\documentclass[a4paper,11pt]{article}

\pdfoutput=1 
\usepackage{jheppub}

\usepackage{graphicx}
\usepackage{epstopdf}

\newcommand{\Dslash}[1] {
\setbox0=\hbox{$#1$}     \dimen0=\wd0        \setbox1=\hbox{/} \dimen1=\wd1
\ifdim\dimen0>\dimen1         \rlap{\hbox to \dimen0{\hfil/\hfil}}       #1
\else           \rlap{\hbox to \dimen1{\hfil$#1$\hfil}}       /         \fi
}

\newcommand{\ba}{\begin{array}}
\newcommand{\ea}{\end{array}}

\newcommand{\be}[1]{
\begin{eqnarray}\label{#1}}
\newcommand{\ee}{\end{eqnarray}}
\newcommand{\bea}{\begin{eqnarray}}
\newcommand{\eea}{\end{eqnarray}}

\newcommand {\nbs}{\Dslash{\bar n}}

\newcommand{\Lin}{\mathcal{L}_{\text{int}}}

\title{\boldmath  Hard exclusive production of two pions in $\gamma\gamma$-collisions
within the SCET factorization approach}
\author{N.
Kivel\note{ On leave of absence from St.~Petersburg Nuclear Physics Institute,
188350, Gatchina, Russia} }

\preprint{HIM-2013-01}

\affiliation{Helmholtz Institut Mainz, Johannes Gutenberg-Universit\"at, D-55099 Mainz, Germany
\\
Institut f\"ur Kernphysik, Johannes Gutenberg-Universit\"at, D-55099 Mainz, Germany
}

\abstract{
We perform an  analysis of  power suppressed contributions  for  large angle  pion pairs production in gamma gamma collisions.   
Using  the soft collinear effective theory (SCET) framework we derive a factorization formula
which is described as a sum of  hard and soft contributions. The soft contribution is an important part  for the consistent description of 
power corrections which can be  described  by matrix elements of  SCET-I operators within the SCET framework. 
For hard power suppressed amplitude  we consider an approximation by  twist-3  chiral enhanced contributions.   
We  define   a physical subtraction scheme  in order  to cancel  endpoint singularities  in  collinear convolution integrals.   
In this case  the subleading correction is described by well defined  expression with the known angular behavior. 
The latter   is  determined by  hard scattering  subprocess  and  can be computed at the leading logarithmic approximation.

The obtained results  are applied  to a  phenomenological analysis of  existing  data. 
Using the leading and subleading power contributions  and fitting  two unknown nonperturbative  amplitudes 
 as a free parameters we can  consistently describe the existing data for  $\pi^{+}\pi^{-}$ and $\pi^{0}\pi^{0}$ production.
We  suggest to measure one more  
hadronic cross section  which can be   accessed   in the unpolarized  $e^{+}e^{-}$ scattering.  
These data  will help  to reduce  theoretical ambiguities  in the phenomenological  analysis  and   to understand   the 
  production mechanism of  the considered  reaction. 
}

\begin{document}
\maketitle
\flushbottom

\section{Introduction}

A factorization theorem for large angle  production of mesons have been
suggested long time ago in Ref.\cite{Brodsky:1981rp}.  The
amplitude of such process is given by a convolution integral of a hard
kernel with  light-cone distribution amplitudes (DAs). The hard kernel  describes  
hard subprocess $\gamma\gamma  \rightarrow\bar q q +\bar q q$  and can be computed systematically in
the perturbation theory. The DA describes the nonperturbative
overlap of the two quarks with  outgoing meson state and can not be computed from the
first principles.

An interesting case  is given by particular  process   $\gamma\gamma\rightarrow\pi\pi$.  The large scattering angles  correspond to a region where
Mandelstam variables are large $s\sim-t\sim-u\gg\Lambda^{2}_{QCD}$.   The asymptotic behavior of the cross section was
obtained in Ref.\cite{Brodsky:1981rp} in the framework of the factorization approach
\begin{align}
\frac{d\sigma^{\gamma\gamma\rightarrow\pi\pi}}{d\cos\theta}\sim\frac
{f^{4}_{\pi}}{s^{3}}\frac{1}{\sin^{4}\theta} .\label{BLas}%
\end{align}
Here $\theta$ is the scattering angle in  center-of-mass frame, $f_{\pi
}=131$MeV is the pion decay constant.  A comprehensive phenomenological
analysis of this reaction was later  carried out in Ref.\cite{Benayoun:1989ng}.

The cross section of  pion  production was already measured
at sufficiently large energy in  several experiments 
\cite{Dominick:1994bw, Heister:2003ae}.  The most precise
measurements  in the region up to $\sqrt{s}=4$GeV were performed by BELLE
collaboration \cite{Nakazawa:2004gu, Uehara:2009cka}, see also review
\cite{Brodzicka:2012jm}.   Comparison of these  accurate data  with the 
theoretical calculations  shows  a significant  underestimate of
 the absolute values of the cross sections \cite{Vogt:2000bz}.  The
largest discrepancy between the theory and experiment is observed for the
neutral pion production.

There are various attempts to find an  explanation of this  problem. 
Within the QCD factorization framework  it is
suggested to use a broad  model for the pion distribution amplitude.  
 In this case  the   virtualities of the hard particles are assumed to be much smaller then the
large external kinematical variables and therefore one can use a relatively
large value of the QCD coupling $\alpha_{s}\simeq0.4$
\cite{Benayoun:1989ng,Chernyak:2006dk,Chernyak:2012pw}.   
Such approach allows to reach a certain qualitative agreement with the $\pi^{+}\pi^{-}$ data
but cannot  explain the discrepancy in the $\pi^{0}\pi^{0}$ channel.  In
Ref.\cite{Chernyak:2012pw} it is proposed that the large contribution in this
channel can arise at higher orders in $\alpha_{s}$ due to the specific three
gluon exchange diagrams.

Other possible scenario which  explains the mismatch  between the leading-order
pQCD predictions and the data implies a different idea about the
underlying QCD dynamics.  It is assumed that the leading-order contribution
 becomes dominant only at very  large  energies  which are considerably larger then the energies
 of existing experiments.   Such scenario suggest a different  shape of the pion DA 
 which yields  small values of the  cross sections.   The model
for  pion DA is obtained from the  process   $\gamma^{*}%
\gamma\rightarrow\pi^{0}$ or using the pion electromagnetic form factor computed in the
QCD sum rule technique  \cite{Khodjamirian:1997tk, Braun:1999uj,
Agaev:2012tm, Bakulev:2011iy, Bakulev:2012nh}.  
The discrepancy in this case must  be explained by a large numerical effect 
of power suppressed corrections. A bulk of this effect is associated with the
so-called soft-overlap mechanism. Such configuration
describes a soft-overlap of  hadronic states and appear  only as a power
correction to the leading asymptotic contribution in Eq.(\ref{BLas}).
Nevertheless it was found that numerically this contribution is large and even
dominant at some moderate values of the hard scale $Q$. The soft-overlap
contribution is especially important if the leading-order approximation  is 
of order $\alpha_{s}$ and therefore  can be suppressed numerically.

This idea was implemented  for  description of  large angle meson production  in 
$\gamma\gamma\rightarrow MM$ process within  the handbag model
in Refs.\cite{Diehl:2001fv, Diehl:2009yi}.  In this model the soft-overlap
contribution is described by the two-pion matrix element which is associated
with the two-pion distribution amplitude. The important feature  is that this
function depends only from the total energy $s$. This allows one to compute
the  angular  behavior of the amplitude because it  is completely defined  by the hard subprocess. 
In Ref.\cite{Diehl:2001fv} it was shown that handbag model gives  $1/\sin^{4}\theta$ behavior
  for the pion cross sections similar to  the leading power  term in Eq.(\ref{BLas}). 
  The unknown normalization of the  two-pion distribution amplitude  was fitted from the data. 
 This automatically  ensures  a large value of the  $\pi^{0}\pi^{0}$ cross
section and gives the ratio $R=d\sigma^{\pi^{0}\pi^{0}}/d\sigma^{\pi
^{+}\pi^{-}}=1/2$.  

Despite interesting  results the handbag model has many
problematic points which have to be better understood. Some critical remarks
are  considered in Refs. \cite{Chernyak:2006dk,Chernyak:2012pw}.  The most
difficult challenge  is to develop  a consistent  formulation of the
soft-overlap configuration in order to describe  it on a systematic way.  
This is important  in order  to avoid a double counting with the power corrections arising
 from the hard power suppressed configurations.  Then such  framework
allows one to reduce a model dependence of  theoretical description to a
minimum. Motivated by this task  we try to develop such approach in
present paper.

Our main  task is to develop the QCD factorization approach beyond the
leading power approximation. Such development  can not be done only by
computation of subleading hard contributions. It is well known that such
corrections are often ill defined because collinear  convolution integrals 
 have the so-called endpoint divergencies.  These
singularities appear due to the overlap of  collinear and soft regions.  The complete
 description in this case can  be  only carried out including a contribution with  soft particles. 
A description of such configurations  can be  performed  in the framework of  effective field theory
which takes into account soft and collinear modes.  Such effective theory
was constructed  recently  and known as soft collinear effective theory (SCET)
 \cite{Bauer:2000ew, Bauer2000,
Bauer:2001ct,Bauer2001,BenCh,BenFeld03}.  

In the SCET  framework  the factorization  is carried out in two steps.  At first step one  factorizes  the hard modes 
( particles with momenta $p_{\mu}\sim Q$).  After integration of  hard modes  
 the full QCD is reduced to the effective theory SCET-I.  This field theory
includes the  hard-collinear,  collinear and soft degrees of freedom. The hard-collinear modes  describe 
particles with the  virtualities $p_{hc}^{2}\sim Q\Lambda$, where the $\Lambda$ is a soft 
scale of order $\Lambda_{QCD}$.  A  further  factorization is possible if the
hard scale $Q$ is sufficiently large so that the hard-collinear scale $\mu
^{2}_{hc}\sim Q\Lambda$ is a good expansion parameter in pQCD. Integrating out
 hard-collinear particles one obtains the effective theory with  soft
and collinear particles which is  called SCET-II.  Such a scheme provide a systematic definition of the soft-overlap 
 configurations and allows one to study the endpoint region in a consistent way.  

The region of \textit{moderate} values of $Q^{2}$ can be defined as  a
region where the hard-collinear scale is still relatively small $\mu^{2}
_{hc}\leq1-1.5$GeV$^{2}$  and further expansion  do not provide a good approximation.  
 The kinematical region of  existing experiments corresponds to $Q\leq 3-4$GeV.
One can easily see  that in this region the hard-collinear scale is
not large enough  $\mu^{2}_{hc}\leq 1.2-1.6$GeV$^{2}$ where we take $\Lambda\simeq
400$MeV.   Therefore in this case one can perform only the factorization of the hard modes.  
 In this case the soft-overlap contributions  can be defined as a  matrix elements of
 SCET-I operators.   Such  contributions must be included into factorization scheme 
 together  with the hard configurations  described by  pure collinear  operators.  We expect 
 that this method will  help  us to obtain a complete and consistent  theoretical description of   power 
 suppressed  corrections.

Our paper is organized as follows. In section \ref{general} we specify
notation and kinematics and briefly review the leading twist results. 
 In section  \ref{toy} we study a scalar integral  using  expansion by momentum regions.  
 The toy integral has contributions associated with  the collinear and soft  regions.  We demonstrate 
 that the overlap of the soft and collinear regions introduce the endpoint divergencies. We show
 that these singularities cancel in the sum of collinear and soft contributions  leaving  a large 
 logarithm.  We also discuss the factorization scheme of this integral in the effective theory
 framework.

 In  section \ref{factSCET} we perform an analysis of  relevant subleading
 operators within the SCET framework.  First  we  consider  the hard contribution and  required  collinear operators.
We  compute the subleading hard  contribution given  by  the chiral enhanced  twist-3  pion distribution amplitude.   
We will show that   factorization  also  include  SCET-I operators  which  describe  the relevant soft  
 contributions.  Using SCET  approach we obtain  that  there is only one  such operator.  
  Using these results we derive  a factorization formula   which  describe the power correction  at order  $1/Q^{2}$.   
 In section \ref{calcofsub} we compute the hard kernels for the soft  contribution. 
  Then we  explain how to define a physical subtraction scheme in order to
avoid  the end-point singularities in the collinear convolution integrals.
Section \ref{phenom} is devoted to a phenomenological
analysis. We compare the obtained results with  data and discuss  
different scenarios  associated with the different models of the pion DA.
A summary and discussion of  obtained  results  is given  in section
\ref{discurs}. In Appendix  we provide  a useful information about
 higher twist distribution amplitudes  and SCET  Lagrangian.

\section{General information about the process $\gamma\gamma\rightarrow\pi\pi
$}

\label{general}

\subsection{Kinematics, amplitudes and cross sections}

In order to describe the process $\ \gamma(q_{1})\gamma(q_{2})\rightarrow
\pi(p)\pi(p^{\prime})$  we choose center-of-mass system (c.m.s.)
$\boldsymbol{p}+\boldsymbol{\bar{p}}^{\prime}=0$ with  pion momenta
directed along $z$-axis.  Mandelstam variables are defined as
\begin{equation}
s=(q_{1}+q_{2})^{2}\equiv W^{2},~\ t=(p-q_{1})^{2},~\ u=(p-q_{2})^{2},
\end{equation}
In c.m.s. the particle momenta read
\begin{align}
p  &  =\frac{W}{2}(1,0,0,\beta),~\ p^{\prime}=\frac{W}{2}(1,0,0,-\beta),\\
q_{1}  &  =\frac{W}{2}(1,\sin\theta,0,\cos\theta),~~q_{2}=\frac{W}{2}%
(1,-\sin\theta,0,-\cos\theta),
\end{align}
where $\theta$ is the scattering angle and $\beta$ is the pion velocity
\begin{equation}
\beta=\sqrt{1-\frac{4m_{\pi}^{2}}{s}}.
\end{equation}
We will also use the auxiliary light-cone vectors
\begin{equation}
n=(1,0,0,-1),~\bar{n}=(1,0,0,1),\ \ (n\cdot\bar{n})=2.
\end{equation}
In this paper we consider the kinematical region where $s\sim-t\sim-u\gg
m^{2}_{\pi}$ therefore we neglect  pion mass. Then the light-cone
decomposition of the momenta read
\begin{equation}
p\simeq W\frac{\bar{n}}{2},~~\ \ \ p^{\prime}\simeq W\frac{n}{2},\label{mom1}%
\end{equation}
\begin{align}
q_{1}  &  =\frac{(1-\cos\theta)}{2}W\frac{n}{2}+\frac{(1+\cos\theta)}{2}
W\frac{\bar{n}}{2}+q_{\bot},~~\\
q_{2}  &  =\frac{(1+\cos\theta)}{2}W\frac{n}{2}+\frac{(1-\cos\theta)}{2}
W\frac{\bar{n}}{2}-q_{\bot},\ \
\end{align}
with
\begin{equation}
q_{\bot}^{2}=\frac{s}{4}(1-\cos^{2}\theta).
\end{equation}

The process $\gamma\gamma\rightarrow\pi\pi$ is described by the matrix element%
\begin{equation}
{\large \langle\pi(p),\pi(p^{\prime})~\text{out}|~\gamma
(q_{1})\gamma(q_{2})\text{ in}\rangle=i(2\pi)^{4}\delta(p_{1}+p_{2}%
-q_{1}-q_{2})~M_{\gamma\gamma\rightarrow\pi\pi},}%
\label{defme}
\end{equation}
where the amplitude
\begin{equation}
M_{\gamma\gamma\rightarrow\pi\pi}=e^{2}~\varepsilon_{\mu}(q_{1})\varepsilon
_{\nu}(q_{2})~M_{\gamma\gamma\rightarrow\pi\pi}^{\mu\nu}~,
\end{equation}
with the following hadronic tensor
\begin{equation}
M_{\gamma\gamma\rightarrow\pi\pi}^{\mu\nu}=~i\int d^{4}x~e^{-i(q_{1}%
x)}\left\langle \pi(p),\pi(p^{\prime})~\left\vert
~T\{~J_{\text{em}}^{\mu}(x),J_{\text{em}}^{\nu}(0)\}\right\vert 0\right\rangle
.
\label{defM}
\end{equation}
Here $~J_{\text{em}}^{\mu}$ denotes the electromagnetic current and
$e^{2}=4\pi\alpha\simeq4\pi/137$. \ It is convenient to pass to the pion
isotopic coordinates $(\pi^{\pm},\pi^{0})\rightarrow(\pi^{1},\pi^{2},\pi^{3})
$ and consider the matrix element describing the process $\gamma
\gamma\rightarrow\pi^{a}\pi^{b}$ 
\begin{equation}
~T_{ab}^{\mu\nu}=~~i\int d^{4}x~e^{-i(q_{1}x)}\left\langle \pi^{a}(p),\pi
^{b}(p^{\prime})\left\vert ~T\{~J_{\text{em}}^{\mu}(x),J_{\text{em}}^{\nu
}(0)\}\right\vert 0\right\rangle ,
\end{equation}
This amplitude can be parametrized as, see e.g. Ref.\cite{Gasser:2006qa}
\begin{align}
T_{ab}^{\mu\nu}  &  =M_{++}^{\mu\nu}\left\{  ~\delta^{ab}T_{++}^{(0)}%
(s,t)+\delta^{a3}\delta^{b3}T_{++}^{(3)}(s,t)\right\} \nonumber\\
&  ~\ \ \ \ \ \ \ \ \ \ \ +M_{+-}^{\mu\nu}\left\{  ~\delta^{ab}T_{+-}%
^{(0)}(s,t)+\delta^{a3}\delta^{b3}T_{+-}^{(3)}(s,t)\right\}  +...~,\label{Tab}%
\end{align}
where dots denote the additional structures which vanish when contracted with
the photon polarization vectors. The two Lorentz tensors in Eq.(\ref{Tab})
read
\begin{equation}
M_{++}^{\mu\nu}=\frac{1}{2}g^{\mu\nu}-\frac{1}{s}q_{1}^{\nu}q_{2}^{\mu},
\end{equation}%
\begin{equation}
M_{+-}^{\mu\nu}=\frac{1}{2}g^{\mu\nu}+\frac{s}{4tu}\left\{  \Delta^{\mu\nu
}-q_{1}^{\nu}q_{2}^{\mu}-\frac{\nu}{s}(q_{1}^{\nu}\Delta^{\mu}-q_{2}^{\mu
}\Delta^{\nu})\right\}  ,
\end{equation}
where we defined $\ \nu=t-u=-2\left(  q_{1}\cdot\Delta\ \right)  $
and$~\ \Delta=p_{1}-p_{2}$. Two tensor structures $M_{+\pm}^{\mu\nu}$
describe the appropriate photon helicity amplitudes $M_{\lambda_{1}\lambda
_{2}}$ and satisfy following relations\footnote{Usually one defines four
helicity amplitudes $M_{++}$, $M_{--}$, $M_{+-}$ and $M_{-+}$. However for the
pion production $M_{++}=M_{--}$ and $M_{+-}=M_{-+}$ and for brevity we do not
write $M_{--}$ and $M_{-+}$.}
\begin{equation}
q_{1}^{\mu}M_{+\pm}^{\mu\nu}=q_{2}^{\nu}M_{+\pm}^{\mu\nu}=0,~\ \ M_{++}%
^{\mu\nu}M_{+-}^{\mu\nu}=0,~\ M_{i}^{\mu\nu}M_{i}^{\mu\nu}=\frac{1}{2}.
\end{equation}
The amplitudes $T_{+\pm}^{(0,3)}$~are symmetric under crossing
$(t,u)\rightarrow(u,t)$. The  pion state in Eq.(\ref{defme})  is $C$-even and can be
decomposed  onto isospin  states  with  $I=0,2$.  For 
physical pion states the amplitude defined  in Eq.(\ref{defM}) we obtain
\begin{equation}
M_{\gamma\gamma\rightarrow\pi^{+}\pi^{-}}^{\mu\nu}=~M_{++}^{\mu\nu}%
~T_{++}^{(0)}+M_{+-}^{\mu\nu}~T_{+-}^{(0)},\label{pi-pm}%
\end{equation}%
\begin{equation}
M_{\gamma\gamma\rightarrow\pi^{0}\pi^{0}}^{\mu\nu}=~M_{++}^{\mu\nu}\left(
T_{++}^{(0)}+T_{++}^{(3)}\right)  +M_{+-}^{\mu\nu}\left(  T_{+-}^{(0)}%
+T_{+-}^{(3)}\right)  .\label{pi-00}%
\end{equation}
The   cross sections  read 
\begin{equation}
\frac{d\sigma^{\pi^{+}\pi^{-}}}{d\cos\theta}=\frac{\pi\alpha^{2}}{16s}~\left(
|T_{++}^{(0)}|^{2}+|T_{+-}^{(0)}|^{2}\right)  ,
\label{dsdcos-pipm}
\end{equation}
\begin{equation}
\frac{d\sigma^{\pi^{0}\pi^{0}}}{d\cos\theta}=\frac{\pi\alpha^{2}}{32s}~\left(
|T_{++}^{(0)}+T_{++}^{(3)}|^{2}+|T_{+-}^{(0)}+T_{+-}^{(3)}|^{2}\right),
\label{dsdcos-pi00}
\end{equation}
where  pion mass is neglected. 

\subsection{Leading twist  approximation}

\label{leading} 

\begin{figure}[h]
\centering
\includegraphics[width=2.0in]{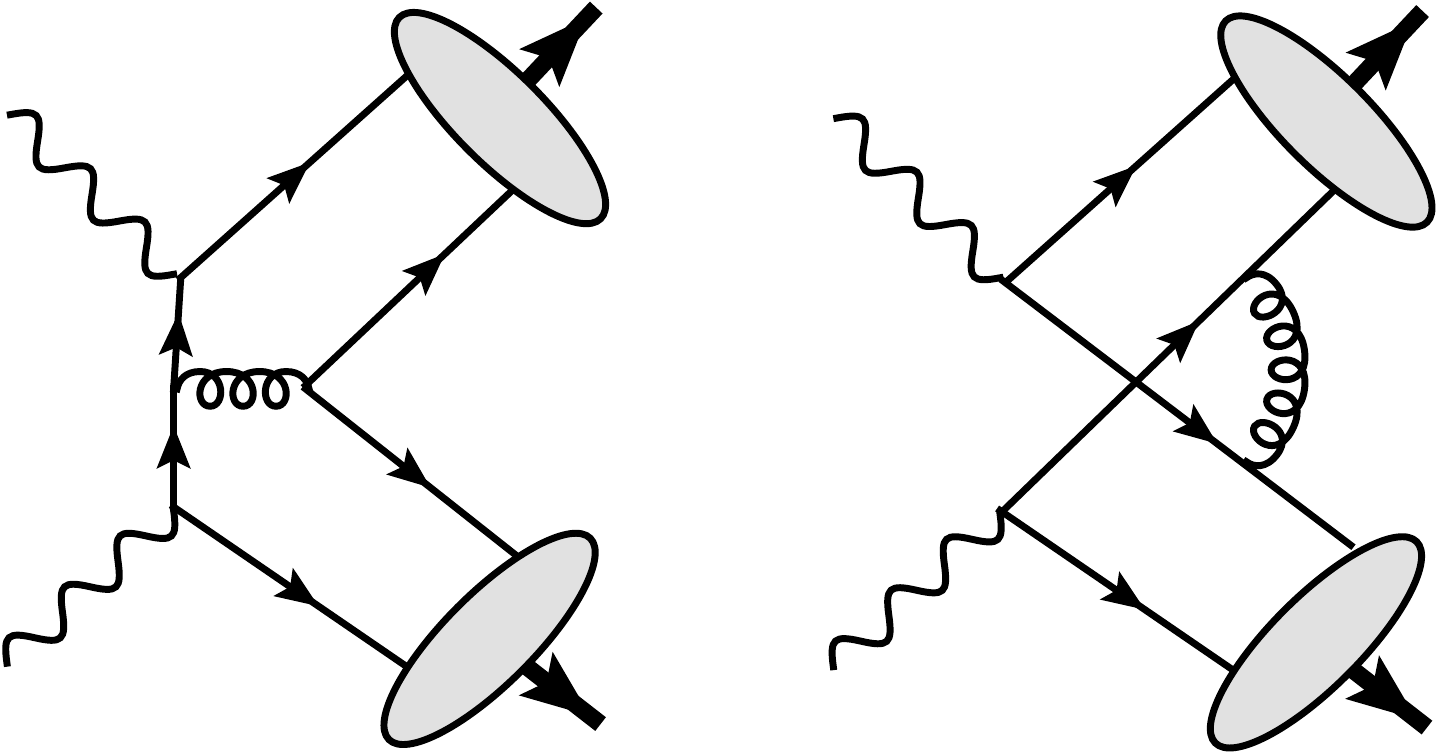}\caption{An example of the
 leading-order  diagrams describing   large-angle
 pion production. The shaded blobs denote  the pion DAs. }%
\label{leading-graphs}%
\end{figure}

In the region where
$s\sim-t\sim-u\gg\Lambda_{\text{QCD}}$  the  amplitude of    process $\gamma\gamma\rightarrow \pi\pi $ can be described  within  the factorization framework. 
The leading order expressions was derived in 
Ref.\cite{Brodsky:1981rp}, see also  Refs.\cite{Lepage:1979zb, Chernyak:1983ej} . 
Let us briefly discuss  these results. Let us write the amplitude as a sum
\begin{equation}
T_{+\pm}^{(i)}(s,\theta)=A_{+\pm}^{(i)}(s,\theta)+B_{+\pm}^{(i)}(s,\theta),~\ \
 \label{defAB}%
\end{equation}
where $A_{+\pm}^{(i)}$  and $B_{+\pm}^{(i)}$ denote the leading and
subleading power  contributions, respectively. The leading-twist and 
leading-order in $\alpha_{s}$  contribution is given by the one gluon exchange diagrams as in
Fig.\ref{leading-graphs}. 

The nonperturbative dynamics is described by the twist-2 pion
distribution amplitude (DA) which is defined as a following matrix element
\begin{equation}
f_{\pi}\varphi_{\pi}(x)=i\int_{-\infty}^{\infty}\frac{d\lambda}{\pi
}e^{-i(2x-1)(p^{\prime}\cdot\bar{n})~\lambda}\left\langle \pi^{-}(p^{\prime
})\left\vert \bar{d}(\lambda\bar{n})~ \setbox0=\hbox{$\bar n$} \dimen0=\wd0
\setbox1=\hbox{/} \dimen1=\wd1 \ifdim\dimen0>\dimen1 \rlap{\hbox to
\dimen0{\hfil/\hfil}} \bar n \else \rlap{\hbox to \dimen1{\hfil$\bar
n$\hfil}} / \fi \gamma_{5}u(-\lambda\bar{n})\right\vert 0\right\rangle
.\label{pionDA}%
\end{equation}
with the pion decay constant~ $f_{\pi}=131$MeV.  Corresponding coefficient functions have been computed in
Ref.\cite{Brodsky:1981rp}. The explicit expressions read 
\begin{equation}
A_{++}^{(0)}(s,\theta)=-A_{++}^{(3)}(s,\theta)=-\frac{\left(  4\pi f_{\pi}\right)  ^{2}}{s}%
\frac{\alpha_{s}}{\pi}\frac{C_{F}}{N_{c}}\frac{1}{1-\cos^{2}\theta
}~\left\langle \frac{1}{x}\right\rangle ^{2},\label{A0pp}%
\end{equation}%
\begin{equation}
A_{+-}^{(0)}(s,\theta)=-\frac{\left(  4\pi f_{\pi}\right)  ^{2}}{s}\frac{\alpha_{s}}{\pi}\frac{C_{F}}{N_{c}}\left\{  \frac{1}{1-\cos^{2}\theta
}\left\langle \frac{1}{x}\right\rangle ^{2}-\frac{1}{9}~J(\theta)\right\}
\label{A0pm}%
\end{equation}%
\begin{equation}
A_{+-}^{(3)}(s,\theta)=\frac{\left(  4\pi f_{\pi}\right)  ^{2}}{s}\frac{\alpha_{s}}{\pi}\frac{C_{F}}{N_{c}}\left\{  ~\frac{1}{1-\cos^{2}\theta
}\left\langle \frac{1}{x}\right\rangle ^{2}-\frac{1}{4}~J(\theta)\right\}
.\label{A3pm}%
\end{equation}
In these equations we used that the pion DA is symmetrical function: $\varphi_{\pi}(1-x)=\varphi_{\pi
}(x)$. We also define  $C_{F}=(N_{c}^{2}-1)/{2N_{c}}$ and  $\alpha_{s}$ denotes  the QCD running coupling.  
 The convolution integrals are defined by 
\begin{equation}
\left\langle \frac{1}{x}\right\rangle =\int_{0}^{1}dx~~\frac{\varphi_{\pi}(x)}{x},\label{inverse}%
\end{equation}
and
\begin{equation}
J(\theta)=\int_{0}^{1}dx \varphi_{\pi}(x) \int_{0}^{1}dy\varphi_{\pi}(y)~\frac{1}{x\bar{x}}\frac{1}{y\bar{y}%
}\frac{~(\bar{x}\bar{y}+xy)(x\bar{x}+y\bar{y})}{(\bar{x}\bar{y}+xy)^{2}%
-(\bar{x}\bar{y}-xy)^{2}\cos^{2}\theta},~\ \bar{x}\equiv1-x\text{.}%
\label{def:J}%
\end{equation}
Using Eqs.(\ref{pi-pm}) and (\ref{pi-00}) one easily finds the physical
amplitudes.  In particuler, for  $\pi^{0}\pi^{0}$ production one obtains
\begin{equation}
A_{++}^{(0)}(s,\theta)+A_{++}^{(3)}(s,\theta)=\mathcal{O}(\alpha_{s}^{2}),\label{App03}%
\end{equation}%
\begin{equation}
A_{+-}^{(0)}(s,\theta)+A_{+-}^{(3)}(s,\theta)=-\frac{\left(  4\pi f_{\pi}\right)  ^{2}}{s}%
\frac{\alpha_{s}}{\pi}\frac{C_{F}}{N_{c}}\frac{5}{36}~J(\theta).\label{Apm03}%
\end{equation}

Notice that the integral $J(\theta)$ is real and therefore all  leading
twist amplitudes at leading-order are real. This is explained by the
absence of any $s$-channel cut in the leading-order diagrams.

A detailed  phenomenological analysis based on the leading twist formulas
(\ref{A0pp})-(\ref{A3pm}) is considered  in Ref.\cite{Benayoun:1989ng}.
Here we briefly discuss  numerical estimates for the  cross  sections. 
In order to apply the leading twist description one has to specify a
model for the pion DA. Following to  standard approach we present these
function as  series over Gebenbauer polynomials 
\begin{equation}
\varphi_{\pi}(x,\mu)=6x\bar{x}\sum_{n}a_{2n}(\mu)~C_{2n}^{3/2}(2x-1)~.
\end{equation}
The coefficients $a_{2n}$ defined by this equation are multiplicatively
renormalizable at the leading logarithmic approximation. We consider few models
which can be defined as following. 
\begin{equation}
\text{model-I:~}~\mu=1\text{GeV},~\ a_{0}=1,~a_{2}=0.25,~a_{2n}%
=0,~n>2.\label{setI}%
\end{equation}
This simple model is based on the estimate suggested in Ref\cite{Mikhailov:1991pt}.  
The following two models
\begin{align}
\text{model-II}   \text{:~}~\mu=1\text{GeV},~ a_{0}=1,~a_{2}=a_{4}%
=a_{6}=0.1,
~a_{8}   =0.034,~a_{2n}=0,~n>8,\label{setBA}%
\end{align}
\begin{align}
\text{model-III}   \mu=2.4\text{GeV},~ a_{0}=1,~a_{2}
=0.157,~a_{4}=-0.192,~a_{6}=0.226,
a_{2n}   =0,~n>4,\label{setBMS}
\end{align}
have been recently suggested in Refs. \cite{Agaev:2012tm} and
\cite{Bakulev:2012nh}, respectively. One more alternative  model  was
suggested in \cite{Chernyak:1981zz}
\begin{equation}
\text{set-CZ:~}\mu=1\text{GeV},~\ a_{0}=1,~a_{2}=2/3,~a_{2n}%
=0,~n>2,\label{setCZ}%
\end{equation}

The  running coupling will be computed at the scale  $\mu_{R}=0.8\, W$GeV  for the models I-III. 
For CZ-model we will use  fixed value $\mu_{R}=1.3$GeV as in Ref.\cite{Benayoun:1989ng}.  
The inverse  moment  defined in Eq. (\ref{inverse}) can be  presented as a sum   
\begin{equation}
\left\langle \frac{1}{x}\right\rangle =3(1+a_{2}+a_{4}+...+a_{2n}+~...).
\end{equation}
Then for the different models  one obtains ($\mu=1$GeV)
\begin{equation}
\left\langle \frac{1}{x}\right\rangle _{\text{I}}=3.75,~\ \left\langle
\frac{1}{x}\right\rangle _{\text{II}}=4.00,~\ \left\langle \frac{1}%
{x}\right\rangle _{\text{III}}=4.05,~~\left\langle \frac{1}{x}\right\rangle
_{\text{CZ}}=5~.\label{1oxnum}%
\end{equation}
The integral $J(\theta)$ defined in Eq.(\ref{def:J}) will be computed
numerically. In Refs. \cite{Brodsky:1981rp, Benayoun:1989ng} it as found that
this integral provides a  small numerical effect. Therefore the main
difference between the different models of pion DA is provided by the inverse
moment (\ref{1oxnum}) and by the value of  running coupling $\alpha_{s}$.
For simplicity, we will also neglect an imaginary part  which appear   in the
timelike kinematics from various large logarithms. 

\begin{figure}[h]
\centering
\includegraphics[width=2.6in]{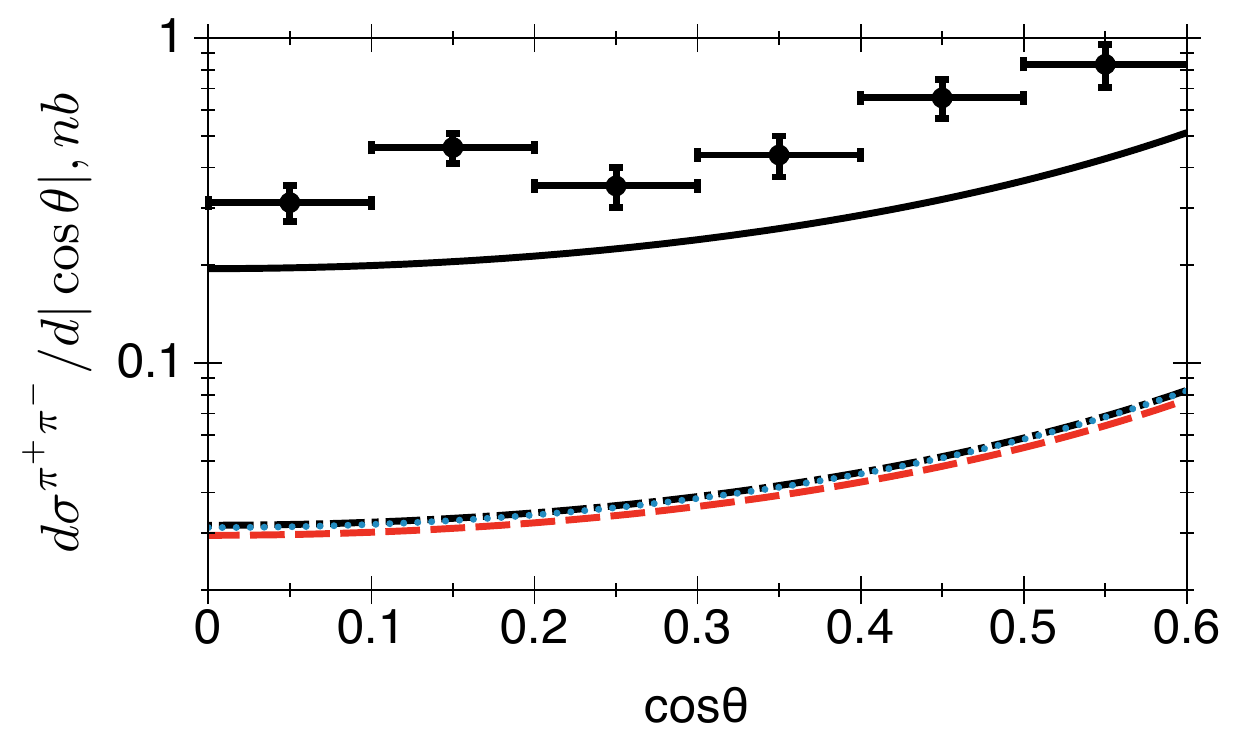}
\includegraphics[width=2.6in]{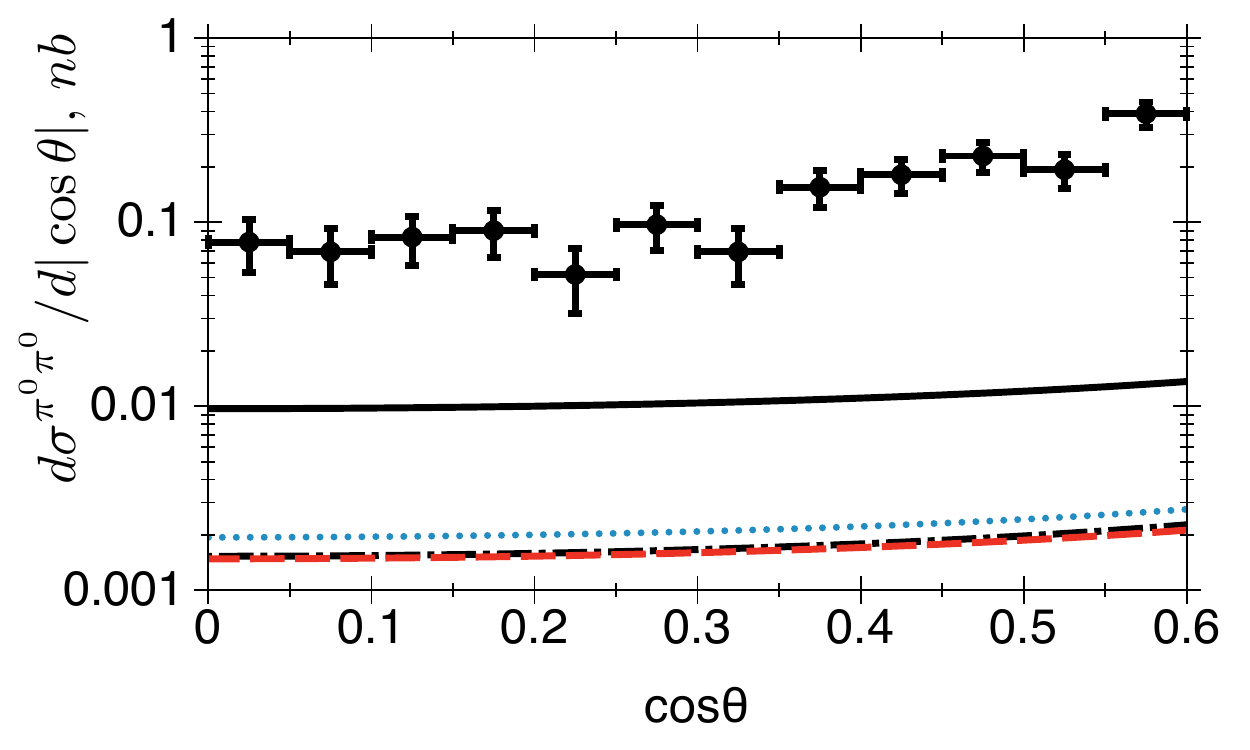}\caption{ The  cross
sections as a function of $\cos\theta$ computed at the leading twist approximation at
fixed $W=3.05$~GeV.  The black solid line corresponds CZ-model, red dashed curve
describes model-I, black dot-dashed and blue dotted curves describe model-II and
III, respectively.  The data are taken  from  Refs.\cite{Nakazawa:2004gu,Uehara:2009cka} }%
\label{plot-lo-pm}%
\end{figure}
In Fig.\ref{plot-lo-pm} we show the leading
twist estimates for differential cross sections in comparison
with  BELLE data \cite{Nakazawa:2004gu,Uehara:2009cka} at $W=3.05$GeV. 
For the pion DA defined by models
I-III  we fix the hard  scale as $\mu=2.4$GeV.  The relatively low value of the
 scale  for  CZ-model  yields $\alpha_{s}(1.7$GeV$^{2})=0.395$.   In  this case such  choice is dictated by a
large role of the endpoint region where $x\sim1$ or $x\sim0$. 
 In this region the virtualities  of  hard partons  are relatively small and
 this leads to  a smaller  value of the hard  scale in phenomenological calculations.

From Fig.\ref{plot-lo-pm} we conclude that  the leading twist approximation  provide  a reasonable description of the
angular behavior  but predicts a very small  absolute
normalization. 
The cross section computed with the models I-III is about  an order of  magnitude below  the data. 

A more realistic  estimate is obtained  only with 
the CZ-model in case of $\pi^{+}\pi^{-}$ production. 
 This model yields  much  larger results because the wide profile of the DA provides a  
 larger value of the inverse momentum $\langle1/x\rangle$ and also due to  larger value of   $\alpha_{s}$. 
In this case one cannot exclude  a sizable contribution from  the higher-order  radiative corrections.   
Some work in this  direction is  presented   in Ref.\cite{Duplancic:2006nv}.

On the other hand a description of the cross section for  $\pi^{0}\pi^{0}$ production remains 
very problematic for all  models of pion DA.  Potentially  large contributions are
compensated  in the expressions (\ref{App03}) and (\ref{Apm03}). Therefore  in this case  any
leading twist estimate  provides
very small numerical value for the cross section  as shown  in Fig.\ref{plot-lo-pm}.
In Ref.\cite{Chernyak:2012pw}  it is suggested that, probably,  some  specific
higher-order radiative corrections  can  help to solve this  problem.

 An alternative  description  can be  developed  if  one assumes that  power suppressed  contributions
are quite large  at  some moderate values of hard scale $Q^{2}$.  
A large numerical contribution  in this case  can be generated by the  soft-overlap mechanism. 
In Refs.\cite{Diehl:2001fv, Diehl:2009yi}  this idea  is used in order to develop the  handbag   model . 
In present work we continue to study the role of the  soft-overlap contribution  using  the SCET factorization framework.

\section{ A  toy integral  with the soft-overlap contribution }
\label{toy}

In this section we consider a specific Feynman integral in order to demonstrate the 
relevance of the hard-collinear modes in description of the soft-overlap mechanism. 
We will show that  the  factorization of the soft
contribution requires  to introduce the hard-collinear and the soft modes.  We will
also see that the overlap of  collinear and soft regions introduces the
endpoint singularities in the collinear convolution integrals. These
divergencies cancel only in the sum of  collinear and soft contributions.

\begin{figure}[h]%
\centering
\includegraphics[height=1.5in]
{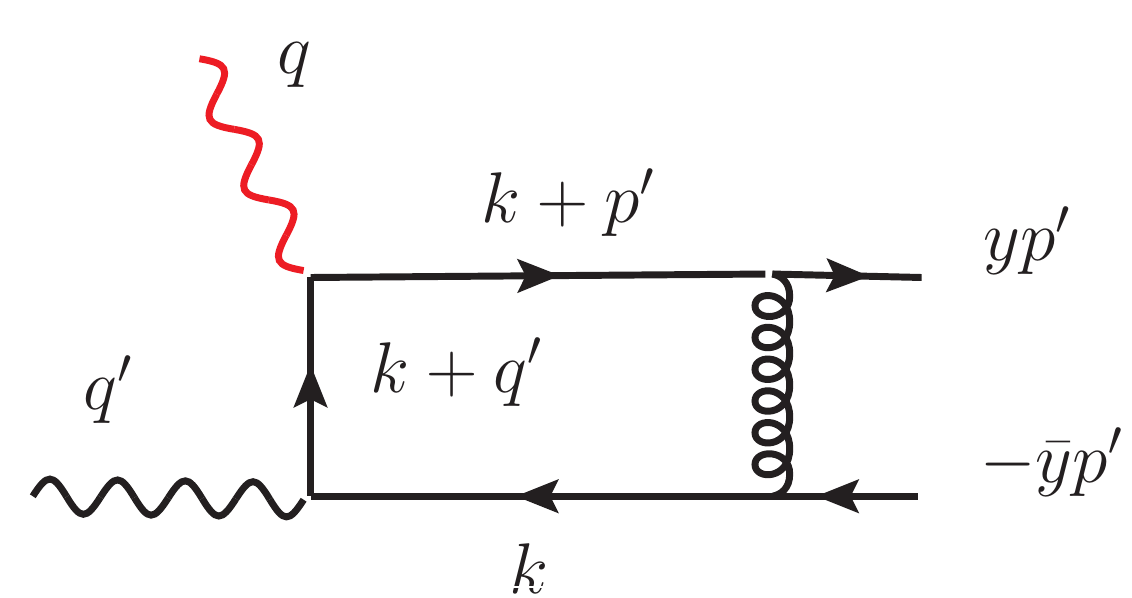}%
\label{toydiagram}%
\caption{The graphical interpretation of the toy integral in Eq.(\ref{Jdef}).  }
\end{figure}

Let us consider the following scalar  integral
\begin{equation}
J=\int dk\frac{m^{2}}{\left[  k^{2}-m^{2}\right]  \left[  (k+p^{\prime}%
)^{2}\right]  \left[  (k+q^{\prime})^{2}\right]  \left[  (k+\bar{y}p^{\prime
})^{2}-m^{2}\right]  }, \label{Jdef}%
\end{equation}
where we assume that~ $p^{\prime}\simeq Qn/2\,$, $q^{\prime}\simeq\bar
Q\bar{n}/2$ so that $p^{\prime2}=q^{\prime 2}=0$ and  $-q^{2}\equiv Q^{2}=2(p'\cdot q')\gg m^{2}$.  For the integral measure we
imply
\begin{equation}
dk=\mu^{2\varepsilon}e^{\varepsilon\gamma_{E}}\frac{d^{D}k}{i\pi^{D/2}%
},~~D=4-2\varepsilon.
\end{equation}
We also assume that all propagators in the square brackets in Eq.(\ref{Jdef}) are defined with
the standard $+i\varepsilon$ prescription. This integral can be associated with
the diagram in Fig.\ref{toydiagram}  where all particles are scalar. 
The dimensionless variable $y$ describes a fraction of the total momentum $p'$ carried by the outgoing  ``quark'', we also use the short notation $\bar y=1-y$.

The integral $J$ (\ref{Jdef}) is ultraviolet (UV) and infrared (IR) finite and can be easily computed in
$D=4$.   Using the standard technique one obtains
\begin{equation}
J=\frac{1}{\bar{y}Q^{2}}\left\{  -\text{Li}(y)+\frac{\pi^{2}}{6}\right\}
+\frac{m^{2}}{\bar{y}Q^{4}}\left(  \ln\frac{m^{2}}{Q^{2}}+~...\right)
+\mathcal{O}(1/Q^{6}), \label{Jexct}%
\end{equation}
where the mass $m$  is considered as a soft scale,  Li$(z)$ denotes the Spence function (or dilogarithm, see definition in
Eq.(\ref{defLi})) and the dots denote  simple non-logarithmic contributions which
will be not considered for brevity.

Let us obtain the expansion in Eq.~(\ref{Jexct}) by identifying the momentum
configurations that give non-vanishing contributions to the integral. For that
purpose we recompute the integral (\ref{Jdef}) using the technique known as
expansion by momentum regions \cite{Beneke:1997zp, Smirnov:1998vk,
Smirnov:2002pj}.

 Below we will use the small  parameter $\lambda\sim
\sqrt{m/Q}$ which is convenient for the estimate of the various terms in the
effective theory. Using this parameter and defining the light-cone coordinates
as $(p\cdot n,p\cdot\bar{n},p_{\bot})\equiv(p_{+},p_{-},p_{\bot})$ one finds
\begin{equation}
q\sim p_{h}\sim Q(1,1,1),~q^{\prime}\sim p_{c}\sim Q(1,\lambda^{4},\lambda
^{2}),~~p^{\prime}\sim p_{c}\sim Q(\lambda^{4},1,\lambda^{2}),
\label{ph/c}%
\end{equation}
where $p_{h}$ and $p_{c}$ denote a generic hard and collinear momenta,
respectively. In addition we will also need the soft $p_{s}$ and
hard-collinear $p_{hc}$ momenta which scale as
\begin{equation}
p_{s}\sim Q(\lambda^{2},\lambda^{2},\lambda^{2}),~~p_{hc}\sim Q(1,\lambda
^{2},\lambda)~\text{or }~p_{hc}\sim Q(\lambda^{2},1,\lambda). \label{ps/hc}%
\end{equation}

\emph{The hard region.} In the hard region  $k\sim p_{h}$, see  Eq.(\ref{ph/c}). 
In this case the expression for the integral reads%
\begin{equation}
J_{h}=\int dk\frac{m^{2}}{\left[  k^{2}\right]  \left[  (k+p^{\prime}%
)^{2}\right]  \left[  (k+q^{\prime})^{2}\right]  \left[  (k+p_{2}^{\prime
})^{2}\right]  }=\frac{m^{2}}{\bar{y}Q^{4}}\left(  \frac{2}{\varepsilon}%
+2\ln\mu^{2}/Q^{2}\right)  .
\label{Jh}
\end{equation}
This integral can be easily computed in dimensional regularization using the Feynman parameters.

\emph{The }$n$\emph{-collinear region.} In this region $k\sim p_{c}\sim
p^{\prime}$, see  Eq.(\ref{ph/c}). Performing the
expansion of the integrand in this region we obtain
\begin{align}
J_{n}  &  \simeq\int dk\frac{m^{2}}{\left[  k^{2}-m^{2}\right]  \left[
(k+p^{\prime})^{2}\right]  \left[  (k+p_{2}^{\prime})^{2}-m^{2}\right]
}\left(  \frac{1}{\left[  2(kq^{\prime})\right]  }-\frac{k^{2}}{\left[
2(kq^{\prime})\right]  ^{2}}\right) =J_{0n}+J_{2n} , \label{Jn-I0}
\end{align}
The first integral $J_{0n}$ is of order $\lambda^{0}$ and UV and IR finite.
Computation this integral  yields
\begin{equation}
J_{0n}=\frac{1}{\bar{y}Q^{2}}\left(  -\text{Li}(y)+\frac{\pi^{2}}{6}\right)  .
\end{equation}
Comparing with the exact answer in Eq.(\ref{Jexct}) we find that this term
reproduces the leading power term of order $1/Q^{2}$.

The second integral in Eq.(\ref{Jn-I0}) is subleading  and scales as $J_{2n}\sim\lambda^{4}$. 
In order to compute this integral we use  a simple trick $k^{2}=[k^{2}-m^{2}]+m^{2}$ in the
numerator (\ref{Jn-I0})  in order to cancel the propagator $[k^{2}-m^{2}]^{-1}$. 
This gives the  sum of the two integrals: one of them is UV-divergent (without the propagator $[k^{2}-m^{2}]^{-1}$).
 Computing these integrals with the help of light-cone variables   in $D=4-2\varepsilon$ one finds 
\begin{equation}
J_{2n}=-\frac{m^{2}}{\bar{y}Q^{4}}\left(  \frac{\mu^{2}}{m^{2}}\right)
^{\varepsilon}\frac{1}{\varepsilon}+\frac{m^{2}}{\bar{y}Q^{4}}
\int_{0}^{p'_{-}}\frac{dk_{-}}{p'_{-}-k_{-}}+~...~. \label{J2div}%
\end{equation}  
 The first term on the {\it rhs} has  pole $1/\varepsilon$ which originates  from the UV-divergent integral.   The second UV finite integral gives the 
 second  contribution on the {\it rhs} (\ref{J2div}) which  is  IR-divergent  and  cannot be defined without  an additional regularization. 
 This singularity can be interpreted as  the endpoint divergency  in the collinear  convolution integral.

\emph{The }$\bar{n}$\emph{-collinear region.}\textit{ } In this case $k\sim
p_{c}\sim q^{\prime}$ and already the leading term in the expansion  is of order $\lambda^{4}$
\begin{equation}
J_{\bar{n}}\simeq\frac{1}{\bar{y}}\int dk\frac{1}{\left[  2(kp^{\prime
})\right]  ^{2}}\frac{m^{2}}{\left[  k^{2}-m^{2}\right]  \left[  (k+q^{\prime
})^{2}\right]  }. \label{Jnb-I0}%
\end{equation}
 The straightforward calculation of this integral  with the help of the light-cone variables gives
\begin{equation}
J_{\bar{n}}=-\frac{1}{2}\frac{m^{2}}{\bar{y}Q^{4}}\frac{1}{\varepsilon}\left(
\frac{\mu^{2}}{m^{2}}\right)  ^{\varepsilon}\int_{0}^{1}\frac{dk_{+}}{\left[
k_{+}\right]  ^{2}}(1-k_{+})^{-\varepsilon}.
\label{Jbn-div}
\end{equation}
In order to obtain this expression we  computed  the integral over $k_{-}$ using residues  and then integrated over the
transverse momentum $k_{\perp}$.   The pole $1/\varepsilon$ is again due to the UV-divergency of the integral over
the transverse momentum $k_{\perp}$. However the remaining  integral over  $dk_{+}$ is power divergent  in the region
$k_{+}\sim 0$.  

In order to resolve this ambiguity let us rewrite the
integrand as a sum of the  following terms
\begin{align}
\frac{1}{\left[  k^{2}-m^{2}\right]  \left[  (k+q^{\prime})^{2}\right]  }  &
=\frac{1}{\left[  k^{2}-m^{2}\right]  }\left(  \frac{1}{\left[  (k+q^{\prime
})^{2}\right]  }-\frac{1}{\left[  2(kq^{\prime})\right]  }\right)  +\frac
{1}{\left[  k^{2}-m^{2}\right]  \left[  2(kq^{\prime})\right]  }, %
\end{align}
This yields   two contributions
\begin{align}
J_{\bar{n}}  &  =\frac{1}{\bar{y}}\int dk\frac{m^{2}}{\left[  2(kp^{\prime
})\right]  ^{2}}\frac{1}{\left[  k^{2}-m^{2}\right]  \left[  2(kq^{\prime})\right]  }
\nonumber \\& 
 -\frac{1}{\bar{y}}\int dk\frac{m^{2}}{\left[  2(kp^{\prime})\right]  ^{2}
}\frac{k^{2}}{\left[  k^{2}-m^{2}\right]  \left[  (k+q^{\prime})^{2}\right]
\left[  2(kq^{\prime})\right]  }
=J_{1\bar{n}}+J_{2\bar{n}}. 
\label{J12nb}
\end{align}

Computing the the first integral $J_{1\bar{n}}$  one obtains   scaleless and power divergent integral
\begin{align}
J_{1\bar{n}}\sim \int_{0}^{\infty}\frac{dk_{+}}{k^{2}_{+}}\int \frac{dk_{\perp}}{k_{\perp}^{2}+m^{2}}=0.
\end{align}%
Therefore we assume  that this integral  vanishes and can be neglected. 
Computation of  the second integral yields
\begin{align}
J_{2\bar{n}}  &  =-\frac{1}{\bar{y}}\int dk\frac{m^{2}}{\left[  2(kp^{\prime
})\right]  ^{2}}\frac{k^{2}}{\left[  k^{2}-m^{2}\right]  \left[  (k+q^{\prime
})^{2}\right]  \left[  2(kq^{\prime})\right]  }\nonumber \\
&  =-\frac{m^{2}}{\bar{y}Q^{4}}
\left( \frac{\mu^{2}}{m^{2}}\right)^{\varepsilon}
+\frac{m^{2}}{\bar{y}Q^{4}}\int_{0}^{q'_{+}}\frac{dk_{+}}{k_{+}^{2}}\ln[1-k_{+}/q'_{+}]~\left(
q'_{+}-2k_{+} \right)  . \label{J2nb-res-1}
\end{align}
The first term on the {\it rhs} has UV-pole $1/\varepsilon$ which appears from the  integration over $k_{\perp}$. 
The second term   has already only the logarithmic  singularity when $k_{+}\rightarrow0$.   Here  we again observe that the UV and IR
singularities in Eq.(\ref{J2nb-res-1})  enter additively.  Again, in order to compute  integral  $J_{2\bar{n}}$ one needs an 
additional regularization prescription. 

\textit{ The soft region.} In this case $k\sim p_{s}$, see  Eq.(\ref{ps/hc}).
Expansion of the integrand in this region yields two contributions of order
$\lambda^{2}$ and $\lambda^{4}$
\begin{equation}
J_{s}\simeq\frac{1}{\bar{y}}\int dk\frac{m^{2}}{\left[  k^{2}-m^{2}\right]
\left[  2(kp^{\prime})\right]  ^{2}\left[  2(kq^{\prime})\right]  }\left\{
1-\frac{k^{2}}{\left[  2(kq^{\prime})\right]  }-\frac{k^{2}}{2(kp^{\prime}%
)}-\frac{k^{2}-m^{2}}{2(kp_{2}^{\prime})}\right\}  . \label{Js}%
\end{equation}
Computation of these integrals in $D=4-2\varepsilon$ yields the scaleless integrals. This is a well known problem  when one has to introduce an additional 
auxiliary regularization.  Some of  the soft integrals in Eq.(\ref{Js}) are even  power divergent.  As a result  such contributions  
can even generate  fictitious subleading terms  which are not presented in the exact answer.  Consider the first contribution  in Eq.(\ref{Js}).      
 It is easy to see that it scales as
\begin{equation}
J_{1s}=\frac{1}{\bar{y}}\int dk\frac{m^{2}}{\left[  k^{2}-m^{2}\right]
\left[  2(kp^{\prime})\right]  ^{2}\left[  2(kq^{\prime})\right]  }\sim
\lambda^{2}, \label{J1s}%
\end{equation}
and gives the power correction of order $1/Q^{3}$ which is not presented in
the exact answer  (\ref{Jexct}).  A more detailed investigation shows that the
corresponding integral is power divergent and therefore must be considered as
scaleless and therefore vanishes.

In order to see this let us  introduce an additional auxiliary regularization. We
 consider the following regularized integral
\begin{equation}
J_{1s}^{\text{reg}}=\frac{m^{2}}{\bar{y}~Q^{2}p_{-}^{\prime}}\frac{1}{2}%
\int\frac{dk_{-}}{\left[  k_{-}\right]  }\int\frac{dk_{+}}{\left[  k_{+}%
-\tau_{+}\right]  ^{2}}\frac{dk_{\bot}}{\left[  k_{+}k_{-}-k_{\bot}^{2}%
-m^{2}\right]  }, \label{Jsreg}%
\end{equation}
where we use the light-cone variables $(k_{+},k_{-},k_{\bot})$ and introduce
  regularization parameter $\tau_{+}$.  We assume that $\tau_{+}$  transforms in the same way as $k_{+}$ 
 under longitudinal boost  $k_{+}\rightarrow \alpha k_{+}$. 
   Notice that without the prefactor $\sim 1/p_{-}^{\prime}$  the integral   is not invariant with respect to longitudinal  boosts because the 
factor $1/[k_{+}+\tau_{+}]^{2}$ in the denominator (\ref{Jsreg}). As a result the computation of this
integral leads to a power divergent contribution when $\tau_{+}\rightarrow 0$.
Taking the integral over $k_{+}$ by residues and integrating over $k_{\perp}$
we obtain
\begin{equation}
J_{1s}^{\text{reg}}=\left( \frac{\mu^{2}}{m^{2}}\right)^{\varepsilon}\frac{m^{2}}{\bar
{y}~Q^{2}}\frac{1}{p_{-}^{\prime}\tau_{+}}.
\label{J1sreg}
\end{equation}
Assuming that $\tau_{+}\sim m$ we obtain the contribution of order $\lambda^{2}$.  
On the other hand   such  contribution can not appear  in the exact  integral $J$ (\ref{Jexct}) because we do not 
have  appropriate external soft  momenta.  Hence we can conclude that the scaleless integrals which are power divergent  can be neglected.  
The main lesson from this consideration  is that  one must consider only such soft integrals which are boost invariant and therefore 
have  only the logarithmic endpoint  singularities.  

Following this argumentation the third and the forth terms on {\it rhs} of Eq.(\ref{Js}) can also be neglected.  Hence only the second term $\sim k^{2}/[2(kq')]^{2}$ in Eq.(\ref{Js}) 
 provides a non-vanishing  contribution of order $\lambda^{4}$. Therefore  we  get
\begin{equation}
J_{s}=\frac{m^{2}}{\bar{y}Q^{4}}\int dk\frac{-k^{2}}{\left[  k^{2}%
-m^{2}\right]  \left[  k_{+}\right]  ^{2}\left[  k_{-}\right]  ^{2}}.
\label{Js-2}%
\end{equation}

The overlap between the collinear and soft regions can be easily established 
taking the soft limit in the collinear integrals in Eqs.(\ref{Jn-I0}) and
(\ref{Jnb-I0}).  It is easy to see that 
\begin{equation}
\left[  J_{n}\right]  _{s}=\left[  J_{\bar{n}}\right]  _{s}=J_{s}.
\end{equation}
This explains that the IR-divergencies in the collinear integrals are related with
the overlap of the soft and collinear regions. 

Notice that   integral $J_{1s}$ in Eq.(\ref{J1s}) also  appears in the soft limit of the well defined  collinear integral  $J_{0n}$ in Eq.(\ref{Jn-I0})
but  this does not provide any IR-divergencies.   On the other hand  the overlap with  the  same  soft integral  $J_{1s}=J_{1\bar n }$  
in the collinear integral $J_{\bar n}$, see Eq.(\ref{J12nb}), gives the  power divergent integral in Eq.(\ref{Jbn-div}).  
Hence we can conclude  that  subtractions of  such spurious soft  integrals    is important in order to clarify  the 
endpoint behavior  of  collinear integrals. 

We find that the contributions of
other possible regions provide scaleless or power suppressed integrals  and
therefore can be neglected. 
For instance, the expansion in the hard-collinear regions $k\sim p_{hc}$ defined in Eq.(\ref{ps/hc}) yields the scaleless integrals because we do
not have external hard-collinear particles.  Hence  we can conclude that 
expansion of the integral $J$ up to order $\lambda^{4}$ must be reproduced by
the sum of the following integrals
\begin{equation}
J=J_{h}+J_{n}+J_{\bar{n}}+J_{s}. \label{Jsum}
\end{equation}
Let us now compute each integral in the {\it rhs} and to check that we reproduce the exact answer in Eq.(\ref{Jexct}). 

Formally the  soft integral (\ref{Js-2}) is  scaleless  in dimensional regularization
and therefore cannot be defined without an additional regularization. In present case it is
convenient to introduce the auxiliary analytic regularization as in  Refs.\cite{Smirnov:1998vk, Smirnov:2002pj}. 
Following this prescription  we introduce  two  regulators $\lambda_{1,2}$ by substituting in Eq.(\ref{Jdef}) 
\begin{equation}
\frac{1}{\left[  (k+p^{\prime})^{2}\right]  \left[  (k+q^{\prime})^{2}\right]
}\rightarrow\frac{\nu^{2(\lambda_{1}+\lambda_{2})}}{\left[  (k+p^{\prime}
)^{2}\right]  ^{1+\lambda_{1}}\left[  (k+q^{\prime})^{2}\right]
^{1+\lambda_{2}}},
\end{equation}
where $\nu$ is the corresponding  regularization scale. Calculation of the regularized  integrals  yields
\begin{align}
J_{2n}^{\text{reg}}  &  =\int dk\frac{\nu^{2(\lambda_{1}+\lambda_{2}
)}~m^{2}}{\left[  k^{2}-m^{2}\right]  \left[  (k+p^{\prime})^{2}\right]
^{1+\lambda_{1}}\left[  (k+\bar y p^{\prime})^{2}-m^{2}\right]  }\frac
{-(1+\lambda_{2})k^{2}}{\left[  2(kq^{\prime})\right]  ^{(2+\lambda_{2})}}\\
&  =\frac{m^{2}}{Q^{4}\bar{y}}\left( -\frac{1}{\varepsilon}+\ln m^{2}/\mu^{2} -\frac{1}{\lambda_{2}-\lambda_{1}}
+\ln Q^{2}/\nu^{2}+~...\right)  ,
\end{align}%
\begin{align}
J_{2\bar{n}}^{\text{reg}}  &  =\frac{1}{Q^{2}\bar{y}}\int dk\frac
{\nu^{2(\lambda_{1}+\lambda_{2})}~m^{2}} {\left[  k^{2}-m^{2}\right]  \left[
2(kp^{\prime})\right]  ^{2+\lambda_{1}} \left[  (k+q^{\prime})^{2} \right]
^{1+\lambda_{2} }}\nonumber\\
&  =\frac{m^{2}}{Q^{4}\bar{y}}\left( -\frac{1}{\varepsilon}+\ln m^{2}/\mu^{2}+ \frac{1}{\lambda_{2}-\lambda_{1}}%
+\ln \nu^{2}/m^{2}+~...\right)  ,
\end{align}%
\begin{equation}
J_{s}^{\text{reg}}=\frac{1}{\bar{y}}\int dk~\frac{\nu^{2(\lambda
_{1}+\lambda_{2})}m^{2}(-k^{2})}{\left[  k^{2}-m^{2}\right]  \left[
2(kp^{\prime})\right]  ^{2+\lambda_{1}}\left[  2(kq^{\prime})\right]
^{2+\lambda_{2}}}=0,
\end{equation}
where dots denote simple non-logarithmic terms as before. The soft
integral in this regularization prescription  remains scaleless and therefore vanishes.
However  additional poles $\sim 1/(\lambda_{2}-\lambda_{1})$ appears in the
collinear integrals.  
These poles and   scale $\nu$  cancel in the sum leaving the large rapidity logarithm  $\ln Q^{2}/m^{2}$
\begin{equation}
J_{2n}^{\text{reg}}+J_{2\bar{n}}^{\text{reg}}=\frac{m^{2}}{Q^{4}\bar{y}}\left(
-\frac{2}{\varepsilon}+2\ln m^{2}/\mu^{2}+\ln Q^{2}/m^{2}+~...\right)  .
\end{equation}
The total expression for the sum of the collinear and soft integral reads
\begin{align}
J_{\bar{n}}^{\text{reg}}+J_{n}^{\text{reg}}+J_{s}^{\text{reg}}  &  =\frac
{1}{Q^{2}}\left\{  -\text{Li}_{2}[y]+\frac{\pi^{2}}{6}\right\}\nonumber \\
&  +\frac{m^{2}}{Q^{4}\bar{y}}\left(  -\frac{2}{\varepsilon}
+2\ln m^{2}/\mu^{2}+\ln Q^{2}/m^{2}+~...\right)  .
\label{sumJnnbs}
\end{align}
Using Eq.(\ref{Jh}) one can easily observe that the pole $1/\varepsilon$ cancel in the sum of the all integrals (\ref{Jsum})
and   the expansion in Eq.(\ref{Jexct}) is reproduced. 
From the structure of different logarithms in Eq.(\ref{sumJnnbs}) we conclude that the simple large logarithm in 
Eq.(\ref{Jexct}) is reproduced  by  the sum of collinear and rapidity logarithms.

\subsection{Interpretation of the result in  soft collinear effective theory }

Let us  perform  interpretation of  the different contributions  in terms of
operators and matrix elements in the effective theory of soft and
collinear particles. 

The hard contribution describes the configuration where all particles in the
loop integral  are hard and therefore the loop diagram in this case can be
interpreted as a next-to-leading  correction to a  hard coefficient $H$. 
 Contracting the hard subdiagram to a "point"  we
obtain the light-cone  leading-order  matrix element constructed from  two fields.  
Therefore we can write
\begin{equation}
J_{h}\simeq\text{FT}\left\langle 0\right\vert B_{c}\left\vert
\gamma\right\rangle ~H_{\text{nlo}}\ast\text{FT}\left\langle
\bar{q}q\right\vert \psi_{c}^{\dag}(\eta_{1}\bar{n})\psi_{c}(\eta_{2}\bar
{n})\left\vert 0\right\rangle =H_{\text{nlo}}\ast\phi_{qq},
\end{equation}
The symbol FT denotes the Fourier transformation with respect to light-cone coordinates $\eta_{i}$.  
 The asterisk denote the convolution  integrals with respect to the corresponding collinear fractions. 
  The fields $\psi_{c}$ and $B_{c}$ denote the  collinear ``quark'' and ``photon'' fields in the effective theory.  
  
The $n$-collinear contribution $k\sim p^{\prime}$  is given by the
sum of two integrals  $J_{0n}\sim\lambda^{0}$ and $J_{2n}\sim\lambda^{4}$, see Eq.(\ref{Jn-I0}).  In  both cases the hard subdiagram  is 
generated  by the ``quark'' propagator $k^{2}+2(kq')$.  
The ``nonperturbative'' subdiagram is given by  remaining collinear  ``quark'' and   ``gluon''  propagators.  
 The leading order contribution $J_{0n}$   can be identified with the convolution of a hard tree kernel with the one-loop contribution 
 to the leading collinear matrix element
\begin{align}
J_{0n}= \text{FT} \left\langle 0\right\vert B_{c}\left\vert \gamma\right\rangle H_{\text{lo}}\ast\text{FT}\left\langle \bar{q}q\right\vert \psi_{c}^{\dag}\psi_{c}\left\vert 0\right\rangle ^{\text{nlo}}=H_{\text{lo}}\ast\phi_{qq}^{\text{nlo}}.
\end{align}
 This  contribution is finite and do not have any large evolution logarithms  because the underlying scalar field theory with   is superrenormalizable.

The subleading  contribution  $J_{2n}$ is given  by the next-to-leading term in the expansion 
of the hard propagator   with respect to the small momentum $k^{2}$.  
Therefore we can also identify it  with the convolution of a hard tree kernel with  the one-loop  correction 
 to a collinear matrix element. But this matrix element must be already associated with the higher twist collinear operator.  
 Corresponding  operator is constructed from  two ``quark''  fields and derivatives in order  to 
 reproduce the factor $k^{2}/[2(kq')]^{2}$ in the expansion (\ref{Jn-I0}). 
 In  position space such subleading operator can be written as 
 \begin{equation}
 \psi_{c}^{\dag}(0) \left\{ \frac{1}{2}(x\bar{n})(n\partial)+\frac{1}
{2}x_{\bot i}x_{\bot j}\partial_{i}\partial_{j} \right \} \psi_{c}(x)\vert_{x=x_{+}}\equiv \psi_{c}^{\dag}{\cal P}(x,\partial)\psi_{c}.
\label{defP}
 \end{equation}
Schematically  the subleading contribution can be represented as 
\begin{equation}
J_{2n}  = C_{\text{lo}}\ast\text{FT}\left\langle \bar{q}q\right\vert
\psi_{c}^{\dag}{\cal P}(x,\partial)\psi_{c}\left\vert 0\right\rangle ^{\text{nlo}}
=C_{\text{lo}}\ast\varphi_{qq}^{\text{nlo}}.
\label{J2n-int}
\end{equation}
 The  UV-divergence  appearing  in this contribution (the pole $1/\varepsilon$  in Eq.(\ref{J2div}) ) can be associated with  the renormalization of the  subleading 
light-cone operator.  The IR-singularity in the  integral in Eq.(\ref{J2div}) can be interpreted as the endpoint divergency in the collinear 
convolution integral in Eq(\ref{J2n-int}) denoted by  asterisk.  

\begin{figure}[ptb]
\centering
\includegraphics[width=4 in]
{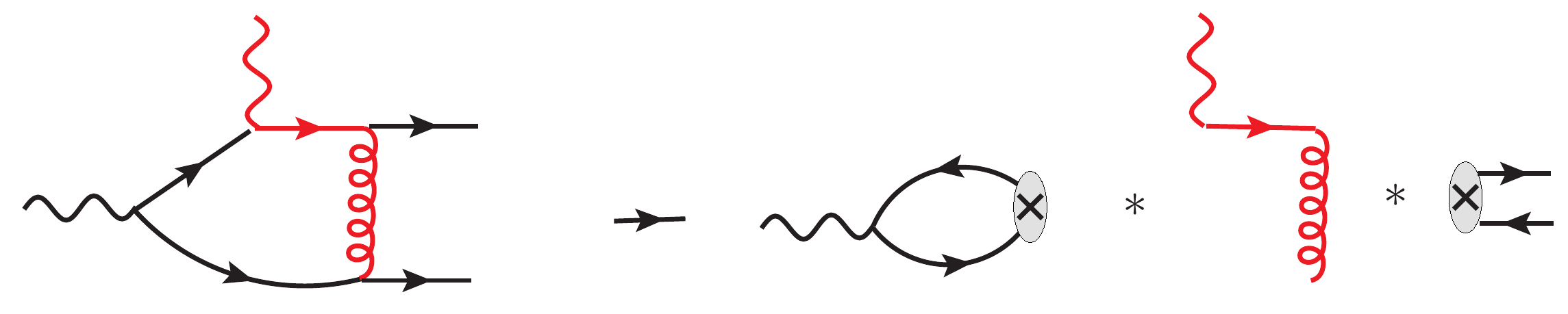}
\caption{Graphical illustration of the factorization in the $\bar{n}
$-collinear region. The hard lines are shown by red color. }
\label{jnb-interpre}
\end{figure}

The  $\bar{n}$-collinear contribution $k\sim q^{\prime}$  defines the 
 configuration which only appears starting  from one-loop.  
 We identify this contribution only with the integral $J_{2\bar n}$ in Eq.(\ref{J12nb}).
  In this  configuration  the hard part  is described by  tree subdiagram with one the ``gluon'' exchange  as illustrated  in
Fig.\ref{jnb-interpre}.
Shrinking  the hard subdiagram  one obtains the four-fermion
collinear operator $\left[  \psi_{c}^{\dag}\psi_{c}\right]  _{\bar{n}}\left[
\psi_{c}^{\dag}\psi_{c}\right]  _{n}$. The $\bar n$-collinear quarks describe  the
loop integral which can be associated with the  matrix element of the operator
$\left[  \psi_{c}^{\dag}\psi_{c}\right]  _{\bar{n}}$ between the real  ``photon''
and vacuum state. This matrix element  can be interpreted as the light-cone
distribution amplitude  of the ``photon''. Therefore  this collinear
contribution can be schematically described  as
\begin{equation}
J_{2\bar n}=\text{FT}\left\langle \bar{q}q\right\vert
\psi_{c}^{\dag}\psi_{c}\left\vert 0\right\rangle \ast T\ast
\text{FT}\left\langle 0\right\vert \psi_{c}^{\dag}\psi_{c}\left\vert \gamma\right\rangle
 =\phi_{qq}^{\text{lo}}\ast T\ast\phi_{\gamma},
\end{equation}
where $T$ denotes the hard coefficient function. This contribution also has
UV-divergency which can be associated with the mixing of the  operators
$\psi_{c}^{\dag}\psi_{c}$ and $B_{c}$   describing  the photon matrix
element.    The IR-singularity in Eq.(\ref{J2nb-res-1}) is interpreted as the endpoint divergency  in 
the convolution integral  with ``photon'' distribution amplitude   $\phi_{\gamma}$. 

We observe  that  both collinear contributions  have the endpoint  divergencies  in  the collinear convolution integrals. 
These divergencies indicate  an overlap between the collinear and soft domains.  
Therefore in order to define  collinear contributions one has to define a regularization which allows one to
 the soft  regions.  This  regularization prescription must be used
uniformly  for  collinear and soft contributions  in order to avoid a  double counting.  

 The soft contribution described by the integral $J_{s}$ in Eq.(\ref{Js-2})  does not have any hard propagator  $\sim 1/p^{2}_{h}$.
  In  this case the hard subgraph can be identified with the tree level  vertex
describing the scattering  of the virtual ``photon'' with the hard-collinear
``quark'' :  $\gamma^{\ast}q_{hc}(q^{\prime}+k)\rightarrow q_{hc}(p+k)$.
 Factorizing  the hard modes one obtains the  matrix element in an effective theory
\begin{equation}
J_{s}= C_{\gamma}\left\langle \bar{q}q\right\vert \psi_{n}^{\dag}\psi_{\bar{n}}\left\vert \gamma\right\rangle _{\text{\tiny SCET}}
\label{scetme}
\end{equation}
where fields  $\psi_{n}^{\dag}$ and $\psi_{\bar{n}}$ describe the hard-collinear
quarks  and  $C_{\gamma}=1$ is the hard kernel. 
\begin{figure}[ptb]
\centering
\includegraphics[width=3.0 in]{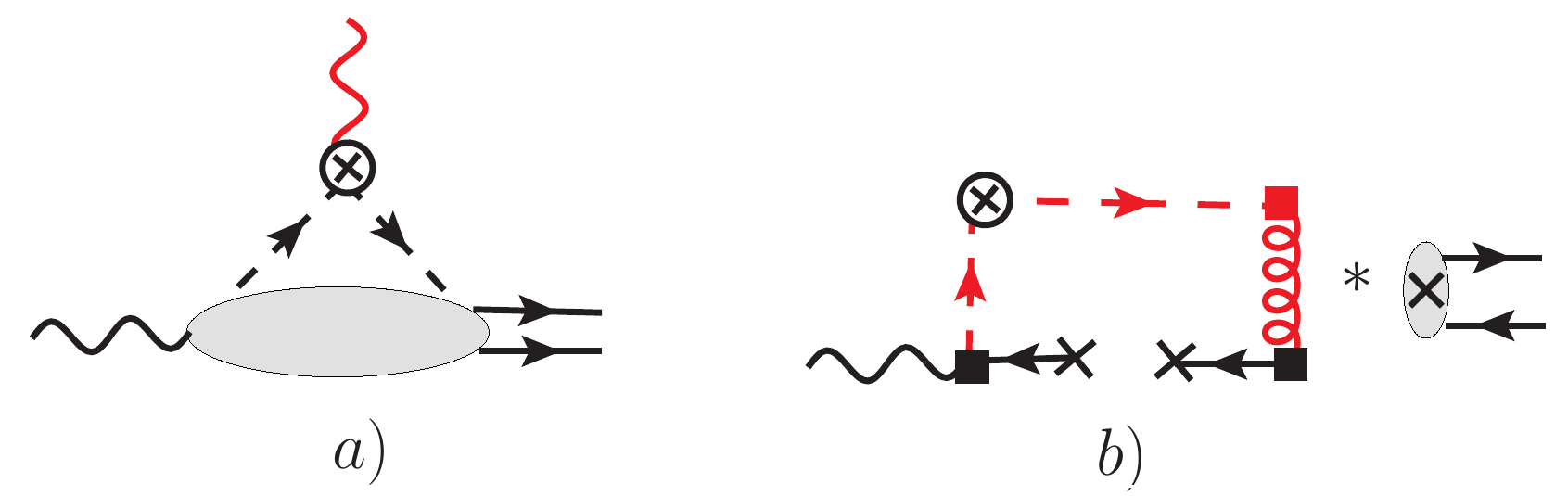}
\caption{The graphical interpretation  of  factorization in the  soft  region. 
The diagram $a)$ illustrates  the factorization of the hard mode. The crossed vertex denote the operator $\psi^{\dagger}_{n}\psi_{\bar n}$, the gray blob corresponds to the matrix element in Eq.(\ref{scetme}).  
The diagram $b)$ describes the factorization of the hard-collinear particles (red lines).  The soft quarks are shown by the solid lines with the crosses. 
The solid squares  denote the effective vertices in the soft-collinear effective theory. }
\label{soft-graph}
\end{figure}

 This matrix element  is described  by  
 hard-collinear, collinear and soft particles. Only this modes define  the soft integral in Eq.(\ref{Js-2}).  Therefore 
this integral  can be understood  in the framework of the soft collinear effective theory that is indicated  by the subscript SCET.  
The factorization of the hard modes is illustrated by the diagram in Fig.\ref{soft-graph}$a$.  The
interactions of  hard-collinear fields and soft fields are described by the
corresponding SCET Lagrangian which we will not define here.   Integrating  over  the hard-collinear fields we 
reduce the  matrix element (\ref{scetme}) to the matrix elements of the soft and  collinear fields. This is illustrated in Fig.\ref{soft-graph}$b$  
and  can be described  as 
\bea
 &\left\langle \bar{q}q\right\vert \psi_{n}^{\dag}\psi_{\bar{n}} \left\vert\gamma\right\rangle _{\text{\tiny SCET}}
 \simeq\text{FT}\left\langle \bar{q}q\right\vert \psi_{c}^{\dag}\psi_{c}\left\vert 0\right\rangle(y) 
 \ast J_{n}(y,Qk_{+})
  \ast\text{FT}\left\langle0\right\vert \psi_{s}^{\dag}\psi_{s}\left\vert 0\right\rangle (k_{+},k_{-})
 \nonumber \\ &
 \ast\, J_{\bar{n}}(Qk_{-})
\text{FT}\left\langle 0\right\vert B_{c}\left\vert \gamma\right\rangle=\phi_{qq}^{\text{lo}}\ast J_{n}\ast S\ast J_{\bar{n}},
 \label{scet-2}
\eea
where   $J_{n}$ and $J_{\bar{n}}$ correspond to the hard-collinear subdiagrams with the
 in each collinear sector.  The soft matrix element is
not local and the asterisks also denote the convolution integral with respect to
the soft fractions $k_{\pm}\sim m$. 

Combining together the  contributions from  all regions one can write the 
factorization formula for the integral $J$ 
\begin{equation}
J =H\ast\phi_{qq}
+\left[  C\ast\varphi_{qq}\right]  _{\text{reg}}
+\left[ \phi_{qq}\ast T\ast\phi_{\gamma}\right]  _{\text{reg}} 
 +C_{\gamma}\left[ \phi_{qq}\ast  J_{n}\ast S\ast J_{\bar{n}}\right]  _{\text{reg}}+{\cal O}(\lambda^{6})
\label{eq1}
\end{equation}
where the brackets $[...]_{\text{reg}}$  indicate the additional regularization prescription 
required for the separation of the collinear and soft modes. This
factorization introduces the additional factorization scale dependence  which must cancel in sum
of the all three contributions. In the calculation carried out above we use
the analytical regularization.  In this case the soft integral remains scaleless and
 vanishes. It is important to keep in mind  that this does not mean the
absence of the soft contribution. In this case the soft contribution
is implicitly included into the collinear integrals and therefore into the
definitions of the collinear matrix elements. This example illustrates that
the definition of the collinear matrix elements is scheme dependent and may
differ from one which is usually  accepted in the collinear factorization
framework (when one does not have any endpoint singularities).  
If one uses a different regularization scheme then the collinear and soft contributions  can  be
different and therefore the soft contribution must be always added into the
factorization formula.  A  one more  approach  to define  the collinear  contributions  is  
to define  subtractions which  remove the soft configurations   from 
the collinear integrals  as suggested in Refs.\cite{Collins:1999dz, Manohar:2006nz}.\footnote{ In SCET  this method  is known  as  zero-bin subtractions. }
 In any case a correct description of the endpoint region can not be
performed without a consistent definition of the soft contribution. 

A  specific feature of the discussed integral is the simple logarithmic structure.   In a renormalizable field
theory, like QCD,  one usually obtains a  large double logarithm. In such case  the auxiliary factorization scale $\nu$ can be used  for
a resummation of  large rapidity logarithms, see e.g. Ref.\cite{Chiu:2012ir} .

One more  important  observation  is related to the interpretation of  the  spurious
power suppressed  contributions   within the SCET approach.  It was demonstrated that in the soft region
 one of such contribution  even predicts power  correction of order  $\lambda^{2}$    
( or  correction suppressed as  $1/Q$ ).  However  corresponding soft  integrals
 are power divergent and must be neglected.  In  the SCET framework 
such spurious   contributions can be generated by appropriate  $T$-products  when matching to SCET-II.   

Consider  the scalar theory discussed in this section.  In this case one can easily find the counting rules for  different fields:
 $\phi_{hc} \sim \lambda$, $\phi_{c}\sim \lambda^{2}$,  $\phi_{s}\sim \lambda^{2}$. Appropriate SCET Lagrangian can be 
 also easily derived and we skip  these details  here.  Then the  soft  integral $J_{1s}$  is obtained from   $T$-product    
 \begin{equation}
 J_{1s}=\left\langle \bar{q}q\right\vert  {\cal O} \left\vert \gamma\right\rangle _{\text{SCET}}
 =\left\langle \bar{q}q\right\vert 
 T\left\{   
 {\cal O},
 \mathcal{L}_{\text{int}}^{(2,\bar{n})},
\mathcal{L}_{\text{int}}^{(2,n)},
\mathcal{L}_{\text{int}}^{(1,n)}
\right\}  
\left\vert \gamma\right\rangle\sim \lambda^{2},
\label{J1s-est}
 \end{equation}
 with the SCET operator 
\begin{eqnarray}
 {\cal O}&=&\psi_{hc,n}^{\dag}(0)\psi_{hc,\bar{n}}(0)\sim {\cal O}(\lambda^{2}),
\end{eqnarray}
and the interaction vertices 
\begin{eqnarray}
\mathcal{L}_{\text{int}}^{(2,\bar{n})}&=&\int d^{4}x \psi_{hc,\bar n}^{\dag}(x)B_{c,\bar n}(x)\psi_{s}(x)\sim {\cal O}(\lambda^{2}) ,
\label{L2nb}
\\
\mathcal{L}_{\text{int}}^{(2,n)}&=& m\int d^{4}x \psi_{s}^{\dag}(x)A_{hc,n}(x)\psi_{c,n}(x)\sim {\cal O}(\lambda^{2}) ,
\\
\mathcal{L}_{\text{int}}^{(1,n)}&=& m\int d^{4}x \psi_{c,n}^{\dag}(x)A_{hc,n}(x)\psi_{hc,n}(x)\sim {\cal O}(\lambda) .
\end{eqnarray}
where we assume $m\sim\lambda^{2}$ and multipole expansion of the arguments for the soft fields.\footnote{Remind that any external collinear state scales as $|p_{c}\rangle\sim \lambda^{-2}$. }   
This  shows that certain $T$-products which  appear within the  SCET framework describe the power divergent integrals which was suggested  to set to zero identically.  Therefore such constructions in the effective theory  are fictitious and  one has to study  various  contractions of the hard-collinear fields  more carefully  in order to  make a conclusion about their relevance. 

 In absence of external soft momenta (as in our  example)  one can obtain a  good  criteria  using  transformation properties of  soft convolution integrals under longitudinal  boosts  
\begin{equation}
n\rightarrow \alpha n,\, \, \bar n\rightarrow \alpha^{-1}\bar  n.
\label{boost}
\end{equation}
If the soft integral is not invariant under (\ref{boost}) then it  is power divergent  and corresponding $T$-product  can be neglected.  For instance,
for the soft convolution integral  $J_{1s}$  in Eq.(\ref{J1s})  one obtains
\begin{equation}
I_{s}=\int\frac{d(\bar n\cdot k)}{ (\bar n\cdot k)  }\int\frac{d(n\cdot k)}{(n\cdot k)^{2}}\frac{dk_{\bot}}{\left[  (n\cdot k)(\bar n\cdot k)-k_{\bot}^{2}
-m^{2}\right]  }\rightarrow \alpha^{-1} I_{s}.
\end{equation}
Then  the answer for this integral must  transform in  the same way   that  leads to the power divergent  expression in  Eq. (\ref{J1sreg}).   

 In case of QCD  the analogous contributions  are described  by the similar integrals but with  a soft function instead of the 
 soft ``quark'' propagator (see more details in Sec.~4.3.2)  
\begin{equation}
I_{s}\sim\int\frac{dk_{+}}{k^{2}_{+}}\int\frac{dk_{-}}{k_{-}}S(k_{+}k_{-}),
\label{Is-QCD}
\end{equation}
The soft function $S(k_{+}k_{-})$ is invariant under longitudinal boosts and therefore depend on the product of the light-cone fractions.  
The power divergency can be easily seen  performing  rescaling  of the one light-cone  variable.  
For instance,  using $k'_{-}=k_{+}k_{-}$ one finds
\begin{equation}
I_{s}\sim\int\frac{dk_{+}}{k^{2}_{+}} \int\frac{dk'_{-}}{k'_{-}}S(k'_{-}).
\label{Is-QCD-power}
\end{equation}
Therefore this property can not be  related with the nonperturbative sector.   We assume   that such integrals 
are similar to  the ``traditional'' scaleless integrals like $\int d^{D}k/k^{2n}$. 
In what follow we assume that such  contributions are fictitious and we will also set them to zero.

The relevant soft contribution  (\ref{Js-2}) is described by
the higher order $T$-product  obtained by expansion of  the argument of
the soft field in the interaction vertex in Eq.(\ref{L2nb}):
\begin{equation}
\mathcal{L}_{\text{int}}^{(\bar{n})}=\int d^{4}x\,   \psi_{hc,\bar n}^{\dag}(x)B_{c,\bar n}(x)
\left\{ \frac{1}{2}(x\bar{n})(n\partial)+\frac{1}{2}x_{\bot i}x_{\bot j}\partial_{i}\partial_{j} \right \}\psi_{s}(x).
\end{equation}

 Therefore  we  find  that certain $T$-products which describe   transition  from SCET-I to SCET-II   are scaleless and power divergent.  
The  given consideration of the toy integral provides  us an evidence that such contribution can be safely excluded  from a consideration.  
 This conclusion will be  very important for an analysis of soft-overlap configurations  which we consider  in the next section.

\section{Factorization of the subleading amplitudes in SCET}

\label{factSCET}

\subsection{ Soft Collinear Effective Theory approach: general remarks}

 In what follow we assume that  it is enough to consider  particles which have hard
$p_{h}$, hard-collinear $p_{hc}$, collinear $p_{c}$ and soft $p_{s}$ momenta.
The light-cone components $\left(  pn,p\bar{n},p_{\bot}\right)  \equiv
(p_{+},p_{-},p_{\bot})$ of the corresponding momenta scale as
\begin{equation}
p_{h}\sim Q\left(  1,1,1\right)  ,~p_{h}^{2}\sim Q^{2},\label{def:ph}%
\end{equation}%
\begin{equation}
p_{hc}\sim Q\left(  1,\lambda^{2},\lambda\right)  \text{ or }p_{hc}^{\prime
}\sim Q\left(  \lambda^{2},1,\lambda\right)  ,~p_{hc}^{2}\sim Q^{2}\lambda
^{2}\ \sim Q\Lambda,\label{def:phc}%
\end{equation}%
\begin{equation}
p_{c}\sim Q\left(  1,\lambda^{4},\lambda^{2}\right)  \text{ or ~}p_{c}%
^{\prime}\sim Q\left(  \lambda^{4},1,\lambda^{2}\right)  ,~\ p_{c}^{2}\sim
Q^{2}\lambda^{4}\sim\Lambda^{2},\label{def:pc}%
\end{equation}%
\begin{equation}
p_{s}\sim Q\left(  \lambda^{2},\lambda^{2},\lambda^{2}\right)  ,~\ p_{s}%
^{2}\sim Q^{2}\lambda^{4}\sim\Lambda^{2}.\label{ps}%
\end{equation}
Here $Q$ and $\Lambda$ denote  generic large and soft scales, respectively.
Further  we will again  use the small dimensionless parameter $\lambda\sim\sqrt{\Lambda/Q}$. 
We assume that we do not need other specific modes  in order to describe factorization of  the 
 power  suppressed amplitudes. Then we expect that  factorization  approach consist of the following two steps. 
 First, we integrate out the hard modes and reduce full QCD to the
effective theory. Corresponding effective Lagrangian is constructed
from the hard-collinear and soft particles.  This effective theory
is denoted as SCET-I. If the hard scale $Q^{2}$ is so large that the hard
collinear scale $\mu_{hc}\sim\sqrt{Q\Lambda}$ is a good parameter for the
perturbative expansion then one can perform the second  step and
integrate out the hard-collinear modes. This reduces  SCET-I to the effective theory 
describing  only collinear and soft particles and known as  SCET-II. 

The different formulation of the SCET can be found in Refs.\cite{Bauer:2000ew,
Bauer2000, Bauer:2001ct,Bauer2001,BenCh,BenFeld03}. In our work we use the
technique developed in the position space in Refs.\cite{BenCh,BenFeld03}. For
the SCET fields we use following notations. The fields $\xi_{n}^{C},$ $A_{\mu
C}^{n}$ and $\xi_{\bar{n}}^{C},~A_{\mu C}^{\bar{n}}~$ denote the
hard-collinear ($C=hc$) or collinear ($C=c$) quark and gluon fields associated
with momentum $p^{\prime}$ and $p$, respectively, see Eq.(\ref{mom1}). As
usually, the hard-collinear and collinear quark fields satisfy to
\begin{equation}
\setbox0=\hbox{$n$} \dimen0=\wd0 \setbox1=\hbox{/} \dimen1=\wd1
\ifdim\dimen0>\dimen1 \rlap{\hbox to \dimen0{\hfil/\hfil}} n
\else \rlap{\hbox to
\dimen1{\hfil$n$\hfil}} / \fi \xi_{n}^{C}=0,~~ \setbox0=\hbox{$\bar n$}
\dimen0=\wd0 \setbox1=\hbox{/} \dimen1=\wd1 \ifdim\dimen0>\dimen1
\rlap{\hbox to \dimen0{\hfil/\hfil}} \bar n \else \rlap{\hbox to
\dimen1{\hfil$\bar n$\hfil}} / \fi \xi_{\bar{n}}^{C}=0.
\end{equation}
The fields $q$ and $A_{\mu}^{s}$ describe  soft quarks and gluons with the
soft momenta (\ref{ps}).  We will use  the standard set  of  convenient notation for the gauge invariant combinations 
\begin{equation}
\chi_{n}^{C}(\lambda\bar{n})\equiv~W_{n}^{\dag}(\lambda\bar{n})\xi_{n}%
^{C}(\lambda\bar{n}),~\ \bar{\chi}_{n}^{C}(\lambda\bar{n})\equiv\bar{\xi}%
_{n}^{C}(\lambda\bar{n})W_{n}(\lambda\bar{n}),\label{Chi}%
\end{equation}%
\begin{equation}
\mathcal{A}_{\mu C}^{n}(\lambda\bar{n})\equiv\left[  W_{n}^{\dag}%
(\lambda\bar{n})D_{\mu C}W_{n}(\lambda\bar{n})\right]  ,\label{AC}%
\end{equation}
where the covariant derivative $D_{\mu C}=i\partial_{\mu}+gA_{ \mu C}^{n}$
acts inside the brackets and the hard-collinear or collinear gluon Wilson line
(WL) read:
\begin{equation}
W_{n}(z)=\text{P}\exp\left\{  ig\int_{-\infty}^{0}ds~\bar{n}\cdot A_{C}%
^{n}(z+s\bar{n})\right\}  .
\end{equation}

In the wide-angle kinematics we have the energetic particles propagating with
large energies in four directions. Therefore it is useful to introduce two
more auxiliary light-cone vectors associated with the photon momenta: $q_{1}$
and $q_{2}$%
\begin{equation}
\bar{v}^{\mu}=\frac{2q_{1}^{\mu}}{\sqrt{s}},~\ v^{\mu}=\frac{2q_{2}^{\mu}%
}{\sqrt{s}},~(\bar{v}\cdot v)=2.
\label{vbv}
\end{equation}
Using the vectors $\bar{v},v$ we also introduce the hard-collinear quark and
gluon fields in the similar way as before just changing $(n,\bar
{n})\rightarrow\left(  v,\bar{v}\right)  $.

The explicit expression for the SCET-I~ Lagrangian in position space can be
found in Refs.\cite{BenCh,BenFeld03}. This Lagrangian being expanded in the
small parameter $\lambda$ is given by the sum
\begin{equation}
\mathcal{L}_{\text{SCET}}^{(n)}=\mathcal{L}_{\xi\xi}^{(0,n)}+\mathcal{L}%
_{\xi\xi}^{(1,n)}+\mathcal{L}_{q\xi}^{(1,n)}+\mathcal{O}(\lambda
^{2}),\label{Lscet}%
\end{equation}
where $\mathcal{L}_{\xi\xi}^{(\lambda,n)}\sim\mathcal{O}(\lambda)$ and the
explicit expressions read
\begin{equation}
\mathcal{L}_{\xi\xi}^{(0, n)}=\bar{\xi}_{n}^{hc}(x)\left(
  i~n\cdot D+gn\cdot A^{s}(x_{-})+i \Dslash{D}_{\bot}(i \bar n\cdot D)^{-1} i\Dslash{D}_{\bot}\right)  \xi_{n}^{hc}(x),\label{Ln0}%
\end{equation}%
\begin{equation}
\mathcal{L}_{q\xi}^{(1,n)}=\bar{\xi}_{n}^{hc}(x)
i \Dslash{D} _{\bot}
W_{n}~q(x_{-})+\bar{q}(x_{-})W_{n}^{\dag}
i  \Dslash{D}_{\bot}
\xi_{n}^{hc}(x).\label{Ln1}%
\end{equation}
where $D_{\mu}=i\partial_{\mu}+gA_{\mu\ hc}^{n}$,~$x_{-}=\frac{1}{2}(x\bar
{n})n$, $A_{\mu}^{s}$ denotes the soft gluon field.  The expression for  the subleading term $\mathcal{L}_{\xi\xi
}^{(1,n)} $ is a bit lengthy and we will not write it here. The similar
expressions are also valid for the other collinear sectors associate with the
directions $\bar{n},v,\bar{v}$.

The matching from SCET-I to SCET-II is performed by substituting  in  SCET-I  Lagrangian 
\be{sub}
\xi^{hc}\rightarrow\xi^{c}+\xi^{hc},\  A_{hc}\rightarrow
A_{c}+A_{hc}
\ee
 and integrating  over  the hard-collinear
fields. A more detailed description of this step can be found in
Refs.\cite{Bauer:2002nz,Bauer:2003mga} in the hybrid representation and in
Ref.\cite{Beneke:2003pa} in the position space formulation. 

The power counting rules for different SCET operators can be fixed using the
power counting of the SCET fields. The counting rules for the SCET fields can
be obtained from the corresponding propagators in momentum space and read (see
for instance Ref.\cite{BenCh})
\begin{equation}
\xi_{n}^{hc}\sim\lambda,~\ \bar{n}\cdot A_{hc}^{n}\sim1,~A_{\bot hc}^{n}\sim\lambda,~ n\cdot A_{hc}^{n}\sim\lambda^{2},
\end{equation}%
\begin{equation}
\xi_{n}^{c}\sim\lambda^{2},~\ \bar{n}\cdot A_{c}^{n}\sim1,~A_{\bot c}^{n}\sim\lambda^{2},~ n\cdot A_{c}^{n}\sim\lambda^{4},\label{colf}
\end{equation}%
\begin{equation}
A^{s}_{\mu}\sim\lambda^{2},~\ q\sim\lambda^{3}.
\end{equation}

In order to determine the counting rules for  physical amplitudes one also needs  to define
the counting for external hadronic states. In c.m.s  frame the outgoing
pions are made of energetic collinear partons therefore assuming the
conventional normalization of hadronic states one obtains
\begin{equation}
\left\langle \pi(p_{c})\right\vert \sim\lambda^{-2}.\label{hadst}%
\end{equation}

\subsection{ The hard contributions within the SCET framework}
\label{hard}
 The hard contributions  are described  by convolution of  a
hard coefficient function with the matrix elements of collinear operators  describing  the overlap with the outgoing pion states.  
Only a  collinear operator in SCET can be matched onto  hadronic states because the invariant mass of  a hadron 
is restricted to order $Q^{2}\lambda^{4}\sim\Lambda^{2}$.

Consider first the leading power contribution discussed in Sec.\ref{leading}.  
In the operator form the  hard contribution can be presented as 
\begin{align}
T\left\{  J_{\text{em}}^{\mu}(0),J_{\text{em}}^{\nu}(x)\right\} _{\text{hard}}  &  =
H^{\mu\nu}\ast  O_{n}^{(4)}O_{\bar n}^{(4)}+\mathcal{O}(\lambda^{9}),
\label{TJJ-hard-LO}%
\end{align}
where asterisk denotes the convolution integrals in position space. 
 The leading twist-2 operator $ O_{n}^{(4)}$  in SCET notation can be written as 
\begin{equation}
O_{n}^{(4)}=\bar{\chi}_{n}^{c} \setbox0=\hbox{$\bar n$} \dimen0=\wd0
\setbox1=\hbox{/} \dimen1=\wd1 \ifdim\dimen0>\dimen1 \rlap{\hbox to
\dimen0{\hfil/\hfil}} \bar n \else \rlap{\hbox to \dimen1{\hfil$\bar
n$\hfil}} / \fi \gamma_{5}\chi_{n}^{c}\sim \mathcal{O}(\lambda^{4})\,.~\label{defO4}%
\end{equation}%
The arguments of the  fields are on the light-cone and not shown for simplicity.  All collinear  operators $O_{n,\bar n}^{(4)}$  are  color singlet
and have appropriate flavor structure.

 In order to compute  the  amplitude  one has to take the matrix element  from  Eq.(\ref{TJJ-hard-LO}). 
 The soft  and collinear modes are decoupled in leading  SCET-II Lagrangian \cite{Bauer:2002nz,
Beneke:2003pa} and therefore the matrix element of the collinear operators  can be factorized
\begin{equation}
\left\langle p,p^{\prime}\right\vert O_{n}^{(4)}O_{\bar{n}}^{(4)}\left\vert
0\right\rangle _{\text{{\tiny SCETII}}}=\left\langle p\right\vert O_{n}%
^{(4)}\left\vert 0\right\rangle _{\text{{\tiny SCETII}}}\left\langle
p^{\prime}\right\vert O_{\bar{n}}^{(4)}\left\vert 0\right\rangle
_{\text{{\tiny SCETII}}}.
\label{fact}
\end{equation}
It is easy to see  that  
collinear operator $O_{n}^{(\lambda)}$ has  the minimal possible  order  $\lambda=4$, i.e.
\begin{equation}
\left\langle \pi(p)\right\vert O_{n}^{(\lambda)}\left\vert 0\right\rangle
_{\text{{\tiny SCETII}}}=0,~\lambda<4\text{. }%
\end{equation}
Therefore  we obtain the first  operator with the nonvanishing matrix element  only at
order $\lambda^{8}$. Substituting in Eq.(\ref{fact}) the parametrization of  matrix elements
(\ref{pionDA}) and performing the Fourier transformation of the hard kernel $H$ we obtain the
factorization formulas discussed in Sec.\ref{leading}. The scaling behavior of
the amplitudes  can be easily obtained using the SCET counting rules that
gives
\begin{equation}
A_{+\pm}^{(i)}\sim\underset{\lambda^{2}}{\underbrace{\left\langle p^{\prime
}\right\vert O_{n}^{(4)}\left\vert 0\right\rangle _{\text{{\tiny SCETII}}}}%
}\ast H^{\mu\nu}\ast\underset{\lambda^{2}}{\underbrace{\left\langle
p\right\vert O_{\bar{n}}^{(4)}\left\vert 0\right\rangle _{\text{{\tiny SCETII}%
}}}}~\sim~\lambda^{4}\sim\Lambda^{2}/Q^{2}.
\label{Apm2}
\end{equation}

 In order  to prove the leading power factorization formula within the SCET framework one
must demonstrate the absence of an appropriate soft-overlap configuration of  the same order  $\sim \lambda^{8}$.  
We will do  this  later  performing an analysis  of the soft  contributions.

The hard power corrections to the leading-order  result (\ref{TJJ-hard-LO}) are defined by  subleading collinear operators. 
The  set of the required  operators $O_{n}^{(i)}\sim\mathcal{O}(\lambda^{i})$  
can be described  by the  two operator subsets  order $\lambda^{6}$ (twist-3) and $\lambda^{8}$ (twist-4). 
Using  the SCET notations these operators can be introduced as following.

The twist-3  operators 
\begin{equation}
O_{n}^{(6)}=\left\{ 
\bar{\eta}^{\alpha}_{n}\nbs\gamma_{\perp} ^{ \beta}\gamma_{5} \chi^{c}_{n},\,  \bar{\chi}^{c}_{n}\nbs \gamma_{\perp}^{ \beta} \gamma_{5} \eta_{n}^{\alpha},\,
\bar{\chi}_{n}^{c}\nbs\Dslash{\mathcal{A}}_{\perp }^{n} \gamma_{5}\chi_{n}^{c}
  \right\},
\label{defO6}%
\end{equation}%
where we use the following notation
\begin{align}
\eta_{n}^{\alpha}(x)=(i\bar{n}\cdot\partial)^{-1}W^{\dagger}_{n}(x)iD^{\alpha}_{\perp c} \xi^{c}_{n}(x)\sim \mathcal{O}(\lambda^{4}), 
\label{etan}
\\
\bar \eta_{n}^{\alpha}(x)= \bar \xi^{c}_{n}(x) i\overleftarrow{D}^{\alpha}_{\perp c} W_{n}(x)(i\bar{n}\cdot\overleftarrow{\partial})^{-1}\sim \mathcal{O}(\lambda^{4}).
\label{eta}
\end{align}

The   set  of the appropriate  twist-4 operators can be schematically introduced  as  
\begin{align}
O_{n}^{(8)}   = & \left\{ 
\bar{\eta}^{\alpha}_{n}\nbs\gamma_{\perp \beta}  \gamma_{5} \Dslash{\mathcal{A}}_{\perp}^{n}  \chi^{c}_{n} ,\, \,
\bar \chi^{c}_{n} \nbs\gamma_{\perp \beta}  \gamma_{5} \Dslash{\mathcal{A}}_{\perp}^{n}{\eta}^{\alpha}_{n},\,  \, 
\bar{\eta}^{\alpha}_{n}\nbs\gamma_{\perp \beta}   \tilde{\Dslash{\mathcal{A}}}_{\perp}^{n}  \chi^{c}_{n}, \,
\right. &\nonumber \\ & \left.
\bar \chi^{c}_{n} \nbs\gamma_{\perp \beta}   \tilde{\Dslash{\mathcal{A}}}_{\perp}^{n} {\eta}^{\alpha}_{n} ,\,\, 
 \bar{\chi}_{n}^{c}\nbs \gamma_{5} (\bar{n}\cdot \mathcal{A}^{n})\chi_{n} ^{c},\,  \,
\bar{\chi}_{n}^{c}\nbs  (\bar{n}\cdot \tilde{\mathcal{A}}^{n})\chi_{n} ^{c},
\right.&
\nonumber \\
&\left.
\bar{\chi}_{n}^{c}\nbs \gamma_{5}\mathcal{A}_{\mu\perp}^{n}\mathcal{A}_{\nu\perp}^{n}\chi_{n} ^{c},\ 
\bar{\chi}_{n}^{c}~
\nbs 
\tilde {\mathcal{A}}^{n}_{\mu\bot} \mathcal{A}_{\nu\perp}^{n} \chi_{n}^{c},\
 \bar{\chi}_{n}^{c}\Gamma_{1}\chi_{n}^{c} ~\bar{\chi}_{n}^{c}\Gamma_{2}\chi_{n}^{c}~\right\} ,& 
 \label{defO8}
\end{align}
where $\tilde {\mathcal{A}}_{\mu\bot} =\frac12 i\varepsilon_{\mu\alpha n\bar n }
\mathcal{A}^{\alpha}_{\bot}$, the symbols $\Gamma_{i}$  in the four-quark operator  denote the appropriate Dirac matrices.

In Appendix~\ref{htDA}  we  also provide  the  QCD definitions well-known in the literature. 
 The QCD operators can be represented   within the SCET framework as the operators listed in Eqs.(\ref{defO6}),(\ref{defO8}). 
More details about this correspondence can be also found  in Ref.\cite{Hardmeier:2003ig}.  Obviously, the analogous  set of
the  operators can also be defined  in the $\bar{n}$-collinear sector. 
Matrix elements of the  operators  in Eqs.(\ref{defO6}), (\ref{defO8}) define the higher twist distribution amplitudes  of  pion.

 Notice that all twist-3 operators are chiral odd.  The first two operators  in set $O_{n}^{(6)}$  in Eq.(\ref{defO6})  are the two-particle operators. They play an important role in phenomenology because their matrix elements  include  the so-called chirally enhanced DAs \cite{Braun:1989iv}.  
 The DAs of these operators  can be  represented  as a sum of the  three-particle DA  defined by the three-particle operator   
 $\bar{\chi}_{n}^{c}(z_{1})\nbs  \Dslash{\mathcal{A}}_{\perp}^{n}(z_{2}) \gamma_{5}\chi_{n}^{c}(z_{3})$ and 
   two-particle contribution  which is proportional to the large numerical factor $m^{2}_{\pi}/(m_{u}+m_{d})$, 
   see the details in Appendix~\ref{htDA}.   

The set of the twist-4 operators $O_{n}^{(8)}$  consists of three-particle (first and second lines) and four-particle operators. 
The QCD definitions of the three-particle  DAs can also be found  in Appendix~\ref{htDA}. \footnote{There are also two-particle twist-4 DAs which  can be expressed through the  twist-2 and  twist-4 DAs. For simplicity we do not introduce them here.  }

Including  all possible subleading contributions  we obtain 
\begin{align}
T\left\{  J_{\text{em}}^{\mu}(0),J_{\text{em}}^{\nu}(x)\right\} _{\text{hard}}  &  =
H^{\mu\nu}\ast  O_{n}^{(4)}O_{\bar n}^{(4)}+ \sum T_{6}^{\mu\nu}\ast  O_{n}^{(6)}O_{\bar n}^{(6)} 
\nonumber \\&
+ \sum T_{8}^{\mu\nu}\ast  \{O_{n}^{(4)}O_{\bar n}^{(8)}+O_{n}^{(8)}O_{\bar n}^{(4)}\}+\mathcal{O}(\lambda^{13}),
\label{TJJ-hard}%
\end{align}
where the sum  over  all operators which enter in the sets  $O_{n}^{(6,8)}$ is implied,  the kernels
$T_{i}^{\mu\nu}$ denote the hard coefficient  function. 
 In Eq.(\ref{TJJ-hard}) we excluded  the contributions of order $\lambda^{10}$ provided by the operators like $O_{n}^{(4)}O_{\bar n}^{(6)}$.  
 Such combinations  can be neglected  because corresponding coefficient functions  vanish  in the massless QCD ($O_{n}^{(4)}$ and $O_{\bar n}^{(6)}$ have different Dirac  structure).  It is easy to see that  subleading  contributions  in  Eq.(\ref{TJJ-hard}) are given by the operators of order $\lambda^{12}$.

The full description of the subleading contribution in Eq.(\ref{TJJ-hard} ) is very complicated  because of  large number  of the various  
subleading operators.  Let us at first step, in order to  simplify the calculations,   to  restrict the following  consideration only by the specific chiral enhanced contribution. 
Such approximation might also be  justified   phenomenologically  due to relatively large normalization coefficients of the corresponding DAs. 
The  consideration of  other contributions  we postpone for future work.  Therefore in what follow we assume  that 
 \begin{align}
T\left\{  J_{\text{em}}^{\mu}(0),J_{\text{em}}^{\nu}(x)\right\} _{\text{hard}}   \simeq
H^{\mu\nu}\ast  O_{n}^{(4)}O_{\bar n}^{(4)}+  T^{\mu\nu}\ast  O_{\chi n}^{(6)}O_{\chi  \bar n}^{(6)}. 
\label{TJJ-hard-chi}
\end{align}
where the operator $O_{\chi n}^{(6)}$  denotes  the chiral enhanced contributions. 

\begin{figure}[ptb]
\centering
\includegraphics[height=1.2437in]{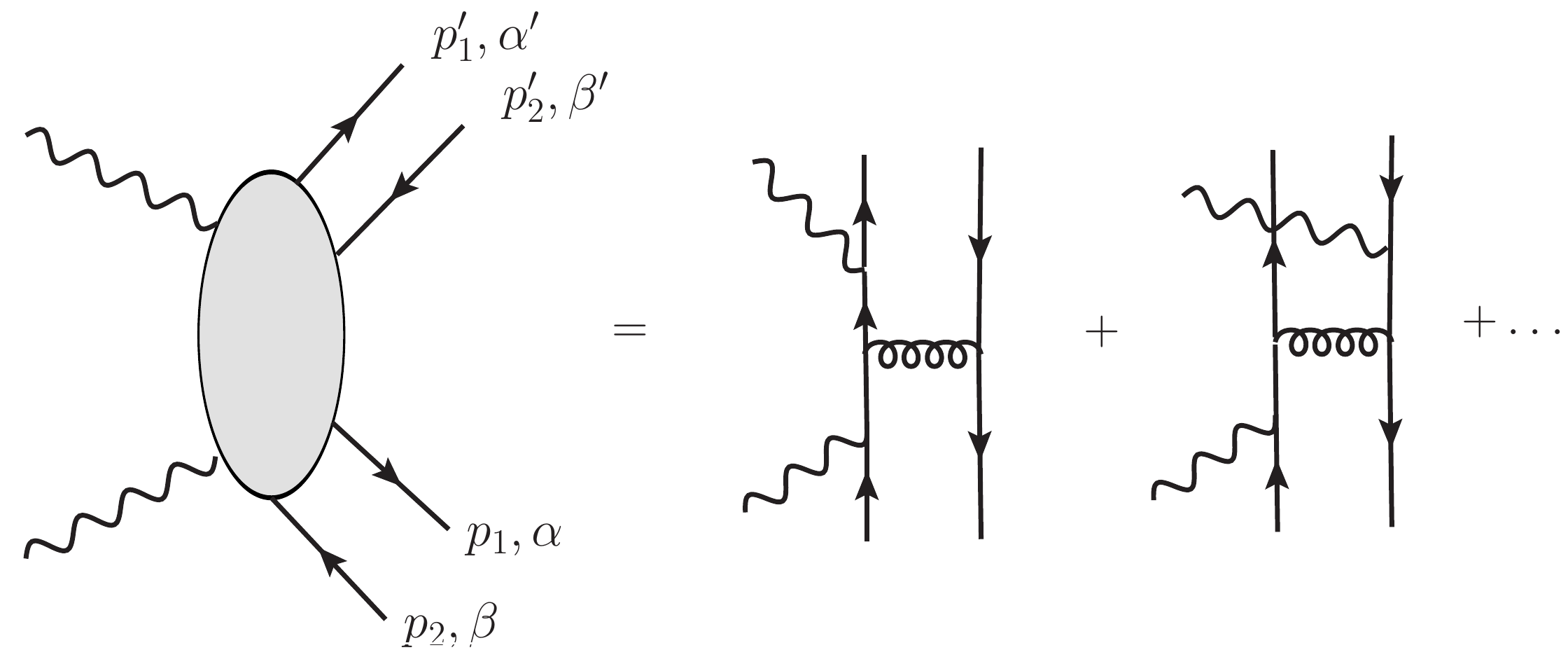}\caption{Graphical
representation of the function $D_{ab}^{\mu\nu}(p_{i}, p_{i}^{\prime})$
defined in Eq.(\ref{hardNLO}).}%
\label{hsa-amplitude}%
\end{figure}

In order to compute the hard kernel $T^{\mu\nu}$  one has to consider   
the diagrams which  are similar to  those in Fig.\ref{leading-graphs} but with the 
appropriate twist-3 projections for pion  DAs.  
We will use a  technique suggested in Refs.\cite{Beneke:2000wa, Beneke:2002bs}.
In this case a compact  expression for the subleading amplitudes $B^{(i,h)}_{+\pm}$ (\ref{defAB} )  can be written as   
\begin{align}
\delta^{ab}B_{+\pm}^{(0,h)}(s,\theta)+\delta^{a3}\delta^{b3}B_{+\pm}^{(3,h)}(s,\theta)  &
=\frac{\left(  f_{\pi}\mu_{\pi}\right)  ^{2}}{16}\int_{0}^{1}dx~\hat{M}
_{\beta^{\prime}\alpha^{\prime}}(x,p^{\prime})\int_{0}^{1}dy~\hat{M}
_{\beta\alpha}(y,p)\nonumber\\
&
~\ \ \ \ \ \ \ \ \ \ \ \ \ \ \ \ \ \times
\left[  2M_{1,2}^{\mu\nu}~D_{ab}^{\mu\nu}(p_{i},p_{i}^{\prime})\right]
_{\alpha^{\prime}\beta;\alpha\beta^{\prime}}, \label{hardNLO}%
\end{align}
where  $D_{ab}^{\mu\nu}(p_{i},p_{i}^{\prime})$ denotes the sum of all 
diagrams describing the hard subprocess $\gamma\gamma\rightarrow(q\bar{q}%
)_{n}+(q\bar{q})_{\bar n}$. The external momenta of the outgoing quarks are
shown in Fig.\ref{hsa-amplitude}. 
The light-cone expansions of  quark momenta  read
\begin{equation}
p_{1}\simeq yp+r_{\perp},~p_{2}\simeq\bar{y}p-r_{\perp},
\end{equation}%
\begin{equation}
p_{1}^{\prime}\simeq xp^{\prime}+k_{\perp},~p_{2}^{\prime}\simeq\bar
{x}p^{\prime}-k_{\perp}.
\end{equation}

Twist-3 quark projectors $\hat{M}$ in Eq.(\ref{hardNLO}) are given by
\begin{equation}
\hat{M}_{\alpha\beta}(y,p)=\phi_{p}(y)~\left[  \gamma_{5}\right]
_{\beta\alpha}-p_{\lambda}\left[  \sigma^{\lambda\rho}\gamma_{5}\right]
_{\beta\alpha}\frac{i}{6}\left[  \frac{n^{\rho}}{(p\cdot n)}\phi_{\sigma
}^{\prime}(y)-\phi_{\sigma}(y)\frac{\partial}{\partial r_{\bot}^{\rho}%
}\right]  ,\label{Myp}%
\end{equation}%
\begin{equation}
\hat{M}_{\beta^{\prime}\alpha^{\prime}}(x,p^{\prime})=\phi_{p}(x)~\left[
\gamma_{5}\right]  _{\beta^{\prime}\alpha^{\prime}} -p^{\prime}_{\lambda
}\left[  \sigma^{\lambda\rho}\gamma_{5}\right]  _{\beta^{\prime}\alpha
^{\prime}}\frac{i}{6}\left[  \frac{\bar{n}^{\rho}}{(p^{\prime}\cdot\bar
n)}\phi_{\sigma}^{\prime}(x)-\phi_{\sigma}(x)\frac{\partial}{\partial k_{\bot
}^{\rho}}\right]  .\label{Mxpf}%
\end{equation}
Here for the DAs $\phi_{p,\sigma}$ we use only the chiral enhanced pieces
\begin{equation}
\phi_{ p}(x)=1,~\ \phi_{ \sigma}(x)=6x\bar{x},~\phi_{\sigma}^{\prime}(x)=6(1-2x).\
\end{equation}
The expressions in Eqs.(\ref{Myp}) and (\ref{Mxpf}) include the
differentiations with respect to the relative transverse momenta $r_{\bot}$
and $k_{\bot}$. After the differentiation
one can put $r_{\bot}=k_{\bot}=0$. In order to compute the traces in  expression
Eq.(\ref{hardNLO}) we used  package \textit{FeynCalc}  \cite{Mertig:1991}.
We obtained the following results
\begin{equation}
B_{++}^{(0,h)}(s,\theta)\simeq B_{++}^{(3,h)}(s,\theta)\simeq 0+\mathcal{O}(\alpha^{2}_{s}),
\label{Bpp0,3h}
\end{equation}%
\begin{equation}
B_{+-}^{(0,h)}(s,\theta)=\frac{\alpha_{s}}{4\pi}\frac{C_{F}}{N_{c}}\frac{\left(
4\pi f_{\pi}\right)  ^{2}}{s}\frac{\mu_{\pi}^{2}}{s}\left\{  (e_{u}^{2}%
+e_{d}^{2})\frac{(3-\eta^{2})}{(1-\eta^{2})}~I_{s}+\frac{2e_{u}e_{d}}%
{(1-\eta^{2})}~I(\eta)\right\}  ,\label{B0hpm}%
\end{equation}%
\begin{equation}
B_{+-}^{(3,h)}(s,\theta)=\frac{\alpha_{s}}{4\pi}\frac{C_{F}}{N_{c}}\frac{\left(
4\pi f_{\pi}\right)  ^{2}}{s}\frac{\mu_{\pi}^{2}}{s}\frac{(e_{u}-e_{d})^{2}%
}{(1-\eta^{2})}I(\eta),\label{B3hpm}%
\end{equation}
where $\eta=\cos\theta$. We also introduced  special  notations for  the  two types of the collinear convolution integrals. 
The singular integrals are given by 
\begin{equation}
I_{s}=\int_{0}^{1} dy \int_{0}^{1} dx\left(  ~\frac{1}{y\bar{x}}+\frac{1}{x\bar{y}}\right).
\end{equation}
The finite collinear integrals can be computed that yields 
\begin{align}
& I(\eta)=8-4\eta\ln\left[  \frac{1+\eta}{1-\eta}\right]  +2(3-\eta
^{2})\left\{  \text{Li}\left[  \frac{{\scriptsize 1+\eta} }{{\scriptsize 2}%
}\right]  +\text{Li}\left[  \frac{{\scriptsize 1-\eta}}{{\scriptsize 2}%
}\right]  \right. \nonumber\\
&  \left.  \phantom{empty space} -\frac{2\pi^{2}}{3}-\ln^{2}\frac
{{\scriptsize 1+\eta}}{{\scriptsize 1-\eta}}+\frac{1}{2}\left(  \ln^{2}
\frac{{\scriptsize 1+\eta}}{{\scriptsize 2}}+\ln^{2}\frac{{\scriptsize 1-\eta
}}{{\scriptsize 2}}\right)  \right\}  ,
\label{defI}
\end{align}
where Li$\left[  z\right]  $ denotes the Spence function  defined  by
\begin{equation}
\text{Li}(z)=-\int_{0}^{z}dt\frac{\ln(1-t)}{t}.
\label{defLi}
\end{equation}
The simple description  of the singular and regular  integrals   is possible due to the different 
isotopic factor in front of the appropriate diagrams.   
The regular integral $I(\eta)$  is only provided by the diagrams where  photons couple to the different quark lines (like the second diagram in Fig.\ref{leading-graphs}).  
Therefore  these diagrams are not singular and corresponding convolution integrals can be
computed explicitly.  At   large values  of scattering angle $\eta\sim0$ ( remind that $\eta=\cos\theta$ hence $\theta\sim90^{o}$) 
we obtain
\begin{equation}
I(\eta\rightarrow0 )=8-3\pi^{2}+\eta^{2}(\pi^{2}-20)+\mathcal{O}(\eta^{4}).
\end{equation}

From obtained  results we also conclude  that   there are no chiral enhanced corrections to the 
amplitudes $B_{++}^{(i,h)}$ at leading order in $\alpha_{s}$.  These amplitudes obtain 
corrections from the twist-3 three-particle DA  which are not considered  in this paper. 

 The end-point singularities  accumulated in the integral $I_{s}$
 originate in the diagrams where both photons couple to the same quark line.     
The integral $I_{s}$ in Eq.(\ref{B3hpm}) is real and  has  the logarithmic  endpoint divergencies. 
For simplicity we do not introduce  any explicit
regularization for $I_{s}$.  The nice feature  is that at leading-order in $\alpha_{s}$  the divergent  integrals does not depend on the  
scattering angle $\theta$.   As we have seen in Sec.~\ref{toy} the endpoint singularities indicate that there is an overlap between the soft and
collinear regions  and in order to develop a consistent description of the power corrections  it is necessary  to include the
appropriate soft-overlap contributions which will be considered  in the next section.

\subsection{The soft-overlap contribution within the SCET framework }
\label{softSCET}

\begin{figure}[ptb]%
\centering
\includegraphics[height=1.3in]{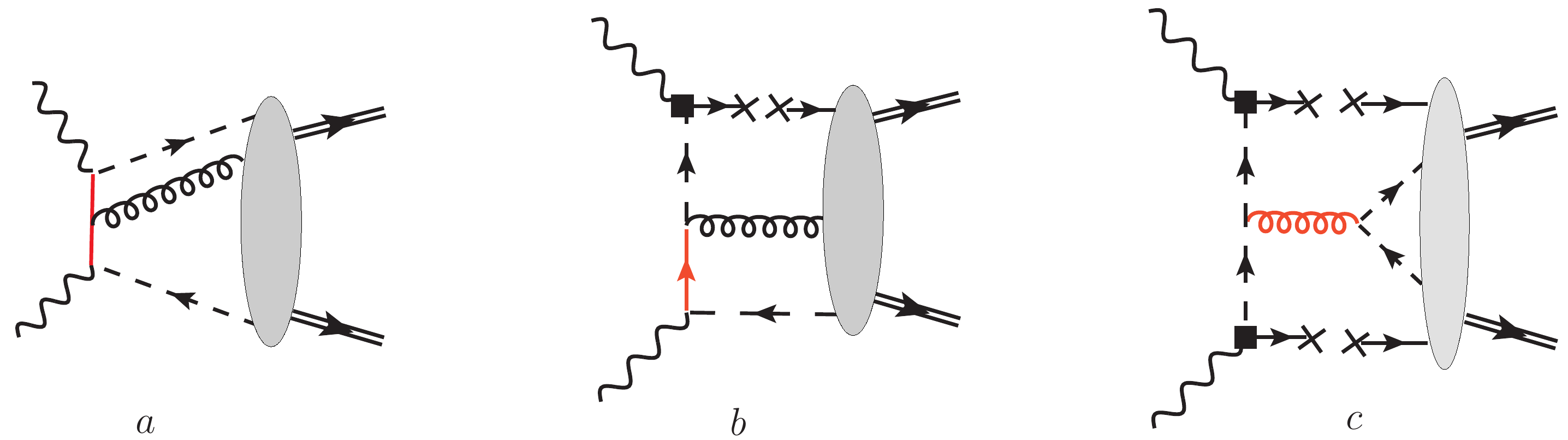}
\caption{The diagrams illustrating the different groups of the SCET operators in Eq.(\ref{TJJ-soft}). The solid red lines  show the hard subdiagrams, the dashed fermion and 
solid gluon lines denote the hard-collinear particles. The solid lines with the crosses show the soft particles.  The black squares denote   vertices  in the effective theory.}
\label{On-groups}%
\end{figure}

We suppose that the complete factorization  is described by the sum of the hard (\ref{TJJ-hard}) and soft contributions
\begin{align}
T\left\{  J_{\text{em}}^{\mu}(0),J_{\text{em}}^{\nu}(x)\right\}=T\left\{  J_{\text{em}}^{\mu}(0),J_{\text{em}}^{\nu}(x)\right\}_{\text{hard}}
+T\left\{  J_{\text{em}}^{\mu}(0),J_{\text{em}}^{\nu}(x)\right\}_{\text{soft}},
\label{TJJ-tot}
\end{align}
The soft-overlap contributions depends on the three scales: hard $\mu_{h}\sim Q^{2}$, hard-collinear $\mu_{hc}\sim \sqrt{\Lambda Q}$,  and soft $\mu_{s}\sim \Lambda$ . 
The factorization in  this case is  performed by integration over  hard and hard-collinear modes.  
Performing the  factorization of the hard modes we reduce $T$-product of the electromagnetic currents  to a set of  SCET-I operators $O^{(k)}\sim \lambda^{k}$ 
constructed from the hard-collinear fields
\begin{align}
T\left\{  J_{\text{em}}^{\mu}(0),J_{\text{em}}^{\nu}(x)\right\}_{\text{soft}}  =\sum_{i}  C_{i}^{\mu\nu}\ast O_{i}^{(k)}.
\label{TJJ-soft}%
\end{align}
 We divide these operators on the three groups  according to their possible structure.   
 The simplest  group  is described by the operators  consist of  hard-collinear fields  from  the $n$- and
$\bar{n}$-collinear sectors. Such configuration can be  illustrated by diagram in Fig.\ref{On-groups}~$(a)$.  
Shrinking the hard lines (hard subgraph) to a ``point''  we obtain the two-jet  operator $\sim \bar \chi^{hc}_{n}{\cal A}^{n}_{\perp hc}\chi^{hc}_{\bar n}$. 
Here we show the hard-collinear gluon just for  illustration, a more detailed analysis is given below.   
Corresponding  contributions can be interpreted as a soft overlap between the outgoing pions.  

 The second group can be associated with the operators which appear if one of the electromagnetic currents on {\it lhs} of  Eq.(\ref{TJJ-soft}) is matched onto 
subleading  SCET  operator with  soft  quark field(s)
 \begin{equation}
 J_{\text{em}}\simeq\bar q  \gamma_{\perp}\chi^{hc}_{v} +{\cal O}(\lambda^{5}).
 \label{Jem-soft}
 \end{equation}
 This possibility is illustrated in Fig.\ref{On-groups}~(b).  Shrinking the hard subgraph we obtain the three-jet hard-collinear operator 
 $\sim \bar{\chi}^{hc}_{v}\gamma \mathcal{A}_{\bot hc}^{{n}}\chi^{hc}_{\bar n}$. 
 These contributions can be interpreted as a soft overlap contributions between photon and outgoing pions.  In the toy integral considered  in Sec.\ref{toy} such contribution can be associated with the soft  $J_{s}$,   see interpretation in  Fig.\ref{soft-graph}.   

The third group SCET operators correspond to the configuration when  the both  electromagnetic currents on {\it lhs} of  Eq.(\ref{TJJ-soft}) are matched onto SCET subleading operators  as in Eq.(\ref{Jem-soft}). Such possibility is illustrated  in Fig.\ref{On-groups}~(c). 
Shrinking the hard subgraph we obtain the four-jet hard-collinear operator $\sim  \bar{\chi}^{hc}_{v}\Gamma\chi^{hc}_{\bar{v}}~\bar{\chi
}^{hc}_{n}\Gamma\chi^{hc}_{\bar{n}}$.  In this case one faces with  the soft overlap configuration 
 between the all external states. 

The main purpose of the following consideration  is   to establish the SCET-I operators which can overlap  with the hard configurations described in Eq.(\ref{TJJ-hard-chi}). 
Then we include corresponding soft-overlap contributions  into the factorization formula in order to have a consistent description of the subleading  corrections.   
In order to establish the  order  and  structure  of the soft-overlap contribution we consider the matching of the  SCET-I operator   $O^{(k)}$ in  Eq.(\ref{TJJ-soft})
onto SCET-II  operators with the appropriate structure. 
Technically this can be done performing the substitution (\ref{sub})  constructing  appropriate 
$T$-products in the intermediate theory with hard-collinear, collinear and soft fields \cite{Bauer:2002nz,Bauer:2003mga,Beneke:2003pa}.

We  start  our analysis  from the  two-jet   operators of the first group which are built from the hard-collinear fields of $n-$ and $\bar n$ sectors. 
 Computation of the relevant  $T$-product gives  the following expression
\begin{equation}
T\left\{  O^{(k)},\mathcal{L}_{\text{int}}^{(l_{1},n)},\mathcal{L}%
_{\text{int}}^{(m_{1},\bar{n})},....\right\}  \simeq O_{n}^{(i)}\ast J_{n}\ast
O_{S}\ast J_{\bar{n}}\ast O_{\bar{n}}^{(j)}, \label{TOk}%
\end{equation}
where the jet-functions $J_{n}$ and $J_{\bar{n}}$ describe the contractions of
the hard-collinear fields. They can be computed from the appropriate SCET
diagrams.  The operators $O_{\bar{n,\bar n}}^{(j)}$ and  $O_{S}$  are built  only  the collinear and soft fields, respectively.  
In what  follow we will call the expression on \textit{rhs} of (\ref{TOk}) by soft-collinear
operator.  If the soft-collinear operator in Eq.(\ref{TOk}) has the same
order  as the collinear operator $O_{n}^{(i)}O_{\bar{n}}^{(j)}\sim \lambda^{i+j}$
describing the hard contribution then we  can conclude that  SCET-I operator  $O^{(k)}$ is relevant for the description
of the soft-collinear overlap and must be included in Eq.(\ref{TJJ-soft}).

Technically  it is convenient to compute the hard-collinear contractions in  each
collinear sector independently considering  soft and collinear fields as
external fields.  Taking into account  that the SCET operator  $O^{(k)}$ from the first group 
can be presented as a simple product of two hard-collinear operators
\begin{equation}
O^{(k)}=O^{(k_{1},n)}O^{(k_{2},\bar{n})}, \label{def:Ok}%
\end{equation}
Therefore we can write 
\begin{equation}
T\left\{  O^{(k)},\mathcal{L}_{\text{int}}^{(l_{1},n)},\mathcal{L}%
_{\text{int}}^{(m_{1},\bar{n})},....\right\}  \simeq T\left\{  O^{(k_{1}%
,n)},\mathcal{L}_{\text{int}}^{(l_{1},n)},....\right\}  T\left\{
O^{(k_{2},\bar{n})},\mathcal{L}_{\text{int}}^{(m_{1},\bar{n})},....\right\}  .
\label{T=T1T2}%
\end{equation}
Computation of  the $T$-products in  each hard-collinear sector  yields
\begin{equation}
T\left\{  O^{(k_{1},n)},\mathcal{L}_{\text{int}}^{(l_{1},n)},....\right\}
\simeq O_{n}^{(i)}\ast J_{n}\ast S, \label{TOk1}%
\end{equation}%
\begin{equation}
T\left\{  O^{(k_{2},\bar{n})},\mathcal{L}_{\text{int}}^{(m_{1},\bar{n}%
)},....\right\}  \simeq\bar{S}\ast J_{\bar{n}}\ast O_{\bar{n}}^{(j)}.
\label{TOk2}%
\end{equation}
where symbols  $S$ and $\bar S$ denote the soft  operators.   
Combining (\ref{TOk1}) and (\ref{TOk2})  we obtain  the  expression in Eq.(\ref{TOk}) with the  soft operator
\begin{equation}
O_{S}=S\bar{S}.
\end{equation}

Combining  $T$-products (\ref{TOk1}) and (\ref{TOk2})  we have impose certain  constraints  on the operators $O^{(k)}$.  
We will exclude the  operators with the odd number of the transverse indices like
\begin{equation}
O_{\bot}^{(k)}[\text{odd}]=\{~\bar{\chi}^{hc}_{n}\gamma_{\bot}^{\mu}\chi^{hc}_{\bar{n}
},~\bar{\chi}^{hc}_{n}~\Dslash{\mathcal{A}}_{\bot hc}^{n}\mathcal{A}_{\bot\mu hc}^{n}
\chi^{hc}_{\bar{n}},~...\}
\end{equation}
 The matrix elements of such operators  vanish  due to the Lorentz invariance
\begin{equation}
\left\langle p,p^{\prime}\right\vert O_{\bot}^{(k)}[\text{odd}]\left\vert
0\right\rangle _{\text{SCET}}=0.
\end{equation}
The  chiral-odd operators $O^{(k)}$  like  $\bar{\chi}^{hc}_{n}\chi^{hc}_{\bar{n}}$,
$\bar{\chi}^{hc}_{n}\gamma_{5}\chi^{hc}_{\bar{n}}$ and so on  can also be neglected because the corresponding  coefficient functions vanish 
in  massless quark limit.

Using that  collinear and soft fields are factorized in  SCET-II  Lagrangian  one finds
\begin{equation}
\left\langle p,p^{\prime}\right\vert O_{n}^{(i)}\ast J_{n}\ast O_{S}\ast
J_{\bar{n}}\ast O_{\bar{n}}^{(j)}\left\vert 0\right\rangle \simeq\left\langle
p^{\prime}\right\vert O_{n}^{(i)}\left\vert 0\right\rangle \ast J_{n}%
\ast\left\langle 0\right\vert O_{S}\left\vert 0\right\rangle \ast J_{\bar{n}%
}\ast\left\langle p\right\vert O_{\bar{n}}^{(j)}\left\vert 0\right\rangle .
\label{soft-fac}%
\end{equation}
This  implies that the soft  operator $O_{S}$  must  have nonvanishing matrix element 
 in Eq.(\ref{soft-fac}).  

In what follows  our task is to find  all  operators Eq.(\ref{TOk}) which provide the overlap 
with the   configurations describing   the hard  contribution in Eq.(\ref{TJJ-hard-chi}).

\subsubsection{The  soft-overlap contribution with   operator  $O_{n}^{(4)}O_{\bar{n}}^{(4)}$}

We start our analysis  from  the  soft-overlap  configuration which is described by the
 collinear operator $O_{n}^{(4)}O_{\bar n}^{(4)}$.  
 In order to obtain the full soft-collinear operator we consider first the $T$-products in the one collinear sector 
\begin{equation}
T\left\{  O^{(k_{1},n)},\mathcal{L}_{\text{int}}^{(l_{1},n)},....\right\}
\simeq O_{n}^{(4)}\ast J_{n}\ast S. \label{O4Os}%
\end{equation}
The simplest possible set of the operators  of order $\lambda$ can be  described  as 
\begin{equation}
O^{(1,n)}=\left\{  \bar{\chi}_{n},\chi_{n}, {\cal A}^{n}_{\bot} \right\},
\label{set1}
\end{equation}
Here and further  we do not write explicitly the label $hc$ for the hard-collinear fields assuming   $\xi^{hc}\equiv \xi$. 
The first and second operators describes the hard-collinear quark  and
antiquark, respectively.  The analysis of the quark and antiquark  configurations are
very similar and   we consider only the quark operator.  The gluon operator can appear only in the next-to-leading in  $\alpha_{s}$ hard coefficient function 
and therefore is subleading. 

It  is not possible to built any $T$-product (\ref{O4Os}) which yields the required soft-collinear operator of
order $\lambda^{4}$ or smaller.  Consider first the case of the quark operator  $\bar{\chi}_{n}$.  
In order  to obtain the two collinear quark fields for $O_{n}^{(4)}$  one needs at least two insertions of
the vertices $\mathcal{L}_{\text{int}}^{(1,n)}$ which include collinear fields $\bar \xi^{c}_{n}$ and  $\xi^{c}_{n}$. Such vertices can be obtained from the leading-order 
Lagrangian  $\mathcal{L}^{(0,n)}$ with the help of the  substitution (\ref{sub}).  In this configuration  only soft quark can appear as a soft field. Hence in order to have 
soft quark  field in \textit{rhs} (\ref{O4Os}) one needs at least one more insertion $\Lin^{(1,n)}$ with the soft quark. 
Therefore one obtains  $T$-products   of order   $\lambda^{4}$.  
Using the SCET Lagrangian we found the  following three possibilities 
\bea
&& T\left\{  \bar{\chi}_{n},\mathcal{L}_{\text{int}}^{(1,n)}[\bar \xi^{c}A_{\bot}A_{\bot} \xi],\mathcal{L}_{\text{int}}^{(1,n)}[\bar \xi A_{\bot}A_{\bot}\xi^{c}],
\mathcal{L}_{\text{int}}^{(1,n)}[ \bar q  A_{\bot} \xi ]\right\}  
\simeq O_{n}^{(4)}\ast J_{n}\ast \bar{q},
\\
&&T\left\{  \bar{\chi}_{n},\mathcal{L}_{\text{int}}^{(1,n)}[\bar \xi^{c}A_{\bot}A_{\bot} \xi],
\mathcal{L}_{\text{int}}^{(2,n)}[\bar q A_{\bot} \xi^{c}]\right\} 
\simeq O_{n}^{(4)}\ast J_{n}\ast \bar{q},
\\
 && T\left\{  \bar{\chi}_{n},\mathcal{L}_{\text{int}}^{(1,n)}[\bar \xi^{c}A_{\bot}A_{\bot} \xi],
\mathcal{L}_{\text{int}}^{(2,n)}[\bar \xi A_{\bot}A^{s}_{\bot} \xi^{c}]\right\}
\simeq O_{n}^{(4)}\ast J_{n}\ast A^{s}_{\bot},
\eea
where in the square brackets $\mathcal{L}_{\text{int}}^{(1,n)}[\dots]$ we show  the field structure of the interaction terms. The
explicit expressions for these SCET interactions  can be found  in Appendix \ref{Lint}. Remind that  field $q$ denotes the soft quark. 
 
 However all these  $T$-products include  the odd number of the transverse hard-collinear 
gluon fields $A_{\bot}$.  In order to contract them one can insert one
more three gluon vertex $\Lin^{(0,n)}[\partial_{\bot}A_{\bot}A_{\bot}A_{\bot}]$. However due to the transverse derivative the obtained  loop integrals
vanish because we do not have external hard-collinear transverse momenta. Hence using the quark (antiquark)  operator $ \bar{\chi}_{n}$ one can not obtain 
the soft-collinear operator (\ref{O4Os}) of order $\lambda^{4}$.

Consider now the gluon operator in Eq.(\ref{set1}). 
In this case  the required  $T$-product can be  described as 
\begin{equation}
T\left\{  \mathcal{A}_{\bot}^{n},\Lin^{(3,n)}\left[  \bar{\xi}_{n}
^{c}~A_{\bot}^{n}A_{\bot}^{s}~\xi_{n}^{c}\right]  \right\}  \simeq O_{n}
^{(4)}\ast J_{n}\ast A_{\bot}^{s}\sim\lambda^{4}.\label{TA-L3}
\end{equation}
However in such contribution  the collinear  operator has isospin zero  while  
we need the operator with isospin one.  Therefore   $T$-product in Eq.(\ref{TA-L3}) 
can be neglected. In this case we did not find any other possibility  to obtain  the soft-collinear operator of order $\lambda^{4}$
with the required  structure. 

Therefore we demonstrated the absence  of the  soft-collinear operator which  can   overlap with  the leading-order collinear operator $O_{\bar n}^{(4)} O_{n}^{(4)}$. 
Therefore the leading order formula (\ref{TJJ-hard-LO}) is valid to all orders in $\alpha_{s}$. 

The hard corrections described in Eq.(\ref{TJJ-hard-LO}) are suppressed  by relative factor  $\lambda^{4}$. 
Is this estimate also  valid for the the soft  corrections in Eq.(\ref{TJJ-soft})?     As we  agreed before we shall 
neglect  the soft-collinear  configurations with  $O_{\bar n}^{(4)} O_{n}^{(4)}$  if they  are also suppressed  by relative factor  $\lambda^{4}$.   
 
In order to  estimate corresponding  power corrections we need to study the higher order  contributions  generated by the 
 soft-collinear operators in Eq.(\ref{O4Os}).  In particular we are interested in   $T$-products in  (\ref{O4Os})   which  can provide the
soft-collinear operator  of order $\lambda^{5}$.  It turns out that such contributions can be easily constructed.  
The simplest possibility can  be  described as
\begin{equation}
T\left\{  \bar{\chi}_{n},\Lin^{(1,n)}\left[  \bar{\xi}^{c}A_{\bot
}A_{\bot}\xi\right]  ,~\Lin^{(3,n)}\left[  \bar{q}A_{\bot}A_{\bot
}\xi^{c}\right]  \right\}  \sim\mathcal{O}(\lambda^{5}),\label{TO1-L13}%
\end{equation}
 Contractions of the gluon fields in (\ref{TO1-L13})  yield the diagram with the hard-collinear loop. One more possibility is given
by the following $T$-product
\begin{equation}
T\left\{  \bar{\chi}_{n},\Lin^{(2,n)}\left[  \bar{\xi}^{c}A_{\bot}A_{\bot}^{s}\xi\right]  ,
~\Lin^{(2,n)}\left[  \bar{q}A_{\bot}\xi^{c}\right]  \right\}  \sim\mathcal{O}(\lambda^{5}),\label{TO1-L22}
\end{equation}
where  the explicit expressions for the Lagrangians $\Lin^{(2,n)}$ are
given in Appendix \ref{Lint}. In this case the soft configuration is more complicate and
includes also the soft gluon field $A_{\bot}^{s}$.

The other  configurations of order $\lambda^{5}$ are related with the higher order operators $O^{(2,n)}\sim \lambda^{2}$
\begin{equation}
O^{(2,n)}=\bar{\chi}_{n}A^{n}_{\bot}. \label{def:O2n}%
\end{equation}
We postpone the discussion of the Dirac and color
structure of the operators $O^{(k,n)}$  until  construction of
the total soft-overlap contribution (\ref{soft-fac}).  The appropriate  $T$-product   
(\ref{O4Os}) which scales as $\lambda^{5}$ are given by
\begin{equation}
T\left\{  \bar{\chi}_{n}  A^{n}_{\bot},\Lin^{(1,n)}\left[  \bar{\xi}^{c}(n\cdot A)\xi \right]  ,\Lin^{(2,n)}\left[  \bar{q}A_{\bot}(\bar
{n}\cdot A)\xi^{c}\right]  \right\}  \sim\mathcal{O}(\lambda^{5}),
\label{TO2-L12}%
\end{equation}
This configuration again describes the digram with  hard-collinear loop.  The  tree
level hard-collinear contribution can be obtained  using the higher order operator
\begin{equation}
O^{(3,n)}=\bar{\chi}_{n}^{c}A^{n}_{\bot}.
\end{equation}
This operator already includes one collinear field  ( it is easy to see  that
$O^{(3,n)}$ is obtained  from $O^{(2,n)}$ using the substitution (\ref{sub})).  In this case 
 the required  $T$-product  read
\begin{equation}
T\{\bar{\chi}_{n}^{c}A^{n}_{\bot},\Lin^{(2,n)}\left[  \bar{q}A_{\bot}
\xi^{c}\right]  ~\}\sim\mathcal{O}(\lambda^{5}), \label{TO3-L2}
\end{equation}
with $\mathcal{L}^{(2,n)}$ shown in Eq.(\ref{L2qAxi}). 
The other higher order operators $O^{(k,n)}$ with $k\geq 3$ provide the $T$-products which have  order $\lambda^{6}$
or higher and therefore we will not consider them now.  For simplicity we also will not consider the $T$-products with the gluon operator $\mathcal{A}_{\bot}^{n}$ 
because it  suppressed by hard $\alpha_{s}$ and therefore can be neglected in our calculation.  

The relevant $T$-products for the $\bar{n}$-collinear sector can be described in the
similar way. Therefore we can construct the total soft-collinear 
operator  following to Eq.(\ref{T=T1T2}).  The  operators $O^{(k_{1},n)}$
and $O^{(k_{2},\bar{n})}$ with $k_{i}\leq3$  can be combined in the operators
$O^{(k)}$ with $k=k_{1}+k_{2}<6$. Building the total operator $O^{(k)}$ in
Eq.(\ref{def:Ok}) we must take into account the  restrictions from the
Lorentz symmetry already  discussed in the previous section.  These operators
are built from  two hard-collinear quark  fields and any number of hard-collinear  gluon fields.  
These are  chiral-even operators which can only have even transverse Lorentz indices.  This allows one to
conclude  that
\begin{equation}
O^{(k)}=\bar{\chi}_{n}\left(  n\cdot\mathcal{A}^{n}\right)  ^{l}\gamma
_{\bot\mu_{1}}~\mathcal{A}^{n}_{\bot\mu_{2}}...\mathcal{A}^{\bar n}_{\bot\mu_{2p}}\left(
\bar{n}\cdot\mathcal{A}^{\bar{n}}\right)  ^{m}\chi_{\bar{n}},
\label{structureOk}%
\end{equation}
where we assume that  all fields in  are hard-collinear. 
Using  $C$-parity  we obtain  the leading order operator
\begin{equation}
O^{(3)}=\bar{\chi}_{n}\left( \Dslash{\mathcal{A}}_{\bot}^{n}+\Dslash{\mathcal{A}}_{\bot}%
^{\bar{n}}\right)  \chi_{\bar{n}}+(n\leftrightarrow\bar{n}), \label{O3}%
\end{equation}%

\begin{figure}[ptb]%
\centering
\includegraphics[height=1.462in]{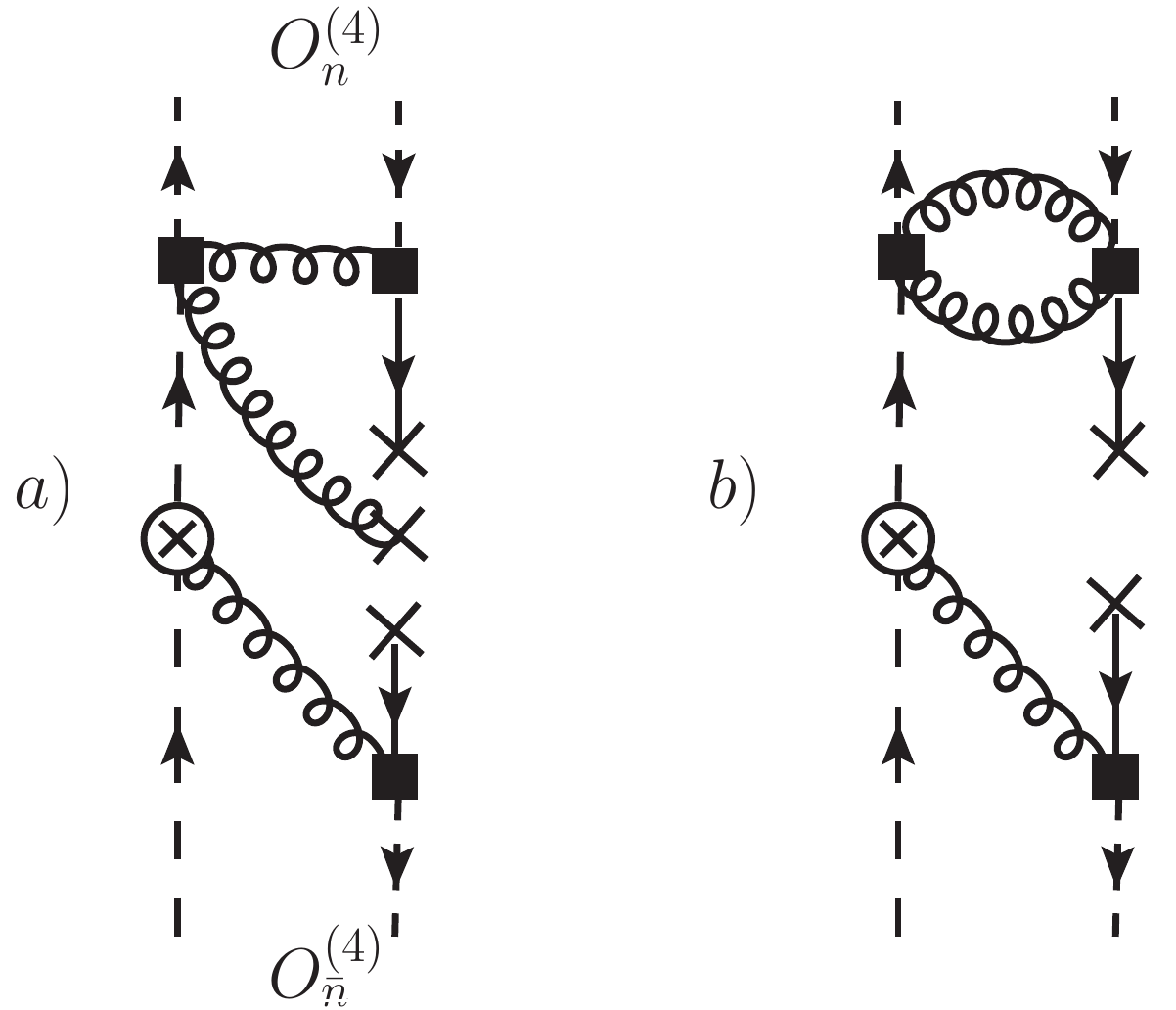}
\caption{The diagrams of order   $\lambda^{10}$  which are provided by  the  $T$-products in Eqs.(\ref{TO3-L22-L2}) and (\ref{TO3-L13-L2}), respectively. }
\label{t-o4}%
\end{figure}

Combining the  $T$-products of order $\lambda^{5}$  described in Eqs.(\ref{TO1-L13}),(\ref{TO1-L22}) and (\ref{TO3-L2})  we obtain the
soft-collinear operators  of order $\lambda^{10}$ which therefore have relative  suppression of order $\lambda^{2}$.  
 All these contributions are described  by the operator $O^{(3)}$ or higher order operators obtained from $O^{(3)}$
 with the help of substitution (\ref{sub}).  There are only   two
$T$-products  which provide the soft-collinear operator of order $\lambda^{10}$.  
Corresponding  contributions  can be illustrated by  diagrams shown in
Fig.\ref{t-o4}.
The diagram $(a)$ is described by the combination  of 
$T$-product as in (\ref{TO1-L22}) and  (\ref{TO3-L2})
\begin{equation}
T\left\{  \bar{\chi}_{n}\Dslash{\mathcal{A}}_{\bot}^{\bar{n}}\chi_{\bar{n}}
^{c},~\Lin^{(2,n)}\left[  \bar{\xi}^{c}A_{\bot}A_{\bot}^{s}\xi\right]  ,~\Lin^{(2,n)}\left[  \bar{q}A_{\bot}\xi^{c}\right]  ,\Lin^{(2,\bar{n})}\left[  \bar{\xi}^{c}A_{\bot}q\right]  \right\}  .
\label{TO3-L22-L2}
\end{equation}
 The diagram $(b)$ is provided by the combination of $T$-products
as in (\ref{TO1-L13}) and  (\ref{TO3-L2})
\begin{equation}
T\left\{  \bar{\chi}_{n}\Dslash{\mathcal{A}}_{\bot}^{\bar{n}}\chi_{\bar{n}}^{c},
~\Lin^{(1,n)}\left[  \bar{\xi}^{c}A_{\bot}A_{\bot}\xi\right]  ,
~\Lin^{(3,n)}\left[  \bar{q}~A_{\bot}\xi^{c}\right],\Lin^{(2,\bar{n})}\left[  \bar{\xi}^{c}A_{\bot}q\right]  \right\}
\label{TO3-L13-L2}
\end{equation}

This is interesting observation which suggest that there are $1/Q$  corrections associated with the soft contributions.   
However as we have already seen in Sec.\ref{toy}  such  contributions can only provide  formal expressions for spurious integrals. 
Therefore in order to make correct conclusion one has to study the SCET diagrams and show that  corresponding
soft integrals are not power divergent.  For that purpose we need to know  their transformation properties under 
longitudinal boost transformations (\ref{boost}).   

Consider the diagram in Fig.\ref{t-o4}~$(a)$.  The soft convolution integral reads 
\begin{align}
J_{n}\ast S\ast J_{\bar{n}}  & \sim\int\frac{d k_{2}^{-}}{  k_{2}^{-}  }\int\frac{dk_{1}^{+}}{ k_{1}^{+} }\frac
{dk_{2}^{+}}{  k_{2}^{+}  }S(k^{+}_{1},k^{+}_{2}, k^{-}_{2}),
\label{JSJa}
\end{align}
where  the soft correlation function (CF)  is defined as\footnote{ Here and further in the text   we do not show the soft Wilson lines for simplicity. } 
\begin{equation}
S(k^{+}_{1},k^{+}_{2}, k^{-}_{2})=\text{FT}
~\left\langle 0\right\vert \bar{q}(\lambda_{1}n)
\Dslash{A}_{\bot}^{s}(\lambda_{2}n) q(\eta_{1}\bar{n})\left\vert 0\right\rangle .
\end{equation}
with the Fourier transformation 
\begin{equation}
\text{FT}\equiv \int d(x\cdot \bar n)~e^{ik^{+}_{1}(x\cdot \bar n)}
\int d(y\cdot \bar n)~e^{i(k^{+}_{1}-k^{+}_{2})(y\cdot \bar n)  }
\int d(x\cdot n) e^{-ik^{-}_{2}(x\cdot n)}.
\end{equation}
From this definition we see that under boost transformations
\begin{equation}
S(k^{+}_{1},k^{+}_{2}, k^{-}_{2})\rightarrow \alpha^{-1}~S(k^{+}_{1},k^{+}_{2}, k^{-}_{2}).
\end{equation}
and therefore the soft integral in Eq.(\ref{JSJa}) is not invariant under boosts. Hence this integral is power divergent and therefore vanishes as the spurious contribution.
 We also checked this conclusion considering the soft limit of the appropriate two-loop QCD diagram.  
 A similar analysis for the diagram in Fig.\ref{t-o4}~$(b)$ also yields that this diagram is associated with the power divergent 
 soft integral and  therefore  can be neglected.

\subsubsection{The  soft-overlap contribution  with  operators  $O_{n}^{(6)}O_{\bar{n}}^{(i)}$}

On order to build the soft-overlap contributions with the twist-3 operators $O_{n}^{(6)}$ we need to 
 consider   $T$-products with the following  structure
\begin{equation}
T\left\{  O^{(k_{1},n)},\mathcal{L}_{\text{int}}^{(l_{1},n)},....\right\}
\simeq O_{n}^{(6)}\ast J_{n}\ast S. \label{TOk->O6n}%
\end{equation}
 We found that there is only  one  such $T$-product  of order  $\lambda^{5}$.
 It can be obtained with the operator $O^{(1,n)}=\bar{\chi}_{n}$ and reads
\begin{equation}
T\left\{  \bar{\chi}_{n},\Lin^{(2,n)}\left[  \bar{\xi}^{c}A_{\bot
}^{c}A_{\bot}\xi\right]  ,\Lin^{(2,n)}\left[  \bar{q}A_{\bot}%
~\xi^{c}\right]  \right\}  \sim\mathcal{O}(\lambda^{5}),
\label{TO1-L22-O6n}%
\end{equation}
where the vertices  $\mathcal{L}^{(2,n)}$  are described  in Eqs.(\ref{L2xiAAxi}) and (\ref{L2qAxi}). 

The other $T$-products  (\ref{TOk->O6n}) scale as $\lambda^{7}$. 
 However combining the contributions  of order $\lambda^{7}$ and  $\lambda^{5}$ 
 we obtain the soft-collinear operators  of order  $\lambda^{12}$ which are suppressed  exactly as the hard subleading contribution
 with $O_{n}^{(6)}O_{\bar{n}}^{(6)}$.   Hence  the $T$-products  of order $\lambda^{7}$ must be also considered. 

Using various operator  $O^{(k,n)}$ we find  the following expressions  of order $\lambda^{7}$
\begin{equation}
T\left\{  \bar{\chi}_{n},\Lin^{(2,n)}\left[  \bar{\xi}^{c}A_{\bot
}^{c}A_{\bot}\xi\right]  ,\Lin^{(4,n)}\left[  \bar{q}A_{\bot}
^{s}A_{\bot}\xi^{c}\right]  \right\},
\label{TO1-L24}
\end{equation}
\begin{equation}
T\left\{  \bar{\chi}_{n}\mathcal{A}_{\bot},
\Lin^{(2,n)}\left[  \bar{q}A_{\bot}\xi^{c}\right]  ,
\Lin^{(3,n)}\left[  \bar{\xi}^{c}A_{\bot}^{c}A_{\bot}^{s}\xi\right]  \right\}, \label{TO2-L23}
\end{equation}
\begin{equation}
T\left\{  \bar{\chi}_{n}^{c}\mathcal{A}_{\bot},
\Lin^{(4,n)}\left[\bar{\xi}^{c}A_{\bot}^{c}A_{\bot}q\right]  \right\},   \label{TO3-L4}
\end{equation}
\begin{equation}
T\left\{  \bar{\chi}_{n}^{c}\mathcal{A}_{\bot}^{c}\mathcal{A}_{\bot
},\Lin^{(2,n)}\left[  \bar{q}A_{\bot}\xi_{n}^{c}\right]  \right\}. \label{TO5-L2}
\end{equation}

\begin{figure}[ptb]%
\centering
\includegraphics[
height=1.7534in]{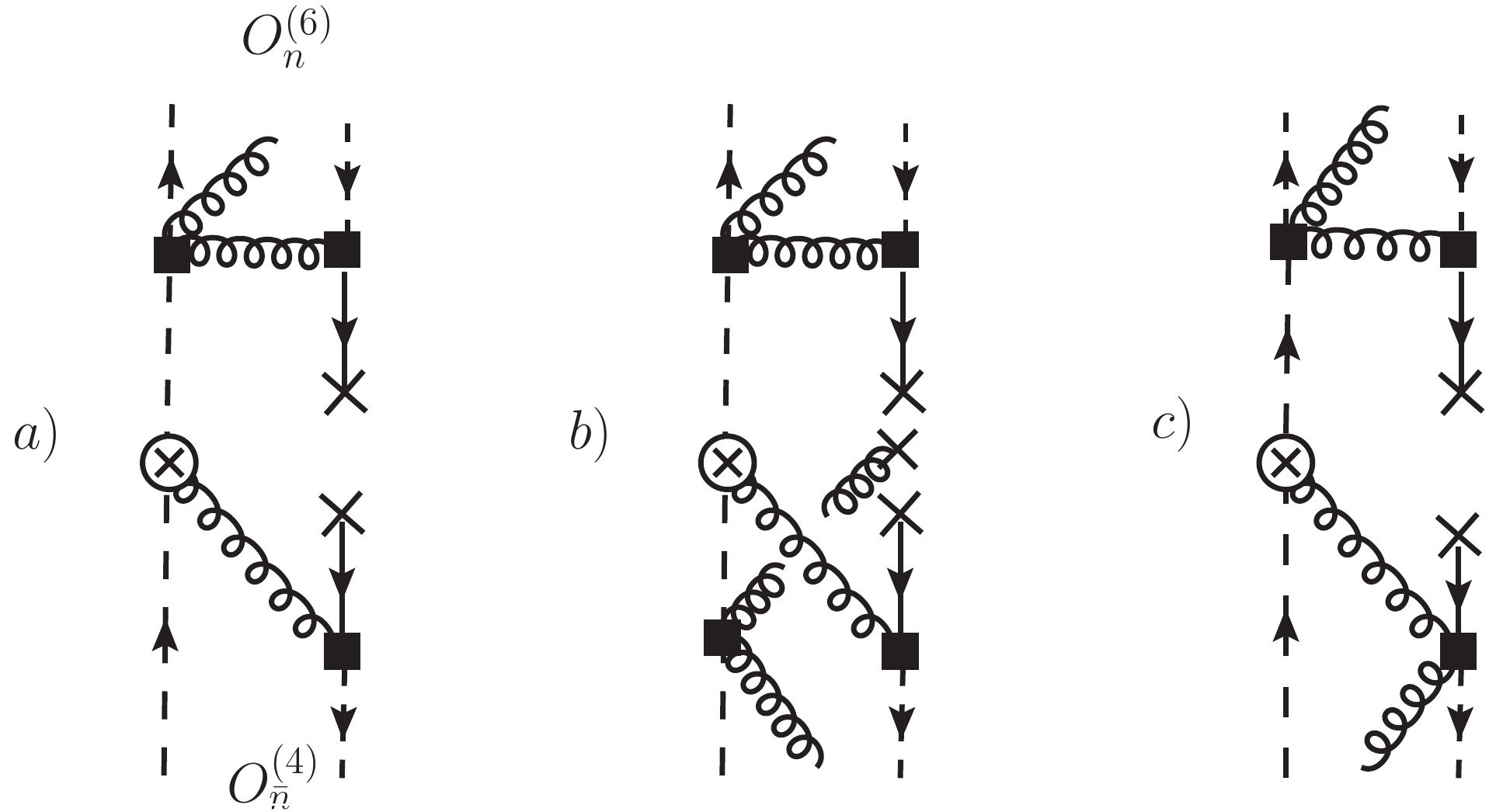}
\caption{The diagrams describing the $T$-products  with the
collinear operators of $O_{n}^{(6)}O_{\bar{n}}^{(i)}$.   }
\label{t-o4o6}%
\end{figure}

One can obtain  the soft-collinear operators of order $\lambda^{10}$   
combining  the  $T$-products (\ref{O4Os}) and (\ref{TOk->O6n}).  
Corresponding soft-collinear operators can not overlap with the hard
configurations because the collinear operators $O_{n}^{(4)}$ and $O_{\bar{n}
}^{(6)}$ have  different chiral structure.  Hence these contributions can
appear only due to the soft-overlap mechanism.  There is only  one possibility to obtain the such contribution at order $\lambda^{10}$  using the combination of
$T$-products in Eq.(\ref{TO3-L2}) and Eq.(\ref{TO1-L22-O6n}) 
\begin{equation}
T\left\{  \bar{\chi}_{n}\Dslash{\mathcal{A}}_{\bot}^{\bar{n}}\chi_{\bar{n}}%
^{c},\Lin^{(2,\bar{n})}\left[  \bar{\xi}^{c}A_{\bot}q\right]
,\Lin^{(2,n)}\left[  \bar{\xi}^{c}A_{\bot}^{c}A_{\bot}\xi \right]  ,\Lin^{(2,n)}\left[  \bar{q}A_{\bot}\xi^{c}\right]
\right\}  \sim\mathcal{O}(\lambda^{10}). \label{TOk4-O4O6}%
\end{equation}
This configuration can be described by the diagram in Fig.\ref{t-o4o6}$(a)$.
The  operator $O^{(k)}=\bar{\chi}_{n}\Dslash{\mathcal{A}}_{\bot}^{\bar{n}}
\chi_{\bar{n}}^{c}$ is  associated with the SCET-I operator $O^{(3)}$   in Eq.(\ref{O3}). 
We again obtain a correction which is suppressed by factor $\lambda^{2}$ comparing to the leading-order contribution.  
The  soft convolution integral reads 
\begin{equation}
J_{n}\ast S \ast J_{\bar{n}}\sim\int\frac{dk_{+}}{k^{2}_{+}}\int\frac{dk_{-}}{k_{-}}S(k_{+},k_{-}),
\label{JSJ}
\end{equation}
where the soft CF  $S(k_{+}, k_{-})$ is defined as 
\begin{equation}
S(k_{+},k_{-})=\int d(\bar n\cdot x)e^{ik_{+}(\bar n\cdot x)}\int d( n\cdot y)e^{-ik_{-}( n\cdot y)}
\left\langle 0\right\vert \bar{q}(\lambda n) q(\eta\bar{n})\left\vert 0\right\rangle. \label{FTSn}
\end{equation}
The soft operator in Eq.(\ref{FTSn}) is chiral-odd and color singlet.  One can easily see that corresponding soft CF is boost invariant and therefore can be written as 
$S(k_{+},k_{-})=S(k_{+}k_{-})$.Hence the corresponding soft integral in Eq.(\ref{JSJ}) is not  invariant under boosts. 
One can easily see that this integral is quite similar to the soft integral $J_{1s}$ in Eq.(\ref{J1s}). 
 Hence the  $T$-product in Eq.(\ref{TOk4-O4O6}) describes the spurious integral and can be neglected.  
 One can expect that the soft contributions with the collinear operator $O_{n}^{(6)}O_{\bar{n}}^{(4)}$ can be obtained from at higher orders   $\sim \lambda^{12}$.
 However  we neglect  such contributions  and therefore we do not consider such operators.

The most important soft-overlap contributions  of order $\lambda^{12}$ are described by the
 soft-collinear operators with   $O_{n}^{(6)}O_{\bar{n}}^{(6)}$  collinear operator.
Corresponding contributions  can be constructed from the $T$-products of $\lambda^{5}$ in
(\ref{TO1-L22-O6n}) and $\lambda^{7}$  described in Eqs.(\ref{TO2-L23}) and
(\ref{TO3-L4})%
\bea
T\left\{ 
 \bar{\chi}_{n}\Dslash{\mathcal{A}}_{\bot}^{\bar{n}}\chi_{\bar{n}}, \right.   &&
\Lin^{(2,\bar{n})}\left[  \bar{\xi}^{c}A_{\bot}q\right],
\Lin^{(3,\bar{n})}\left[  \bar{\xi}A_{\bot}^{c}A_{\bot}^{s}\xi^{c}\right] ,
    \nonumber \\  &&  \left.
\Lin^{(2,n)}\left[  \bar{\xi}^{c}A_{\bot}^{c}A_{\bot}\xi \right] ,
\Lin^{(2,n)}\left[  \bar{q}A_{\bot}\xi^{c}\right]
\right\}\sim\mathcal{O}(\lambda^{12}),
 \label{TO3-O6O6}
\eea
\begin{equation}
T\left\{  \bar{\chi}_{n}\Dslash{\mathcal{A}}_{\bot}^{\bar{n}}\chi_{\bar{n}}^{c},
~\Lin^{(4,\bar{n})}\left[  \bar{\xi}^{c}A_{\bot}^{c}A_{\bot}q\right] ,
\Lin^{(2,n)}\left[  \bar{\xi}^{c}A_{\bot}^{c}A_{\bot}\xi\right]  ,
\Lin^{(2,n)}\left[  \bar{q}A_{\bot}\xi^{c}\right] 
 \right\}
 \sim\mathcal{O}(\lambda^{12}).
  \label{TO4-O6O6}
\end{equation}
Appropriate diagrams are shown in Fig.\ref{t-o4o6} $(b)$ and $(c)$,
respectively. Both diagrams describe the suitable soft convolution integrals which have only the logarithmic singularities.  

We can not find  other contributions  with the more complicate operators $O^{(k)}$  which can  provide the
soft-overlap operator as in Eq.(\ref{structureOk}) at order $\lambda^{12}$.   Let us also observe  that  all  operators $O^{(k)}$ in
Eqs.(\ref{TO3-O6O6}) and (\ref{TO4-O6O6}) are related to the leading SCET-I   operator $O^{(3)}$  in Eq.(\ref{O3}). Therefore  corresponding 
soft-overlap  contribution  is described by the matrix element of the SCET operator $O^{(3)}$ in  Eq.(\ref{O3}).
This matrix element also describes  the  soft-collinear operators associated with  other possible collinear 
configurations like  $O_{n}^{(4)}O_{\bar{n}}^{(4)}$,  $O_{n}^{(4)}O_{\bar{n}}^{(6)}$, $O_{n}^{(8)}O_{\bar{n}}^{(4)}$ 
which we  do not consider in this paper for simplicity.

\subsubsection{The  soft-overlap contributions  with  photon states  }

Some examples of the soft-overlap contributions with  photons have been provided earlier, see Fig.\ref{On-groups}.
The specific feature of these configurations is that they are described by the SCET-I operators associated with more 
that two light-like directions (three and four-jet operators).  As a result  the analysis of such contributions is more complicate 
then the corresponding analysis of the two-jet operators carried in the previous sections. 
 
In general, real photon has the nonperturbative component of the wave function and therefore
one can define the matrix element which defines the photon DA. The leading
twist DA is defined by the chiral-odd operator Ref.\cite{Balitsky:1997wi}
\begin{equation}
~O_{v}^{(4)}=\bar{\chi}_{v}^{c}~\Dslash{\bar v} \gamma_{T}\chi_{v}^{c}\,\text{ }
\label{O4g}
\end{equation}
where, remind,  the auxiliary  light-like vectors are $v$ and $\bar{v}$ are defined in Eq.(\ref{vbv}). The 
$\gamma_{T}$ denotes the suitable transverse projection.\footnote{This definition implies a choice  $v$ and $\bar v$ as the basic light-like vectors.}    
The hard subleading contribution with the nonperturbative  photon can be associated with the following collinear operator
\begin{equation}
 O_{v}^{(4)} O_{\bar{n}}^{(6)}O_{n}^{(4)}\sim\lambda^{14},
 \label{O464}
\end{equation}
and the similar  operators obtained by appropriate permutation of the collinear indices $v, \bar v$ and $n, \bar n$.  
In order to obtain the nontrivial matrix element we need at least
two operators with the chiral-even Dirac structure and therefore we need the subleading  operator $O_{\bar{n}}^{(6)}$ in Eq.(\ref{O464}). 
As a result this contribution suppressed  as 
${\cal O}(\lambda^{14})$ and therefore can be neglected.    If we use  the  twist-3 chiral-odd operator  for description of the  photon DA
\begin{equation}
 O_{v}^{(6)} =\bar{\chi}_{v}^{c}~\Dslash{\bar v}{\cal A}^{v}_{\bot \alpha}\chi_{v}^{c} \sim\lambda^{6}.
\end{equation}
we also obtain the contribution of order $\lambda^{14}$. 
Therefore  the contribution of order $\lambda^{12}$    can  only  be obtained  from the  soft-overlap contributions as suggested in Fig.\ref{On-groups}~$(b,c)$.

Appropriate  SCET  operators include the hard-collinear fields associated with the photon and pion
momenta. The lowest order operators read
\begin{equation}
O_{\gamma}^{(3)}=\left\{  \bar{\chi}_{n}\gamma_{\alpha}\mathcal{A}_{\beta\bot
}^{\bar{n}}\chi_{v},\bar{\chi}_{n}\gamma_{\alpha}\mathcal{A}_{\beta\bot}%
^{\bar{n}}\chi_{\bar{v}},~\left\{  n\leftrightarrow\bar{n}\right\}
,...\right\}  , \label{O3g}%
\end{equation}%
\begin{equation}
O_{\gamma\gamma}^{(4)}=\left\{  \bar{\chi}_{v}\Gamma\chi_{\bar{v}}~\bar{\chi
}_{n}\Gamma\chi_{\bar{n}},...\right\}  . \label{O4gg}%
\end{equation}
where dots denote the other  suitable  combinations with similar field
structure, the matrix $\Gamma$ denotes  appropriate Dirac and color structures.
 From Fig.\ref{On-groups}~$(b,c)$  one can observe that only the
operators $O_{\gamma}^{(3)}$ have  hard coefficient functions at  leading order in $\alpha_{s}$.

Suppose that we consider a configuration where one of
the photons interacts with the soft quark, like in Fig.\ref{On-groups}~$(b)$. 
 Assume that  the  coupling of the collinear  photon to the hard-collinear and soft quarks can be  described by the
leading order SCET Lagrangian 
\begin{equation}
\Lin^{(2,v)}[\bar \xi B_{c} q]=ee_{q}\int d^{4}x~\bar{\chi}_{v}\Dslash{B}  _{c}^{(v)}q,
\label{L2em}%
\end{equation}
where $B_{\mu c}^{(v)}$ describes the collinear photon field.   Inserting  such contribution to the SCET matrix element we obtain
\begin{align}
&  \left\langle p,p^{\prime}\right\vert T\left\{  O_{\gamma}^{(3)},\dots,\Lin^{(l_{i},n)}, \Lin^{(2,v)}[\bar \xi B_{c} q]\right\}  \left\vert q_{2}\right\rangle
\nonumber\\
&  \simeq ee_{q}\varepsilon_{\nu}(q_{2})~\int d^{4}x~e^{-i(q_{2}
x)}\left\langle p^{\prime},p\right\vert T\left\{  O_{\gamma}^{(3)},\dots,\Lin^{(l_{i},n)},\bar{\chi}_{v}\gamma_{T}^{\nu}q\right\}  \left\vert 0\right\rangle ,
\label{<pp|Okg}%
\end{align}
where dots denote the other  SCET interactions. The external
collinear photon state yields the factor $\lambda^{-2}$. Then inserting
interaction (\ref{L2em}) which is of order $\lambda^{2}$ we compensate this
factor. Hence in order to estimate  the relative
order of  the such contribution one needs to  estimate of the
remaining set of the $T$-product in Eq.(\ref{<pp|Okg})
\begin{equation}
\dim\left\{  O_{\gamma}^{(k)},\dots,\Lin^{(l_{i},n)}\right\}  \sim\lambda^{3+...+l_{i}}.
\label{estOkg}%
\end{equation}
This allows one   to compare this configuration with  the two-jet contributions
discussed in the previous sections.

Each  operator $O_{\gamma}^{(3)}$ from the set in Eq.(\ref{O3g})  consists of  the three different hard-collinear
fields.   As a result  the  SCET matrix element   $\langle p,p'\vert O_{\gamma}^{(3)}\vert \gamma\rangle$  defines an  amplitude
depending on the energy $s$ and scattering angle $\theta$.  As a result in this case the hard factorization does not allow one to get any restrictions on
  the  structure of the amplitude.

The further analysis is the same as before: we  need to find  suitable  $T$-products (\ref{TOk})  giving the 
 required soft-collinear operators of order $\lambda^{12}$ or smaller.     In order to be specific we consider  
 $O_{\gamma}^{(3)}= \bar{\chi}_{n}\gamma_{\alpha}\mathcal{A}_{\beta\bot}^{\bar{n}}\chi_{v}$.   
 The construction of the soft-collinear operator can be done in the same way as before combining the contributions from each
 hard-collinear sector.  
In order to describe the electromagnetic interaction we use the subleading
SCET  vertex (\ref{L2em}).  This yields 
\begin{equation}
T\{\chi_{v},\bar{\chi}_{v}\gamma^{\nu}q \}\simeq J^{\nu}_{v}\ast q. 
\label{Jq}
\end{equation}
The appropriate  $T$-products with the hard-collinear operator $O^{1,n}= \bar{\chi}_{n}$ were already discussed before, see Eqs.(\ref{TO1-L13}),(\ref{TO1-L22}) and (\ref{TO1-L22-O6n}).
These terms provide the contributions  of order  $\lambda^{5}$
\begin{equation}
T\{ \bar{\chi}_{n},\dots \}\simeq O_{n}^{(4,6)}*J_{n}\ast \bar{q} \sim \lambda^{5}. 
\end{equation}

\begin{figure}[ptb]
\centering
\includegraphics[height=1.9352in]{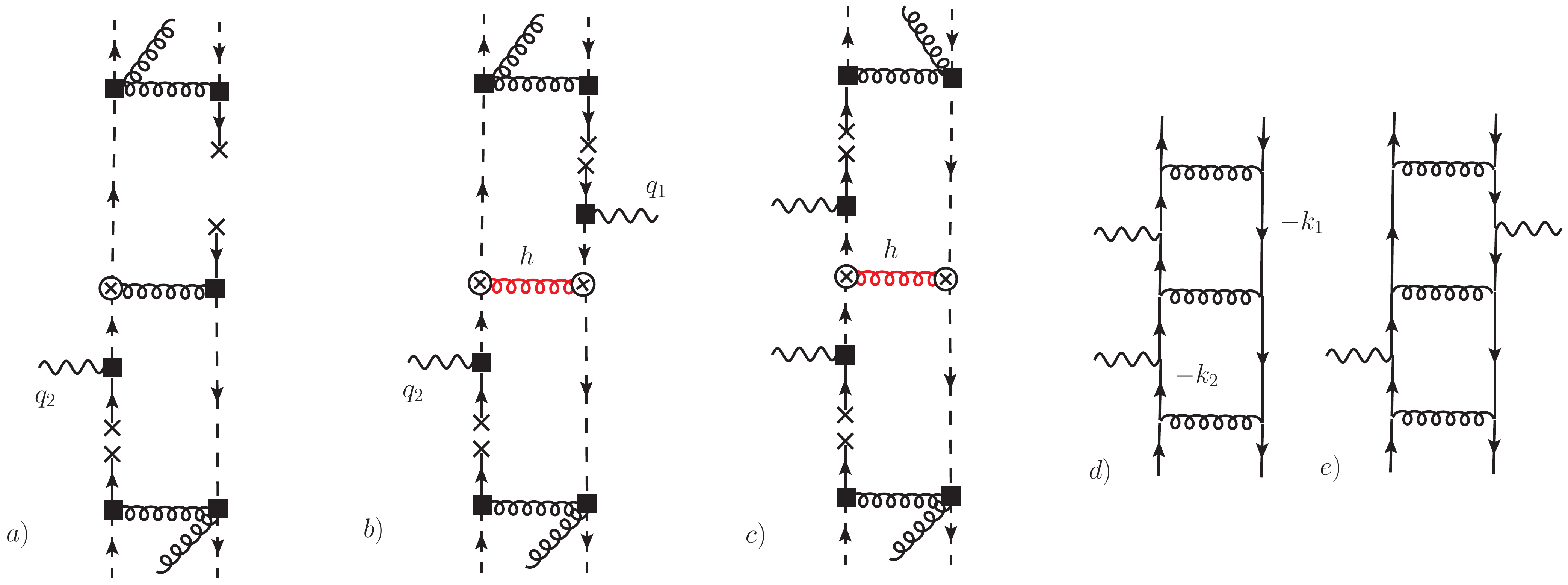}
\caption{The diagrams illustrating the $T$-products of the operators  
$O_{\gamma}^{(3)}$ and $O_{\gamma\gamma}^{(4)}$.  The four-fermion vertices of the operator  $O_{\gamma\gamma}^{(4)}$ in diagrams $(b)$ and $(c)$ are shown by two crossed circles connected by red
gluon line  with index  ``h''.  }
\label{og-diagrams}
\end{figure}

The only new element now  is the $T$-product with the gluon field which must have the following structure
\begin{equation}
T\left\{  \mathcal{A}_{\beta\bot}^{\bar{n}}\right\}  \simeq\bar{S}\ast
J_{\bar{n}}\ast O_{\bar{n}}^{(i)}.
\label{TA}
\end{equation}
We  find the following lowest order possibilities 
\begin{equation}
T\left\{  \mathcal{A}_{\beta\bot}^{\bar{n}},\Lin^{(2,\bar{n})}\left[
\bar{q}A_{\bot}\xi^{c}\right]  ,\Lin^{(3,\bar{n})}\left[  \bar{\xi}%
^{c}A_{\bot}A_{\bot}q\right]  ~\right\}  \simeq [\bar{q}q]\ast J_{\bar{n}}\ast
O_{\bar{n}}^{(4)}\sim\lambda^{6},
\end{equation}%
\begin{equation}
T\left\{  \mathcal{A}_{\beta\bot}^{\bar{n}},\Lin^{(1,\bar{n})}\left[
\bar{\xi}A_{\bot}q\right]  ,\Lin^{(2,\bar{n})}\left[  \bar{\xi}
^{c}A_{\bot}^{c}A_{\bot}\xi\right]  ,\Lin^{(2,\bar{n})}\left[  \bar
{q}A_{\bot}\xi^{c}\right]  ~\right\}  \simeq [\bar{q}q]\ast J_{\bar{n}}\ast
O_{\bar{n}}^{(6)}\sim\lambda^{6}.
\label{TA-O6}
\end{equation}
Combining  together the contributions in Eqs.(\ref{Jq})-(\ref{TA}) we obtain only two possible soft-collinear operators 
of order  $\lambda^{12}$ as described  in Eq.(\ref{estOkg}). 
These  operators  contain $O_{n}^{(4)}O_{\bar{n}}^{(4)}$ and $O_{n}^{(6)}O_{\bar{n}}^{(6)}$.  We now focus  on the term with  the twist-3 collinear  
operators $O_{n,\bar n}^{(6)}$.    Corresponding $T$-product is combined from Eqs. (\ref{TO1-L22-O6n}) and (\ref{TA-O6}) that gives
\begin{equation}
\dim\left\{  O_{\gamma}^{(3)},\Lin^{(1,\bar{n})},\Lin^{(2,\bar{n})},\Lin^{(2,\bar{n})},\Lin^{(2,n)},\Lin^{(2,n)}\right\}
\sim\lambda^{12}.
\label{To3g-b}
\end{equation}
 The full $T$-product  can  be illustrated by the diagram  in Fig.\ref{og-diagrams}~$(a)$.
Therefore from this result we  conclude that this contribution may also be relevant for description of the endpoint singularities  in hard term (\ref{TJJ-hard-chi}).

The analysis of the operator $O_{\gamma\gamma}^{(4)}$ in Eq.(\ref{O4gg}) is
quite similar.    Using the same arguments as before one can write%
\begin{align}
&\left\langle p,p^{\prime}\right\vert T\left\{  O_{\gamma\gamma}^{(k)}
, \dots,\Lin^{(l_{i},n)}, \Lin^{(2,v)}[\bar \xi B_{c} q],\Lin^{(2,\bar{v})}[\bar \xi B_{c} q]\right\}
\left\vert q_{1}, q_{2}\right\rangle
 \nonumber\\
&  \simeq\varepsilon_{\mu}(q_{1})\varepsilon_{\nu}(q_{2})~\left\langle
p,p^{\prime}\right\vert O_{n}^{(i)}\ast J_{n}\ast J_{v}^{\nu}\ast O_{S}\ast
J_{\bar{v}}^{\mu}\ast J_{\bar{n}}\ast O_{\bar{n}}^{(j)}\left\vert
0\right\rangle , \label{TO4gg}%
\end{align}
where jet functions $J_{i}$ describe the hard-collinear interactions
associated with the different light-cone vectors. Then the relative order of the corresponding 
contribution  is defined by the order of the 
\begin{equation}
\dim\left\{   O_{\gamma\gamma}^{(4)},\dots,\Lin^{(l_{i},n)}\right\}  \sim\lambda^{4+...+l_{i}}.
\label{estO4g}
\end{equation}
The construction of the required $T$-products  follows the same line as before.  We use  expression  (\ref{TO1-L22-O6n}) in order
to  convert the hard-collinear quark field to the required soft-collinear combination.  
Combining the  known $T$-products  we obtain the contributions of order $\lambda^{12}$
\begin{equation}
\dim\left\{  O_{\gamma\gamma}^{(4)},\Lin^{(2,\bar{n})},\Lin^{(2,\bar{n})},\Lin^{(2,n)},\Lin^{(2,n)}\right\}
\sim\lambda^{12}.
\label{To4gg}
\end{equation}
 Then the total expression in Eq.(\ref{TO4gg})  can be described by diagrams shown in Fig.\ref{og-diagrams}~$(b,c)$.
These diagrams  describe  the configurations  when colliding photons interact with the same or different  quarks.  
Hence, at least formally,   we again obtain the contributions which  can overlap with the hard configuration in Eq.(\ref{TJJ-hard-chi}) and therefore 
can be relevant for description of the endpoint region. 

However there are some observations which  indicate that these   $T$-product  can  describe the spurious integrals similar to  $J_{1s}$  in the toy model.   
Indeed, the configurations as in  Fig.\ref{og-diagrams}~$(b)$ implies that the hard diagram with photons attached to the different spinor lines can also produce the singular
endpoint contributions. However from the calculation in Sec.\ref{hard}  we obtain the different result: such diagrams provide the regular contributions described by factors  $I(\eta)$,  see Eq.(\ref{defI}).     Next, the structure of the hard-collinear diagrams in Fig.\ref{og-diagrams}~$(a,b,c)$  is quite similar to   diagram in Fig.\ref{t-o4o6}~$(a)$ which  vanishes.  In addition, the angular  dependence of the photon soft-overlap contributions is not  fixed  by the hard subdiagram. 
The soft convolution integral also depends on $\theta$ and this can potentially provide a more complicate function of $\theta$ than one  in front of the singular integral $I_{s}$ in Eq.(\ref{B0hpm}).  Therefore let us study these diagrams in detail. 

 The common feature of the photon soft-overlap contributions is related with the four-quark soft CFs. For instance for the diagram in Fig.\ref{og-diagrams}~$(a)$ the soft  CF is given by the matrix element  (remind that we do not show explicitly  the soft Wilson lines)
\begin{equation}
\left\langle 0\right\vert \bar{q}(\lambda_{1}n)\Gamma_{1}q(\eta_{1}\bar{n})~\bar{q}(\eta_{2}\bar{n})\Gamma_{2}q(\sigma v)\left\vert 0\right\rangle,
\label{4qsoft}
\end{equation}
where matrices $\Gamma_{1}\otimes\Gamma_{2}$ describe Dirac and color structure. It is clear that this  matrix element describes a complicate  soft  CF.  
Our consideration can be simplified if we consider the large-$N_{c}$ limit \cite{'tHooft:1973jz,  Witten:1979kh}  for the  diagrams in Fig.\ref{og-diagrams}. In this limit  the  soft gluon exchanges between the soft quark field in the different loops (or between the different soft quark ``propagators'' )  give always  non-planar  diagrams which are suppressed by $1/N_{c}$  comparing to planar diagrams according to large-$N_{c}$ counting rules \cite{'tHooft:1973jz, Witten:1979kh}.  Therefore  in the large-$N_{c}$ picture  the soft matrix element in Eq.(\ref{4qsoft}) simplifies and  can be described  as  product of the two-quark soft CFs. In  case of diagram in Fig.\ref{og-diagrams}~$(a)$  this  gives
 \begin{equation}
\left\langle 0\right\vert \bar{q}(\lambda_{1}n)\Gamma_{1}q(\eta_{1}\bar{n})~\bar{q}(\eta_{2}\bar{n})\Gamma_{2}q(\sigma v)\left\vert 0\right\rangle
\approx 
\left\langle 0 \right\vert \bar{q}(\lambda_{1}n)q(\eta_{1}\bar{n}) \left\vert 0\right\rangle  
\left\langle 0\right\vert \bar{q}(\eta_{2}\bar{n})q(\sigma v)\left\vert 0\right\rangle,
\label{2x2qsoft}
\end{equation}
where each two-quark CF  is defined as in Eq.(\ref{FTSn}).  In this case the corresponding soft convolution integral  reads
 \begin{eqnarray}
J_{n}\ast S\ast J_{\bar{n}}\ast J_{v}\sim 
\int_{0}^{\infty}\frac{dk_{2}^{-}}{\left[  -k_{2}^{-}\right]  ^{2}}
\int_{0}^{\infty}\frac{d(k_{2}\cdot v)}{[-(k_{2}\cdot v)]}
~S(k_{2}^{-}(k_{2}v))
\nonumber
 \\  \times 
\int_{0}^{\infty}\frac{dk_{1}^{+}}{\left[-k_{1}^{+}\right]  ^{2}}
\int_{0}^{\infty}\frac{dk_{1}^{-}}{\left[  k_{1}^{-}-k_{2}^{-}\right]  }
~S(k_{1}^{+}k_{1}^{-}),
\label{softSS}
\end{eqnarray}
where momenta $k_{1}$ and $k_{2}$ can be associated with the momenta of the soft quarks in the loops of diagram in Fig.\ref{og-diagrams}~$(a)$.  
Each soft CF  in Eq.(\ref{softSS}) is invariant under longitudinal boosts and therefore depends only on the products of the appropriate light-cone fractions. Notice that the factorization of the soft  CF (\ref{2x2qsoft}) allows one to consider the longitudinal integrals as independent  and  to compute  them  using  different basis of the light-cone vectors in each loop. In such situation  one can consider     different types of the longitudinal boosts in each sector.  In order to analyze these integrals let us introduce the two auxiliary regulators $\tau_{+}$ and $\tau_{-}$ which transforms as plus and minus components under longitudinal boosts.      
Consider now the integrals over $k_{1}^{\pm}$ in Eq.(\ref{softSS}).  
\begin{align}
\int_{0}^{\infty}\frac{dk_{1}^{+}}{\left[  -k_{1}^{+}-\tau_{+}\right]  ^{2}}\int_{0}^{\infty}
\frac{dk_{1}^{-}}{k_{1}^{-}-k_{2}^{-}}S(k_{1}^{+}k_{1}^{-})  &
\label{tau+}\\
 =\int_{0}^{\infty}~dk_{1}^{-}S(k_{1}^{-})~\int_{0}^{\infty}\frac{dk_{1}^{+}
}{\left[  -k_{1}^{+}-\tau_{+}\right]  ^{2}}\frac{1}{\left[  k_{1}^{-}
-k_{2}^{-}k_{1}^{+}\right]  }  \sim a\tau_{+}^{-1}+b~k_{2}^{-},
\end{align}
where $a$ and $b$ are some constants and $k^{+}_{i}=(n\cdot k_{i})$, $k^{-}_{i}=(\bar n\cdot k_{i})$. 
We observe that the integral is power divergent providing  $\sim \tau_{+}^{-1}$.   
 Computing     two  remaining integrals in Eq.(\ref{softSS})  we obtain
\begin{align}
& \int_{0}^{\infty}\frac{dk_{2}^{-}}{\left[  k_{2}^{-}+\tau_{-}\right]  ^{2}}\int_{0}^{\infty}
\frac{d(k_{2}\cdot v)}{(k_{2}\cdot v)}~S(k_{2}^{-}(k_{2}\cdot v))\left[  a\tau_{+}^{-1}
+~b~k_{2}^{-}\right] \label{tau-} \\
& =\int_{0}^{\infty}\frac{d(k_{2}\cdot v)}{(k_{2}\cdot v)}~S((k_{2}\cdot v))\int_{0}^{\infty} dk_{2}^{-}\frac{a\tau
_{+}^{-1}+~b~k_{2}^{-}}{\left[  k_{2}^{-}+\tau_{-}\right]  ^{2}}\sim\frac
{a}{\tau_{+}\tau_{-}}+~...~.
\label{1/tau}
\end{align} 
Hence we  obtain  that  the soft convolution integral obtained  from the diagram in  Fig.\ref{og-diagrams}~$(a)$    is power divergent and therefore vanishes. 
 The similar  consideration  allows one to obtain the same  results   for the two diagrams in  Fig.\ref{og-diagrams}~$(b,c)$. 
This allows us to conclude that the $T$-products   associated with diagrams in Fig.\ref{og-diagrams}~$(a,b,c)$  describe the spurious contributions at 
leading-order of $1/N_{c}$ expansion.   

Consider now the  hard contribution   $B^{0,h}_{+-}$  which has integral with the endpoint singularities.  
 At large-$N_{c}$ limit  it can be easily estimated using the  Eq.(\ref{B0hpm})  
\begin{align*}
B^{0,h}_{+-}\sim f^{2}_{\pi}\mu^{2}_{\pi}\,  \alpha_{s} C_{F}/N_{c}\sim f^{2}_{\pi}\mu^{2}_{\pi}/N_{c}. 
\end{align*} 
 The fictitious contributions of  diagrams in  Fig.\ref{og-diagrams}~$(a,b,c)$ at large-$N_{c}$ limit  are of the same order. 
 Hence  suppressed  by $1/N_{c}$  configurations which were neglected in Eq.(\ref{softSS})   are irrelevant
  for the matching  of the endpoint singularities of the amplitude  $B^{0,h}_{+-}$ in Eq.(\ref{B0hpm} ).   The  subleading in $1/N_{c}$ configurations might 
  be important for descriptions of the endpoint  singularities which only appear at the next-to-leading order in  $\alpha_{s}$   in  the hard amplitudes $B^{i,h}_{+\pm}$, 
i.e.  suppressed by  $1/N_{c}$ contributions  can be associated with the subleading  logarithms.   In our analysis we will not consider  such contributions. 
 Let us also remind that   our  conclusion is only  valid for the chiral enhanced contributions.

The obtained  conclusions   can be verified  by  investigating  the appropriate two-loop QCD diagrams  shown in Fig.\ref{og-diagrams}~$(d,e)$ and using the expansion 
by momentum regions.  Within  this technique one can see  that the soft regions of these diagrams yield the SCET diagrams in Fig.\ref{og-diagrams}~$(a,b,c)$.  
The two-quark soft CFs in this case are described by  integrals from the propagators of the soft quarks with mass $m$ which plays the role of the soft scale. 

In order  to be concrete let us consider  the diagram $D_{d}$ in  Fig.\ref{og-diagrams}~$(d)$  in such  configuration when  the soft limit  corresponds to the SCET diagram in Fig.\ref{og-diagrams}~$(a)$.  
Expansion of the corresponding  QCD expression in  the  soft limit  $k_{1,2}\sim m$  yields  
\begin{align}
D_{d}^{s,s} &  \sim 
\int dk_{2}\frac{m}{\left[
k_{2}^{2}-m^{2}\right]  \left[ - (v \cdot k_{2}) \right]  
\left[    - k^{-}_{2}  \right]  ^{2}}
\int dk_{1}\frac{m}{\left[  k_{1}^{2}-m^{2}\right]
\left[  -k^{+}_{1} \right]  ^{2}  \left[    k^{-}_{1} -   k^{-}_{2}\right]  }
\label{Dss}
 \end{align}
Notice that  contribution of each soft quark propagator is  given by chiral-odd term that  provides the mass factor $m$ in the numerator.  This expression reproduces the  structure of the integral  in  Eq.(\ref{softSS}). Computing the soft  integral (\ref{Dss}) with the regulators $\tau_{\pm}$ as in Eqs.(\ref{tau+}) and (\ref{tau-})  we confirm the qualitative result 
obtained in Eq.(\ref{1/tau}).

In order to see the  overlap of the collinear and soft regions one can also consider  the contributions from  the collinear  regions.  
The collinear contributions with 
$ k_{1}\sim p',\, k_{2}\sim p $  can be interpreted as the convolution of the tree level hard kernel with the one-loop collinear matrix elements  for which   we assume the appropriate twist-3 projections.  Taking the soft limit for  collinear contributions  and comparing with the soft contribution in Eq.(\ref{Dss}) we can  see a  possible overlap with the soft region.  Following this line  we find  that  the soft limit of the collinear contributions  does not match exactly  the soft expression  for these two loop diagrams.
 The soft limit of the collinear contribution   yields
\begin{align}
D_{d}^{n/s,\bar{n}/s}   \sim &
\int  dk_{1}\frac{m}{\left[  k_{1}^{2}-m^{2}\right]  \left[  - k^{+}_{1}\right]^{2}\left[  k^{-}_{1} \right]  } 
 \int dk_{2}\frac{m}{\left[  k_{2}^{2}-m^{2}\right]  \left[  - k^{-}_{2} \right]  ^{2}\left[  - (v\cdot k_{2})
\right]  },
\label{Dnsnbs}
\end{align}
We see that  the  soft integrals in Eq.(\ref{Dnsnbs})  are completely factorized. 
One immediately see that each soft integral is similar to the spurious integral $J_{1s}$ in Eq.(\ref{J1s}).  
Therefore   $D_{d}^{n/s,\bar{n}/s} =0$ and  we do not have an overlap between the collinear and soft  regions.   
In  cases when only one of the momenta $k_{1}$ or $k_{2}$ is taken to be soft one obtains  $D_{d}^{n,\bar{n}/s} =D_{d}^{n/s,\bar{n}} =0$  up to power suppressed
contributions. 

 The analogous results are also  valid for the second diagram   Fig.\ref{og-diagrams}~$(e)$  and  for the configurations  associated 
 with the diagrams in  Fig.\ref{og-diagrams}~$(a,b,c)$.   Therefore  we confirm  that in the perturbation theory the formal SCET  $T$-products associated with the diagrams 
 in Fig.\ref{og-diagrams}~$(a,b,c)$  also describe the  inessential  power divergent integrals.

One more  possibility to obtain the  photon soft-overlap contribution is  provided by  the operators of order $\lambda^{8}$
\begin{equation}
O_{\gamma}^{(8)}=\bar{\chi}_{v}\Gamma\chi_{n}~O_{\bar{n}}^{(6)}, \dots,
\label{O8g}
\end{equation}
 where $\Gamma$ denotes a chiral-odd  Dirac matrix $\Gamma=\gamma_{5},\sigma\gamma_{5}$ and dots denote 
 similar operators with the different collinear labels.  
 \begin{figure}[h]
\centering
\includegraphics[ height=1.4753in,width=1.1606in]{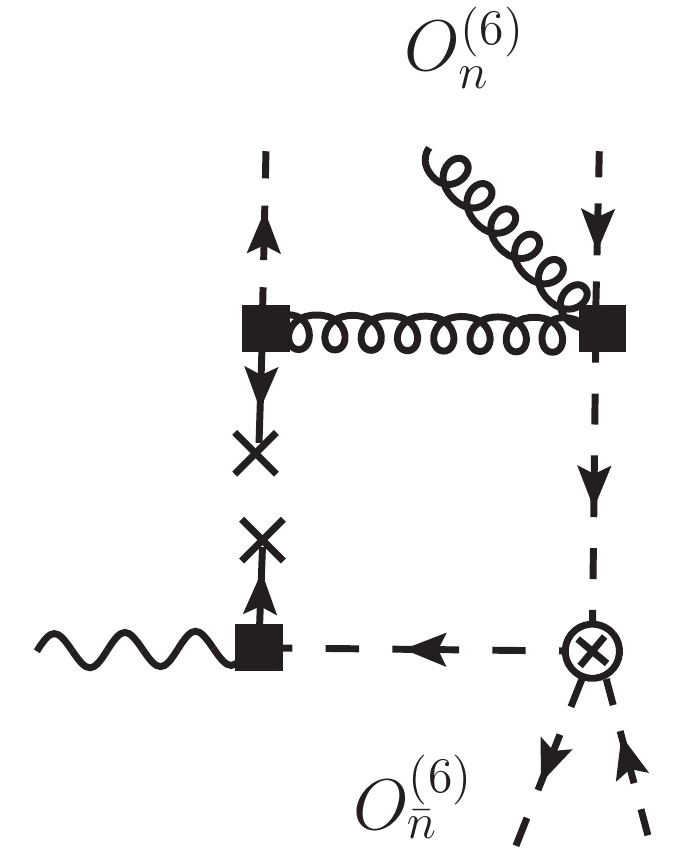}
\caption{The SCET  diagram describing the photon soft-overlap
contribution in Eq(\ref{TO8g}). The crossed circle denotes the hard-collinear
SCET operator $O_{\gamma}^{(8)}$. }
\label{o8g-o6n}
\end{figure}
The  appropriate  contribution   can be  obtained    
 combining  the $T$-product of  order $\lambda^{5}$ for the  hard-collinear field $\chi_{n}$, see Eq.(\ref{TO1-L22-O6n}), 
 with  the photon interaction vertex $\Lin^{(2,v)}$
\begin{align}
\left\langle p,p^{\prime}\right\vert O_{\gamma}^{(6)}\left\vert q_{2}
\right\rangle  &  =\left\langle p,p^{\prime}\right\vert O_{\bar{n}}
^{(4)}T\left\{  \bar{\chi}_{v}\gamma_{\sigma}\chi_{n},\Lin^{(2,v)}[\bar q B_{\perp} \xi_{v}],...\right\}  \left\vert q_{2}\right\rangle \nonumber \\
&  =    \left\langle p^{\prime}\right\vert O_{\bar{n}}^{(6)} \left\vert 0\right\rangle \ast J_{n}\ast O_{S}\ast J_{v}\ast \left\langle p\right\vert O_{n}^{(6)}\left\vert 0\right\rangle
 \sim \lambda^{8},
 \label{TO8g}
\end{align}
In Fig.\ref{o8g-o6n} we show the SCET diagram
generated by  the $T$-product in Eq.(\ref{TO8g}).  The  SCET vertices $\Lin^{(2,v)}$ and $\Lin^{(2,n)}$ describing the interaction with the soft quark  fields   restrict
 the  structure of the soft  operator in the matrix element.  This can be only chiral-odd  operator   $\sim \langle0| \bar q q|0 \rangle$.  Then the  hard-collinear diagram in Fig.\ref{o8g-o6n} has  three chiral-odd vertices:  hard-collinear vertex $\bar{\chi}_{v}\Gamma\chi_{n}$,  collinear  operator $O_{n}^{(6)}$  and the soft operator.   As a result this diagram generates a trace with odd number of the $\gamma$-matrices  and  this contribution vanishes.

\subsection{Summary of the SCET analysis}

Let us briefly summarize the obtained results.   We 
confirm  the  structure of  leading power contribution. It  is only  described by the hard
contribution  associated with the leading-twist collinear operator $O_{n}^{(4)}O_{\bar{n}}^{(4)}
\sim\mathcal{O}(\lambda^{8})$   defined in Eq.(\ref{defO4}). 

The hard power suppressed contributions are described by suitable  collinear operators of order $\lambda^{12}$. 
 There are  many appropriate collinear  operators at this order.   In order to simplify our consideration   
 we take into account   only the specific chiral enhanced contributions $O_{\chi n}^{(6)}O_{\chi  \bar n}^{(6)}$ associated with the
twist-3 DA of pion, see Eq.(\ref{defO6}) and discussion in Sec.~\ref{hard}.  The specific feature of this contribution is that corresponding DA  is  known 
exactly in QCD  and  it is numerically enhanced comparing to  other higher-twist corrections.  
However the corresponding hard coefficient function has the endpoint singularities, see Eq.(\ref{B0hpm}),  that can be explained by 
the overlap of  collinear and soft regions.   We expect  that 
the endpoint  singularities  must cancel in the sum of hard and a suitable soft-overlap contribution.     
In SCET-II such soft-overlap contribution is described by a  soft-collinear operator of order  $\lambda^{12}$   
which is constructed  from the same collinear operators  $O_{\chi n}^{(6)}O_{\chi  \bar n}^{(6)}$  and soft fields. 
 
After hard factorization   the soft-overlap contribution is  described by a set of  SCET-I  operators.  
We demonstrated  that there is only one SCET-I operator which   provides the soft-collinear 
operator with  required properties  in  SCET-II.  Corresponding  SCET-I operator $O^{(3)}$ is  given in Eq.(\ref{O3})  and can be 
associated with the  soft-overlap contribution between the outgoing pions.  
We also obtain  that  more complicate soft-overlap configurations with  pion  
and photon states  are power suppressed at least to a leading logarithmic accuracy.  
This is enough in order  to obtain a consistent description  in our case.  
   
These results  allows us  to write the  following  relatively simple formula 
\begin{align}
T\left\{  J^{\mu}(x),J^{\nu}(0)\right\}   &  \simeq
 H^{\mu\nu}\ast O_{n}^{(4)}O_{\bar{n}}^{(4)}
 \nonumber \\
&+[ T^{\mu\nu}\ast  O_{\chi n}^{(6)}O_{\chi  \bar n}^{(6)}]_{\text{reg}}
+  C_{3}^{\mu\nu}\ast [O^{(3)}]_{\text{reg}},
\label{TJJ-res}
\end{align}
where  brackets  $[...]_{\text{reg}}$   symbolically  denote a specific  regularization and subtractions scheme  which allow one  to  separate
 the collinear and soft modes.  More detailed discussion of this  point will be presented below.   Let us  note  that  the  hard convolution  integral 
 in the third term  on {\it rhs} of Eq.(\ref{TJJ-res}) is well defined  and  the corresponding convolution integral 
 does  not depend on the  specific regularization  and therefore    the  coefficient function   $C_{3}^{\mu\nu}$ is shown outside  brackets  $[...]_{\text{reg}}$.

\section{ Calculation of the amplitude in the  physical subtraction scheme  }

\label{calcofsub}

In this section we  compute the hard coefficient function $C_{3}^{\mu\nu}$ which appears in  the soft-overlap contribution  in
Eq.(\ref{TJJ-res}).  Using this result we define the  physical subtraction scheme  which allows one  to separate   unambiguously
the regularized contributions in Eq.(\ref{TJJ-res}) and to define the hard subleading in $\lambda$ contribution  without  the endpoint singularities. 
Within this framework  we obtain the well defined expressions for the physical amplitudes  which can be used for a phenomenological analysis. 

\subsection{The leading-order hard coefficient functions of the soft contribution}

Let us clarify the arguments of  fields in the required  SCET operator $O^{(3)}$ (\ref{O3}).
 We assume that the fields are multipole expanded
in the position space.  We define this  operator as
\begin{equation}
O^{(3)}(\lambda)=\bar{\chi}_{n}(0)\left(  \mathcal{\ \setbox0=\hbox{$A $}
\dimen0=\wd0 \setbox1=\hbox{/} \dimen1=\wd1 \ifdim\dimen0>\dimen1
\rlap{\hbox to \dimen0{\hfil/\hfil}} A \else \rlap{\hbox to
\dimen1{\hfil$A $\hfil}} / \fi }_{\bot}^{(n)}(\lambda\bar{n}%
)+\mathcal{\ \setbox0=\hbox{$A $} \dimen0=\wd0 \setbox1=\hbox{/} \dimen1=\wd1
\ifdim\dimen0>\dimen1 \rlap{\hbox to
\dimen0{\hfil/\hfil}} A \else \rlap{\hbox to \dimen1{\hfil$A
$\hfil}} / \fi }_{\bot }^{(\bar{n})}(\lambda n)\right)  \chi_{\bar{n}}(0),
\end{equation}
where hard-collinear fields  $\bar{\chi}_{n}$ and $\mathcal{\ A }_{\bot }$ are defined
in Eqs.(\ref{Chi}) and (\ref{AC}). This operator depends on the
relative light-cone distance $\lambda$ between the hard-collinear quark and
gluon. Performing the Fourier transformation with respect to $\lambda$ one
introduces the conjugate variable $\tau$ which can be interpreted as a
fraction of the total hard-collinear momentum carried by gluon. It is
convenient to introduce the following momentum space representation
\begin{equation}
\mathcal{A}_{\mu}^{(n)}(\tau)=\int\frac{d\lambda}{2\pi}~P_{+}^{\prime
}~e^{-i\lambda P_{+}^{\prime}\tau}\mathcal{A}_{\mu }^{(n)}(\lambda\bar{n}),
\end{equation}
where $P^{\prime}$ denote the momentum operator in the $n$-collinear sector.
The similar expression holds also for the $\mathcal{A}_{\mu }^{(\bar{n})}$.
Using this compact notation one can define the matrix element directly in the
momentum space. Hence we define the SCET-I operator as
\begin{equation}
O^{(3)}(\tau)=\bar{\chi}_{n}(0)\left(  
\Dslash{{ \cal A}}_{\bot}^{(n)} (\tau)
+\Dslash{{ \cal A}}
_{\bot }^{(\bar{n})}(\tau)\right)  \chi_{\bar{n}}(0).
\end{equation}
The parametrization of the corresponding SCET-I matrix element can be defined  as
\begin{equation}
\left\langle \pi^{a}(p),\pi^{b}(p^{\prime})\left\vert O^{(3)}(\tau)\right\vert
0\right\rangle _{\text{{\tiny SCET-I}}}=\delta^{ab}~(4\pi f_{\pi} )^{2}~f_{\pi\pi}(\tau,s),
 \label{fpi}
\end{equation}
where, just for convenience,  we used the dimensional factor$~(4\pi f_{\pi})^{2}$,  in this section we also do not write explicitly the  regularization 
symbol  $[...]_{\text{reg}}$   as in  Eq.(\ref{TJJ-res}). 
The dimensionless SCET amplitude  $f_{\pi\pi}(\tau,s)$ depends from  the collinear
fraction $\tau$ and the total energy  $s$ and  from the factorization scale which is not shown for simplicity.\footnote{Let us emphasize  in order to avoid misunderstanding  that  here we assume the factorization scale  associated with  the factorization of hard modes while the additional regularization denoted as $[...]_{\text{reg}}$  is introduced for a separation of the collinear and soft  modes. }
The evolution of the operators like $O^{3}(\tau)$ has been studied in  Refs.\cite{Chay:2010hq, Kivel:2010ns}.

 In Eq.(\ref{fpi}) we assume that the
operator $O^{3}(\tau)$ is the singlet in the flavor space
\begin{equation}
O_{B}(\tau)\sim\bar{u}u+\bar{d}d.
\end{equation}
The operator with isospin $I=1$ cannot contribute in  this case  because of  $C$-parity. 
This allows one to conclude that such soft contribution is  relevant only for the isoscalar amplitudes $B_{+\pm}^{(0)}$. 
Therefore the  factorization of the soft contributions can be written as
\begin{equation}
B_{+\pm}^{(0,s)}(s,\theta)=(4\pi f_{\pi})^{2}\int_{0}^{1}d\tau~C_{+\pm}
^{(0)}(s,\theta,\tau)~f_{\pi\pi}(\tau,s),\label{B0s}%
\end{equation}
\begin{equation}
B_{+\pm}^{(3,s)}(s,\theta)=\mathcal{O}(\alpha_{s}).
\label{B3s}
\end{equation}
We see that the angular dependence  of the soft amplitudes is defined by the hard subprocess and therefore can be computed in perturbation theory. 
 
\begin{figure}[ptb]
\centering
\includegraphics[height=1.0909in,width=4.0257in]{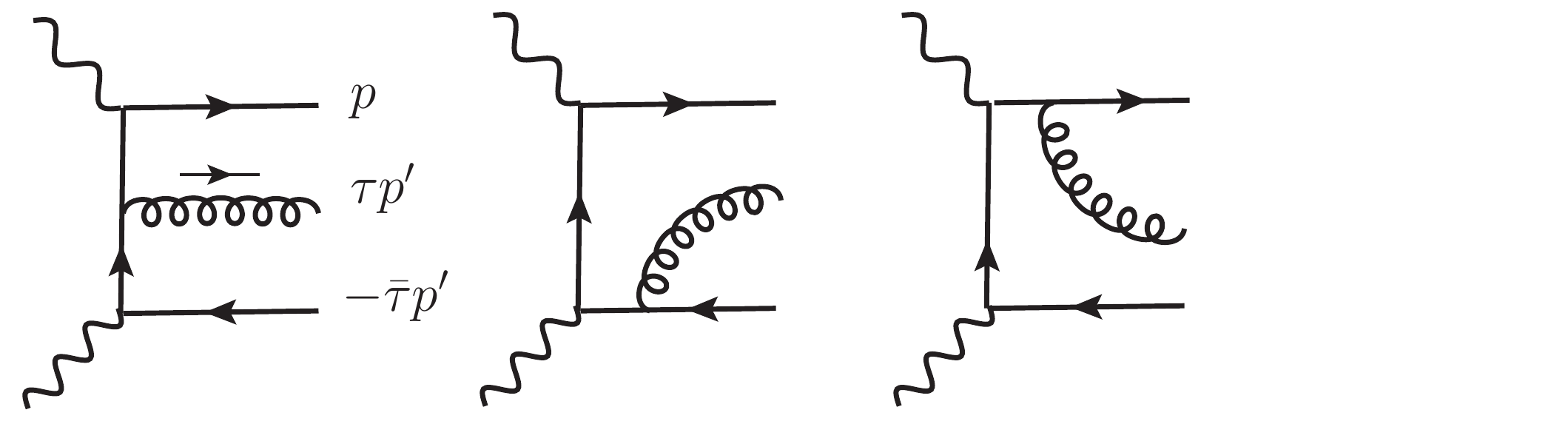}
\caption{The
set of  diagrams required for the matching of T-product of the
electromagnetic currents onto SCET-I operator $O^{(3)}(\tau)$.  We only  show the diagrams for the
case $\bar{\chi}_{n} \Dslash{{\cal A}}_{\bot}^{(\bar{n})}\chi_{\bar{n}}$. The crossed
diagrams are not shown for simplicity.  The diagrams describing the other
configurations are similar,  we also used $\bar{\tau}\equiv1-\tau$. }
\label{pi-pi-soft-lo}
\end{figure}

In order to obtain  tree level expressions for the  coefficient functions $C_{+\pm}^{(0)}$ 
one has to compute the  diagrams shown in Fig.\ref{pi-pi-soft-lo}.
The computation is quite standard therefore let us provide the resulting
expressions  for the soft amplitudes
\begin{equation}
B_{++}^{(0,s)}(s,\theta)=~-(e_{u}^{2}+e_{d}^{2})\frac{(4\pi f_{\pi})^{2}}
{s}\frac{4}{1-\eta^{2}}\int_{0}^{1}d\tau~\frac{\tau}{1-\tau}~f_{\pi\pi}
(\tau,s),\label{AppDc}%
\end{equation}
\begin{equation}
B_{+-}^{(0,s)}(s,\theta)=~-2(e_{u}^{2}+e_{d}^{2})~\frac{(4\pi f_{\pi})^{2}}
{s}\frac{3-\eta^{2}}{1-\eta^{2}}\int_{0}^{1}d\tau~~f_{\pi\pi}(\tau
,s),\label{ApmDc}%
\end{equation}
where we again used notation $\eta=\cos\theta$.
The  both helicity amplitudes $B_{+\pm}^{(0,s)}$ are defined by
the same SCET amplitude  $f_{\pi\pi}$ but through the different  convolution integrals with respect to $\tau$.
 We also observe that  the angular behavior (associated with the variable $\eta$)  in Eqs.(\ref{AppDc}) and (\ref{ApmDc}) is different. 
 For definiteness let us  assume that  the renormalization scale in the amplitude $f_{\pi\pi}$ is fixed  to be large $\mu_{F}\simeq s$. 
 
The  obtained expressions in Eqs.(\ref{AppDc}) and (\ref{ApmDc})  demonstrate the one important property:  
the angular behavior of the soft amplitudes 
is defined by the simple factors $1/(1-\eta^{2})$  and $(3-\eta^{2})/(1-\eta^{2})$  which are factorized from the 
convolution integrals over $\tau$.  This allows one to define the  following two SCET amplitudes as
\begin{equation}
\Phi_{++}(s)=\int_{0}^{1}d\tau~\frac{\tau}{1-\tau}~f_{\pi\pi}(\tau,s),\quad 
\Phi_{+-}(s)=\int_{0}^{1}d\tau~f_{\pi\pi}(\tau,s).
\label{def:Phi}
\end{equation}
The factorization of the angular dependence from the convolution integrals in  Eqs.(\ref{AppDc}) and (\ref{ApmDc})  provides a very important  check  of the 
suggested formalism.   In this case  the  endpoint singularities in the soft term (remind that they are regularized by the special regularization  denoted by  
$[...]_{\text{reg}}$)  аре  and provided by the  amplitudes   $\Phi_{+\pm}(s)$ which does not depend on the scattering angle $\theta$.  
 On the other hand the compensation  of  the endpoint divergencies  between the hard and soft contributions in  Eq.(\ref{TJJ-res})   requires a strong correlation of the angular
 dependence  in these contributions.  This correlation can be used in order  to define a specific subtraction  procedure of the endpoint singularities from the hard subleading term  $[ T^{\mu\nu}\ast  O_{\chi n}^{(6)}O_{\chi  \bar n}^{(6)}]_{\text{reg}}$ in Eq.(\ref{TJJ-res}) and therefore to  obtain  well defined expressions for the  physical amplitudes.

\subsection{ Subleading amplitude in the physical subtraction scheme}

Combining the results for the hard subleading amplitudes  (\ref{Bpp0,3h})-(\ref{B3hpm})  and for 
the soft contributions  (\ref{AppDc})-(\ref{def:Phi})  
we  obtain the following expressions
\begin{align}
B_{+-}^{(0)}(s,\theta)  &  \simeq-2(e_{u}^{2}+e_{d}^{2})~\frac{(4\pi f_{\pi
})^{2}}{s}\frac{3-\eta^{2}}{1-\eta^{2}}
[\Phi_{+-}(s)]_{\text{reg}}\nonumber\\
&  +\frac{\alpha_{s}}{4\pi}\frac{C_{F}}{N_{c}}\frac{\left(  4\pi f_{\pi}\right)  ^{2}}{s}\frac{\mu_{\pi}^{2}}{s}
\left[ \left\{  
(e_{u}^{2}+e_{d}^{2})\frac{(3-\eta^{2})}{(1-\eta^{2})}~I_{s}+\frac{2e_{u}e_{d}}{(1-\eta^{2}
)}~I(\eta)
\right\} \right]_{\text{reg}}
\label{Bpmdiv}
\end{align}
\begin{align}
B_{++}^{(0)}(s,\theta) & \simeq -(e_{u}^{2}+e_{d}%
^{2})\frac{(4\pi f_{\pi})^{2}}{s}\frac{4}{1-\eta^{2}}[\Phi_{++}(s)]_{\text{reg}}
+\left[   \langle p,p'| O_{n}O_{\bar n}\ast T_{++}|0\rangle \right]_{\text{reg}}(\theta, s),
\label{B0ppdiv}
\end{align}
\begin{equation}
B_{++}^{(3)}(s,\theta)\simeq0,~
\label{B3pp}
\end{equation}%
\begin{equation}
B_{+-}^{(3)}(s,\theta)\simeq B_{+-}^{(3,h)}(s,\eta)=\frac{\alpha_{s}}{4\pi}%
\frac{C_{F}}{N_{c}}\frac{\left(  4\pi f_{\pi}\right)  ^{2}}{s}\frac{\mu_{\pi
}^{2}}{s}\frac{(e_{u}-e_{d})^{2}}{(1-\eta^{2})}I(\eta).
\label{B3pm}
\end{equation}
The expression for the amplitude  $B_{+-}^{(0)}$ in Eq.(\ref{Bpmdiv}) includes
the divergent integral $I_{s}$  in the hard contribution  and therefore  we use  the 
specific regularization  indicated by square brackets.   The same situation  must 
take place  in  the  expression in Eq.(\ref{B0ppdiv}). However in this case the endpoint
singularities  are related to a hard  configuration which is not described  by the chiral enhanced 
contributions and therefore have not been computed in our  consideration. 
In order to emphasize  this point we  denoted in Eq.(\ref{B0ppdiv})   the hard contributions  as  the matrix element 
 $\left[   \langle p,p'| O_{n}O_{\bar n}\ast T_{++}|0\rangle \right]_{\text{reg}}$.  
The amplitudes $B_{+\pm}^{(3)}$ do not obtain any  soft-overlap contributions and therefore we  do not need any special regularization in this case. 

In order to proceed further  we must define explicitly the  regularization    and subtraction scheme indicated as $[...]_{\text{reg}}$. 
 In order to solve this problem we are going to use the observation made in the previous section: the factorization of the   angular  dependent factors in the 
 expressions for the soft contributions.  The definition of the  subtraction scheme given below  is  very close to the idea which was suggested  in  
 description of $B$-decays  in Refs.\cite{Beneke:2000ry,Beneke:2000wa} therefore  we will also refer to this receipt  as  physical subtraction scheme. 
 Consider for instance the amplitude $B_{+-}^{(0)}$.  Using that the amplitude  $\Phi_{+-}$ defined  in  Eq.(\ref{def:Phi})  does not depend on the scattering 
 angle $\theta$  we can define it  at some fixed angle as following  expression  
 $\theta_{0}$  as 
\begin{align}
[\Phi_{+-}(s)]_{\text{reg}} &  =
\left(
-\frac{10}{9} \frac{(4\pi f_{\pi})^{2}}{s}\frac{3-\eta_{0}^{2} }{1-\eta_{0}^{2}}
\right)^{-1} 
\nonumber \\ &
\times \left(
B_{+-}^{(0)}(s,\theta_{0})
-\frac{\alpha_{s}}{4\pi}
\frac{\left(  4\pi f_{\pi}\right)  ^{2}}{s}
\frac{\mu_{\pi}^{2}}{s}\frac43 
\left[
\left\{ \frac59\frac{3-\eta^{2}_{0}}{1-\eta^{2}_{0}}~I_{s}-\frac49\frac{1}{1-\eta^{2}_{0}}~I(\eta_{0})\right\} 
\right]_{\text{reg}}
 \right)
 \label{Phipm}
\end{align}
where $\eta_{0}=\cos\theta_{0}$ and we also substituted  $N_{c}=3$, $e_{u}=2/3$, $e_{d}=-1/3$ in order to simplify the  analytical expression. 
This equation  defines
the soft factor  $[\Phi_{+-}(s)]_{\text{reg}}$  through the physical amplitude   $B_{+-}^{(0)}(s,\theta_{0})$ and the regularized hard contribution.  
The {\it rhs} in Eq.(\ref{Phipm}) does not depend  on the subtraction angle $\theta_{0}$  therefore the value  $\theta_{0}$  can be   fixed  
using  phenomenological arguments.   It is clear that the best choice  $\theta_{0}$ corresponds  to a  region where our description  is expected to be most accurate. 
In what follow we choose the value $\theta_{0}=90^{o}$  which is equidistant  from the forward and backward regions.  
 Then  substituting  Eq.(\ref{Phipm}) into  expression with arbitrary $\theta$  (\ref{Bpmdiv})  
and using  $\theta_{0}=90^{o}$ ($\eta_{0}=0$)
we obtain
\begin{equation}
B_{+-}^{(0)}(s,\theta)=~\frac{1-\eta^{2}/3}{1-\eta^{2}}~B_{+-}^{(0)}(s,90^{o})+ \Delta_{+-}^{(0)}(s,\eta),
\label{B0pm}
\end{equation}
with 
\begin{align}
\Delta_{+-}^{(0)}(s,\eta) =\frac{\alpha_{s}}{4\pi}
\frac{\left(  4\pi f_{\pi}\right)  ^{2}}{s}\frac{\mu_{\pi}^{2}}{s}
\frac{16}{27}
\frac{1}{(1-\eta^{2} )}\left\{  \left(  1-\frac{\eta^{2}} 3 \right)
~I(0)-I(\eta) \right\}.
 \label{Dlt0pm}%
\end{align}
  From this result we observe  that the divergent integral  $I_{s}$ cancel.  Formally this is  the direct consequence of the simple  fact:  
 the angular  factor  $(3-\eta^{2} )/(1-\eta^{2})$ in front of   divergent integral $I_{s}$  in Eq.(\ref{Bpmdiv})  
 is exactly the same as in front of the  soft  amplitude $\Phi_{+-}(s)$.   Let us stress  that this coincidence  is  obtained from the two
different matching  calculations.  More generally,  the  singular terms on {\it rhs } of Eq.(\ref{B0pm}) enter only in the combination  $\Delta_{+-}^{(0)}$  and 
must  cancel  if our   factorization formula (\ref{Bpmdiv}) is complete, i.e. it  describes the all required  configurations (or regions).  Then the  regularization $[...]_{\text{ reg}}$
 on {\it rhs }  of Eq.(\ref{B0pm}) can be omitted. Therefore  the cancellation of the singular term $I_{s}$ in the expression (\ref{Dlt0pm}) can be considered  as a  direct  confirmation of the SCET analysis   carried  out  in Sec.~\ref{softSCET}.   
 Let us also remind, that  the hard  contribution $\Delta_{+-}^{(0)}$ in Eq.(\ref{Dlt0pm}) is not complete and can also include  other contributions which are related 
to the different higher twist DAs. 

Application of  the same scheme for the amplitude $B_{++}^{(0)}$ yields
\begin{equation}
B_{++}^{(0)}(s,\theta)\simeq~\frac{B_{++}^{(0)}(s,90^{o})}{1-\eta^{2}} + \Delta_{++}^{(0)}(s,\eta)\approx \frac{B_{++}^{(0)}(s,90^{o})}{1-\eta^{2}}.
\label{B0pp-fin}
\end{equation}
where the hard correction $\Delta_{++}^{(0)}$ is given by the appropriate combination of the contributions
$\left[   \langle p,p'| O_{n}O_{\bar n}\ast T_{++}|0\rangle \right](s, \theta)\sim \alpha_{s}$   described by the higher twist pion DAs associated with the 
 operators $O_{n}$ and $O_{\bar n}$.
 We neglected  these terms  assuming that their numerical values  are  smaller comparing to the chiral enhanced corrections.  
 In Eq.(\ref{B0pp-fin})  we just indicate this possible contribution for clarity.

Let us remind that in expressions given by Eqs.(\ref{B0pm}) and (\ref{B0pp-fin}) we consider only the power suppressed  amplitudes $B_{+\pm}^{(i)}$  as defined in 
Eq.(\ref{defAB}).  The physical subtractions  reorganize  the  formal  expansions  (\ref{Bpmdiv}) and (\ref{B0ppdiv})  in such way that the soft-overlap contributions are
accumulated in the power suppressed amplitudes  $B_{+\pm}^{(0)}(s,\theta=90^{o})$ which we consider as nonpertubative quantities.  
The angular behavior of the power suppressed amplitudes $B_{+\pm}^{(i)}$ in the   Eqs.(\ref{B0pm}) and (\ref{B0pp-fin})    is obtained  from  the hard coefficient 
functions and   therefore  can be considered as a model independent result.   The  effect of the power suppressed hard contribution can be associated in our approximation  with the corrections described  by $\Delta_{+-}^{(0)}$ and $B_{+-}^{(3,h)}$ in Eqs.(\ref{B0pp-fin})  and  (\ref{B3pm}), respectively.

\begin{figure}[ptb]
\centering
\includegraphics[width=2.8in]{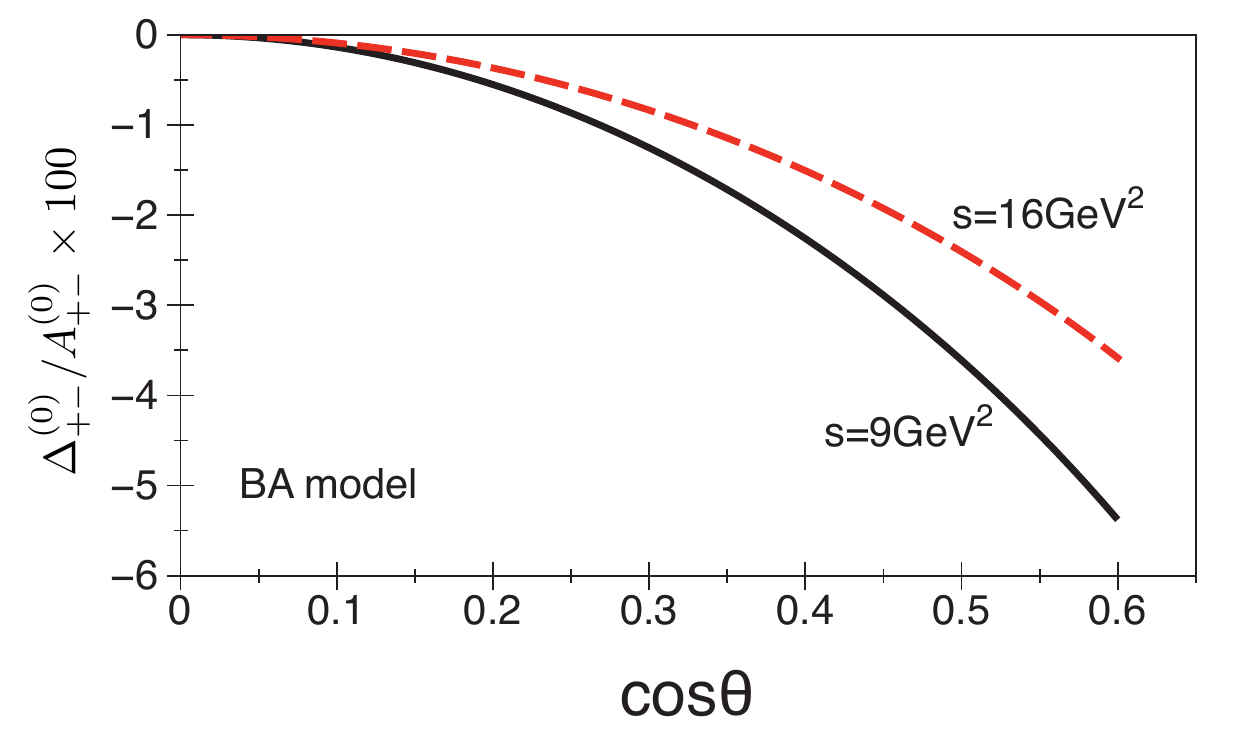}
\includegraphics[width=2.8in]{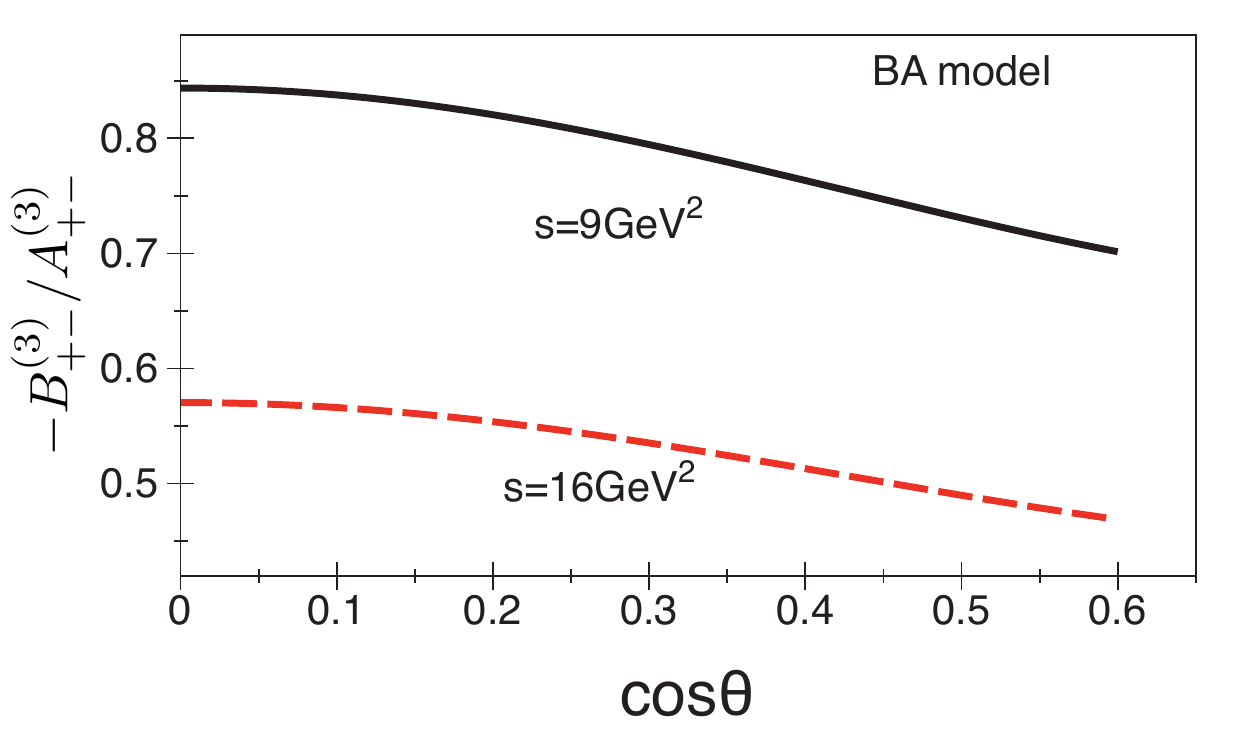} \newline%
\includegraphics[width=2.8in]{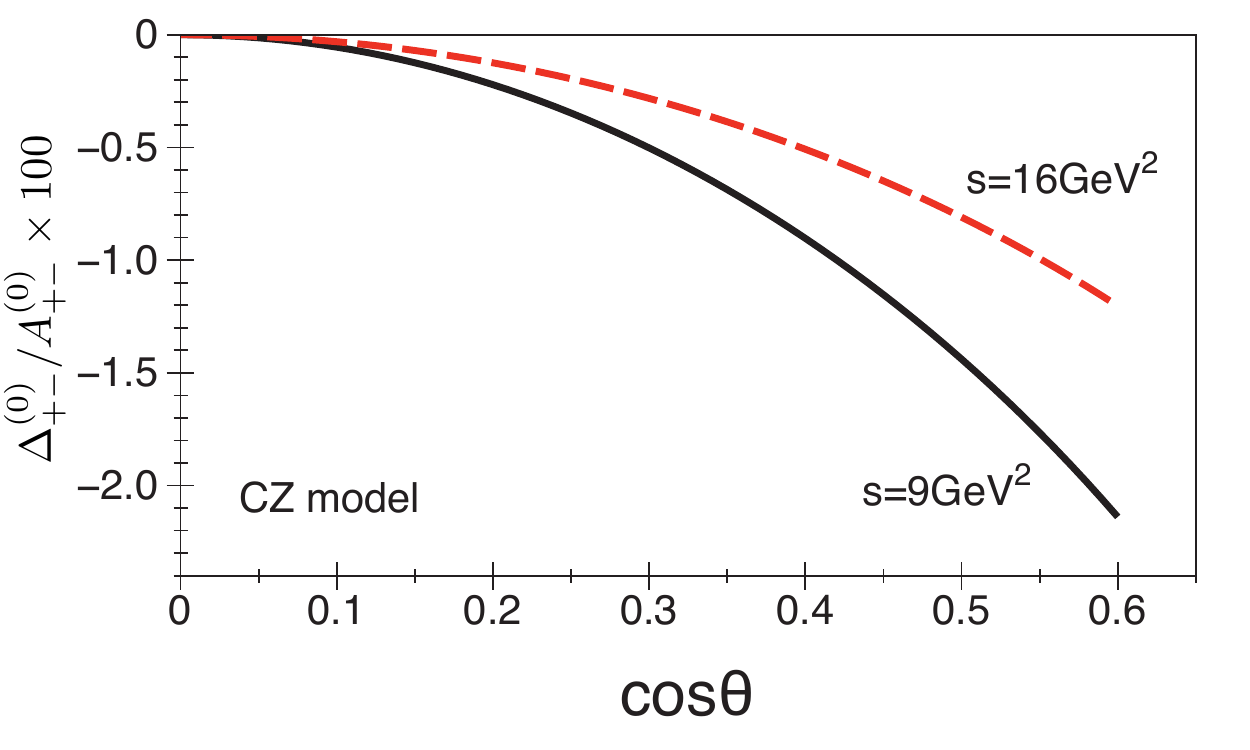}
\includegraphics[width=2.8in]{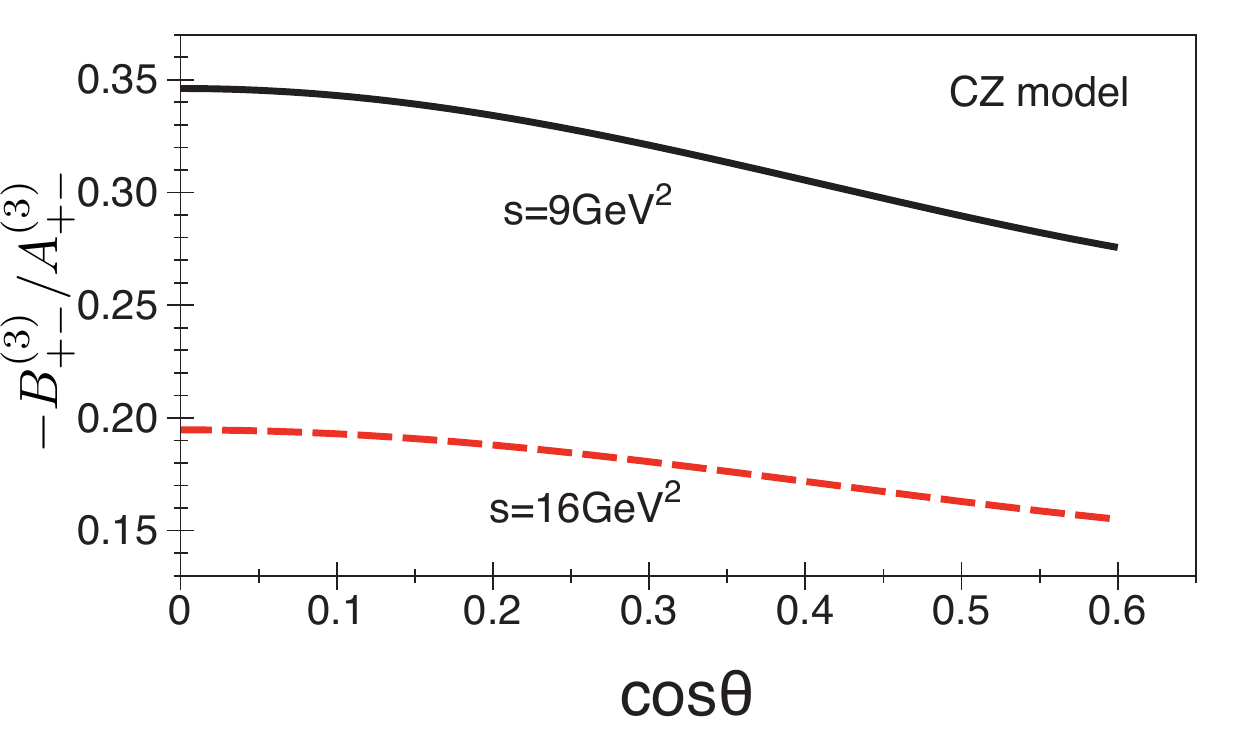}\caption{The ratios of the hard
subleading power amplitude to the leading power contribution. The fist
(second) line shows the plots for the BA(CZ)-model of pion DA. }%
\label{r3pm}%
\end{figure}

It is interesting to compare the hard leading-order  contributions with  the
hard  power suppressed  corrections. In Fig.\ref{r3pm} we
show the ratios of the $\Delta_{+-}^{(0)}/A_{+-}^{(0)}$ (left) and $B_{+-}
^{(3)}/A_{+-}^{(3)}$ (right) as functions of $\cos\theta$ for two different
values of $s$. In order to compute these ratios we used the leading
twist pion DA defined by  BA-set, see Eq.(\ref{setBA}). The renormalization scale $\mu_{R}$ for
the running coupling and for the quark masses is fixed to be $\mu_{R}=0.8W$
GeV. 

The ratio $\Delta_{+-}^{(0)}/A_{+-}^{(0)}$ is  relatively small (about few percent) and 
 tends to zero when  $\cos\theta\rightarrow0$. This behavior  is the consequence of the our subtraction
scheme with subtractions at $\theta_{0}=90^{o}$ that ensures 
$\Delta_{+-}^{(0)}(s,\eta\rightarrow0)\rightarrow0$.  Therefore we see that  numerical
effect from the hard chiral enhanced power correction in the amplitude $T_{+-}^{(0)}$ is very small.    
The result for  ratio $B_{+-}^{(3)}/A_{+-}^{(3)}$ is different  different. In this
case the absolute vale of the subleading amplitude is comparable to the
leading one providing a larger numerical impact. Even at $s=16$GeV$^{2}$ the effect
from the chiral enhanced power correction is of order $40-50\%$. The 
amplitudes  $B_{+-}^{(3)}$ and $A_{+-}^{(3)}$  enter with opposite signs therefore 
 the value of the combination  $B_{+-}^{(3)}+A_{+-}^{(3)}$ is strongly
reduced. Hence the effect of the chiral power correction for this
amplitude is numerically significant  and must be taken into account.

For the  CZ-model of the pion DA ($\mu_{R}=1.3$GeV)  the relative
value of the hard subleading corrections is considerably smaller because the
absolute value of the leading  contribution is considerably larger in this case. 
The results for the corresponding  ratios  are shown in the bottom plots  in Fig.\ref{r3pm}.

Summarizing,  in (\ref{B0pm}) and (\ref{B0pp-fin}) we  provide the  expressions  for the power suppressed
 amplitudes which  depend on the two unknown functions $B_{+-}^{(0)}(s,90^{o})$ and $B_{++}
^{(0)}(s,90^{o})$  describing  the soft-overlap contribution in the physical
subtraction scheme. Now we can combine  these results  with the leading-twist contributions $A_{+\pm}^{(i)}$
 in order to perform a phenomenological analysis  of the existing data.

\section{Phenomenological analysis of BELLE data}

\label{phenom}

The expressions for the cross sections can be written in the  relatively simple form if we present the total amplitudes in the following form
\begin{align}
T^{(0)}_{++}(s,\theta)& = A^{(0)}_{++}(s,\theta) +   \frac{1}{1-\eta^{2}}B_{++}^{(0)}(s),
\label{T0pp-fin}
\\
T^{(0)}_{+-}(s,\theta)& = A^{(0)}_{+-}(s,\theta)+\frac{1-\eta^{2}/3}{1-\eta^{2}}~B_{+-}^{(0)}(s)+ \Delta_{+-}^{(0)}(s,\eta),
\label{T0pm-fin}
\\
T^{(3)}_{++}(s,\theta)& = A^{(3)}_{++}(s,\theta),
\\
T^{(3)}_{+-}(s,\theta)& = A^{(3)}_{+-}(s,\theta)+B^{(3,h)}_{+-}(s,\theta)
\label{T3pm-fin}
\end{align}
where $\eta=cos\theta$  and we  used that the amplitude $B^{(3)}_{++}$ is small, see (\ref{B3pp}).   The expressions describing the leading
power contributions $A_{+\pm}^{(i)}$ are given in Eqs.(\ref{A0pp})-(\ref{A3pm}).  The hard  subleading  amplitudes 
$B_{+-}^{(3,h)}$ and $\Delta_{+-}^{(0)}$ can be fond  in Eqs.(\ref{B3pm}) and (\ref{Dlt0pm}), respectively.   
In Eqs.(\ref{T0pp-fin}) and (\ref{T0pm-fin}) we used for the subleading amplitudes as given in  Eqs.(\ref{B0pm}) and (\ref{B0pp-fin}). 
For the unknown soft-overlap contributions  we  introduced  short  notations  
\begin{equation}
B_{++}^{(0)}(s,\theta=90^{o})\equiv B_{++}^{(0)}(s), \quad B_{+-}^{(0)}(s,\theta=90^{o})\equiv B_{+-}^{(0)}(s).  
\end{equation}

Using  equations (\ref{T0pp-fin})-(\ref{T3pm-fin})  we can write  expressions for the cross sections (\ref{dsdcos-pipm}) and (\ref{dsdcos-pi00}) as  following
\begin{align}
\frac{d\sigma^{\pi^{+}\pi^{-}}}{d\cos\theta}  &  =\frac{\pi\alpha^{2}}%
{16s}~\left(  \frac{\left\vert B_{++}^{(0)}\right\vert ^{2}}{\left(
1-\eta^{2}\right)  ^{2}}+\left(  \frac{1-\eta^{2}/3}{1-\eta^{2}}\right)
^{2}\left\vert B_{+-}^{(0)}\right\vert ^{2}\right.  +\left\vert
A_{++}^{(0)}\right\vert ^{2}+\left\vert A_{+-}^{(0)}+\Delta_{+-}^{(0)}\right\vert ^{2}\nonumber\\
&  \left.  +2A_{++}^{(0)}\frac{\operatorname{Re}[B_{++}^{(0)}]}%
{1-\eta^{2}}+2\frac{1-\eta^{2}/3}{1-\eta^{2}}\left(  A_{+-}^{(0)}+\Delta
_{+-}^{(0)}\right)  \operatorname{Re}[B_{+-}^{(0)}]\right)  ,\label{dspm}%
\end{align}%
\begin{align}
\frac{d\sigma^{\pi^{0}\pi^{0}}}{d\cos\theta}  &  =\frac{\pi\alpha^{2}}%
{32s}~\left(  \frac{\left\vert B_{++}^{(0)}\right\vert ^{2}}{\left(
1-\eta^{2}\right)  ^{2}}+\left(  \frac{1-\eta^{2}/3}{1-\eta^{2}}\right)
^{2}\left\vert B_{+-}^{(0)}\right\vert ^{2}
\right. 
 +\left\vert A_{+-}^{(0)}+\Delta_{+-}^{(0)}+A_{+-}^{(3)}+B_{+-}^{(3,h)}\right\vert ^{2}
\nonumber\\
&  \left.  +2\frac{1-\eta^{2}/3}{1-\eta^{2}}\left(  A_{+-}^{(0)}+\Delta
_{+-}^{(0)}+A_{+-}^{(3)}+B_{+-}^{(3,h)}\right)  \operatorname{Re}[B_{+-}^{(0)}]\right)  .
\label{ds00}%
\end{align}
In these expressions we used that all hard contributions $A^{(i)}_{+\pm}$, $B_{+-}^{(3,h)}$ and $\Delta_{+-}^{(0)}$  are real and therefore 
only the real parts  $\operatorname{Re} [B^{(0)}_{+\pm}]$   appear in the interference.

Using the known angular behavior of the  cross sections in Eqs.(\ref{dspm}) and   (\ref{ds00})  one can try to 
describe BELLE  data  \cite{Nakazawa:2004gu, Uehara:2009cka}  in   the region $W=3-4$GeV 
accepting   the  amplitudes $B_{+\pm}^{(0)}$ as free parameters.  
 In the subsequent analysis we assume that  the   amplitudes $B_{+\pm}^{(0)}$ are dominated by real parts and  
\begin{equation}
\operatorname{Im}B_{+\pm}^{(0)}(s)\approx 0. 
\end{equation}
Then at  fixed energy $s$  we only have  two unknown real parameters $B_{+\pm}^{(0)}(s)$ 
in order to describe the two differential cross sections.    

The part of  the nonperturbative input  is  given by  the pion DA. 
In  our numerical estimates we will use  two different  models described  in Eqs. (\ref{setBA}) and
(\ref{setCZ}) ( BA- and CZ-model, respectively).   The numerical results  obtained with the other  
models defined  in Eqs.(\ref{setI}) and  (\ref{setBMS})  are very close to one obtained with the BA-model.

At the beginning  let us consider some qualitative properties  
of the cross sections described by   Eqs.(\ref{dspm}),(\ref{ds00}).  
These expressions  allows one  to confront the  theoretical 
predictions for the angular behavior  at  fixed energy $s$ against the data. 
The angular behavior is described by the known coefficients in front of the soft amplitudes $B_{+\pm}^{(0)}(s)$ 
and by the different hard amplitudes. 

Let us consider a soft approximation obtained  neglecting  the \textit{all} hard contributions in Eqs.(\ref{dspm}) and (\ref{ds00}).
Then the cross sections are described only by the quadratical terms $|B_{++}^{(0)}|^{2}$ and $|B_{+-}^{(0)}|^{2}$. 
Their coefficients  differ  by relatively small  factor  $\left( \eta^{2}/3\right)/(1-\eta^{2})^{2}$ 
and therefore we can conclude that  such ``soft''  cross sections  behave as approximately as  $(1-\eta^{2})^{-2}$.  
From Eqs.(\ref{dspm}) and (\ref{ds00}) it is  easily to  see  that the ratio of the cross sections  in this case is fixed
\begin{equation}
\left[  \frac{d\sigma^{\pi^{0}\pi^{0}}}{d\cos\theta}/\frac{d\sigma^{\pi^{+}
\pi^{-}}}{d\cos\theta}\right]  _{\text{soft}}=\frac{1}{2}.\label{Ratio}%
\end{equation}
The deviation from this value in the present formalism can be explained only
by the presence of the interference of the computed hard contributions
$A_{+\pm} ^{(i)},~B_{+-}^{(3,h)}$ with the unknown soft amplitudes $B_{+\pm
}^{(0)}(s,0)$.

Therefore  relatively small deviation from $1/(1-\eta^{2})$
behavior can only be visible if the value of the amplitude $B_{+-}^{(0)}$ is
quite large.

\begin{figure}[ptb]
\centering
\includegraphics[width=2.7in]{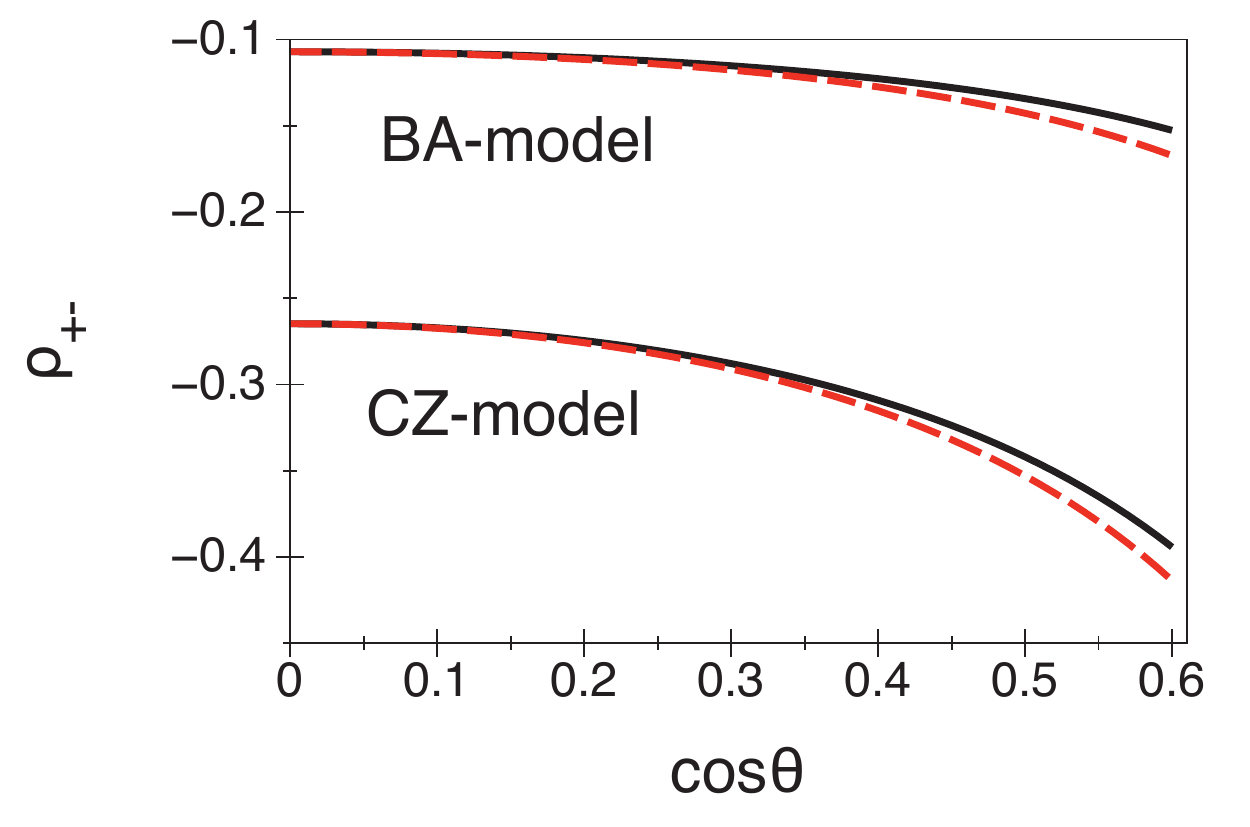}
\includegraphics[width=2.7in]{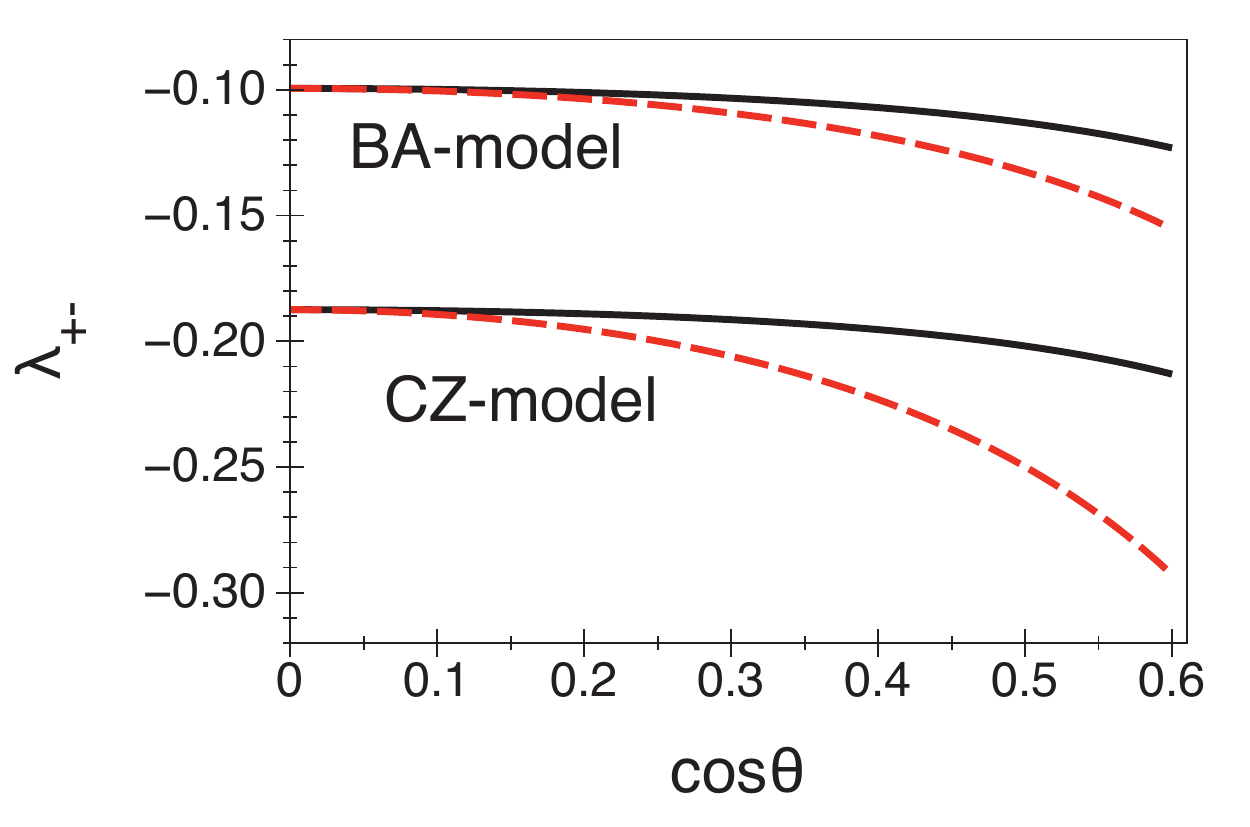}\caption{The linear
coefficients $\rho_{+-}$ defined in Eq.(\ref{rho-exc}) (left figure) and
$\lambda_{+-}$ defined in Eq.(\ref{lam}) (right figure) as a functions of
$\eta=\cos\theta$ at fixed energy $W=3$GeV. The exact values of these
coefficients are shown by solid black lines. The approximations $\tilde{\rho
}_{+-}$ (\ref{rho-app}) and $\tilde{\lambda}_{+-} $ (\ref{lamtilde}) are given
by dashed red lines. For the numerical calculations of the amplitudes we used
the same input parameters as in Figs.\ref{r3pm}}%
\label{linear-bpm}%
\end{figure}

The linear in $B_{+\pm}$ terms in Eqs.(\ref{dspm}),(\ref{ds00}) can also
provide a significant numerical effect. Let us consider their behavior in
$\cos\theta$. From Eq.(\ref{A0pp}) one finds that $A_{++}^{(0)}\sim
1/(1-\eta^{2})$. The combination $(1-\eta^{2}/3)\left(  A_{+-}^{(0)}%
+\Delta_{+-}^{(0)}\right)  $ which inter in Eq.(\ref{dspm})\ also behaves as
$1/(1-\eta^{2}).$ In Fig.\ref{linear-bpm} we plot the corresponding  linear term
\begin{equation}
\rho_{+-}(s,\eta)=(1-\eta^{2}/3)\left(  A_{+-}^{(0)}(s,\eta)+\Delta_{+-}%
^{(0)}(s,\eta)\right)  ,\label{rho-exc}%
\end{equation}
and  the approximation which is given by
\begin{equation}
\tilde{\rho}_{+-}(s,\eta)\simeq\frac{\rho_{+-}(s,0)}{1-\eta^{2}}%
.\label{rho-app}%
\end{equation}
at fixed $W=3$GeV. From this figure one can see that the difference between the
two expressions is quite small for all models of pion DA. This picture
is not changed in the region where $W=3-4$GeV. Therefore to a very good
accuracy one can expect that
\begin{equation}
\frac{d\sigma^{\pi^{+}\pi^{-}}}{d\cos\theta}\sim\frac{1}{(1-\eta^{2})^{2}%
}.\label{dspm-cos}%
\end{equation}
This simple observation   allows us to conclude that one can not perform a good 
extraction of the two amplitudes $B_{+\pm}^{(0)}$  by  fitting the differential
cross section $d\sigma^{\pi^{+}\pi^{-}}/d\cos\theta$ .

The qualitative observation (\ref{dspm-cos}) is in agreement with experimental
data \cite{Nakazawa:2004gu}. The behavior of the cross section as in
Eq.(\ref{dspm-cos}) was also obtained in handbag model approach
Ref.\cite{Diehl:2001fv}. However within this framework $B^{(i)}_{++}=0$ that
is not agree with the our result in Eq.(\ref{dspm}).

The linear (with respect to $B_{+\pm}^{(0)}$) contribution in  the  cross sections for neutral pions in Eq.(\ref{ds00}) 
depends only from the amplitude $B_{+-}^{(0)}$. The corresponding hard  coefficients reads
\begin{equation}
\lambda_{+-}(s,\eta)=A_{+-}^{(0)}(s,\eta)+\Delta_{+-}^{(0)}(s,\eta
)+A_{+-}^{(3)}(s,\eta)+B_{+-}^{(3,h)}(s,\eta).\label{lam}%
\end{equation}
The angular behavior  of this expression deviates from a simple 
 behavior like $(1-\eta^{2})^{-1}$. In Fig.\ref{linear-bpm} we show the exact value
$\lambda_{+-}(s,\eta)$ in comparison with the approximation
\begin{equation}
\tilde{\lambda}_{+-}(s,\eta)=\frac{\lambda_{+-}(s,0)}{(1-\eta^{2}%
)}.\label{lamtilde}%
\end{equation}
We observe that in this case the deviation from the profile $1/(1-\eta^{2})$
reaches $30\%$ for the CZ-model. Therefore if the absolute value of the amplitude
$B_{+-}^{(0)}$ is relatively large then the corresponding linear term in the
 cross section can already provide a sizable numerical effect. In this
case the separation of the two amplitudes using the angular behavior can be
performed in a better way. 

The boundary value $\eta_{\max}=0.6$ in  the
energy interval $W=3-4$~GeV  corresponds  to the values of $|t|=|u|=1.8-3.2$~GeV$^{2}$. 
Such values of $t$ and $u$ are still far from the asymptotic domain  and we can not
exclude a substantial numerical corrections associated with the different
subleading  contributions. Therefore we admit  that 
our description can be less accurate  in the  region $\eta \sim0.6$ for 
$W=3-4$~GeV.

\subsection{Phenomenological analysis  using  model-II for  pion DA  }

 We start  our  phenomenological analysis  using  the pion
DA defined by set-II,  see Eq.(\ref{setBA}).  
The value of  the hard  scale is fixed to be  $\mu_{R}=0.8W$. This  scale  is used in order to compute the
running coupling $\alpha_{s}(\mu_{R})$, the moments $a_{2n}(\mu_{R})$ and
quark masses in expression for $\mu_{\pi}$ in Eq.(\ref{mupi}).

In order to see an  effect from the subleading corrections let us consider
numerical values of  different contributions in the cross sections. Taking
$W=3.05$GeV, $\cos\theta=0.05$ we obtain
\begin{align}
&  \left(  \frac{d\sigma^{\pi^{+}\pi^{-}}}{d\cos\theta}\right)  _{\text{LT}}
=0.0283, \quad\left(  \frac{d\sigma^{\pi^{+}\pi^{-}}}{d\cos\theta}\right)
_{\text{exp}}=0.312\pm0.039,\label{LTpmW30-005}\\
&  \left(  \frac{d\sigma^{\pi^{0}\pi^{0}}}{d\cos\theta}\right) _{\text{LT}%
}=0.0014, \quad\left(  \frac{d\sigma^{\pi^{0}\pi^{0}}}{d\cos\theta}\right)
_{\text{exp}}=0.078\pm0.025,\label{LT00W30-005}\\
\frac{d\sigma^{\pi^{+}\pi^{-}}}{d\cos\theta}  &  =0.879|B_{++}^{(0)}
|^{2}+0.878|B_{+-}^{(0)}|^{2}-0.259B_{++}^{(0)}-0.180B_{+-}^{(0)}
+0.0283,\label{pmW30-005}\\
\frac{d\sigma^{\pi^{0}\pi^{0}}}{d\cos\theta}  &  =0.440|B_{++}^{(0)}
|^{2}+0.439|B_{+-}^{(0)}|^{2}-0.083B_{+-}^{(0)}+0.004.\label{00W30-005}%
\end{align}
Here the subscript ``LT'' denotes the leading-twist approximation. For
comparison we also show in Eqs.(\ref{LTpmW30-005}) and (\ref{LT00W30-005})
the experimental values from Refs.\cite{Nakazawa:2004gu,Uehara:2009cka}.

Comparing  Eqs.(\ref{LT00W30-005}) and (\ref{00W30-005}) in the limit $B_{+pm}^{(0)}=0$
one can see  that the hard subleading contribution $B_{+-}^{(3,h)}$  provides a large numerical  contribution 
( $\pi^{0}\pi^{0}$ channel).   However the absolute value of this correction  is still very
small in order to obtain the experimental value in Eq.(\ref{LT00W30-005}). This
observation  is also valid  for other values of the scattering angle $\theta$. Hence we
conclude that  the data  can be described only if  the soft amplitudes
$B_{+\pm}^{(0)}$ are quite large.

\begin{figure}[ptb]
\centering
\includegraphics[width=1.8in]{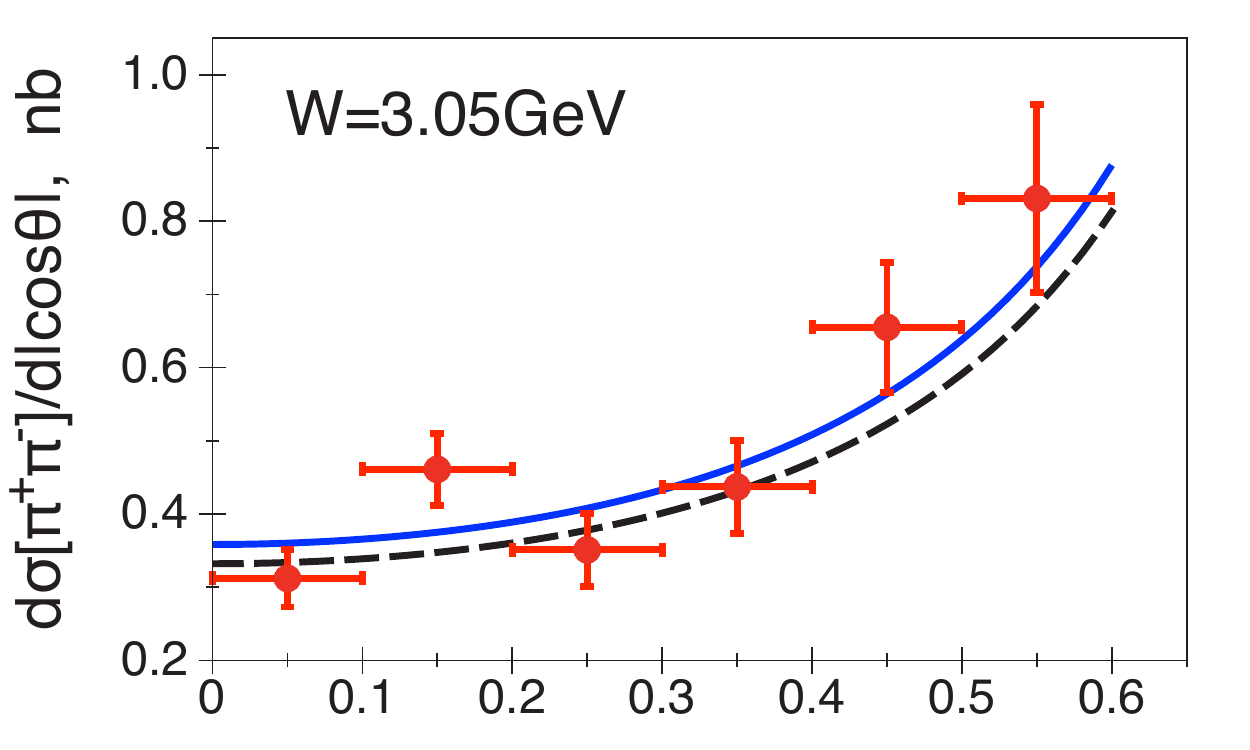}
\includegraphics[width=1.8in]{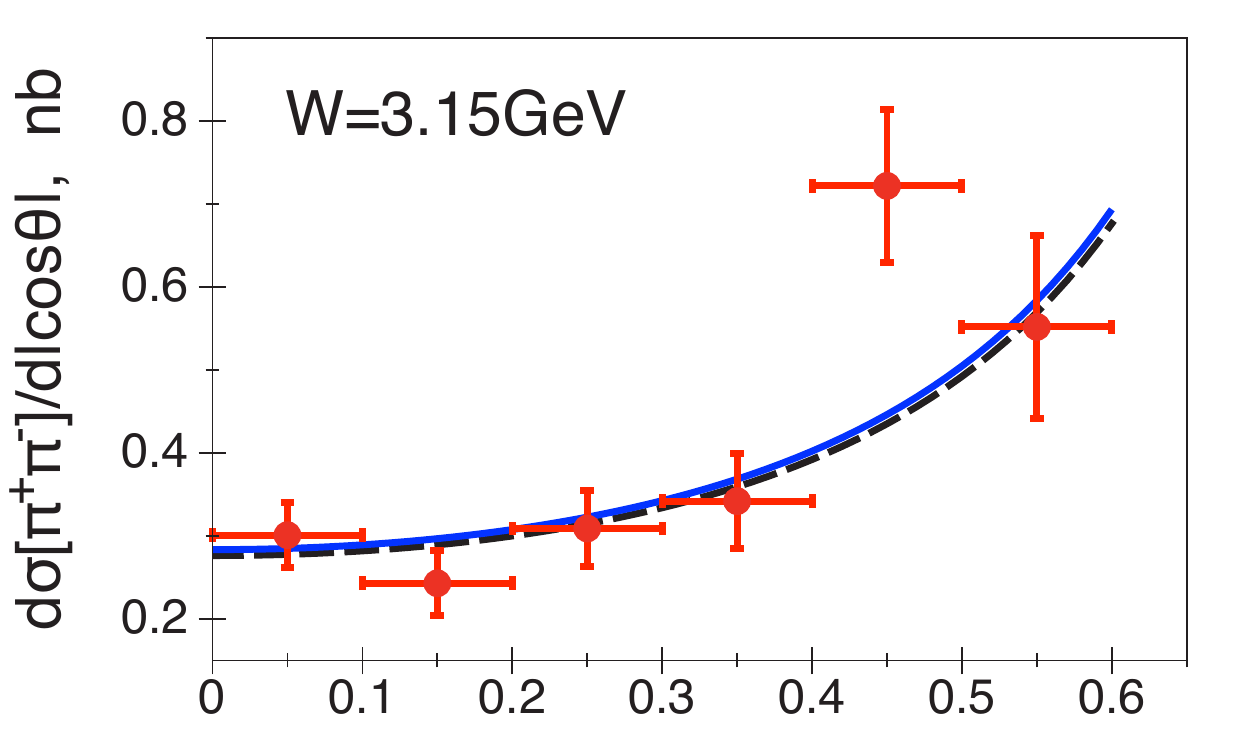}
\includegraphics[width=1.8in]{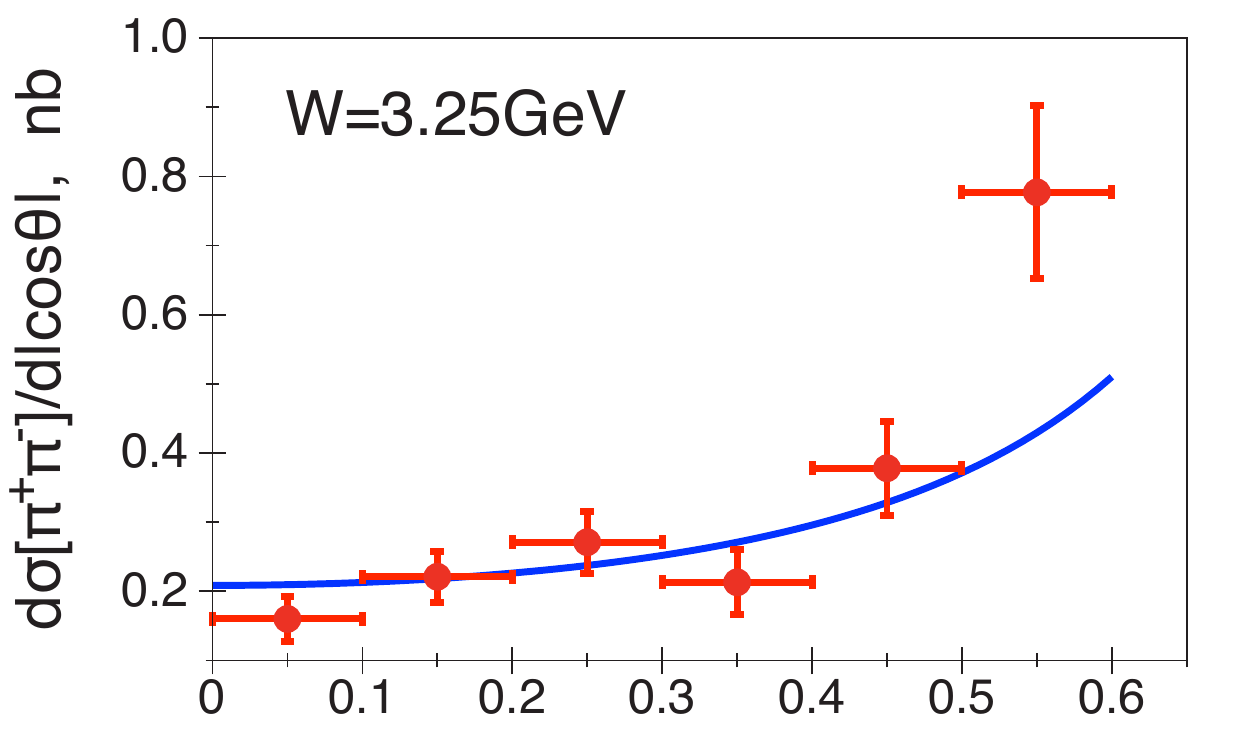} \newline%
\includegraphics[width=1.8in]{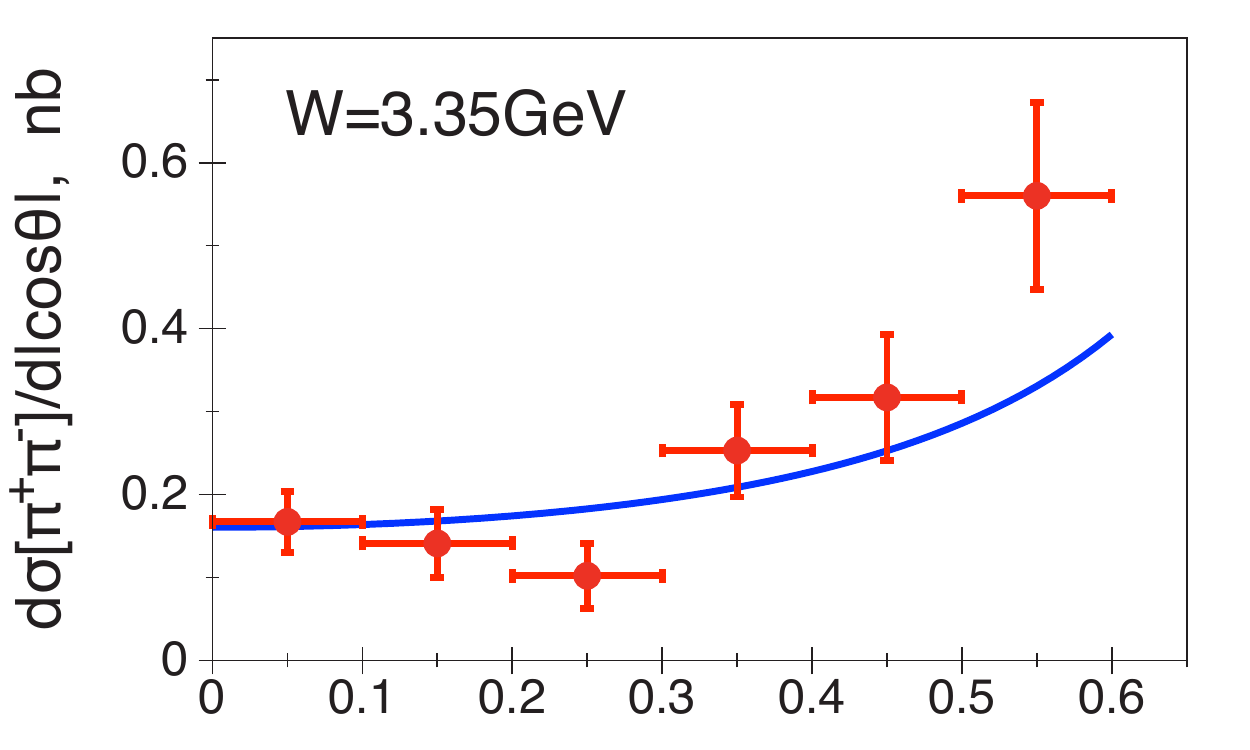}
\includegraphics[width=1.8in]{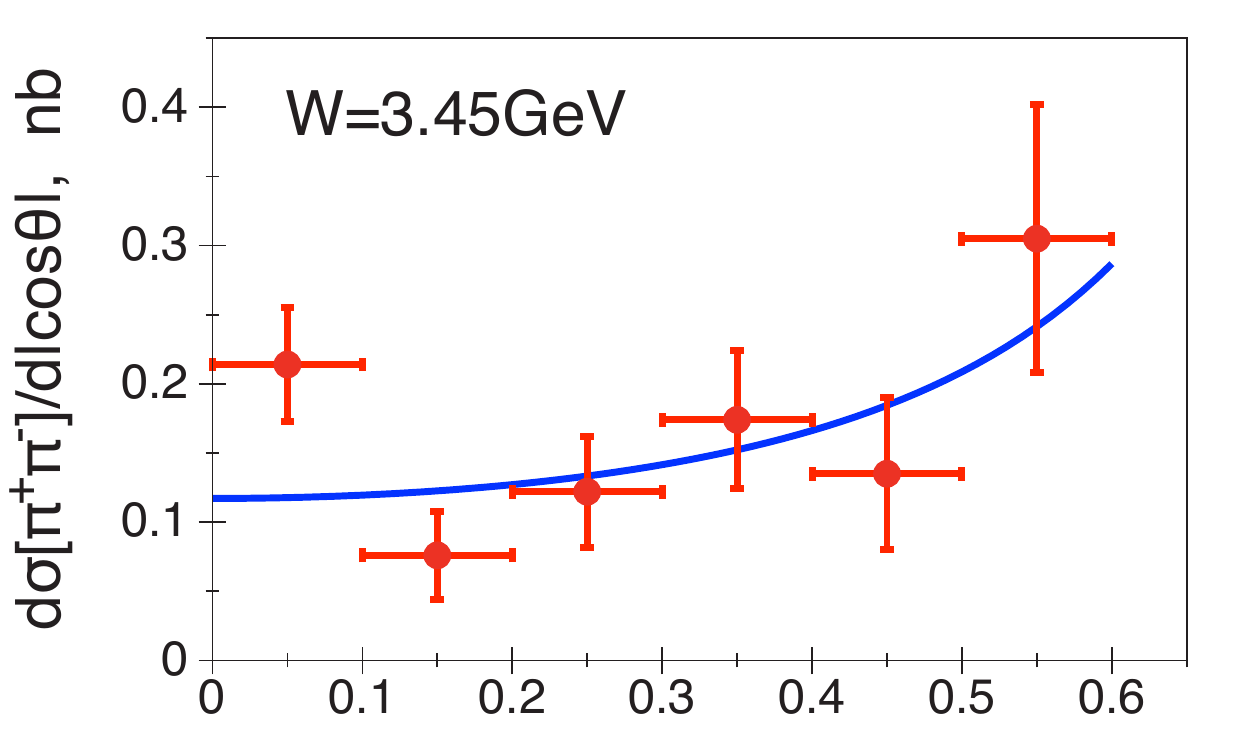}
\includegraphics[width=1.8in]{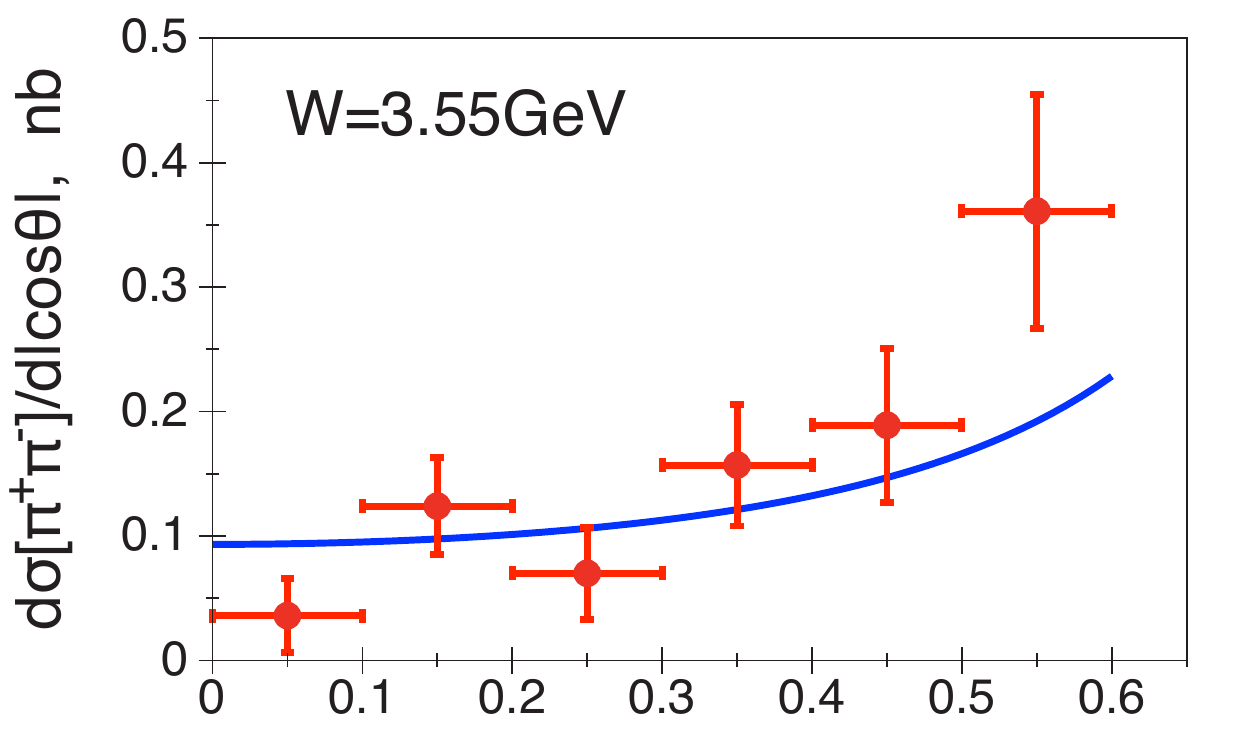} \newline%
\includegraphics[width=1.8in]{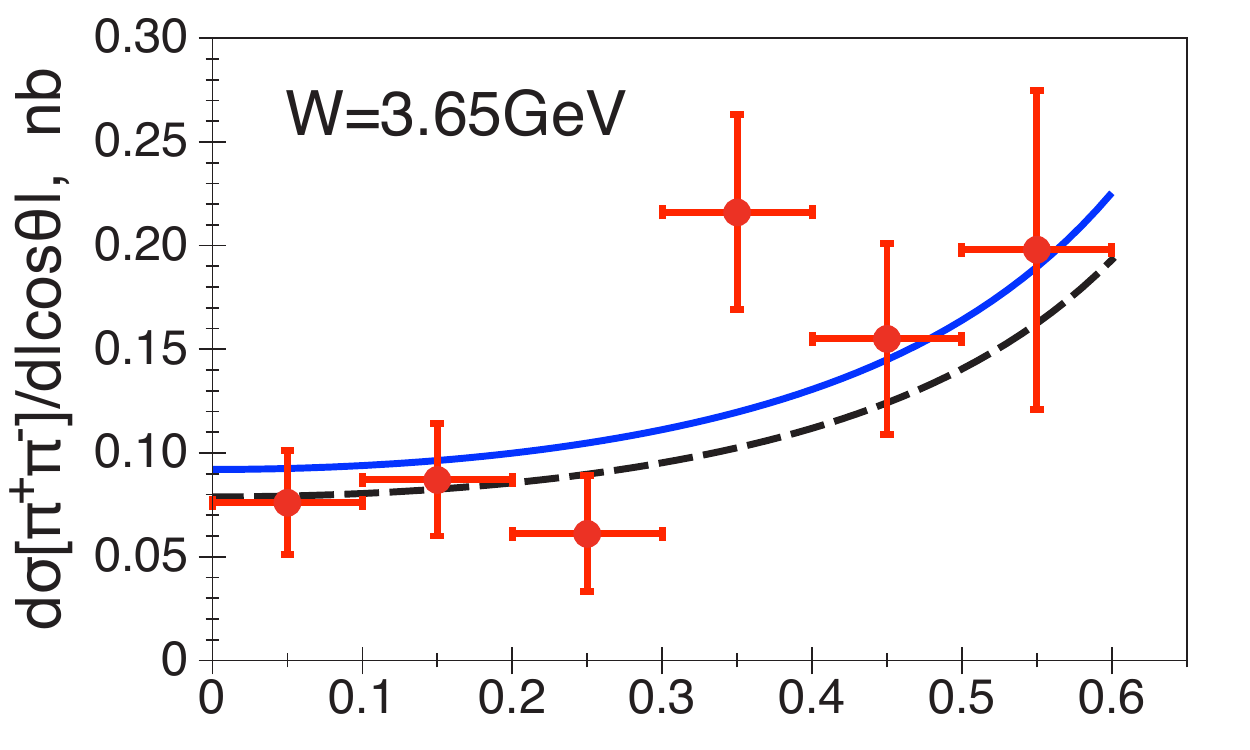}
\includegraphics[width=1.8in]{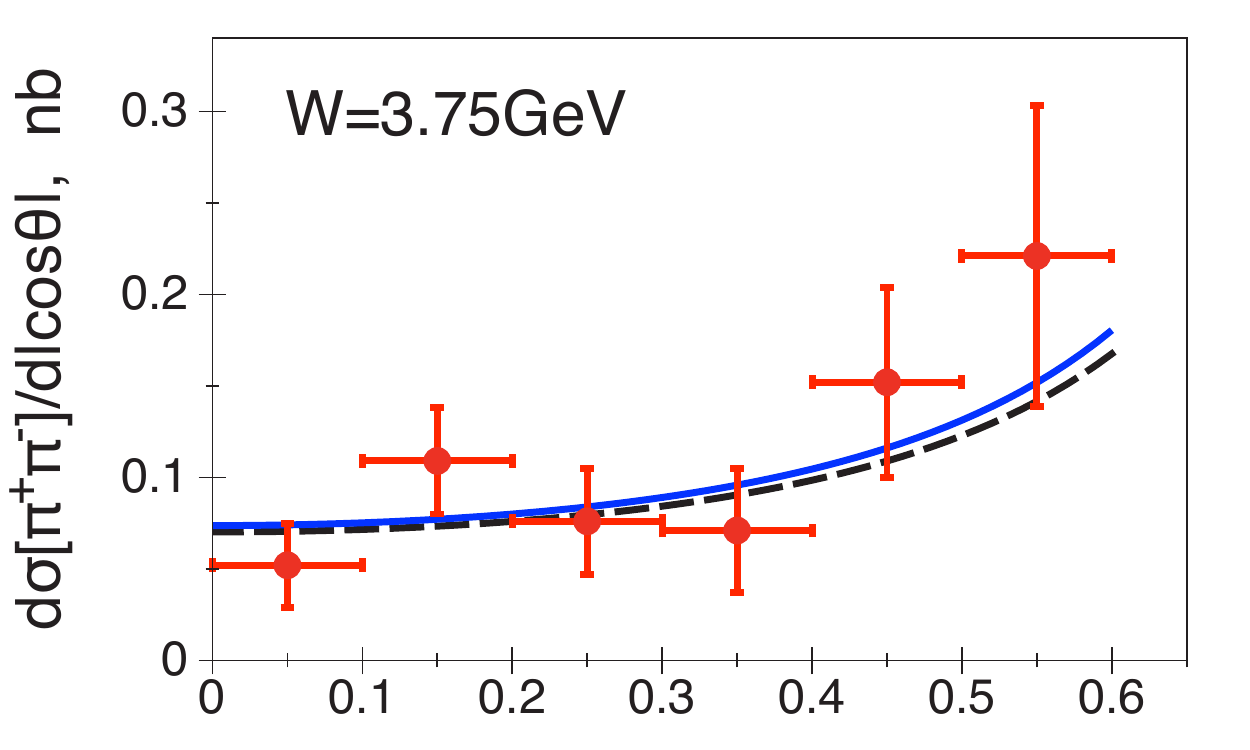}
\includegraphics[width=1.8in]{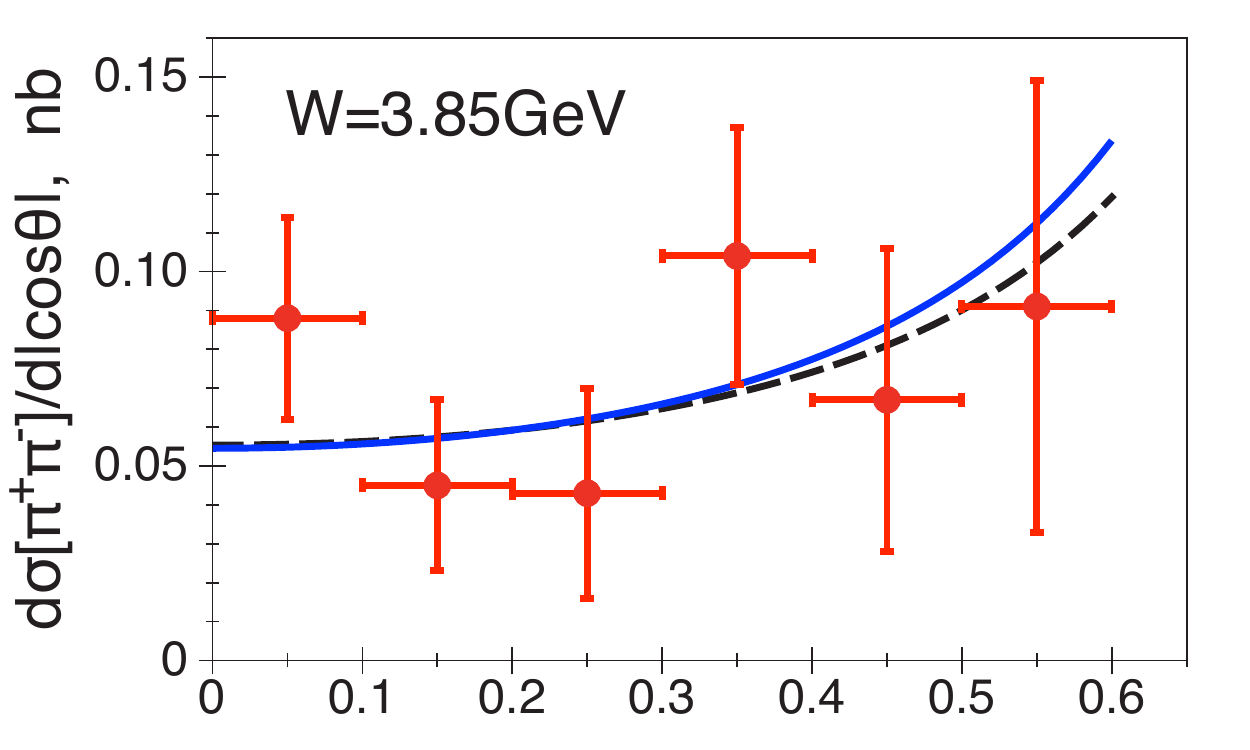} \newline%
\includegraphics[width=1.8in]{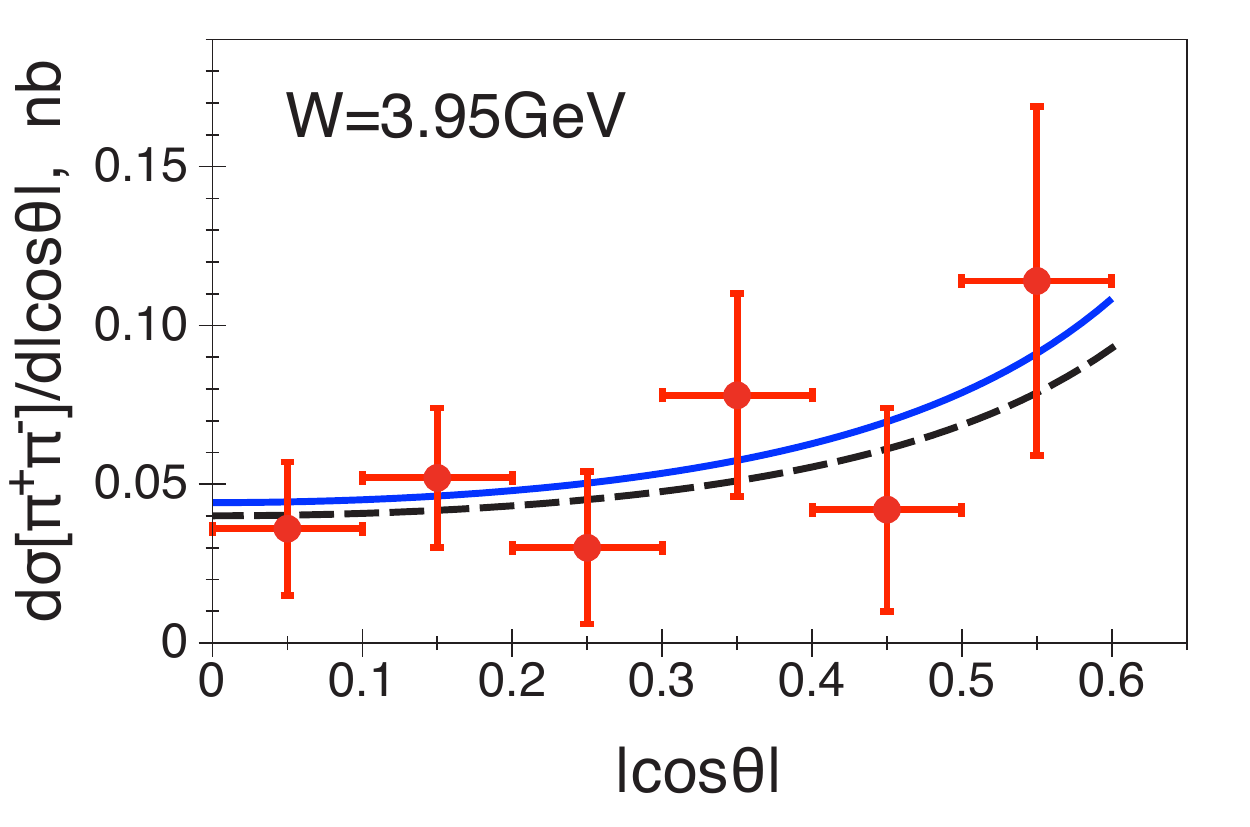}
\includegraphics[width=1.8in]{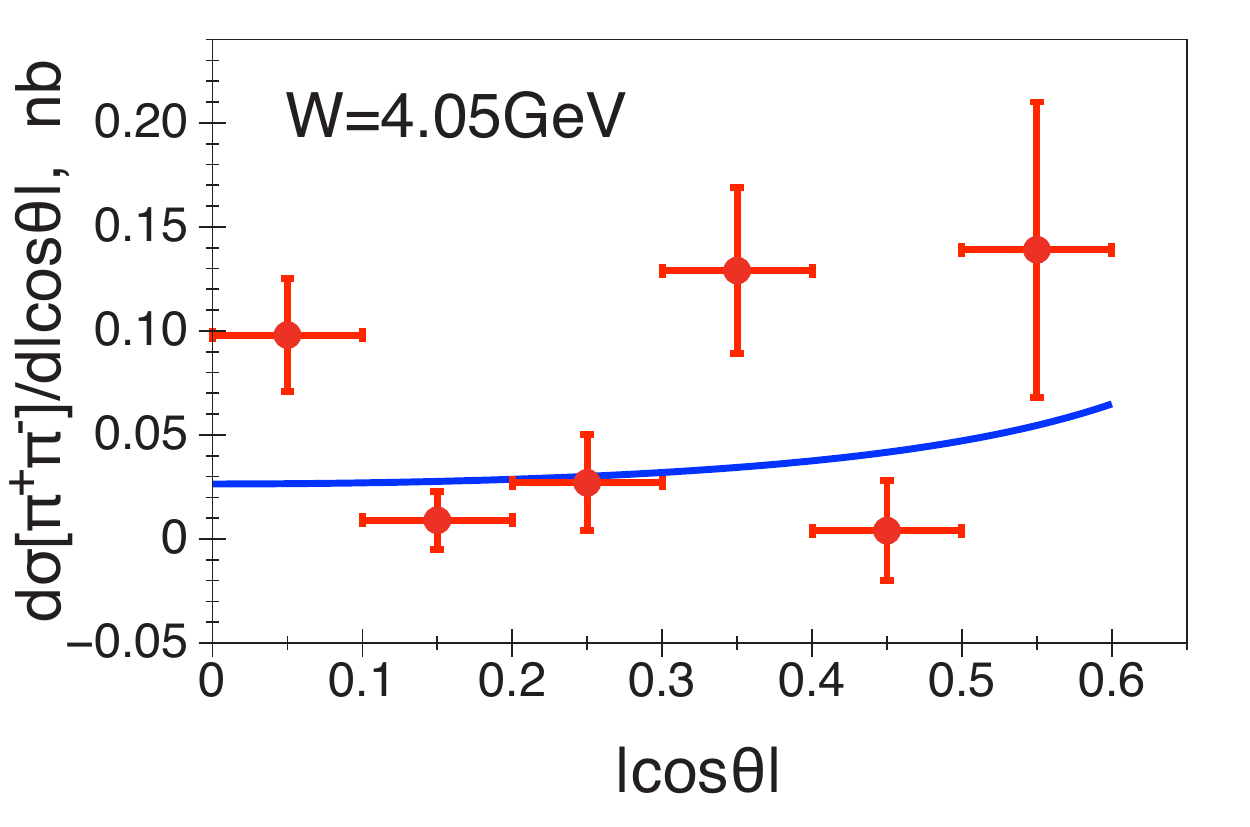}
\caption{Results of the fit for the
charged pion cross section, see discussion in the text. The data are results
from BELLE \cite{Nakazawa:2004gu}. }%
\label{dspmBA}%
\end{figure}

\begin{figure}[ptb]%
\begin{tabular}
[c]{ccc}%
\includegraphics[width=1.8in]{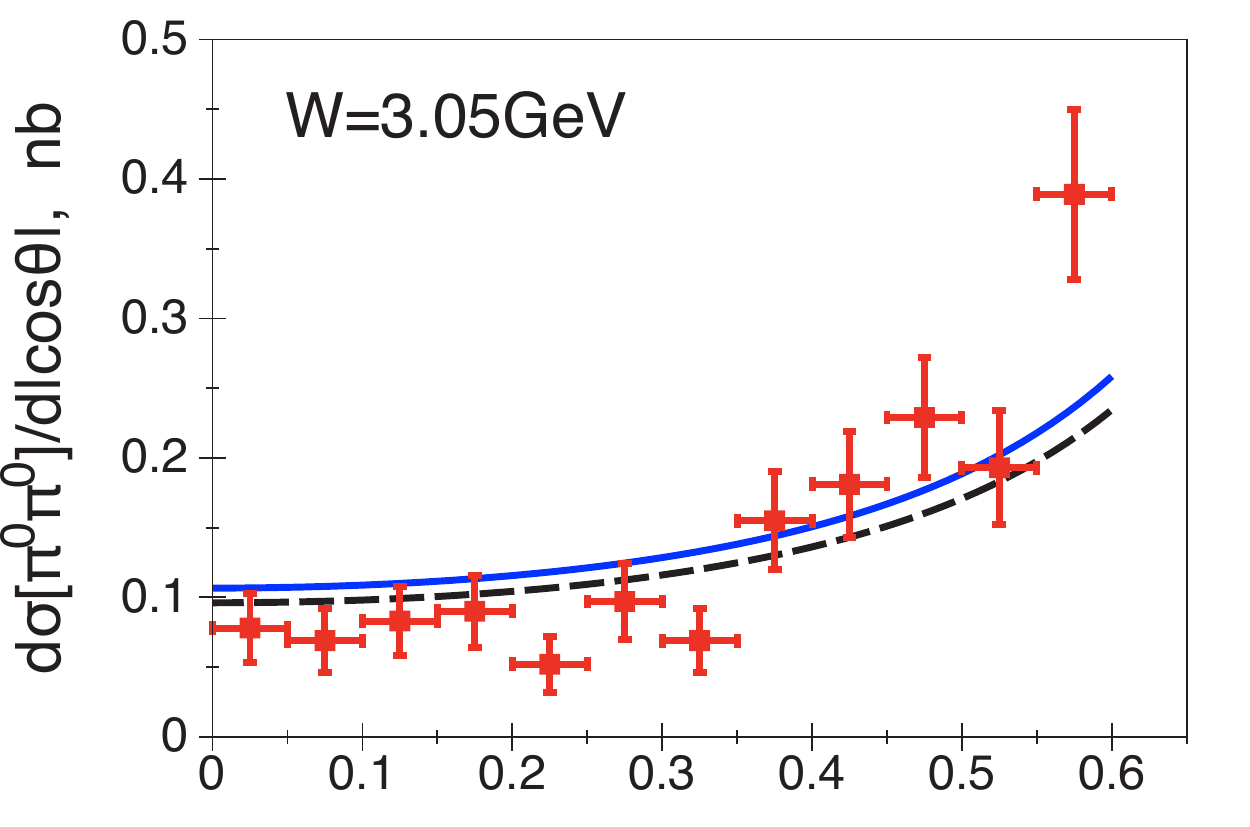} &
\includegraphics[width=1.8in]{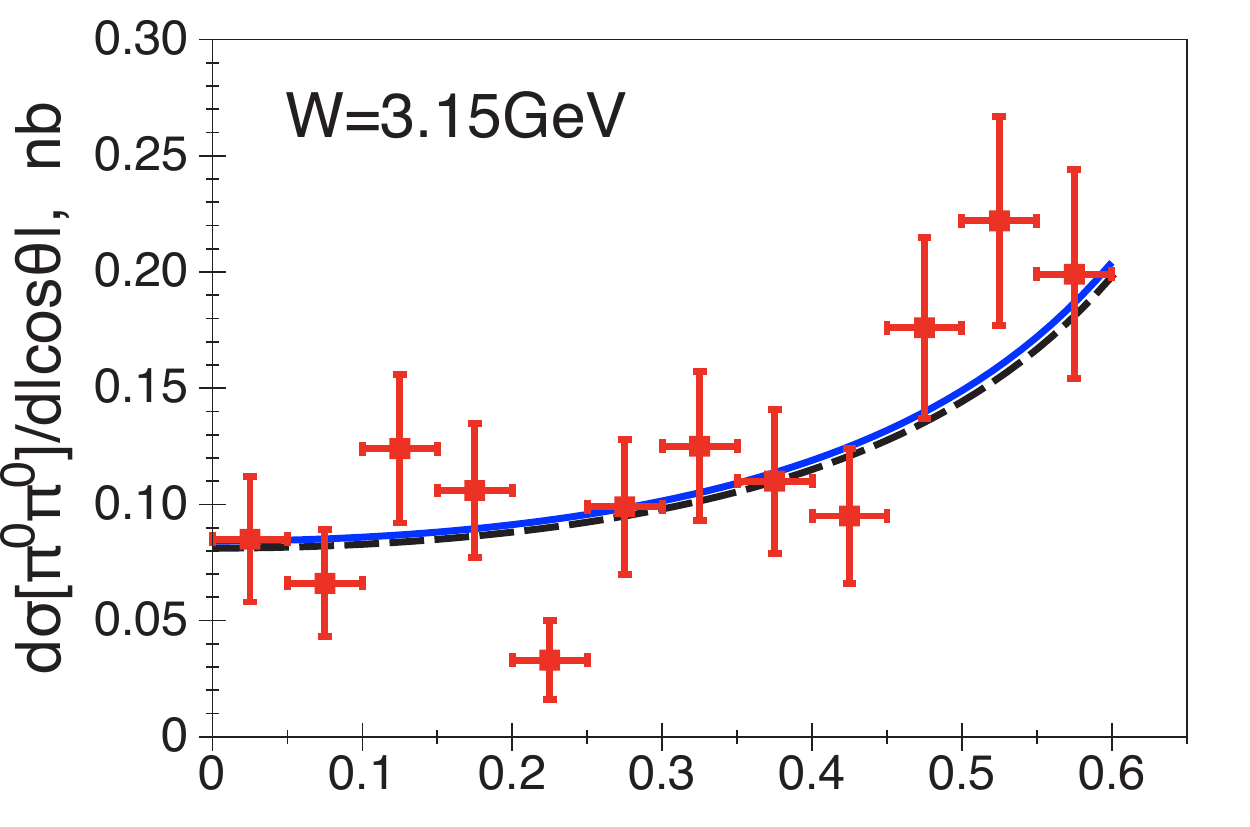} &
\includegraphics[width=1.8in]{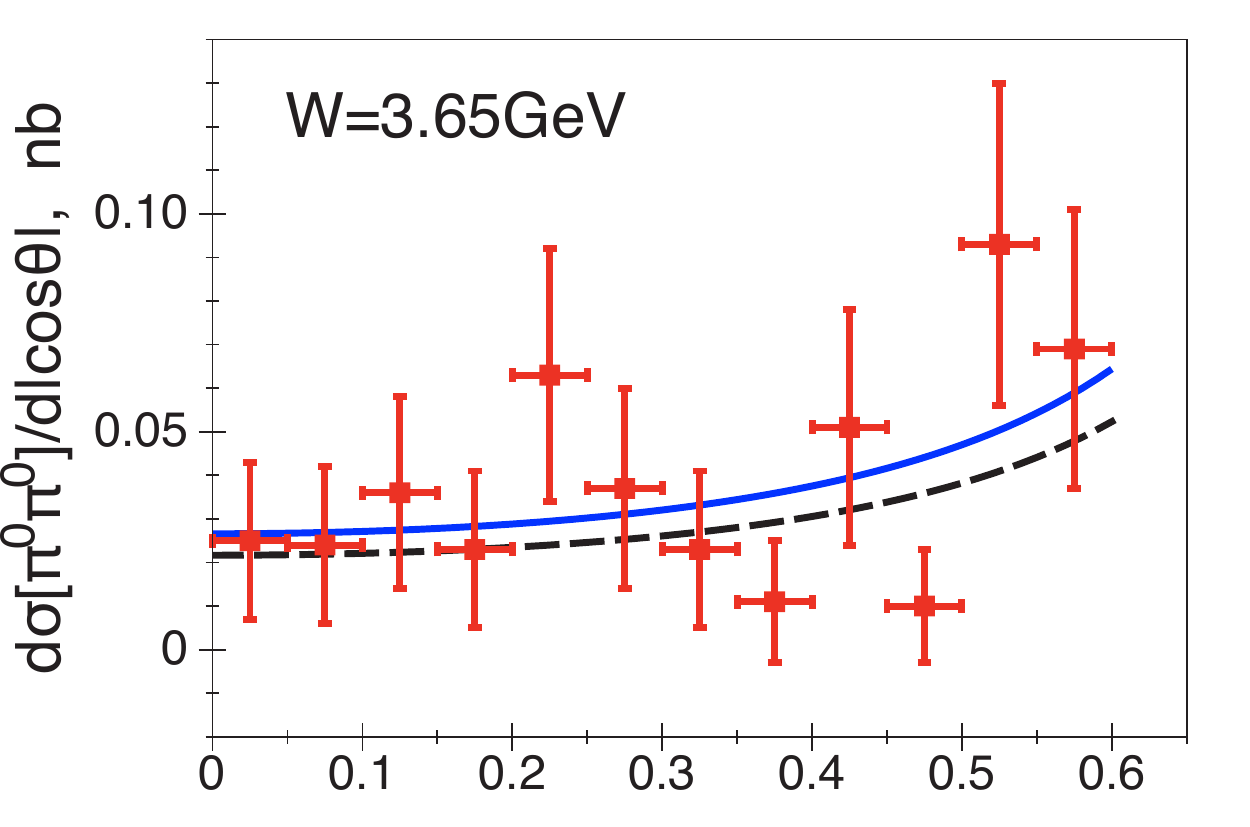}\\
\includegraphics[width=1.8in]{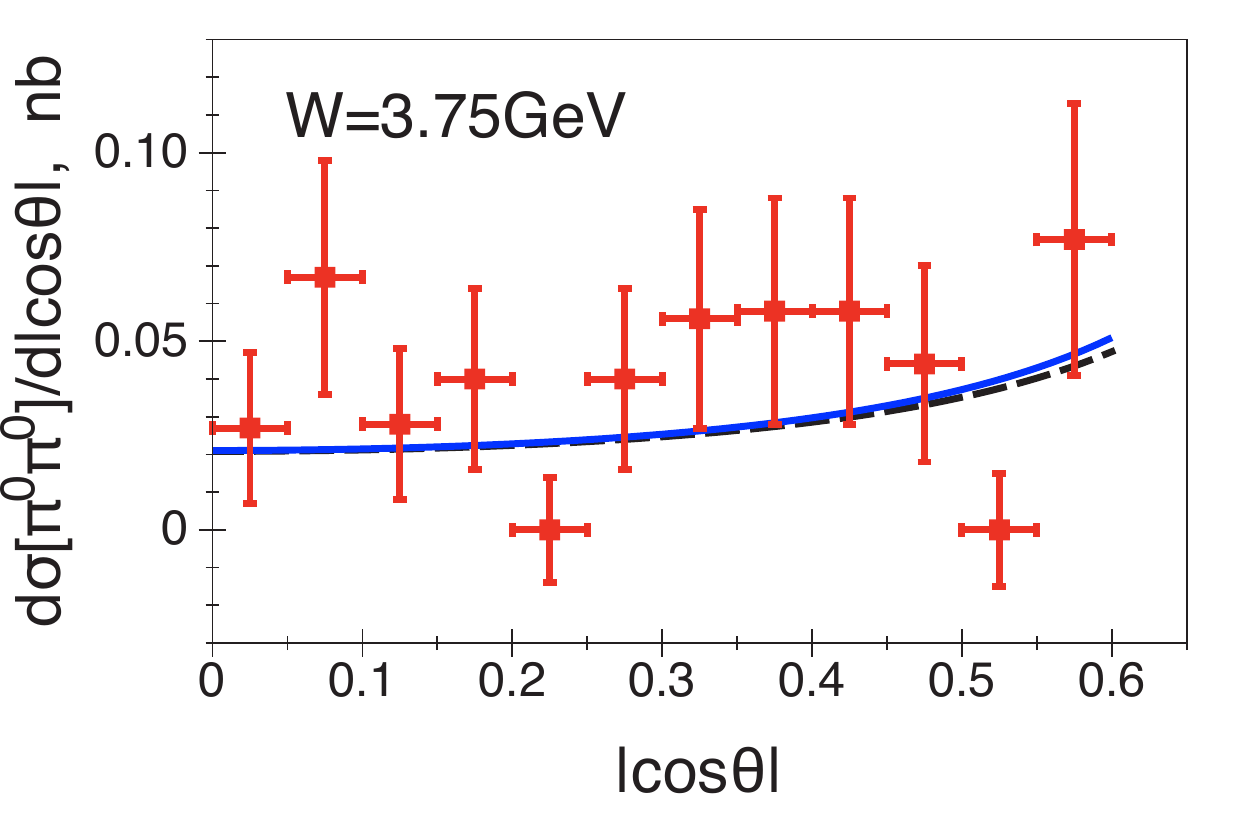} &
\includegraphics[width=1.8in]{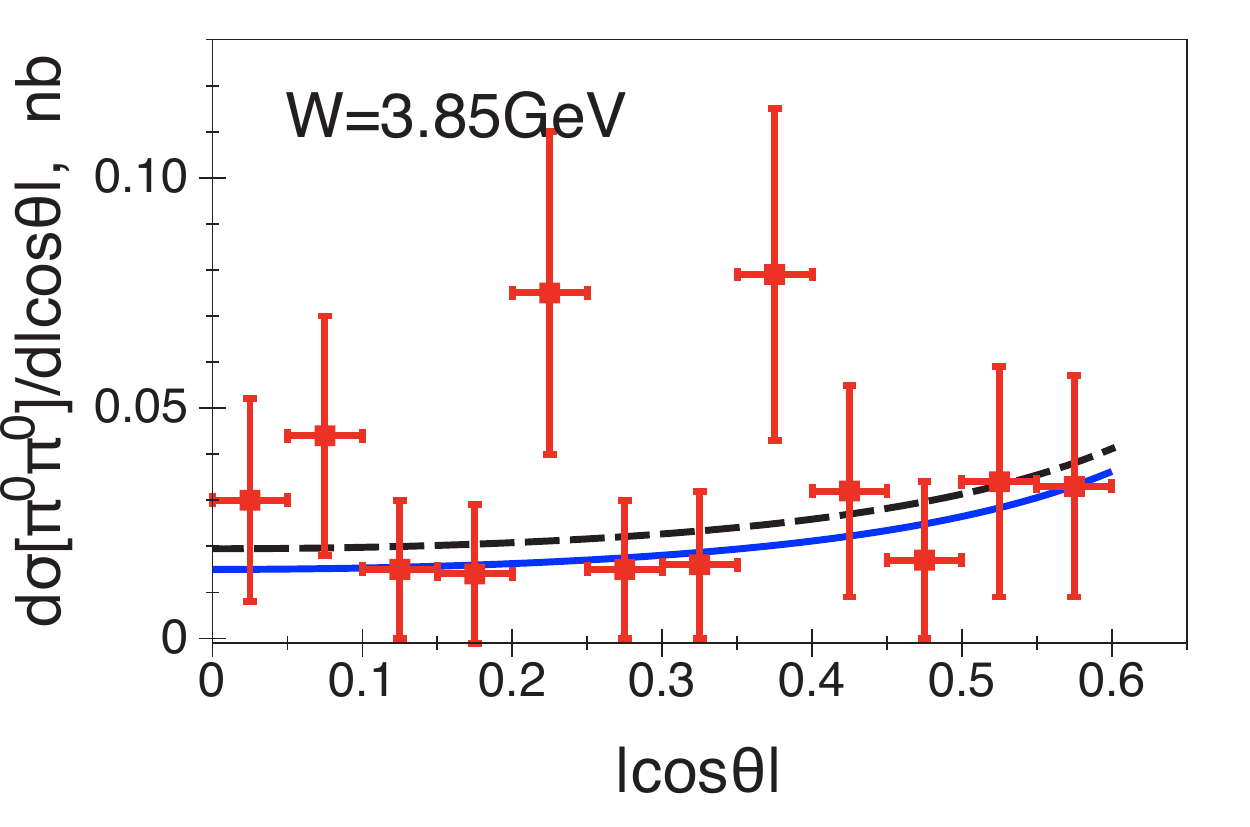} &
\includegraphics[width=1.8in]{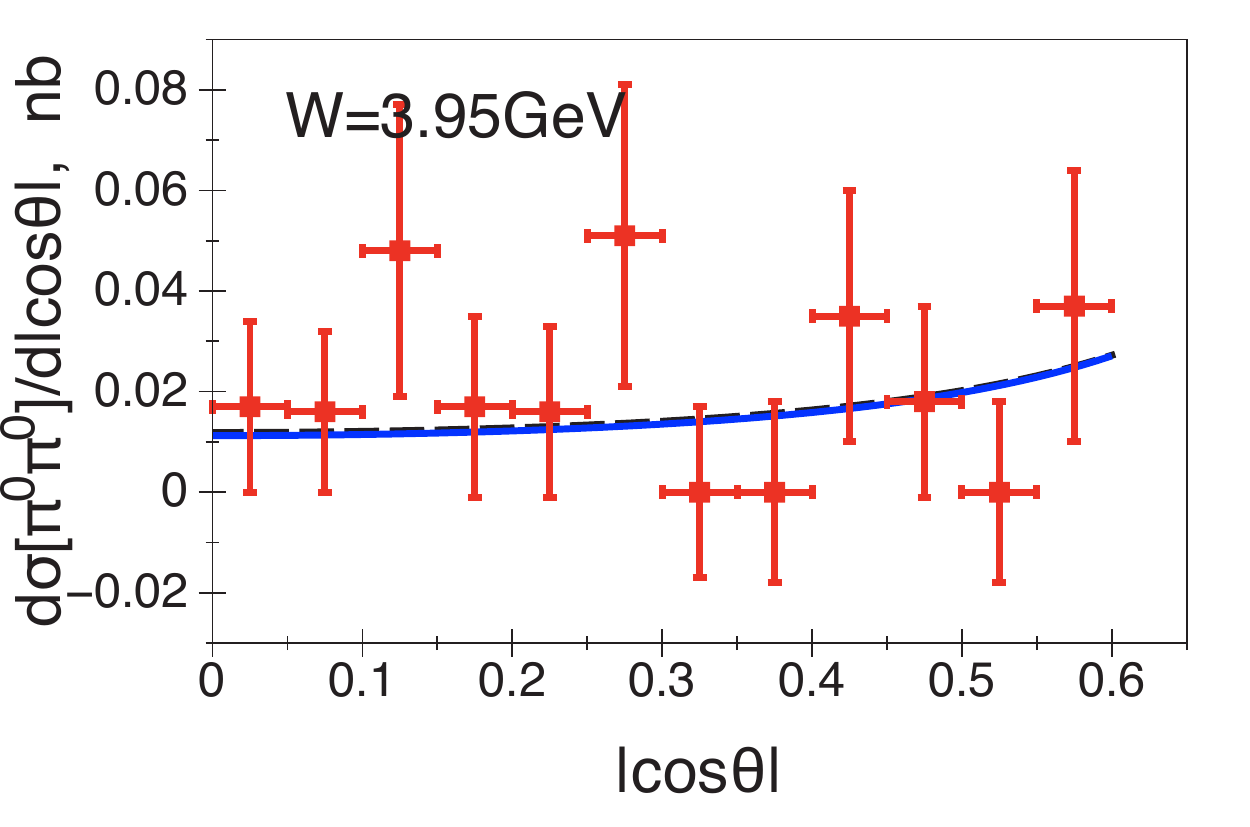}\\
&  &
\end{tabular}
\caption{Results of the fit for the neutral pion cross section, see discussion
in the text. The data are taken from Ref.\cite{Uehara:2009cka}. }%
\label{ds00BA}%
\end{figure}

If our description of the angular behavior  is consistent with the data then we can determine  
the  values of the amplitudes $B_{+\pm}^{(0)}$.
In Figs.~\ref{dspmBA} and  \ref{ds00BA} we present our  results for  the fit of
the BELLE data \cite{Nakazawa:2004gu,Uehara:2009cka} for  charged and
neutral pions, respectively.  
The results of the two-parameter fit of  both
data sets  are shown  in
Figs.\ref{dspmBA} and \ref{ds00BA} by dashed lines. 
In Table \ref{tabBppm} we present the numerical values of the  amplitudes $B_{+\pm}^{(0)}$ in for each energy $W$. 
  The data for $\pi^{0}\pi^{0}$ production  in the region $W=3.3-3.6$GeV  have a gap because of charmonium production.
In this region we can not perform the two-parameter fit of the amplitudes $B_{+\pm}^{(0)}$  and therefore corresponding 
values can not be obtained using this method.
\begin{table}[h]
\caption{ The values of the amplitudes  $B_{+\pm}^{(0)}(s)$ obtained from the
two-parameter fit of the cross sections provided  by BELLE collaboration
\cite{Nakazawa:2004gu,Uehara:2009cka}.  The reduced values of $\chi^{2}$
are computed with dof$=16$. }%
\label{tabBppm}
\begin{center}
\begin{tabular}
[c]{|l|l|l|l|}\hline
$W$,~GeV & $B_{++}^{(0)}(s,0)$ & $B_{+-}^{(0)}(s,0)$ & $\chi^{2}/$dof\\\hline
$3.05$ & $-0.47\pm0.16$ & $0.068\pm2.96$ & $2.32$\\\hline
$3.15$ & $-0.44\pm0.17$ & $0.057\pm2.68$ & $1.89$\\\hline
$3.65$ & $-0.260\pm0.34$ & $-0.003\pm1.7$ & $0.98$\\\hline
$3.75$ & $-0.22\pm0.21$ & $-0.1\pm0.38$ & $1.28$\\\hline
$3.85$ & $-0.17\pm0.12$ & $0.25\pm0.14$ & $0.67$\\\hline
$3.95$ & $-0.18\pm0.11$ & $0.15\pm0.26$ & $0.58$\\\hline
\end{tabular}
\end{center}
\end{table}
We obtain  that in  the region $W=3.05-3.65$GeV the absolute value of
$B_{+-}^{(0)}$ is  quite small.  The angular separation of the amplitudes 
 cannot  be done accurately   in this case  and  this  leads to  large   errors for  $B_{+-}^{(0)}$.  It
demonstrates that  application of the expressions in
Eqs.(\ref{dspm}) and (\ref{ds00}) for the fit of the data does not allow one
to determine  the unknown amplitudes  with a reasonable accuracy  without  some additional information.
\begin{table}[h]
\caption{Results for the amplitude $B_{++}^{(0)}(s)$ obtained from the
one-parameter fit of  BELLE data \cite{Nakazawa:2004gu} for charged pions.}%
\label{tabBpp}
\begin{center}
\begin{tabular}
[c]{|c|c|c|c|c|c|}\hline
$W,$ GeV & $3.05$ & $3.15$ & $3.25$ & $3.35$ & $3.45$\\\hline
$B_{++}^{(0)}(s,0)$ & $\scriptstyle -0.48\pm0.02$ & $\scriptstyle -0.44\pm
0.02$ & $\scriptstyle -0.39\pm0.02$ & $\scriptstyle -0.35\pm0.03$ &
$\scriptstyle -0.30\pm0.03$\\\hline
$\chi^{2}/$dof & $1.5$ & $2.3$ & $2.5$ & $2.0$ & $1.8$\\\hline
\end{tabular}
\\[0pt]%
\begin{tabular}
[c]{|c|c|c|c|c|c|c|}\hline
$W,$ GeV & $3.55$ & $3.65$ & $3.75$ & $3.85$ & $3.95$ & $4.05$\\\hline
$B_{++}^{(0)}(s,0)$ & $\scriptstyle -0.27\pm0.03$ & $\scriptstyle -0.29\pm
0.04$ & $\scriptstyle -0.27\pm0.03$ & $\scriptstyle -0.23\pm0.03$ &
$\scriptstyle -0.20\pm0.03$ & $\scriptstyle -0.15\pm0.04$\\\hline
$\chi^{2}/$dof & $1.9$ & $0.8$ & $0.8$ & $0.8$ & $0.4$ & $3.6$\\\hline
\end{tabular}
\end{center}
\end{table}

The result of the two-parameter fit allow us  to conclude that $|B_{++}^{(0)}|\gg|B_{+-}^{(0)}|$. For
larger values of energy $W\geq3.75$GeV the  values  of $B_{+-}^{(0)}$ in Table \ref{tabBppm} are
already quite large but the error bars  are also large.  Guided by these observations  we consider a simple model  assuming
\begin{equation}
B_{+-}^{(0)}\simeq0.
\end{equation}
In this case one can perform a more simple one-parameter fit of the data for  charged pion
 in order to define the values of $B_{++}^{(0)}$.  Then  these
values  can be used for computation of the  cross section for neutral pions.  Comparison of  the  cross section  with the
data allows one to check the consistency of the model.   
The obtained results are also shown in Figs.~\ref{dspmBA} and \ref{ds00BA}  by solid line. 
We observe that the difference between one- and two-parameter fits
in this case is small.  

In Table~\ref{tabBpp} we present the results for the values $B_{++}^{(0)}$
defined by the one-parameter fit. One can see that the quality of the fit is
much better.  Such fit  has better $\chi^{2}$ and obtained values have relatively small  error bars.
However this estimates are no longer unbiased and the small error bars can
also arise due to the effect of underfitting\footnote{The author thanks to
M.Distler for the discussion of this moment.}.  Nevertheless we consider this 
this model is interesting providing a simple scenario which is consistent with the
data. 

Substituting the values $B_{++}^{(0)}=-0.48$, $B_{+-}^{(0)}=0$  in Eq.~(\ref{pmW30-005}) we obtain
\begin{align}
\frac{d\sigma^{\pi^{+}\pi^{-}}}{d\cos\theta}(3.05\text{GeV},\cos\theta &
=0.05)=0.879|B_{++}^{(0)}|^{2}-0.259B_{++}^{(0)}+0.0283\\
&  =0.19_{B_{++}^{2}}+0.12_{B_{++}}+0.0283\simeq0.34,
\end{align}
where the subscripts indicate the corresponding contribution in the upper
line.  We observe that the interference 
contribution (linear in $B_{++}^{(0)}$) provides quite sizable numerical effect of order $30\%$. 
 The similar calculation  for the neutral channel yields
\begin{equation}
\frac{d\sigma^{\pi^{0}\pi^{0}}}{d\cos\theta}=0.440|B_{++}^{(0)}|^{2}%
+0.004=0.096+0.004=0.1.
\end{equation}
In this case the interference  depends only from the small amplitude $B_{+-}^{(0)}$
and therefore corresponding numerical  effect is negligible.  Hence  we  conclude that the sizable
deviation from the simple value $1/2$ for the cross section ratio (\ref{Ratio}) can
be only obtained due to the  linear  contribution with the amplitude $B_{++}^{(0)}$  in Eq.(\ref{dspm}).

\begin{figure}[ptb]
\centering
\includegraphics[width=2.7in]{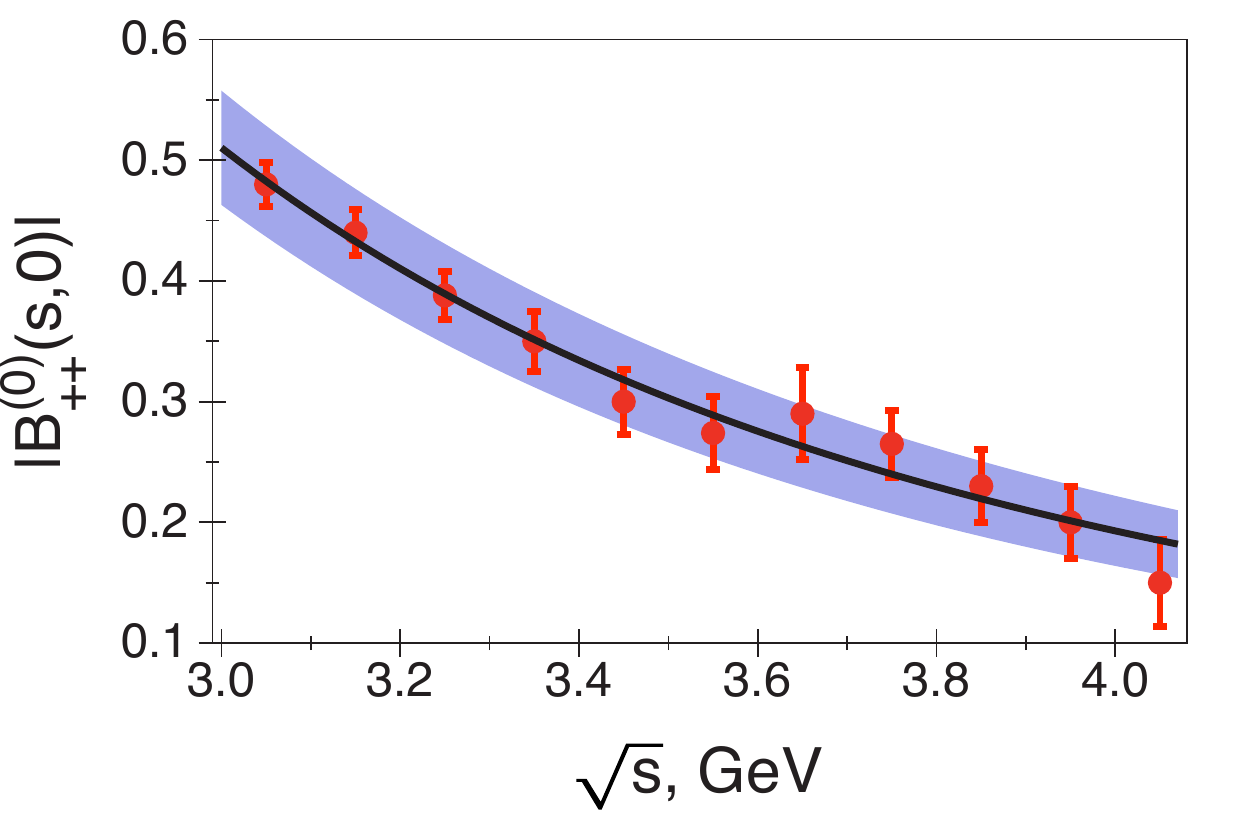}\caption{The amplitude $B_{++}%
^{(0)}(s,0)$ as a function of energy $s$. }%
\label{B0ppfit}%
\end{figure}

Performing an empirical  power fit of results in Table \ref{tabBpp} we obtain
\begin{equation}
B_{++}^{(0)}(s,0)=\left(  \frac{s_{0}}{s}\right)  ^{a},~\text{with }%
\ s_{0}=6.0\pm0.3,~a=1.7\pm0.15.\label{Bppfit}%
\end{equation}
This result is shown in Fig.\ref{B0ppfit}. 
Using  SCET framework we obtained  in Sec.\ref{factSCET}   that this amplitude is suppressed in the limit $s\rightarrow\infty$ as  
$B_{++}^{(0)}(s){\sim}{\Lambda^{4}}/{s^{2}}$.   The obtained empirical value of the power exponent $a$ is smaller but not far from  
this  expectation.  The large 
value of the effective scale  $s_{0}$ in Eq.(\ref{Bppfit})  is necessary  in order  to  have a large normalization of the cross section.   
We expect  the  nonperturbative dynamical   scale defining  the behavior of the amplitude  $B_{++}^{(0)}(s)$ is  
the hard-collinear scale $\mu_{hc}\sim \Lambda Q$ where  $\Lambda$ can be interpreted as a typical value of  soft particles momenta.
We can rewrite the empirical formula in Eq.(\ref{Bppfit})  as  
$B_{++}^{(0)}(s,0)=\left(  \frac{s'_{0}}{\Lambda W }\right)  ^{2a}$ where  $W\equiv \sqrt{s}$. 
Taking  $\Lambda\simeq 300-400$MeV  we obtain that the value of the intrinsic scale  is $s'_0=\Lambda\sqrt{s_{0}}\simeq 0.73-0.98$~GeV$^{2}$. 

\begin{figure}[ptb]
\centering
\includegraphics[width=2.7in]{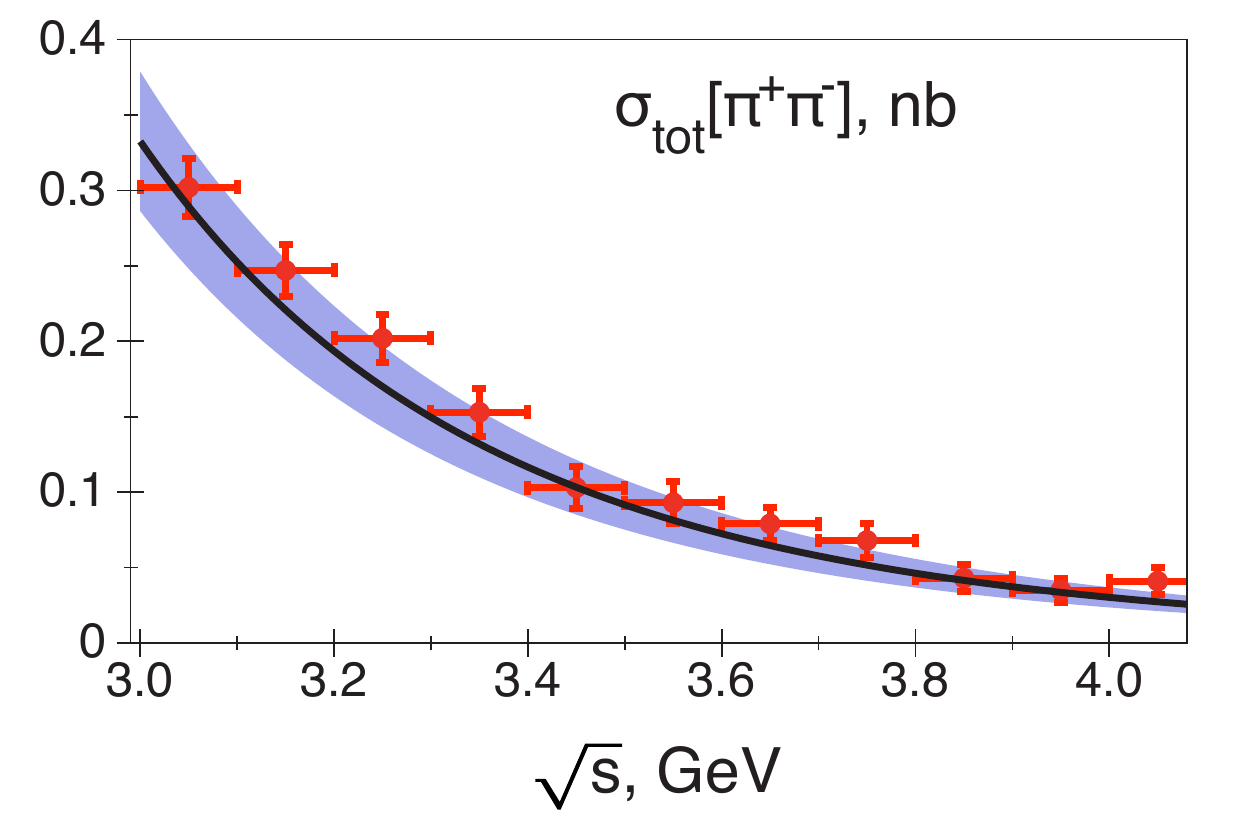}
\includegraphics[width=2.7in]{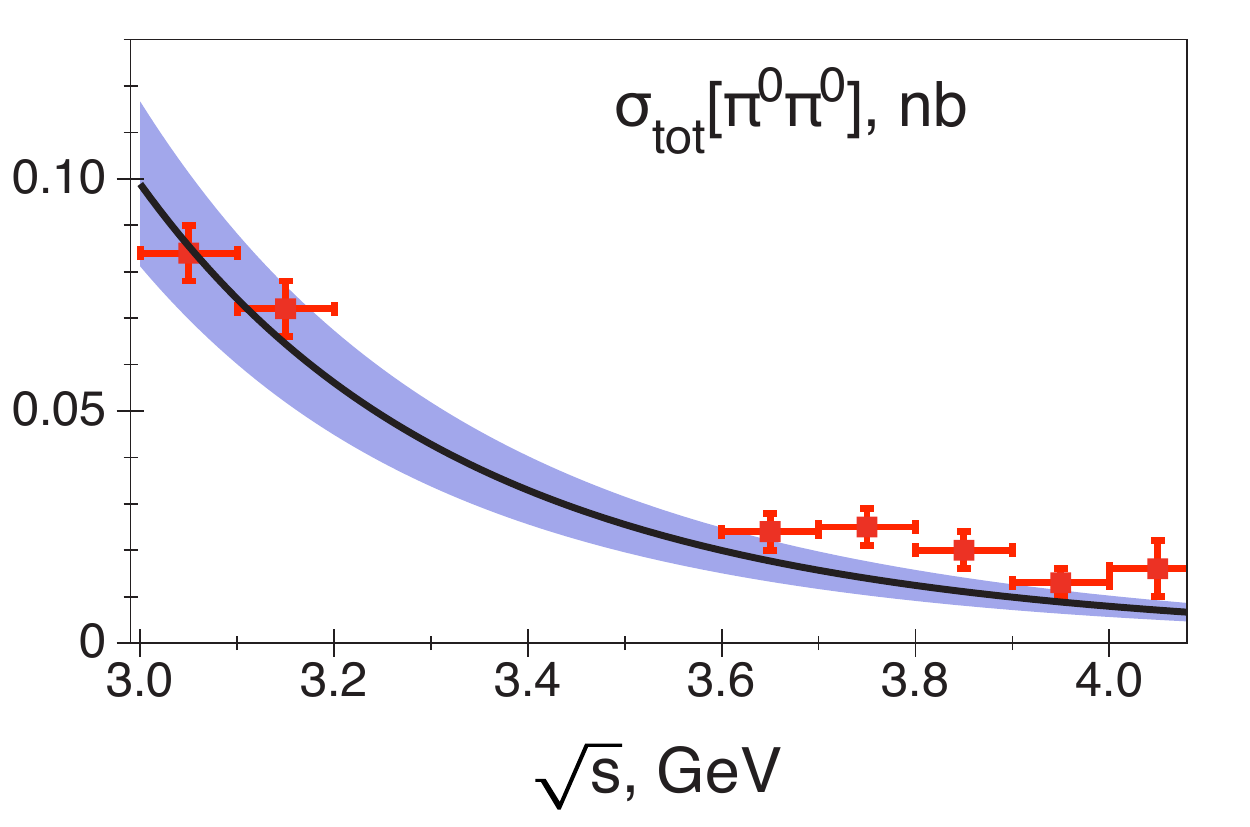}\caption{The total cross
sections as a functions of energy. }%
\label{sigma-tot}%
\end{figure}

\begin{figure}[ptb]
\centering
\includegraphics[width=3.5in]{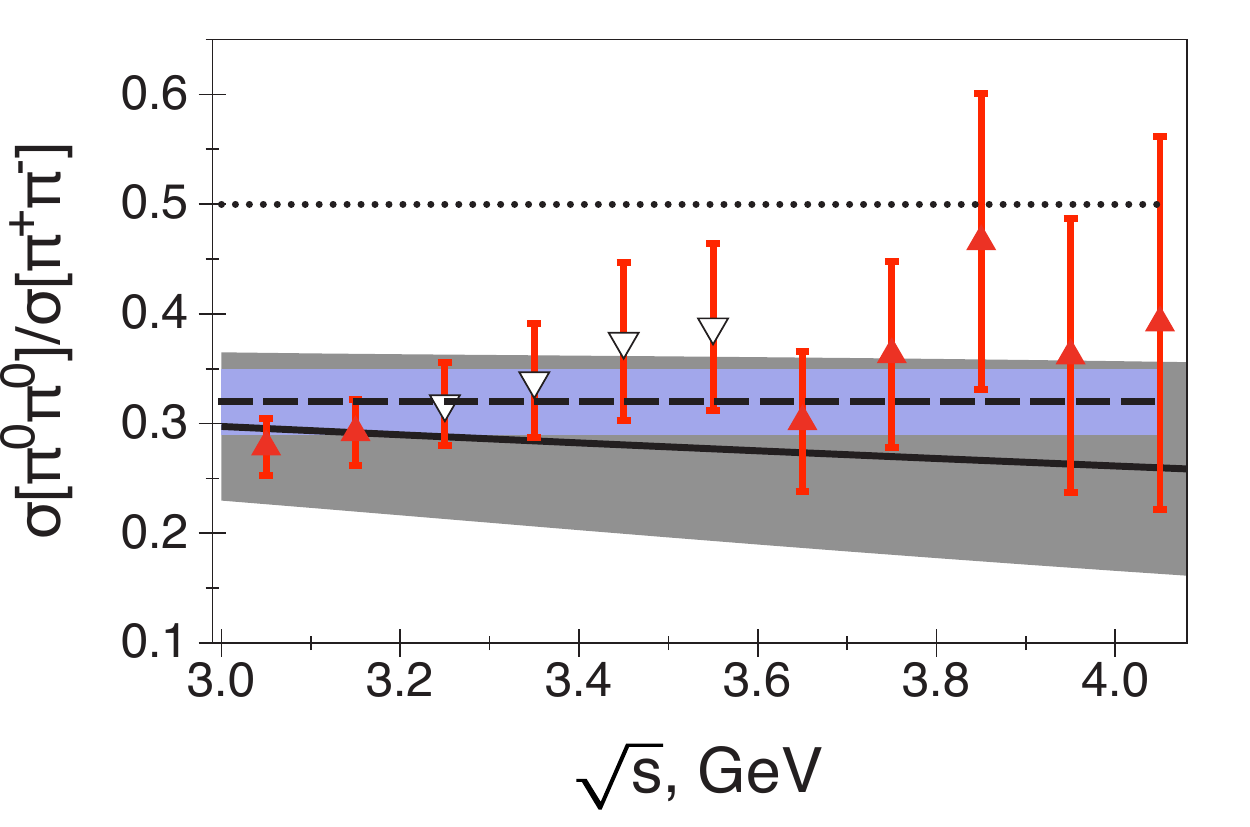}\caption{The ratio of the cross
sections $\sigma_{tot}^{\pi^{0}\pi^{0}}/\sigma_{tot}^{\pi^{+}\pi^{-}}$ as a
function of the energy. The data are taken from Ref.\cite{Uehara:2009cka} .
The dashed line shows the experimental fit with 1$\sigma$ error bands. The
solid line shows the computed ratio with the 1$\sigma$ error bands. }%
\label{R002pm}%
\end{figure}

 Eq. (\ref{Bppfit}) for $B_{++}^{(0)}(s)$ allows one one to compute the total
cross sections and their ratio. Corresponding results are  are shown in Figs.~\ref{sigma-tot} and \ref{R002pm}. 
 From  Fig.~\ref{sigma-tot} it is
seen that  obtained $\sigma_{tot}^{\pi^{0}\pi^{0}}$ is somewhat smaller than
the experimental values. As a result the  ratio (solid line in Fig.\ref{R002pm}) slightly decreases  for larger energy $W$. 
The gray area around the solid line shows $1\sigma$ bands
obtained from  the errors of the parameters $s_{0}$ and $a$ in
Eq.(\ref{Bppfit}).  The  fit of the experimental data yields
$R=0.32\pm0.03$ \cite{Uehara:2009cka}.\footnote{The results in the  region
$W=3.3-3.6$~GeV (plotted with open triangles) are not used for the fit in
Ref\cite{Uehara:2009cka}.} In Fig.\ref{R002pm} this result is shown by dashed
line with 1$\sigma$ error bands.  

\subsection{Phenomenological analysis  using  CZ-model of  pion DA}

Using  CZ-model  defined in Eq.(\ref{setCZ}) and fixing a small value of the
renormalization scale $\mu_{R}=1.3$GeV one obtains a larger contribution from
the leading-twist part. In that case the numerical values  of  the differential
cross sections at $W=3.05$~GeV and $\cos\theta=0.05$  read (c.f. with the
corresponding Eqs.(\ref{LTpmW30-005})-(\ref{00W30-005}))
\begin{align}
&  \left(  \frac{d\sigma^{\pi^{+}\pi^{-}}}{d\cos\theta}\right)  _{\text{LT}}
=0.18, \quad\left(  \frac{d\sigma^{\pi^{+}\pi^{-}}}{d\cos\theta}\right)
_{\text{exp}}=0.312\pm0.039,\label{LTpmCZ}\\
&  \left(  \frac{d\sigma^{\pi^{0}\pi^{0}}}{d\cos\theta}\right) _{\text{LT}%
}=0.009, \quad\left(  \frac{d\sigma^{\pi^{0}\pi^{0}}}{d\cos\theta}\right)
_{\text{exp}}=0.078\pm0.025,\label{LT00CZ}\\
\frac{d\sigma^{\pi^{+}\pi^{-}}}{d\cos\theta}  &  =0.88|B_{++}^{(0)}%
|^{2}+0.88|B_{+-}^{(0)}|^{2}-0.65B_{++}^{(0)}-0.45B_{+-}^{(0)}%
+0.18.\label{dspmCZ}\\
\frac{d\sigma^{\pi^{0}\pi^{0}}}{d\cos\theta}  &  =0.44|B_{++}^{(0)}%
|^{2}+0.44|B_{+-}^{(0)}|^{2}-0.16B_{+-}^{(0)}+0.014.\label{ds00-CZ}%
\end{align}

\begin{figure}[ptb]
\begin{center}%
\begin{tabular}
[c]{cc}%
\includegraphics[width=2in]{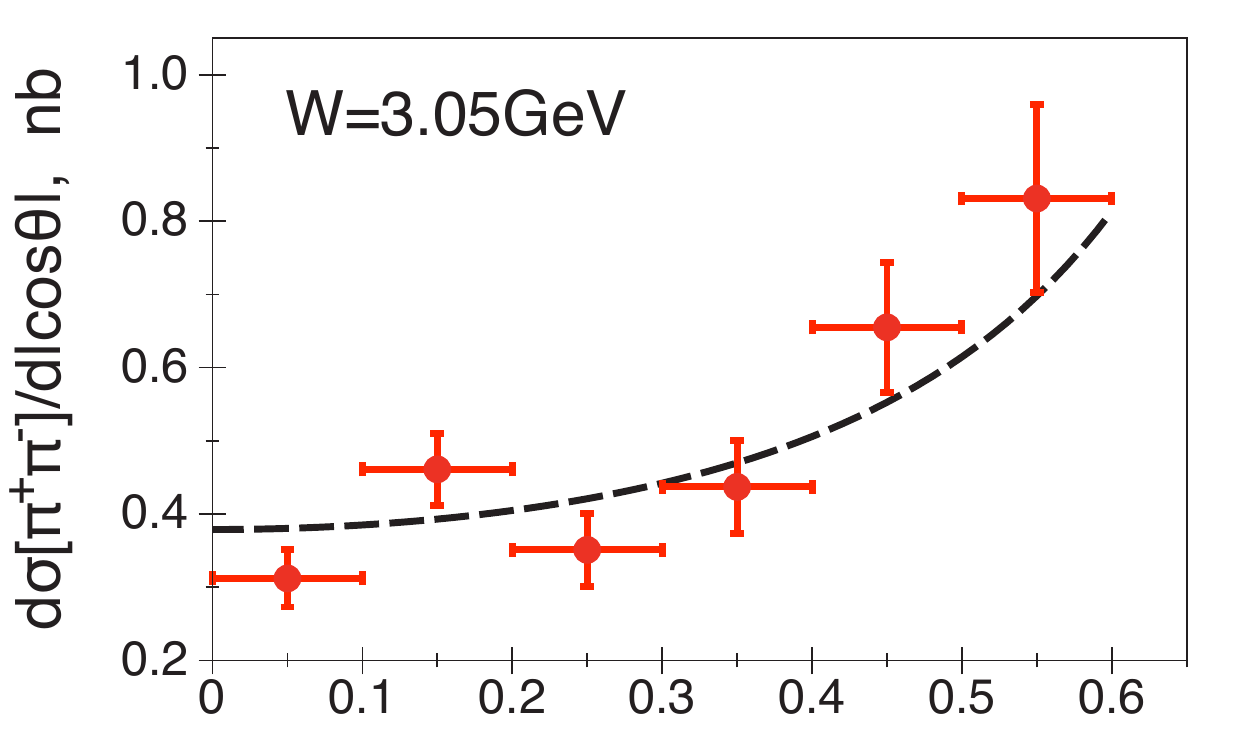} &
\includegraphics[width=2in]{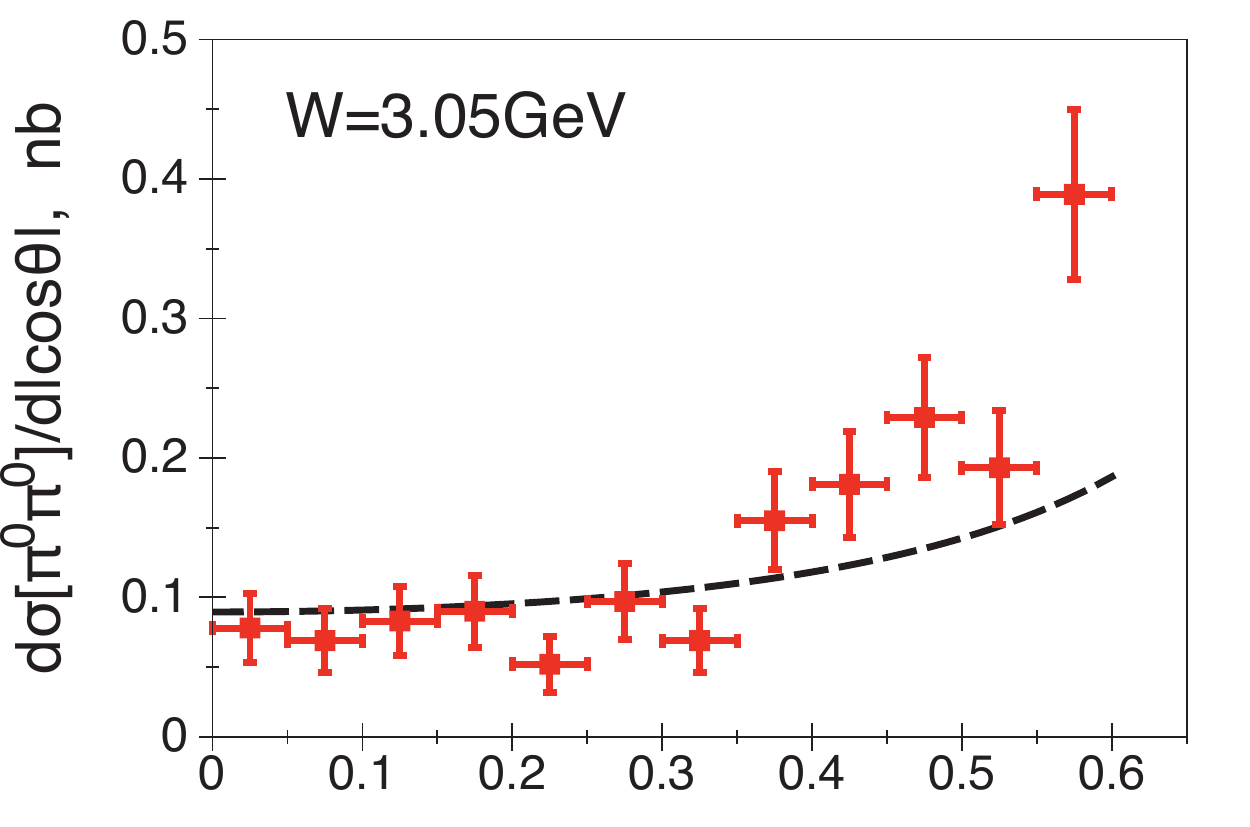}\\
\includegraphics[width=2in]{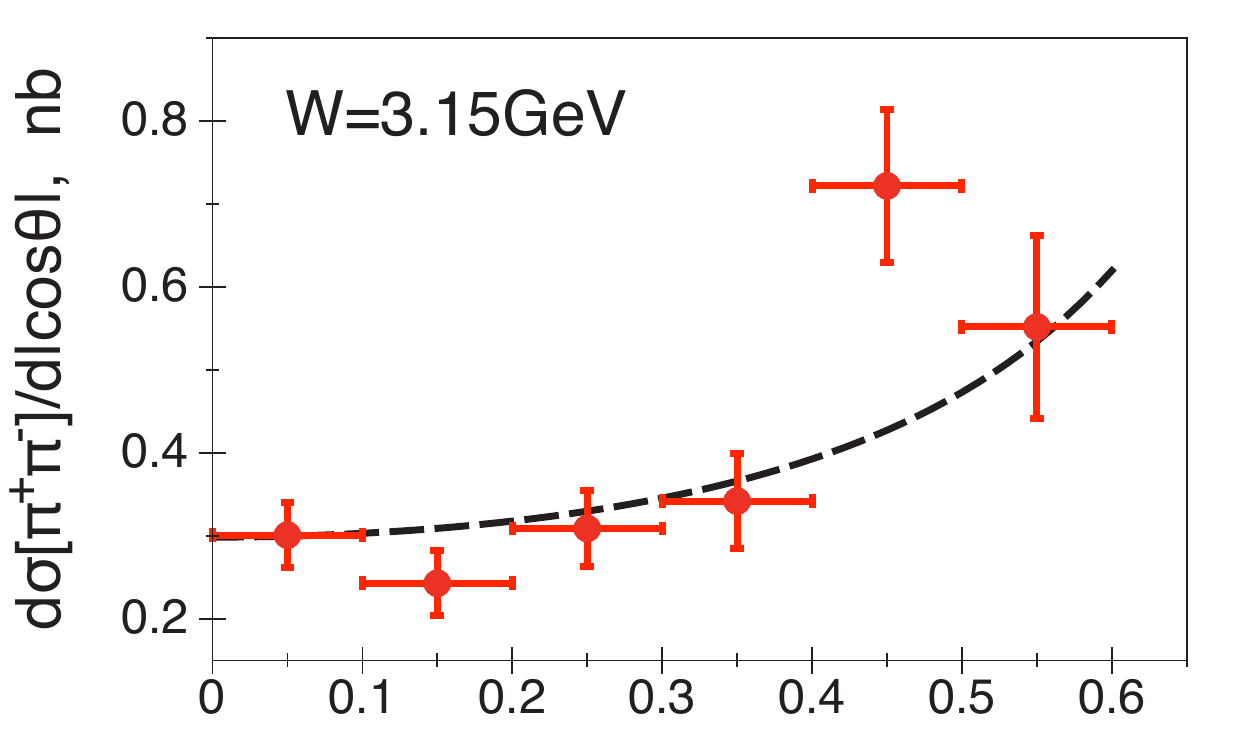} &
\includegraphics[width=2in]{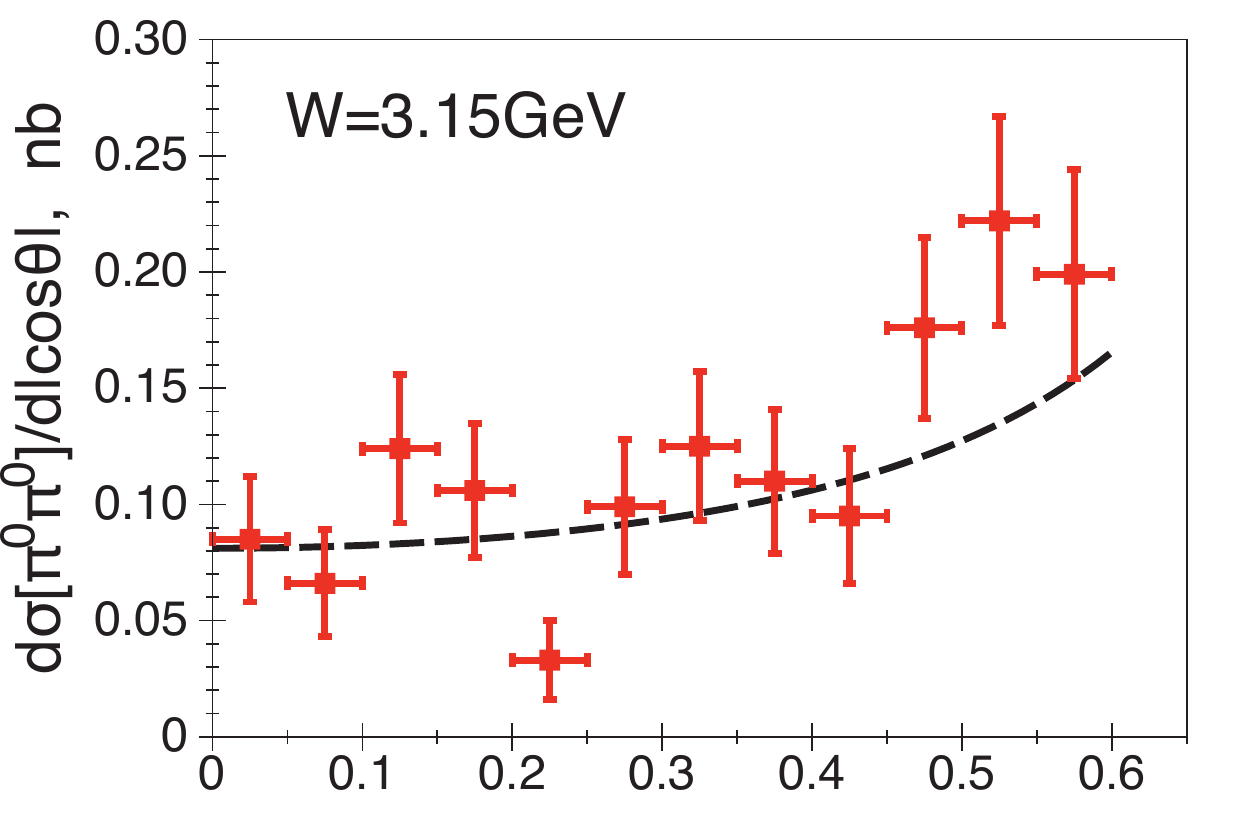}\\
\includegraphics[width=2in]{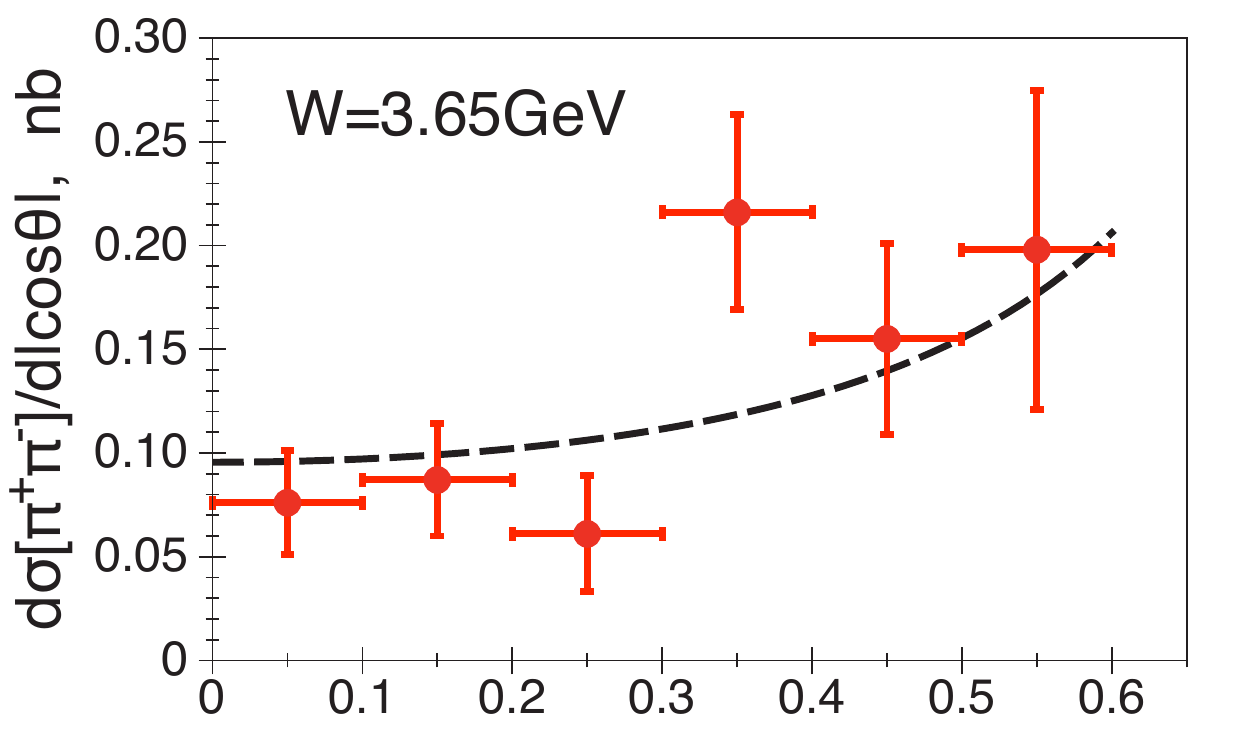} &
\includegraphics[width=2in]{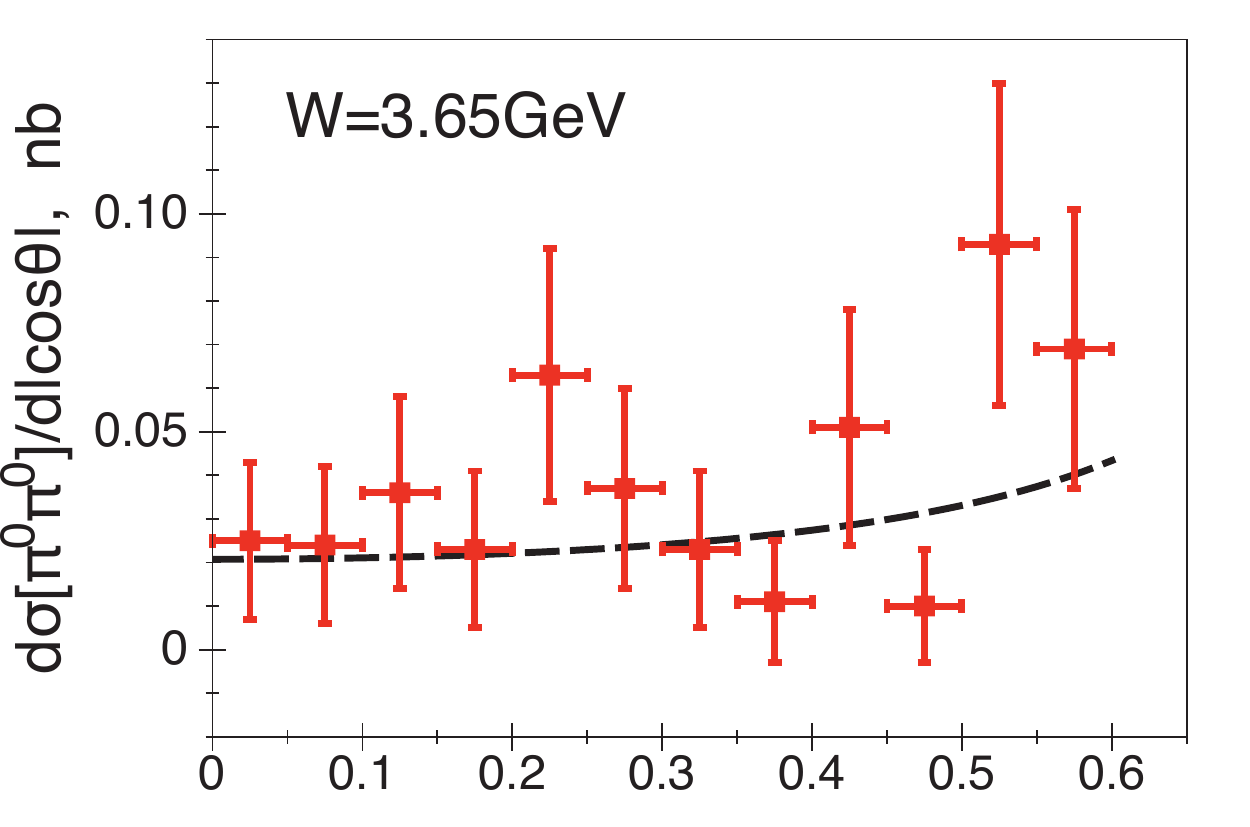}\\
\includegraphics[width=2in]{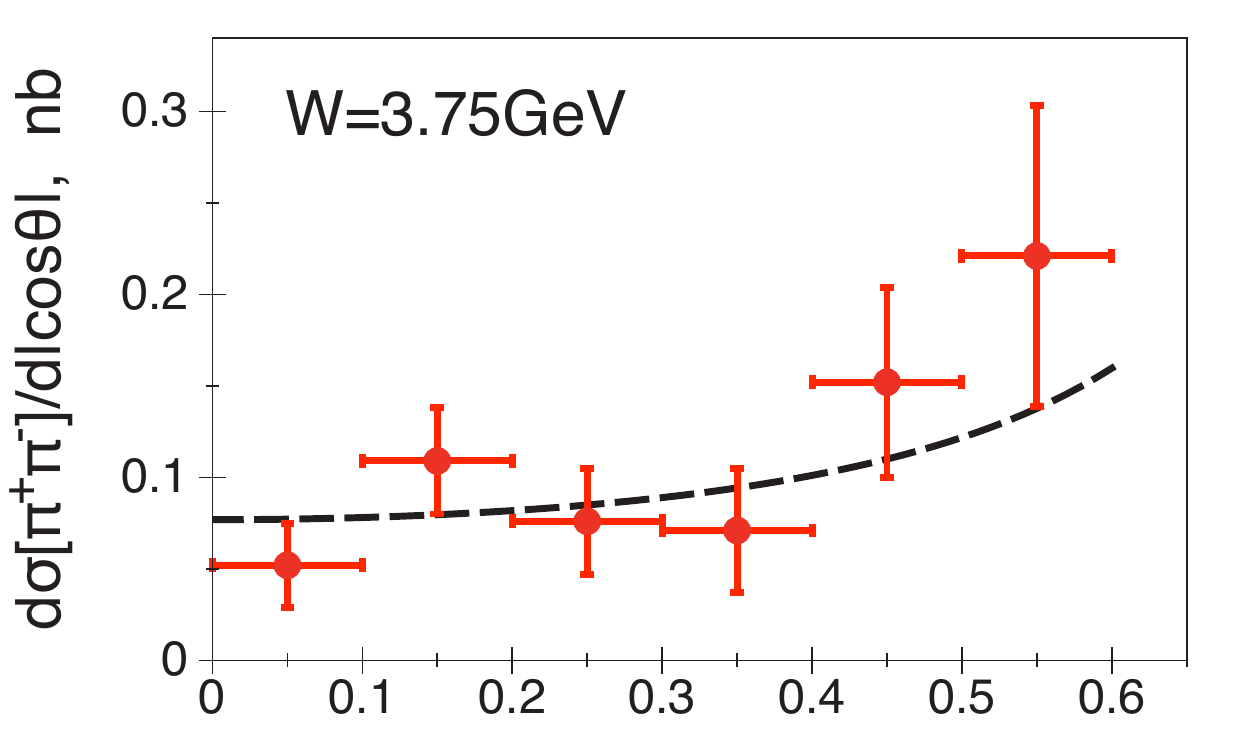} &
\includegraphics[width=2in]{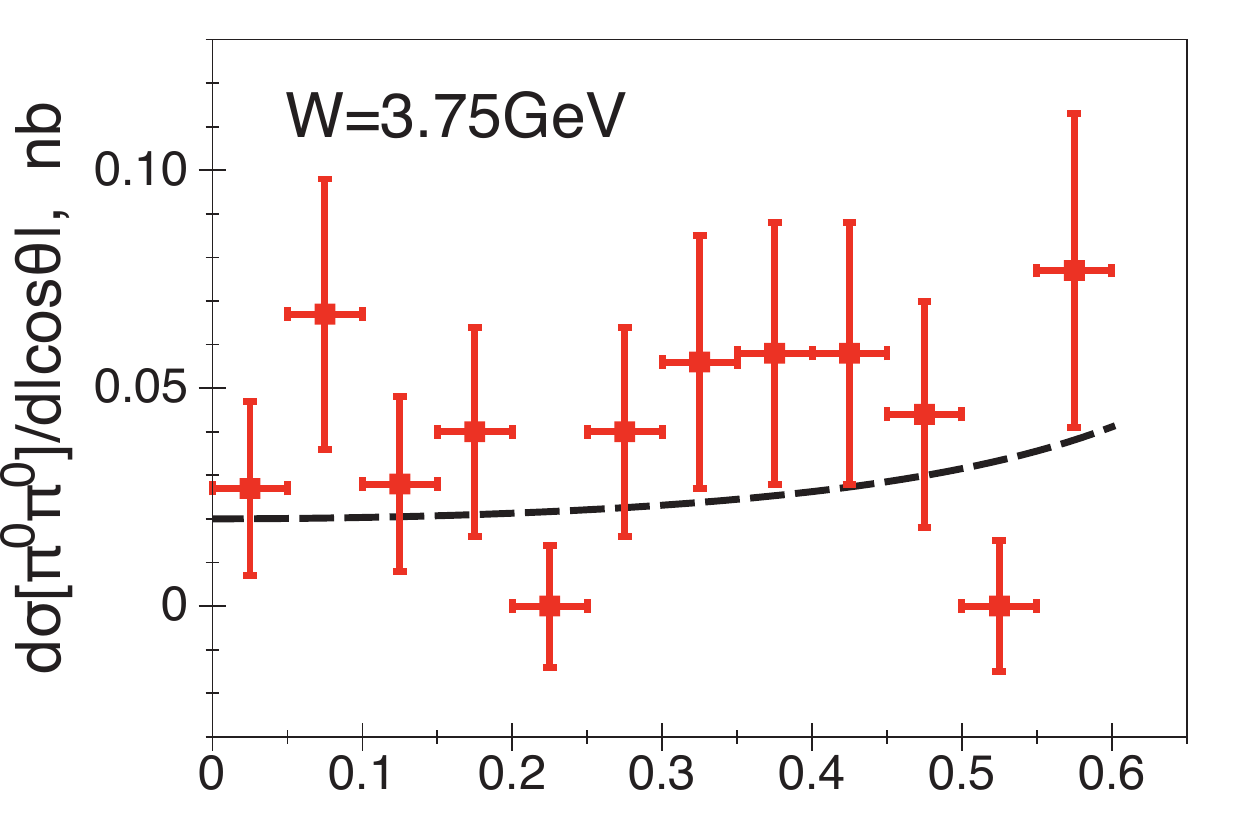}\\
\includegraphics[width=2in]{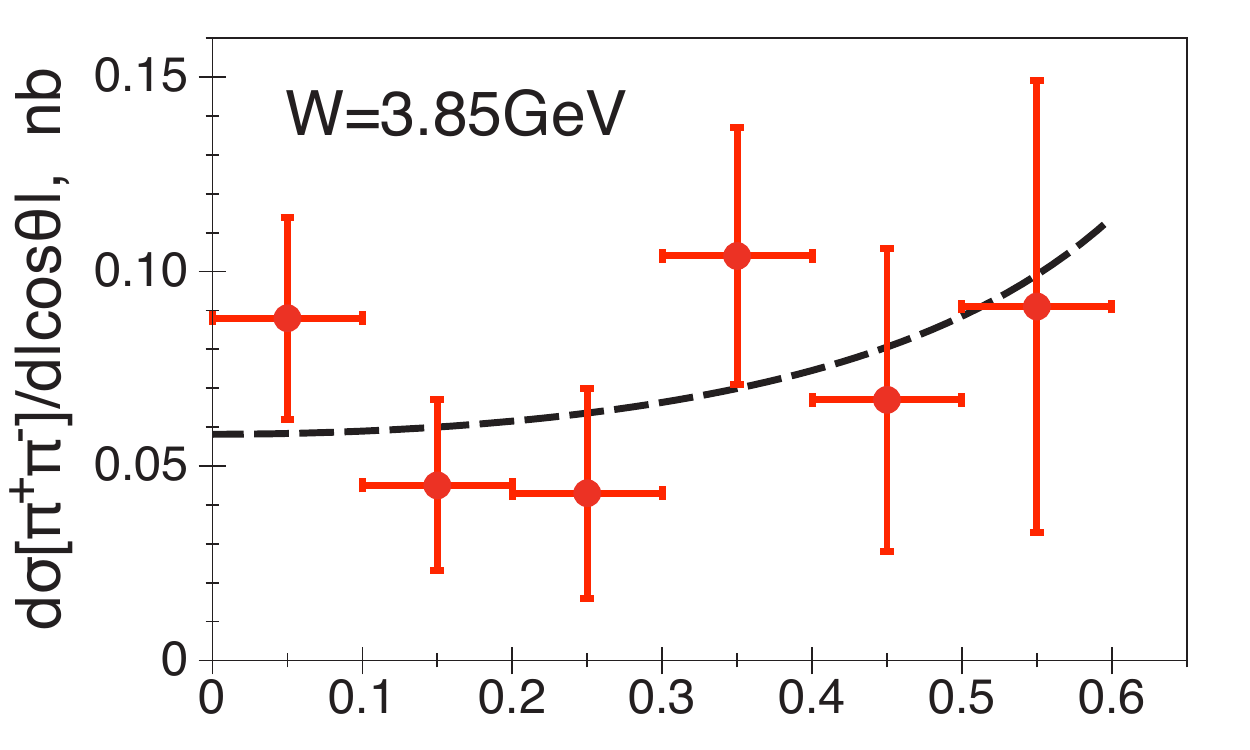} &
\includegraphics[width=2in]{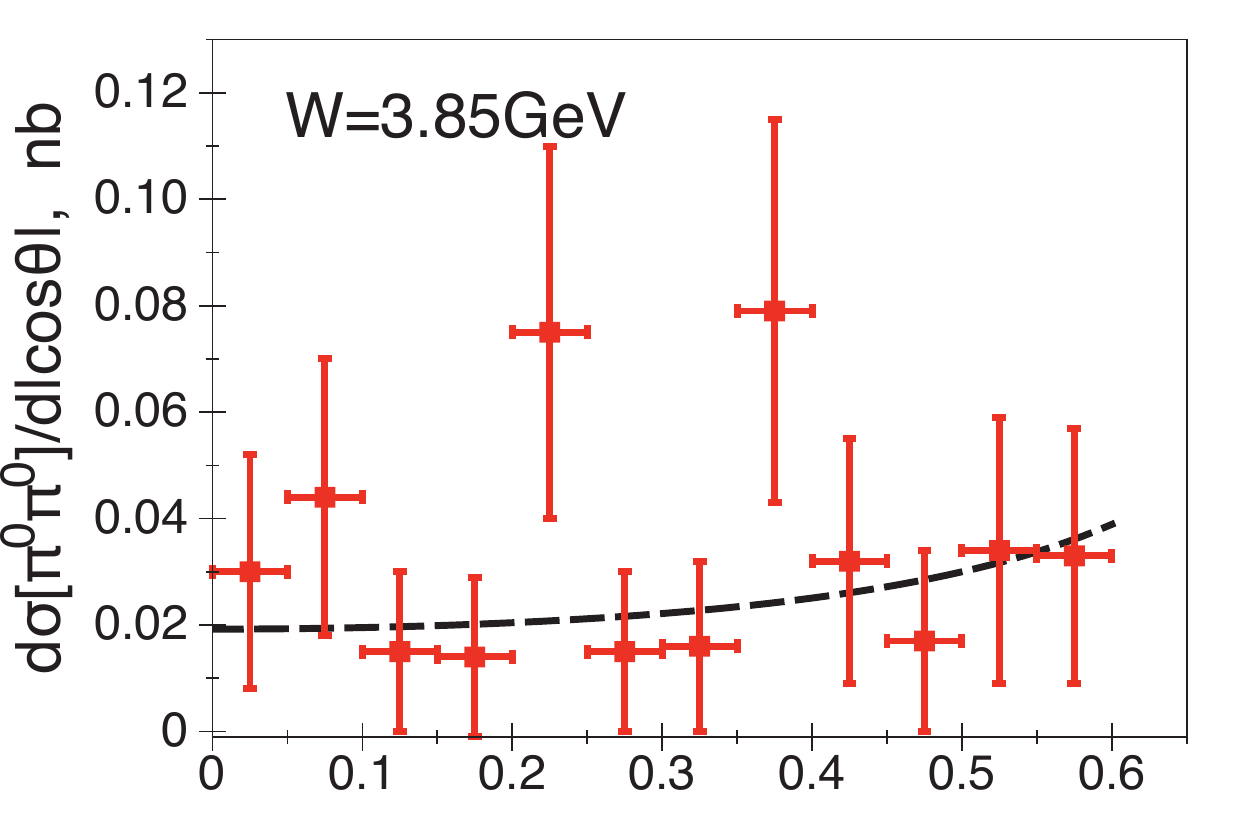}\\
\includegraphics[width=2in]{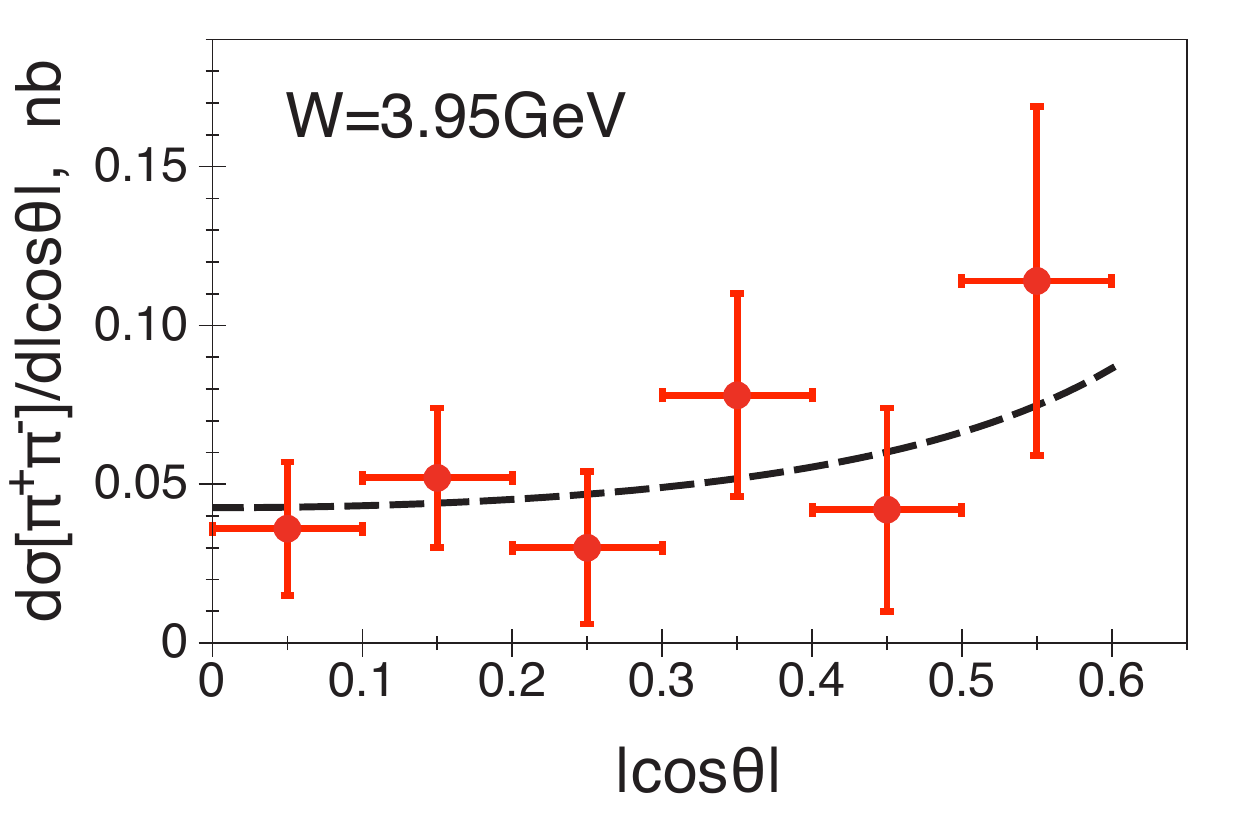} &
\includegraphics[width=2in]{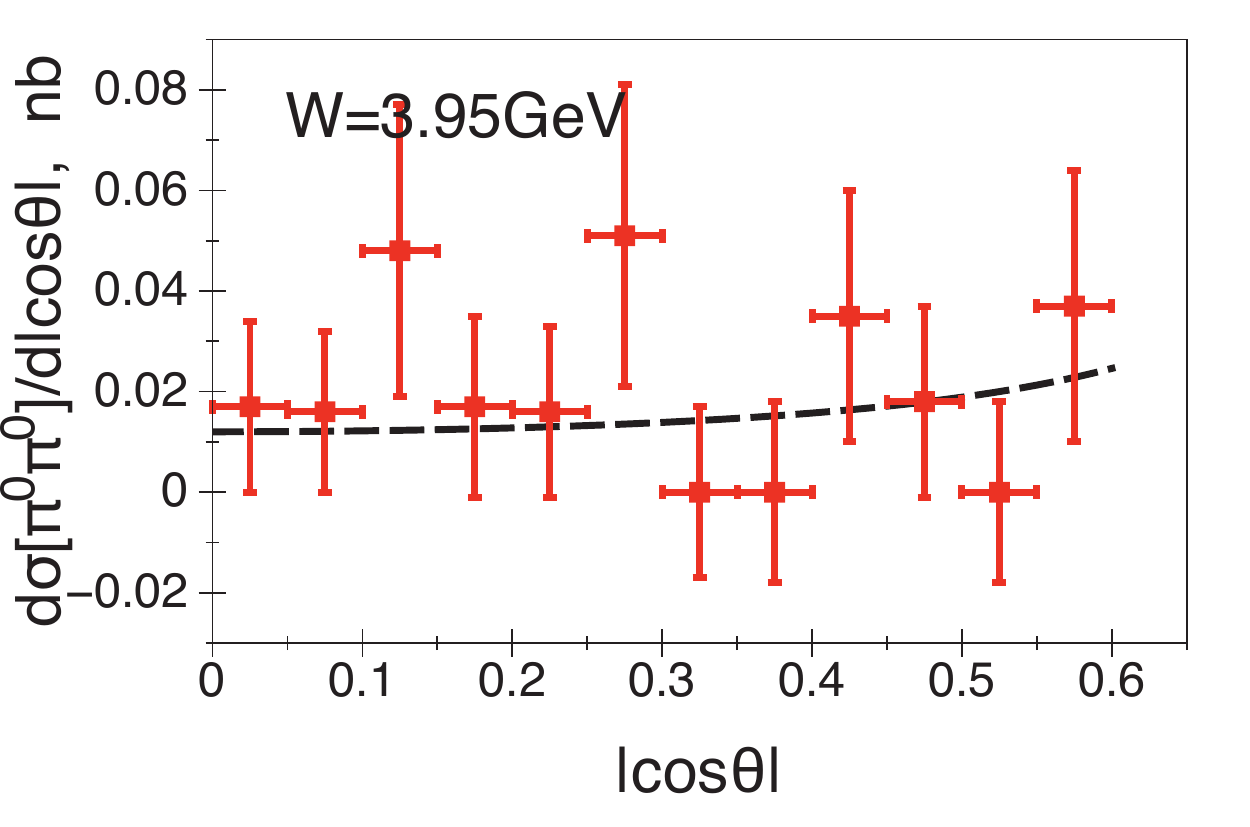}
\end{tabular}
\end{center}
\caption{Results of the fit of BELLE data
\cite{Nakazawa:2004gu,Uehara:2009cka} for the differential cross sections at
different energies. }%
\label{ds-CZ}%
\end{figure}

\begin{figure}[ptb]
\begin{center}
\includegraphics[width=2.6in]{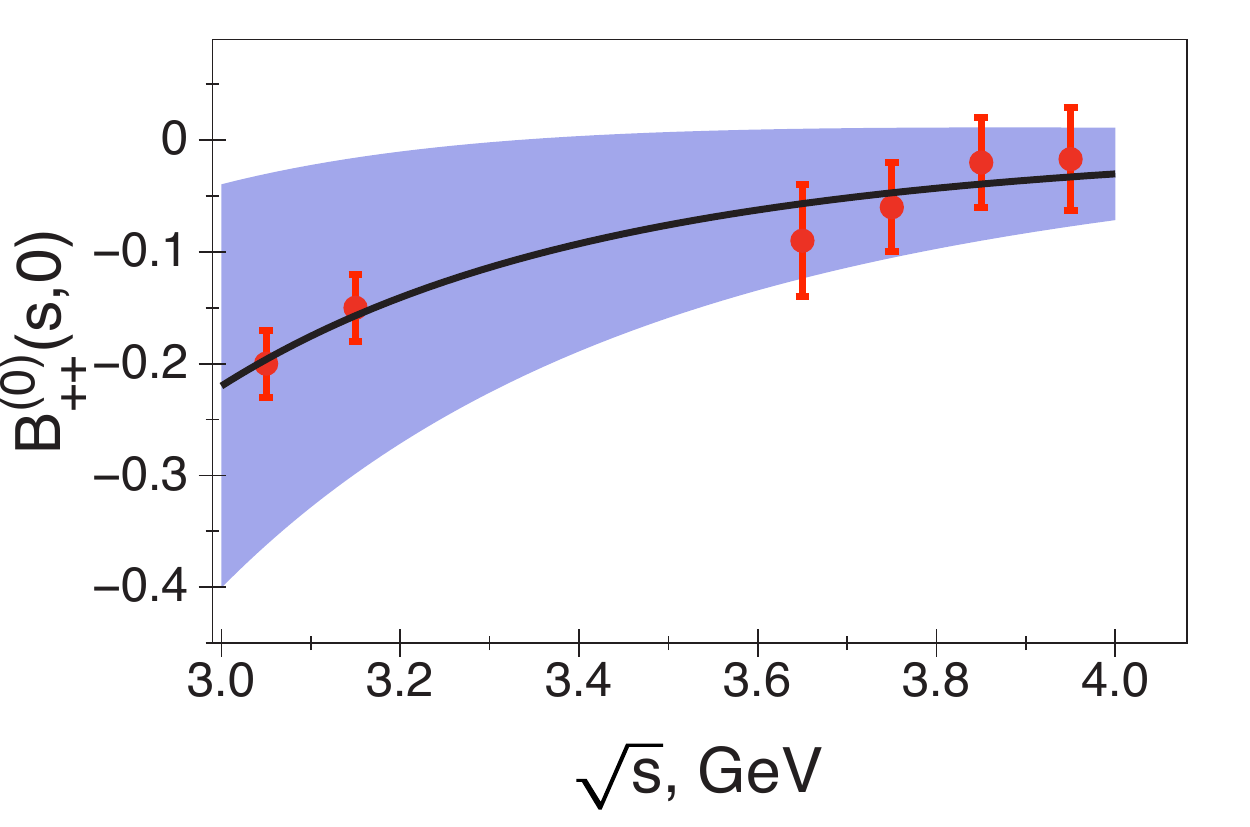}
\includegraphics[width=2.6in]{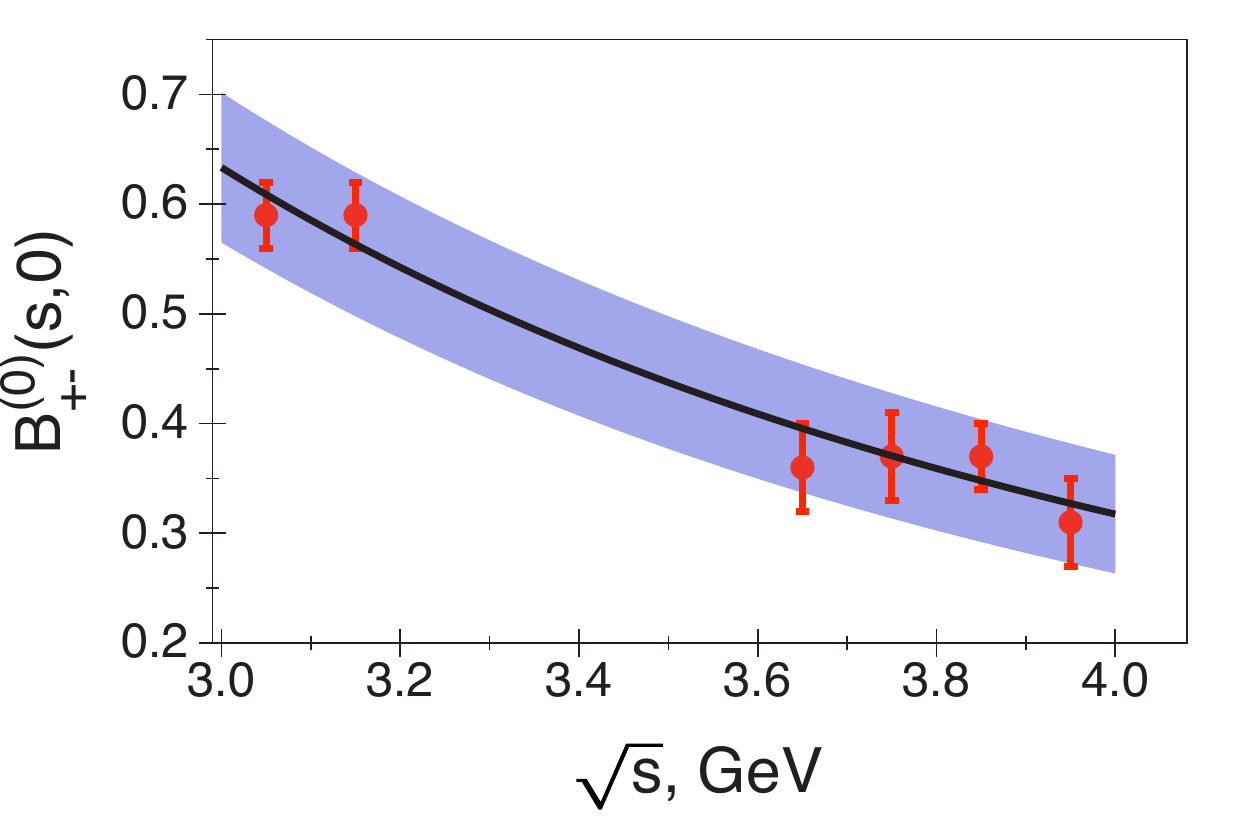}
\end{center}
\caption{The amplitudes $B_{++}^{0}(s,0)$ and $B_{+-}^{0}(s,0)$ as a functions
of energy obtained from the fit of the differential cross sections. }%
\label{B0ppmCZ}%
\end{figure}

\begin{figure}[ptb]
\begin{center}
\includegraphics[width=2.6in]{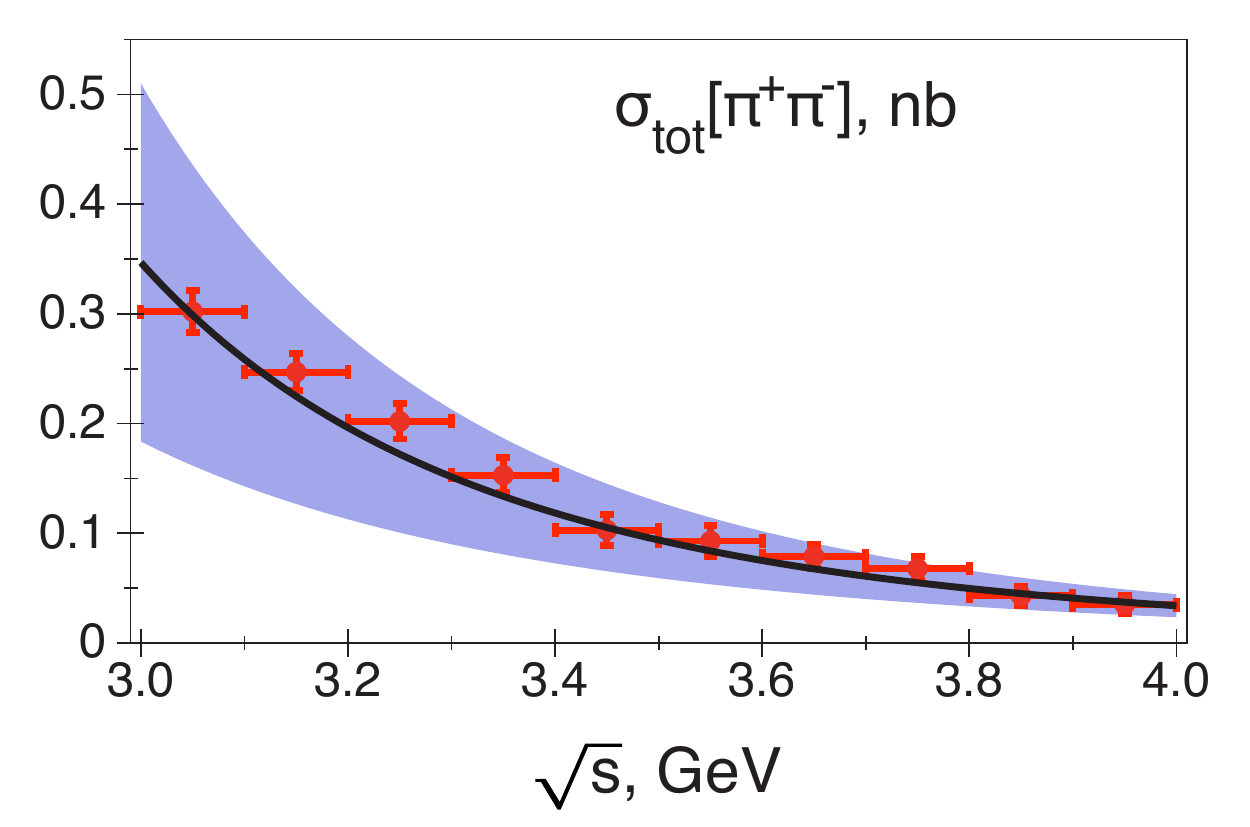}
\includegraphics[width=2.6in]{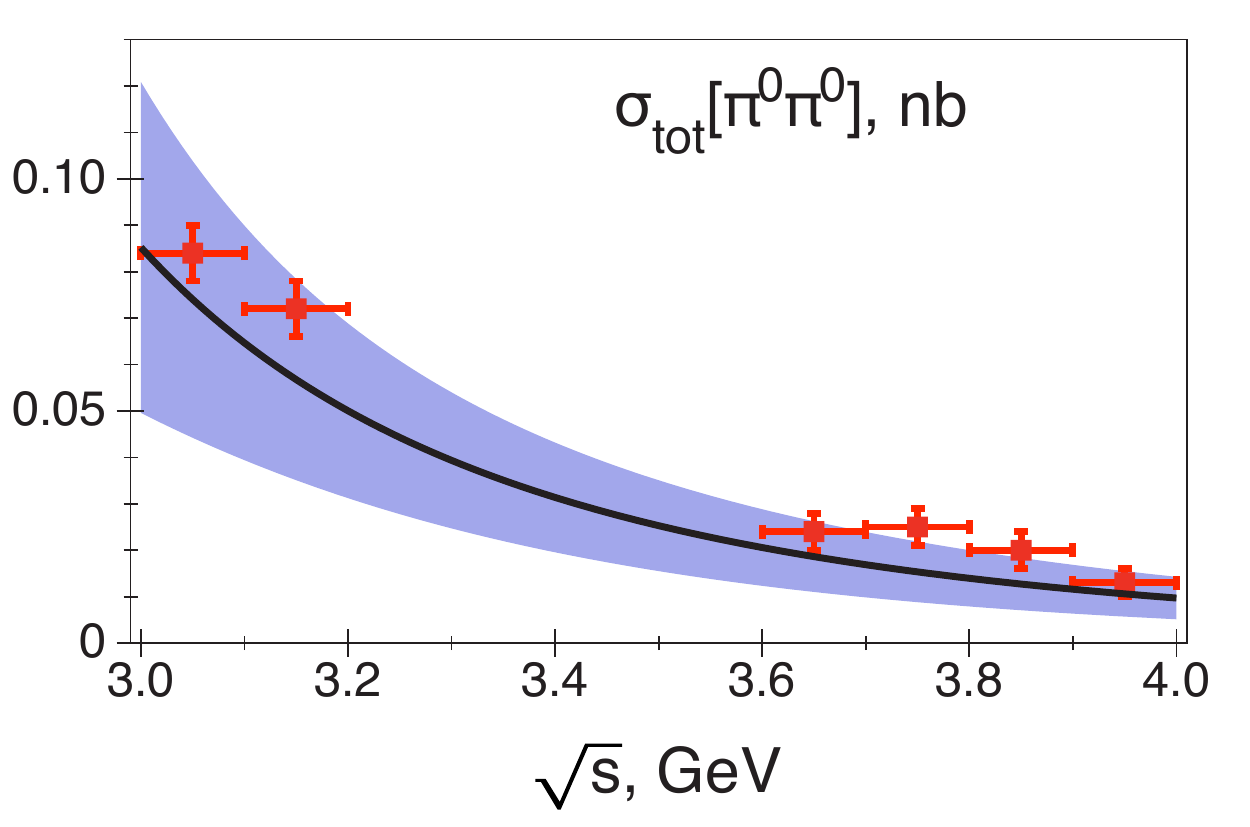}
\end{center}
\caption{The total cross sections as functions of energy. }%
\label{sigmaCZ-tot}%
\end{figure}

\begin{figure}[ptb]
\begin{center}
\includegraphics[width=3.5in]{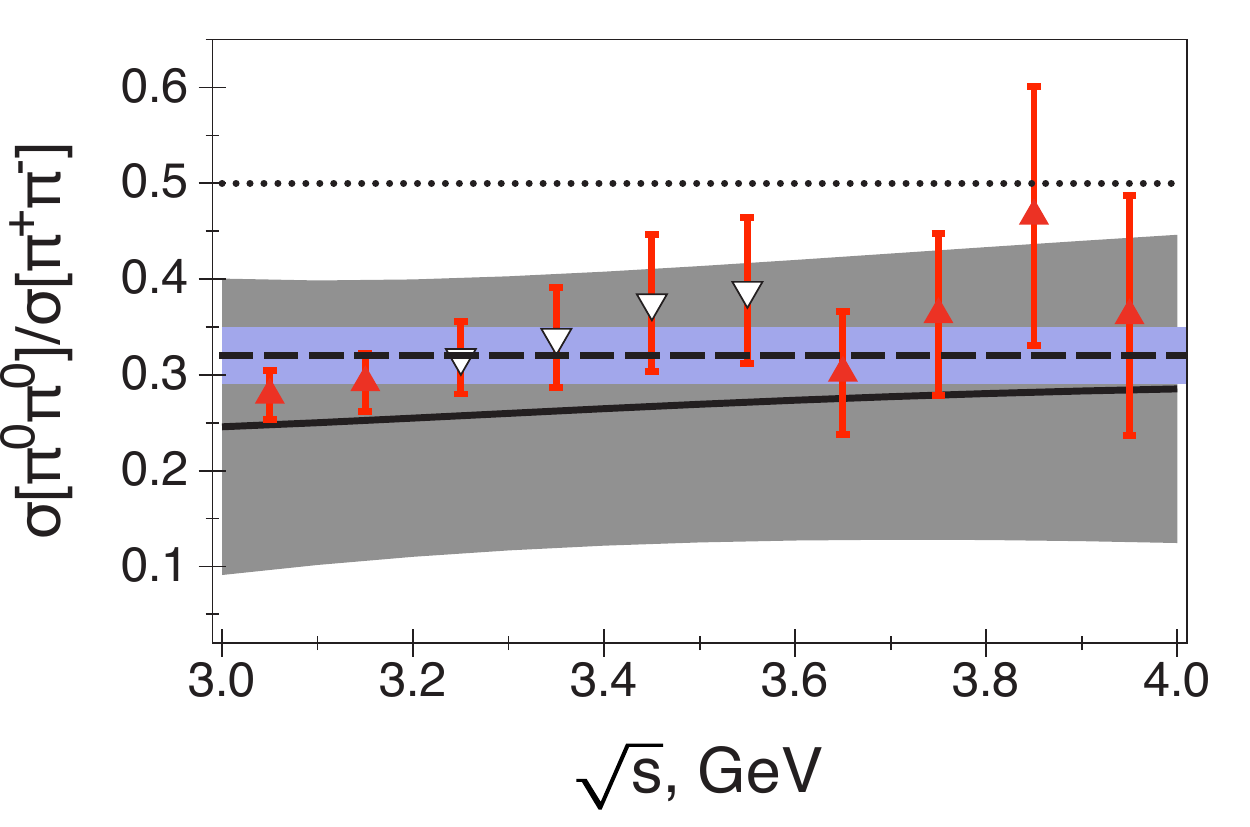}
\end{center}
\caption{The ratio $\sigma^{\pi^{0}\pi^{0}}/\sigma^{\pi^{+}\pi^{-}}$ as
functions of energy. The notations are the same as in Fig.\ref{R002pm}. }%
\label{R002pmCZ}%
\end{figure}

In this case the quality of the two-parameter fit is better: the
obtained error bars are smaller. The results for the differential cross
sections are shown in Fig.\ref{ds-CZ} and the numerical values of the
amplitudes  for the different energies $W$ are summarized in Table \ref{tabBppmCZ}. 
One can see  that in this  case the solution is given by large $B_{+-}^{(0)}$ and relatively small
$B_{++}^{(0)}$.  The coefficient of the linear contribution $\lambda_{+-}$
defined in (\ref{lam}) is quite large and therefore the angular separation works better 
 that explains the better results of the two-parameter fit. 
The results are presented in  Fig.\ref{ds-CZ}.  The power fit of the obtained points  yields
\begin{equation}
B_{++}^{(0)}(s)=\left(  \frac{s_{0}}{s}\right)  ^{a},\text{ with~\ }%
s_{0}=5.8\pm1\text{GeV}^{2},\text{ }a=3.4\pm1.2,
\end{equation}%
\begin{equation}
B_{+-}^{(0)}(s)=-\left(  \frac{s_{1}}{s}\right)  ^{b},\text{ with~\ }%
s_{1}=6.1\pm0.5\text{GeV}^{2},\text{ }b=1.2\pm0.15.
\end{equation}
Corresponding figures with  $1\sigma$ error bands are shown in  Fig.\ref{B0ppmCZ}. 
\begin{table}[h]
\caption{The amplitudes $B_{+\pm}^{(0)}(s)$ obtained from the
two-parameter fit of  BELLE data \cite{Nakazawa:2004gu,Uehara:2009cka}  with CZ-model of pion DA (dof=$16$).}%
\label{tabBppmCZ}
\begin{center}
\begin{tabular}
[c]{|c|c|c|c|c|c|c|}\hline
$W,$ GeV & $3.05$ & $3.15$ & $3.65$ & $3.75$ & $3.85$ & $3.95$\\\hline
$B_{++}^{(0)}(s,0)$ & $\scriptstyle -0.20\pm0.03$ & $\scriptstyle -0.15\pm
0.03$ & $\scriptstyle -0.09\pm0.05$ & $\scriptstyle -0.06\pm0.04$ &
$\scriptstyle -0.02\pm0.04$ & $\scriptstyle -0.017\pm0.046$\\\hline
$B_{+-}^{(0)}(s,0)$ & $\scriptstyle 0.59\pm0.03$ & $\scriptstyle 0.59\pm0.03$
& $\scriptstyle 0.36\pm0.04$ & $\scriptstyle 0.37\pm0.04$ &
$\scriptstyle 0.37\pm0.03$ & $\scriptstyle 0.31\pm0.04$\\\hline
$\chi^{2}/$dof & $2.7$ & $2.2$ & $0.94$ & $1.3$ & $0.7$ & $0.60$\\\hline
\end{tabular}
\end{center}
\end{table}
The absolute value of $B_{++}^{(0)}$  is smaller comparing to $B_{+-}^{(0)}$ and much stronger suppressed with the energy $s$.
The power behavior of the dominant amplitude $B_{+-}^{(0)}$ is much smaller than  
the  asymptotic prediction $\sim1/s^{2}$.   An  attempt  to apply  the 
one-parameter model with $B_{++}^{(0)}\simeq0$ gives  a bad 
description indicating that $B_{++}^{(0)}$ can not be neglected for the whole energy interval. 
The total cross sections and their ratio are shown in
Figs.~\ref{sigmaCZ-tot} and \ref{R002pmCZ}.  All  notations and the error bands are the same  as  in
Figs.\ref{sigma-tot} and \ref{R002pm}, respectively.


In case of CZ-model one also needs to take into account the  numerical
effect provided by the the linear  contributions  with the amplitudes $B_{+\pm}^{(0)}$ in Eqs.(\ref{dspm}) and (\ref{ds00}). 
Substituting the numerical values for the amplitude in Eqs.(\ref{dspmCZ}) and (\ref{ds00-CZ}) 
we obtain ($W=3.05\text{GeV}$ and $\cos\theta=0.05$)
\begin{align}
\frac{d\sigma^{\pi^{+}\pi^{-}}}{d\cos\theta}=0.035_{B_{++}^{2}}+0.3_{B_{+-}%
^{2}}+0.13_{B_{++}}-0.26_{B_{+-}}+0.18=0.38,\label{dspnnumCZ}%
\end{align}
\begin{align}
\frac{d\sigma^{\pi^{0}\pi^{0}}}{d\cos\theta} =0.018_{B_{++}^{2}}%
+0.15_{B_{+-}^{2}}-0.09_{B_{+-}}+0.014=0.09.
\label{ds00numCZ}%
\end{align}
Here the subscripts again show the numerical contribution of the appropriate
quadratical or linear terms in Eq's.(\ref{dspmCZ}) and (\ref{ds00-CZ}).
Numerical  values   in (\ref{dspnnumCZ}) and (\ref{ds00numCZ}) demonstrate that
  all terms  are  significant for  description of the cross sections.

We observe that it is possible to fit the data using very different models of  pion DA.
Hence our consideration shows  that  our approach cannot help to constrain pion DA and 
as a result we can obtain  different solutions for the amplitudes $B_{+\pm}^{(0)}$. 
 In this case  one needs more information  in order to  constrain the input parameters. 
 Potentially a precise measurement  of the cross section $e^{+}e^{-}\rightarrow e^{+}e^{-}\pi\pi$ with
unpolarized electron beams  allows one to obtain  more information about  the 
amplitudes of the hadronic subprocess.

The corresponding  cross section  is described by the
two contributions, see the details in Ref.\cite{Budnev:1974de}. Schematically
the expression for the cross section reads
\begin{equation}
d\sigma^{e^{+}e^{-}\rightarrow e^{+}e^{-}\pi\pi}=\left\{  A~\frac{1}{2}\left(
\sigma_{\Vert}^{\gamma\gamma\rightarrow\pi\pi}+\sigma_{\bot}^{\gamma
\gamma\rightarrow\pi\pi}\right)  +B~\left(  \sigma_{\Vert}^{\gamma
\gamma\rightarrow\pi\pi}-\sigma_{\bot}^{\gamma\gamma\rightarrow\pi\pi}\right)
\cos2\varphi\right\}  \frac{d^{3}p_{1}^{\prime}}{E_{1}}\frac{d^{3}%
p_{2}^{\prime}}{E_{1}},\label{dsee}%
\end{equation}
where the $\sigma_{\Vert,\bot}^{\gamma\gamma\rightarrow\pi\pi}$ denotes cross
sections for the scattering of photons with the parallel $(\sigma_{\Vert})$
and orthogonal $(\sigma_{\bot})$ linear polarizations. Here the coefficients
$A$ and $B$ denote the functions of the kinematical variables  and their explicit expressions  can be found  in Ref.\cite{Budnev:1974de}. 
The azimuth angle $\varphi$ is defined as
the angle between the electron scattering planes in the colliding electron
c.m.s.,  $E_{i}$ and   $p_{i}^{\prime}$ denote the scattered electron
energies and momenta. In the present work we only investigated the first
combination of the cross sections which is proportional to the sum of the
helicity amplitudes:
\begin{equation}
\sigma_{\Vert}^{\gamma\gamma\rightarrow\pi\pi}+\sigma_{\bot}^{\gamma
\gamma\rightarrow\pi\pi}\sim|T_{++}^{\gamma\gamma\rightarrow\pi\pi}%
|^{2}+|T_{+-}^{\gamma\gamma\rightarrow\pi\pi}|^{2}.
\end{equation}
Using the angular dependence given by the factor $\cos2\varphi$ one can also
access the second contribution which is sensitive to the difference of the
cross sections in Eq.(\ref{dsee}). This combination is proportional only to
the one helicity amplitude
\begin{equation}
\sigma_{\Vert}^{\gamma\gamma\rightarrow\pi\pi}-\sigma_{\bot}^{\gamma
\gamma\rightarrow\pi\pi}\sim|T_{++}^{\gamma\gamma\rightarrow\pi\pi}|^{2}.
\label{dsTpp}
\end{equation}
Hence such data can provide an additional  information which allows one to perform  a  better separation of the
helicity amplitudes and to perform  extraction of the  soft-overlap form factors
$B^{(0)_{+\pm}}(s)$.

In our consideration the values of the amplitudes $B^{(0)}_{+\pm}(s)$
depends on the model of  pion DA. For model defined by set-II   we
obtain that $|T_{++}^{\gamma\gamma\rightarrow\pi\pi}|\gg|T_{+-}
^{\gamma\gamma\rightarrow\pi\pi}|$ and contrariwise for the CZ-model.
The consideration within the  handbag model in Ref.\cite{Diehl:2001fv}
predicts  $T_{++}^{\gamma\gamma\rightarrow\pi\pi}\simeq0$.

\begin{figure}[ptb]
\begin{center}
\includegraphics[width=3.5in]{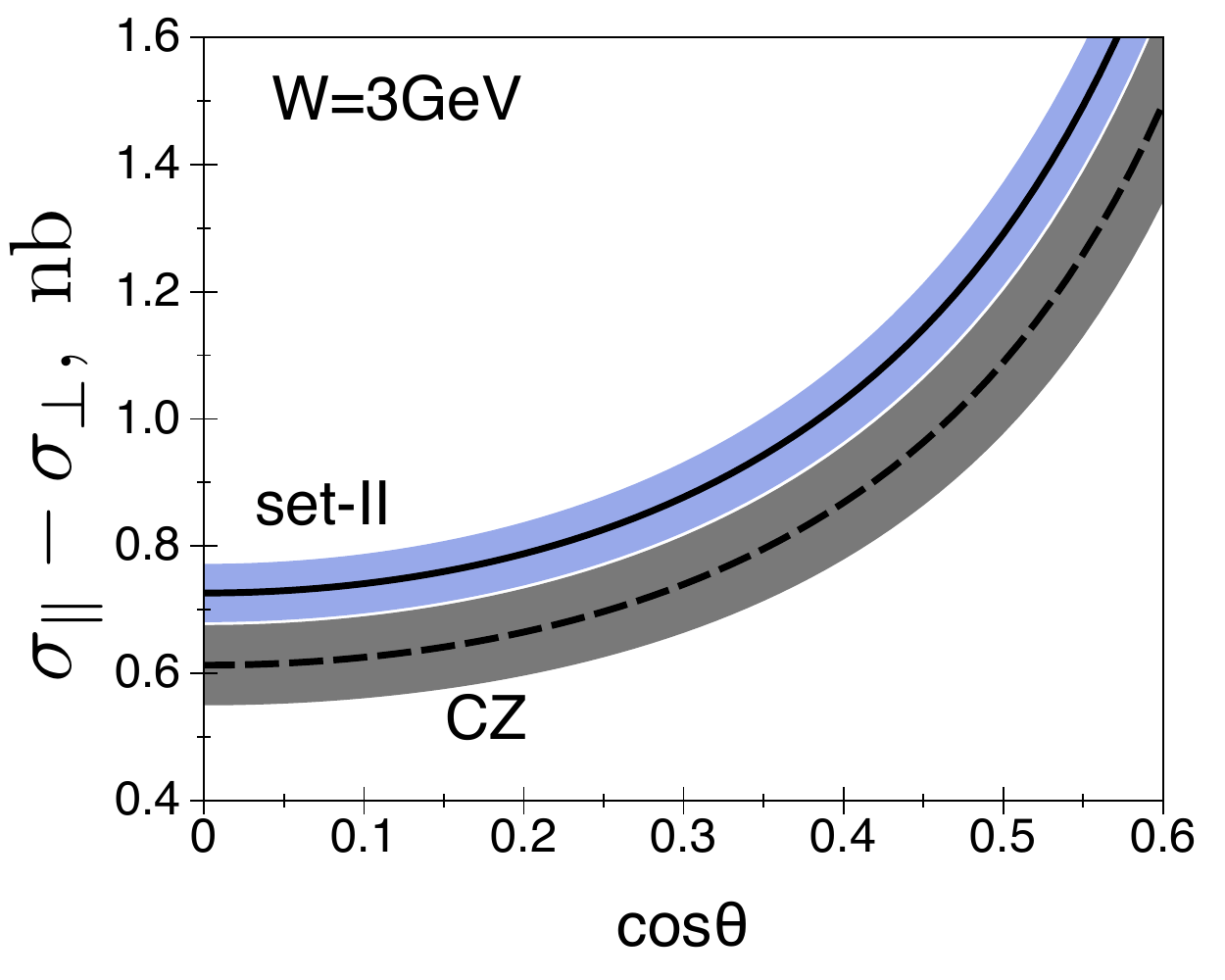}
\end{center}
\caption{Predictions of the cross sections $\sigma_{\Vert
}^{\gamma\gamma\rightarrow\pi\pi}-\sigma_{\bot}^{\gamma\gamma\rightarrow\pi
\pi}$  for the two different model of pion DA. See discussion in the text. }
\label{dsLT}
\end{figure}

 Our estimations  of the cross section  $\sigma_{\Vert}-\sigma_{\bot}$ at $W=3.05$GeV
 are shown in Fig.\ref{dsLT}.   We consider  two different scenarios associated with the two different  models of pion DA as discussed above.  
 In order to  draw these plots we use  and the  values  $B^{(0)}_{++}(s)$  obtained in Tables \ref{tabBpp} and \ref{tabBppmCZ}. 
The shaded area shows the $1\sigma$ error bands which corresponds  to the uncertainties in the  determination of $B^{(0)}_{++}(s)$. 
From this figure we see that  the  difference between the two  calculations   is not very
large.  Hence one needs  precise data in order  to  distinguish between the different scenarios.  
 Nevertheless we expect that  new  data for the 
cross sections $\sigma_{\Vert}^{\gamma\gamma\rightarrow\pi\pi}-\sigma_{\bot
}^{\gamma\gamma\rightarrow\pi\pi}$  can be very helpful in order to understand
the reaction mechanism of the large angle  pion  production at
intermediate values of the energy and momentum transfer.

\section{Discussion}

\label{discurs}

We have discussed a contribution of  the subleading power
corrections in  the process $\gamma\gamma\rightarrow\pi\pi$ in
the region where all Mandelstam variables are large $s\sim-t\sim-u\gg
\Lambda^{2}$ (large angle scattering).  The leading power behavior of the
corresponding amplitude is described by the hard scattering mechanism
suggested in Ref.\cite{Brodsky:1981rp}.  
The factorization of  power suppressed corrections  can be  described as a  
sum of the hard and soft-overlap contributions.  We develop a systematic approach 
for  description of such configurations within  the SCET framework. 

Factorizing the hard modes in SCET  we obtain  that the soft-overlap contribution can be described by  
matrix elements of suitable SCET-I operators. 
 We  present  an analysis of the appropriate SCET-I operators which are required for description
 $1/Q^{2}$ corrections to the leading power approximation.
We  demonstrate  that the $T$-products of the SCET-I operators  mix  in SCET-II  with the
pure collinear operators  describing   hard subleading contributions.  Such
mixing is possible due  to an overlap of  collinear and soft domains and leads
to the endpoint singularities in  soft and collinear  convolution integrals.
In order to factorize consistently  the hard and soft-overlap  contributions one has to
define a specific regularization which allows one  to define the integrals in the endpoint region.  
In present  case  such scheme can be avoided  if one uses  the so-called physical 
subtraction scheme. 



The subleading hard contributions are described by the matrix elements of the different higher twist 
collinear operators.  All such  contributions provide the power correction of order $1/s^{2}$.
 In present  work  we took into account only the specific twist-3 collinear operators which can
provide the largest numerical corrections. Such terms are described by the
 two-particle  operators and known  as chiral enhanced corrections. Such  approximation  also  allows us 
 to simplify the theoretical consideration.  

We compute the corresponding  leading-order  hard  kernels  and  obtain that the  hard  amplitudes have
the endpoint singularities.   Therefore in order to be consistent we must add  the
SCET-I operators  which can overlap with this subleading hard contribution.  We obtain that
 such contribution can be only described  by one suitable hard-collinear operator. 
Corresponding SCET-I matrix element defines  the one soft-overlap amplitude which depends only from the 
total energy $s$.   The angular dependence in this case is given by the hard subprocess.  
Combining  the hard and soft-overlap contributions we show that the  endpoint  singularities    
are cancel in the physical subtraction scheme.  This allows us to 
compute the angular dependence of the amplitudes in a model independent way.

The  final expression for the  power suppressed amplitude  depends
on the two unknown amplitudes  $B^{0}_{+\pm}(s)$ which are some
functions of the total energy $s$.  These amplitudes are defined by subprocess  associated with
the hard-collinear scale $\mu_{hc}\sim \Lambda \sqrt{s}$ where the soft scale $\Lambda$ is of order
of the typical momenta of  soft particles.      The  hard-collinear scale is still small  in the kinematical region $\sqrt{s}=3-4$GeV where we have data
 and we consider the amplitudes  $B^{0}_{+\pm}(s)$ as the nonperturbative quantities.

The obtained results have been used in order to perform a phenomenological analysis of existing  data. 
We obtain that to a very good  accuracy the angular behavior of the cross section for the production of
charged pions is proportional to the $\sin^{-4}\theta$ that is in the reasonable 
agreement with the  data.  This allows us  to fix the nonperturbative functions  $B^{0}_{+\pm}(s)$  by fitting 
the differential cross sections in the region $\sqrt{s}=3-4$GVe and  $|\cos\theta|<0.6$.  
Combining the leading and subleading  amplitudes we obtain a reasonable description of the cross sections 
and their ratio $R=d\sigma^{\pi^{0}\pi^{0}}/d\sigma^{\pi^{+}\pi^{-}}$.   

We find that the results for the functions $B^{(0)}_{+\pm}(s)$ are sensitive to the
models of pion distribution amplitude used in the numerical calculations. For
the CZ-model the best fit is obtained if $|B^{(0)}_{+-}(s,0)|\gg|B^{(0)}_{++}(s,0)|$. 
For the other models  the angular separation of the
amplitudes $B^{(0)}_{+\pm}(s)$ can not be performed  with a good precision indicating 
 the qualitative estimate $|B^{(0)}_{++}(s)|\gg|B^{(0)}_{+-}(s)|$.  Following this observation we consider  the
model with $B^{(0)}_{+-}(s)\approx 0$.   We find that in this case the data can be  described quite well.

We expect that additional   experimental information can help us to reduce the model dependence in the 
phenomenological analysis.    More accurate data,
especially for larger values of energy $s$  will be very helpful but one has to
remember that our description so far  is also restricted only by the leading-order
accuracy in $\alpha_{s}$.   The other possibility  to improve the phenomenological  analysis is to consider  an additional observable
which is sensitive to a different combination of the helicity amplitudes. 
A required cross section can also be measured in the  process $e^{+}e^{-}
\rightarrow e^{+}e^{-}\pi\pi$  using the modulation with respect 
to a angle between the electron scattering planes in the colliding electron
c.m.s., see e.g. Ref.\cite{Budnev:1974de}.   In this case one can can access the
combination 
$\sigma_{\Vert}^{\gamma\gamma\rightarrow\pi\pi}-\sigma_{\bot
}^{\gamma\gamma\rightarrow\pi\pi}\sim|T_{++}^{\gamma\gamma\rightarrow\pi\pi
}|^{2}$, where the $\sigma_{\Vert,\bot}^{\gamma\gamma\rightarrow\pi\pi}$
denotes cross sections for the scattering of photons with parallel
$(\sigma_{\Vert}) $ and orthogonal $(\sigma_{\bot})$ linear polarizations.
This observable is only sensitive to the amplitude $B^{(0)}_{++}(s)$ that 
 allows one to perform  an accurate extraction of the amplitudes
$B^{(0)}_{+\pm}(s)$.  We expect that this will  be very helpful  for understanding  of the dominant  mechanism in the
large angle meson  production.   

\begin{appendix}


\section{ Higher twist  distribution amplitudes}
\label{htDA}
For convenience in this Appendix we briefly discuss  higher twist pion distribution amplitudes.  More  detailed
discussion can be fond in Ref.\cite{Braun:1989iv}.  Below we present a
standard QCD formulation. In order to obtain the equivalent SCET  description  one has to
split the quark fields into large $\xi_{n}$ and small $\eta_{n}$ components
and use the equation of motion in order to eliminate the field $\eta_{n}$.  A discussion
of the some higher twist collinear matrix elements in the SCET  framework can be found
in Ref.\cite{Hardmeier:2003ig}.

 In the  QCD formulation there are three  twist-3 collinear operators.
Two of them are  2-particle operators and their light-cone matrix  elements are  defined as
\begin{equation}
~\left\langle p^{\prime}\right\vert \bar{q}(\lambda_{1}\bar{n})W_{n}%
(\lambda_{1}\bar{n})i\gamma_{5}W_{n}^{\dag}(\lambda_{1}\bar{n})q(\lambda
_{2}\bar{n})\left\vert 0\right\rangle =~f_{\pi}\mu_{\pi}\int_{0}%
^{1}du~e^{iu\lambda_{1}p_{-}^{\prime}+i\bar{u}\lambda_{2}p_{-}^{\prime}}%
~\phi_{p}(u), \label{Op}%
\end{equation}%
\begin{align}
~\left\langle p^{\prime}\right\vert \bar{q}(\lambda_{1}\bar{n})W_{n}%
(\lambda_{1}\bar{n})\sigma^{\alpha\beta}\gamma_{5}W_{n}^{\dag}(\lambda_{2}%
\bar{n})q(\lambda_{2}\bar{n})\left\vert 0\right\rangle  &  =i~f_{\pi}\mu_{\pi
}\left(  p^{\alpha}z^{\beta}-p^{\beta}z^{\alpha}\right) \\
&  ~\ \ \ \ \int_{0}^{1}du~e^{iu\lambda_{1}p_{-}^{\prime}+i\bar{u}\lambda
_{2}p_{-}^{\prime}}\frac{\phi_{\sigma}(u)}{6},
\end{align}
with $z^{\alpha}=(\lambda_{1}-\lambda_{2})\bar{n}^{\alpha},~p_{-}^{\prime
}\equiv(p^{\prime}\cdot\bar{n})$ and
\begin{equation}
\mu_{\pi}=\frac{m_{\pi}^{2}}{m_{u}+m_{d}},~\ p^{\prime}\simeq p_{-}^{\prime
}\frac{\bar{n}}{2}.~\ \label{mupi}%
\end{equation}
For the sum of the quark masses we use following estimate
\bea
(m_{u}+m_{d})(2\text{GeV})=8.5 \text{MeV}.
\eea
There is only one twist-3 three-particle operator and its matrix element
defines the three-particle DA $\phi_{3\pi}(\alpha_{i})$%
\begin{equation}
\left\langle \pi^{+}(p^{\prime})\right\vert \bar{u}(\lambda\bar{n}%
)\sigma^{\bar{n}\alpha_{\bot}}\gamma_{5}~gG_{\bar{n}\alpha_{\bot}}%
(v\lambda\bar{n})d(-\lambda\bar{n})\left\vert 0\right\rangle =if_{3\pi}%
p_{-}^{\prime2}~2~\int\mathcal{D}\alpha_{i}~e^{i\lambda p_{-}^{\prime}%
(\alpha_{u}-\alpha_{d}+v\alpha_{g})}\phi_{3\pi}(\alpha_{i}),
\end{equation}
Here $\mathcal{D}\alpha_{i}=d\alpha_{u}d\alpha_{d}d\alpha_{g}\delta(\alpha
_{u}+\alpha_{d}+\alpha_{g}-1)$, $G_{\bar{n}\alpha_{\bot}}=G_{\mu\alpha_{\bot}%
}\bar{n}^{\mu}$. Using QCD equations of motion one can show ( see details in
Refs.\cite{Braun:1989iv,Ball:2006wn}) that 2-particles DAs $\phi_{p,\sigma}$ can
be presented as
\begin{align}
\phi_{p}(u)  &  =1+R~V_{p}(u,\alpha_{i})\ast\phi_{3\pi}(\alpha_{i}%
),~\label{phi-p}\\
\ \phi_{\sigma}(u)  &  =\ 6u\bar{u}+R~V_{\sigma}(u,\alpha_{i})\ast\phi_{3\pi
}(\alpha_{i}), \label{phi-sgm}%
\end{align}
where $V_{p,\sigma}$ denotes a certain dimensionless kernel, the asterisks
denote the convolution integrals with respect to the fractions $\alpha_{i}$ and
$R=f_{3\pi}/f_{\pi}\mu_{\pi}\simeq0.014$ \cite{Braun:1989iv,Ball:2006wn}.
One can assume that corrections associated with the admixture $\phi_{3\pi}$ in
Eqs.(\ref{phi-p}) and (\ref{phi-sgm}) are relatively small because the
factor $R$ is numerically small comparing to the constant $\mu_{\pi}$  defined in Eq.(\ref{mupi}).
Therefore   neglecting  the 3-particle contributions  with $\phi_{3\pi}$ in Eqs. (\ref{phi-p}) and (\ref{phi-sgm}) one finds
\begin{equation}
\phi_{p}(u)\simeq1,~\ \phi_{\sigma}(u)\simeq\ 6u\bar{u}.
\end{equation}

The discussion of the twist-3 matrix elements within  the SCET framework can also be found in  Ref.\cite{Hardmeier:2003ig}.

The twist-4 operators can be divided on the following  groups: two-, three- and four-particle operators.  
The two-particles operator are not independent and can be expressed  through the  other DAs, see Ref.\cite{Braun:1989iv}. 
In phenomenological applications the four-particle twist-4 contributions are assumed to be small and as a rule neglected. 
Following to Ref.\cite{Braun:1989iv}   one can define  four three-particle DAs of twist four (for simplicity we do
not write the collinear Wilson lines $W_{n}$)
\begin{equation}
\left\langle \pi^{+}(p^{\prime})\right\vert \bar{u}(\lambda\bar{n}%
)\gamma_{\bot}^{\alpha}\gamma_{5}~gG_{\alpha_{\bot}\bar{n}}(v\lambda\bar
{n})~d(-\lambda\bar{n})\left\vert 0\right\rangle =2f_{\pi}m_{\pi}^{2}%
p_{-}^{\prime}~A_{\bot}(v,\lambda p_{-}^{\prime}),
\end{equation}%
\begin{equation}
\left\langle \pi^{+}(p^{\prime})\right\vert \bar{u}(\lambda\bar{n}%
)\gamma_{\bot}^{\alpha}~g\tilde{G}_{\alpha_{\bot}\bar{n}}(v\lambda\bar
{n})~d(-\lambda\bar{n})\left\vert 0\right\rangle =2f_{\pi}m_{\pi}^{2}%
p_{-}^{\prime}V_{\bot}(v,\lambda p_{-}^{\prime}),
\end{equation}
\begin{equation}
\left\langle \pi^{+}(p^{\prime})\right\vert \bar{u}(\lambda\bar{n})\nbs
\gamma_{5}~gG_{\bar{n}n}(v\lambda\bar{n})~d(-\lambda\bar{n})\left\vert
0\right\rangle =f_{\pi}m_{\pi}^{2}p_{-}^{\prime}~A_{\Vert}(v,\lambda
p_{-}^{\prime}),
\end{equation}%
\begin{equation}
\left\langle \pi^{+}(p^{\prime})\right\vert \bar{u}(\lambda\bar{n})\nbs
g\tilde{G}_{\bar{n}n}(v\lambda\bar{n})~d(-\lambda\bar{n})\left\vert
0\right\rangle =f_{\pi}m_{\pi}^{2}p_{-}^{\prime}V_{\Vert}(v,\lambda
p_{-}^{\prime}),
\end{equation}
with the
\begin{equation}
\left\{  V_{i},A_{i}\right\}  (v,\lambda p_{-}^{\prime})=\int\mathcal{D}%
\alpha_{i}~e^{i\lambda p_{-}^{\prime}(\alpha_{u}-\alpha_{d}+v\alpha_{g}%
)}\left\{  V_{i},A_{i}\right\}  (\alpha_{i}).
\end{equation}
and  we used
\begin{equation}
\tilde{G}_{\alpha\beta}=\frac{1}{2}\varepsilon_{\alpha\beta\mu\nu}G^{\mu\nu}.
\end{equation}
We will  not describe  in detail the structure of these DAs  because we neglect corresponding  contributions in the our calculations.

\section{The list of  SCET  interactions}
\label{Lint}
\begin{equation}
\mathcal{L}^{(1,n)}\left[  \bar{\xi}_{n}^{c}~A_{\bot}A_{\bot}\xi_{n}\right]
\simeq\int d^{4}x~\bar{\xi}_{n}^{c}~\Dslash{A}_{\bot}(\bar{n}\partial)^{-1}\Dslash{A}_{\bot}\frac{\nbs}{2}\xi_{n}, \label{L1AA}
\end{equation}%
\begin{equation}
\mathcal{L}^{(2,n)}\left[  \bar{\xi}_{n}^{c}~A_{\bot}A_{\bot}^{(s)}\xi
_{n}\right]  \simeq\int d^{4}x~\bar{\xi}_{n}^{c}~\left\{  \Dslash{A}_{\bot}
^{s}(\bar{n}\partial)^{-1}\Dslash{A}_{\bot}+\Dslash{A}_{\bot}
(\bar{n}\partial)^{-1}\Dslash{A}_{\bot}^{s}\right\}  \frac{\nbs}{2}\xi_{n},
\label{L2AAs}%
\end{equation}%
\begin{equation}
\mathcal{L}^{(2,n)}\left[  \bar{q}~A_{\bot}~\xi_{n}^{c}\right]  \simeq\int
d^{4}x~\bar{q}~\Dslash{A}_{\bot}\xi_{n}^{c}. \label{L2qAxi}
\end{equation}%
\begin{equation}
\mathcal{L}^{(2,n)}\left[  \bar{\xi}_{n}^{c}~A_{\bot}^{c}A_{\bot}\xi
_{n}\right]  \simeq\int d^{4}x~\bar{\xi}_{n}^{c}~\Dslash{A}_{\bot}^{c}
(\bar{n}\partial)^{-1}\Dslash{A}_{\bot}\frac{\nbs}{2}\xi_{n}, \label{L2xiAAxi}
\end{equation}%
\begin{equation}
\mathcal{L}^{(3,n)}\left[  \bar{q}~A_{\bot}A_{\bot}~\xi_{n}^{c}\right]
\simeq\int d^{4}x~\bar{q}~\Dslash{A}_{\bot}(\bar{n}\partial)^{-1}\Dslash{A}
_{\bot}\frac{\nbs}{2}\xi_{n}^{c}. \label{L3qAAxi}
\end{equation}%
\begin{equation}
\mathcal{L}^{(4,n)}\left[  \bar{q}~A_{\bot}^{s}A_{\bot}\xi_{n}^{c}\right]
=\int d^{4}x~\bar{q}\left\{  ~\Dslash{A}_{\bot}^{s}(\bar{n}\partial)^{-1}
\Dslash{A}_{\bot}+\Dslash{A}_{\bot}(\bar{n}\partial)^{-1}~\Dslash{A}_{\bot}
^{s}\right\}  \frac{\nbs}{2}\xi_{n}^{c}
\end{equation}

\end{appendix}

\section*{Acknowledgments}

This work was supported by the Helmholtz Institute Mainz. The author is
grateful to M. Vanderhaeghen, M. Distler and L.Tiator for many useful
discussions and to S. Eydelman and H. Nakazava for the helpful correspondence.

\label{gobib}


\begin{thebibliography}{99}                                                                                               %




\bibitem {Brodsky:1981rp}S.~J.~Brodsky and G.~P.~Lepage,
Phys.\ Rev.\ D \textbf{24} (1981) 1808.


\bibitem {Benayoun:1989ng}M.~Benayoun and V.~L.~Chernyak,
Nucl.\ Phys.\ B \textbf{329} (1990) 285.


\bibitem {Dominick:1994bw}J.~Dominick \textit{et al.} [CLEO Collaboration],
Phys.\ Rev.\ D \textbf{50} (1994) 3027 [hep-ph/9403379].


\bibitem {Heister:2003ae}A.~Heister \textit{et al.} [ALEPH Collaboration],
Phys.\ Lett.\ B \textbf{569} (2003) 140.


\bibitem {Nakazawa:2004gu}H.~Nakazawa \textit{et al.} [BELLE Collaboration],
Phys.\ Lett.\ B \textbf{615} (2005) 39 [hep-ex/0412058].


\bibitem {Uehara:2009cka}S.~Uehara \textit{et al.} [BELLE Collaboration],
Phys.\ Rev.\ D \textbf{79} (2009) 052009 [arXiv:0903.3697 [hep-ex]].


\bibitem {Brodzicka:2012jm}J.~Brodzicka \textit{et al.} [Belle
Collaboration],
PTEP \textbf{2012} (2012) 04D001 [arXiv:1212.5342 [hep-ex]].




\bibitem {Vogt:2000bz}C.~Vogt,
hep-ph/0010040.




\bibitem {Chernyak:2006dk}V.~L.~Chernyak,
Phys.\ Lett.\ B \textbf{640} (2006) 246 [hep-ph/0605072].




\bibitem {Chernyak:2012pw}V.~L.~Chernyak,
arXiv:1212.1304 [hep-ph];
V.~L.~Chernyak,
Chin.\ Phys.\ C \textbf{34} (2010) 822 [arXiv:0912.0623 [hep-ph]];




\bibitem {Khodjamirian:1997tk}A.~Khodjamirian,
Eur.\ Phys.\ J.\ C \textbf{6} (1999) 477 [hep-ph/9712451].


\bibitem{Mikhailov:1991pt}
  S.~V.~Mikhailov and A.~V.~Radyushkin,
  Phys.\ Rev.\ D {\bf 45} (1992) 1754.



\bibitem {Agaev:2012tm}S.~S.~Agaev, V.~M.~Braun, N.~Offen and F.~A.~Porkert,
Phys.\ Rev.\ D \textbf{86} (2012) 077504 [arXiv:1206.3968 [hep-ph]].




\bibitem {Bakulev:2011iy}A.~P.~Bakulev, S.~V.~Mikhailov, A.~V.~Pimikov and
N.~G.~Stefanis,
Nucl.\ Phys.\ Proc.\ Suppl.\ \textbf{219-220} (2011) 133 [arXiv:1108.4344
[hep-ph]].




\bibitem {Bakulev:2012nh}A.~P.~Bakulev, S.~V.~Mikhailov, A.~V.~Pimikov and
N.~G.~Stefanis,
Phys.\ Rev.\ D \textbf{86} (2012) 031501 [arXiv:1205.3770 [hep-ph]].




\bibitem {Braun:1999uj}V.~M.~Braun, A.~Khodjamirian and M.~Maul,
Phys.\ Rev.\ D \textbf{61} (2000) 073004 [hep-ph/9907495].




\bibitem {Diehl:2001fv}M.~Diehl, P.~Kroll and C.~Vogt,
Phys.\ Lett.\ B \textbf{532} (2002) 99 [hep-ph/0112274].




\bibitem {Diehl:2009yi}M.~Diehl and P.~Kroll,
Phys.\ Lett.\ B \textbf{683} (2010) 165 [arXiv:0911.3317 [hep-ph]].




\bibitem {Bauer:2000ew}C.~W.~Bauer, S.~Fleming and M.~E.~Luke,
Phys.\ Rev.\ D \textbf{63}, 014006 (2000).


\bibitem {Bauer2000}C.~W.~Bauer, S.~Fleming, D.~Pirjol and I.~W.~Stewart,
Phys.\ Rev.\ D \textbf{63}, 114020 (2001).


\bibitem {Bauer:2001ct}C.~W.~Bauer and I.~W.~Stewart,
Phys.\ Lett.\ B \textbf{516}, 134 (2001).


\bibitem {Bauer2001}C.~W.~Bauer, D.~Pirjol and I.~W.~Stewart,
Phys.\ Rev.\ D \textbf{65}, 054022 (2002).


\bibitem {BenCh}
M.~Beneke, A.~P.~Chapovsky, M.~Diehl and T.~Feldmann,
Nucl.\ Phys.\ B \textbf{643}, 431 (2002).


\bibitem {BenFeld03}M.~Beneke and T.~Feldmann,
Phys.\ Lett.\ B \textbf{553}, 267 (2003).






\bibitem {Gasser:2006qa}J.~Gasser, M.~A.~Ivanov and M.~E.~Sainio,
Nucl.\ Phys.\ B \textbf{745} (2006) 84 [hep-ph/0602234].


\bibitem {Lepage:1979zb}G.~P.~Lepage and S.~J.~Brodsky,
Phys.\ Lett.\ B \textbf{87}, 359 (1979).


\bibitem {Chernyak:1983ej}V.~L.~Chernyak and A.~R.~Zhitnitsky,
Phys.\ Rept.\ \textbf{112}, 173 (1984).


\bibitem {Isgur:1984jm}N.~Isgur and C.~H.~Llewellyn Smith,
Phys.\ Rev.\ Lett.\ \textbf{52}, 1080 (1984).


\bibitem {Isgur:1988iw}N.~Isgur and C.~H.~Llewellyn Smith,
Nucl.\ Phys.\ B \textbf{317}, 526 (1989).


\bibitem {Isgur:1989cy}N.~Isgur and C.~H.~Llewellyn Smith,
Phys.\ Lett.\ B \textbf{217}, 535 (1989).




\bibitem {Nesterenko:1982gc}V.~A.~Nesterenko and A.~V.~Radyushkin,
Phys.\ Lett.\ B \textbf{115} (1982) 410.




\bibitem {Bakulev:2000uh}A.~P.~Bakulev, A.~V.~Radyushkin and N.~G.~Stefanis,
Phys.\ Rev.\ D \textbf{62} (2000) 113001 [hep-ph/0005085].






\bibitem {Chernyak:1981zz}V.~L.~Chernyak and A.~R.~Zhitnitsky,
Nucl.\ Phys.\ B \textbf{201} (1982) 492 [Erratum-ibid.\ B \textbf{214} (1983)
547].


\bibitem {Duplancic:2006nv}G.~Duplancic and B.~Nizic,
Phys.\ Rev.\ Lett.\ \textbf{97} (2006) 142003 [hep-ph/0607069].


\bibitem{Beneke:1997zp}
  M.~Beneke and V.~A.~Smirnov,
  Nucl.\ Phys.\ B {\bf 522} (1998) 321
  [hep-ph/9711391].

\bibitem{Smirnov:1998vk}
  V.~A.~Smirnov and E.~R.~Rakhmetov,
  Theor.\ Math.\ Phys.\  {\bf 120} (1999) 870
   [Teor.\ Mat.\ Fiz.\  {\bf 120} (1999) 64]
  [hep-ph/9812529].


\bibitem{Smirnov:2002pj}
  V.~A.~Smirnov,
  Springer Tracts Mod.\ Phys.\  {\bf 177} (2002) 1.

\bibitem{Collins:1999dz}
  J.~C.~Collins and F.~Hautmann,
  Phys.\ Lett.\  B {\bf 472}, 129 (2000)
  [arXiv:hep-ph/9908467].

\bibitem{Manohar:2006nz}
  A.~V.~Manohar and I.~W.~Stewart,
  Phys.\ Rev.\  D {\bf 76}, 074002 (2007)
  [arXiv:hep-ph/0605001].


\bibitem{Chiu:2012ir}
  J.~-Y.~Chiu, A.~Jain, D.~Neill and I.~Z.~Rothstein,
  JHEP {\bf 1205} (2012) 084
  [arXiv:1202.0814 [hep-ph]].



\bibitem {Bauer:2002nz}C.~W.~Bauer, S.~Fleming, D.~Pirjol, I.~Z.~Rothstein and
I.~W.~Stewart,
Phys.\ Rev.\ D \textbf{66} (2002) 014017 [hep-ph/0202088].




\bibitem {Bauer:2003mga}C.~W.~Bauer, D.~Pirjol and I.~W.~Stewart,
Phys.\ Rev.\ D \textbf{68} (2003) 034021 [hep-ph/0303156].




\bibitem {Beneke:2003pa}M.~Beneke and T.~Feldmann,
Nucl.\ Phys.\ B \textbf{685} (2004) 249 [hep-ph/0311335].




\bibitem {Hardmeier:2003ig}A.~Hardmeier, E.~Lunghi, D.~Pirjol and D.~Wyler,
Nucl.\ Phys.\ B \textbf{682} (2004) 150 [hep-ph/0307171].




\bibitem {Balitsky:1997wi}Y.~Y.~Balitsky, V.~M.~Braun and
A.~V.~Kolesnichenko,
Sov.\ J.\ Nucl.\ Phys.\ \textbf{48} (1988) 348 [Yad.\ Fiz.\ \textbf{48} (1988)
547];
Nucl.\ Phys.\ B \textbf{312} (1989) 509.

\bibitem{'tHooft:1973jz}
  G.~'t Hooft,
  Nucl.\ Phys.\ B {\bf 72} (1974) 461.
  


\bibitem{Witten:1979kh}
  E.~Witten,
  Nucl.\ Phys.\ B {\bf 160} (1979) 57.





\bibitem {Beneke:2000ry}M.~Beneke, G.~Buchalla, M.~Neubert and
C.~T.~Sachrajda,
Nucl.\ Phys.\ B \textbf{591} (2000) 313 [arXiv:hep-ph/0006124].




\bibitem {Beneke:2000wa}M.~Beneke and T.~Feldmann,
Nucl.\ Phys.\ B \textbf{592} (2001) 3 [arXiv:hep-ph/0008255].




\bibitem {Braun:1989iv}V.~M.~Braun and I.~E.~Filyanov,
Z.\ Phys.\ C \textbf{48} (1990) 239 [Sov.\ J.\ Nucl.\ Phys.\ \textbf{52}
(1990) 126] [Yad.\ Fiz.\ \textbf{52} (1990) 199].




\bibitem {Kivel:2010ns}N.~Kivel and M.~Vanderhaeghen,
Phys.\ Rev.\ D \textbf{83} (2011) 093005 [arXiv:1010.5314 [hep-ph]].




\bibitem {Chay:2010hq}J.~Chay and C.~Kim,
Phys.\ Rev.\ D \textbf{82} (2010) 094021 [arXiv:1007.4395 [hep-ph]].




\bibitem {Beneke:2002bs}M.~Beneke,
Nucl.\ Phys.\ Proc.\ Suppl.\ \textbf{111} (2002) 62 [hep-ph/0202056].




\bibitem {Mertig:1991}R.~Mertig, M.~Bohm and A.~Denner,
Comput.\ Phys.\ Commun.\ \textbf{64} (1991) 345; R.~Mertig and R.~Scharf,
Comput.\ Phys.\ Commun.\ \textbf{111} (1998) 265 [hep-ph/9801383];
see also web-page \verb| http://www.feyncalc.org |



\bibitem {Ball:2006wn}P.~Ball, V.~M.~Braun and A.~Lenz,
JHEP \textbf{0605} (2006) 004 [hep-ph/0603063].



\bibitem {Budnev:1974de}V.~M.~Budnev, I.~F.~Ginzburg, G.~V.~Meledin and
V.~G.~Serbo,
Phys.\ Rept.\ \textbf{15} (1975) 181.



\end{thebibliography}
\end{document}